\documentclass{fac}

\usepackage{graphicx}
\usepackage{amssymb}
\usepackage{amsfonts}
\usepackage{amsmath}
\usepackage{dsfont}
\usepackage{MnSymbol}
\usepackage{mathtools}
\usepackage{epsfig}
\usepackage{epstopdf}
\usepackage{tikz}
\usepackage{amsthm}
\ifprodtf \else \usepackage{latexsym}\fi

\usepackage{tikz}
\usetikzlibrary{calc,decorations.pathmorphing,shapes,graphs}

\newcounter{sarrow}

\newcounter{sarrow1}
\newcommand\xnrsquigarrow[1]{%
\stepcounter{sarrow1}%
\mathrel{\begin{tikzpicture}[baseline= {( $ (current bounding box.south) + (0,-0.5ex) $ )}]
\node[inner sep=.5ex] (\thesarrow) {$\scriptstyle #1$};
\path[draw,<-,decorate,
  decoration={zigzag,amplitude=0.7pt,segment length=1.2mm,pre=lineto,pre length=4pt}]
    (\thesarrow1.south east) -- (\thesarrow1.south west);
    $\slashedarrowfill@\relbar\relbar/$
    \end{tikzpicture}}%
}

\makeatletter
\def\slashedarrowfill@#1#2#3#4#5{%
  $\m@th\thickmuskip0mu\medmuskip\thickmuskip\thinmuskip\thickmuskip
   \relax#5#1\mkern-7mu%
   \cleaders\hbox{$#5\mkern-2mu#2\mkern-2mu$}\hfill
   \mathclap{#3}\mathclap{#2}%
   \cleaders\hbox{$#5\mkern-2mu#2\mkern-2mu$}\hfill
   \mkern-7mu#4$%
}
\def\rightslashedarrowfillb@{%
  \slashedarrowfill@\relbar\relbar/\rightarrow}
\newcommand\xnrightarrow[2][]{%
  \ext@arrow 0055{\rightslashedarrowfillb@}{#1}{#2}}

\def\rightslashedarrowfille@{%
  \slashedarrowfill@\relbar\relbar/\twoheadrightarrow}
\newcommand\xntworightarrow[2][]{%
  \ext@arrow 0055{\rightslashedarrowfille@}{#1}{#2}}

\def\rightslashedarrowfillg@{%
  \slashedarrowfill@\relbar\relbar{\raisebox{.12em}{}}\twoheadrightarrow}
\newcommand\xtworightarrow[2][]{%
  \ext@arrow 0055{\rightslashedarrowfillg@}{#1}{#2}}

\def\rightslashedarrowfillx@{%
  \slashedarrowfill@\Relbar\Relbar/\rightrightarrows}
\newcommand\xnTworightarrow[2][]{%
  \ext@arrow 0055{\rightslashedarrowfillx@}{#1}{#2}}

\def\rightslashedarrowfilly@{%
  \slashedarrowfill@\Relbar\Relbar{\raisebox{.12em}{}}\rightrightarrows}
\newcommand\xTworightarrow[2][]{%
  \ext@arrow 0055{\rightslashedarrowfilly@}{#1}{#2}}

\pgfdeclareshape{slash underlined}
{
  \inheritsavedanchors[from=rectangle] 
  \inheritanchorborder[from=rectangle]
  \inheritanchor[from=rectangle]{north}
  \inheritanchor[from=rectangle]{north west}
  \inheritanchor[from=rectangle]{north east}
  \inheritanchor[from=rectangle]{center}
  \inheritanchor[from=rectangle]{west}
  \inheritanchor[from=rectangle]{east}
  \inheritanchor[from=rectangle]{mid}
  \inheritanchor[from=rectangle]{mid west}
  \inheritanchor[from=rectangle]{mid east}
  \inheritanchor[from=rectangle]{base}
  \inheritanchor[from=rectangle]{base west}
  \inheritanchor[from=rectangle]{base east}
  \inheritanchor[from=rectangle]{south}
  \inheritanchor[from=rectangle]{south west}
  \inheritanchor[from=rectangle]{south east}
  \inheritanchorborder[from=rectangle]
  \foregroundpath{
    \southwest \pgf@xa=\pgf@x \pgf@ya=\pgf@y
    \northeast \pgf@xb=\pgf@x \pgf@yb=\pgf@y
    \pgf@xc=\pgf@xa
    \advance\pgf@xc by .5\pgf@xb
    \pgf@yc=\pgf@ya
    \advance\pgf@xc by -1.3pt
    \advance\pgf@yc by -1.8pt
    \pgfpathmoveto{\pgfqpoint{\pgf@xc}{\pgf@yc}}
    \advance\pgf@xc by  2.6pt
    \advance\pgf@yc by  3.6pt
    \pgfpathlineto{\pgfqpoint{\pgf@xc}{\pgf@yc}}
    \pgfpathmoveto{\pgfqpoint{\pgf@xa}{\pgf@ya}}
    \pgfpathlineto{\pgfqpoint{\pgf@xb}{\pgf@ya}}
 }
}
\tikzset{nomorepostaction/.code=\let\tikz@postactions\pgfutil@empty}

\makeatother

\newcommand\black{\ensuremath{\blacktriangleright}}
\newcommand\white{\ensuremath{\vartriangleright}}

  \newcommand\whbl{\white\kern-.1em--\kern-.1em\black}
  \newcommand\blwh{\black\kern-.1em--\kern-.1em\white}
  \newcommand\blbl{\black\kern-.1em--\kern-.1em\black}
  \newcommand\whwh{\white\kern-.1em--\kern-.1em\white}

\newtheorem{theorem}{Theorem}[section]
\newtheorem{definition}[theorem]{Definition}
\newtheorem{proposition}[theorem]{Proposition}

\title[Draft of An Algebra of Reversible Computation]
      {An Algebra of Reversible Computation}

\author[Yong Wang]
    {Yong Wang\\
     College of Computer Science and Technology,\\
     Faculty of Information Technology,\\
     Beijing University of Technology, Beijing, China\\
     }

\correspond{Yong Wang, Pingleyuan 100, Chaoyang District, Beijing, China.
            e-mail: wangy@bjut.edu.cn}

\pubyear{2018}
\pagerange{\pageref{firstpage}--\pageref{lastpage}}

\begin{document}
\label{firstpage}

\makecorrespond

\maketitle

\begin{abstract}
Process algebra ACP based on the interleaving semantics can not be reversed. We design a reversible version of APTC called RAPTC. It has algebraic laws of reversible choice, sequence, parallelism, communication, silent step and abstraction, and also the soundness and completeness modulo strongly forward-reverse truly concurrent bisimulations and weakly forward-reverse truly concurrent bisimulations.
\end{abstract}


\section{Introduction}{\label{int}}

Reversible computation \cite{RCCS2} \cite{TCSR} \cite{CR} is an interesting topic, and in \cite{RCCS2}, an algebraic way of reversible computation used communication key was proposed. In \cite{RCCS2}, a reversible version of process algebra CCS \cite{CC} \cite{CCS} was presented, however, properties of this reversible CCS based on the so-called forward-reverse interleaving bisimulation semantics were not discussed.

We tried to do some work on reversible ACP \cite{ALNC} \cite{ACP} in the former version of this paper, but, there were some errors that can not be remedied, just because ACP based on the interleaving semantics can not be reversed. Until we found algebraic laws for true concurrency called $APTC$ \cite{APTC}, we recall the axiomatization of reversible process algebra, and design a reversible version of $APTC$ called $RAPTC$.

This paper is organized as follows. In section \ref{bg}, we give a background of $APTC$. In section \ref{ftc}, we interpret the concepts of the so-called forward-reverse truly concurrent bisimulations on which $RAPTC$ is based. We introduce Basic Reversible Algebra for True Concurrency ($BRATC$) in section \ref{bratc} and Reversible Algebra for Parallelism in True Concurrency ($RAPTC$) in section \ref{raptc}, and abstraction in section \ref{abs}. Finally, in section \ref{con}, we conclude this paper.

\section{Backgrounds}\label{bg}

In this subsection, we introduce the preliminaries on truly concurrent process algebra $APTC$ \cite{APTC}, which is based on the truly concurrent bisimulation semantics. $APTC$ has an almost perfect axiomatization to capture laws on truly concurrent bisimulation equivalence, including equational logic and truly concurrent bisimulation semantics, and also the soundness and completeness bridged between them.

\subsection{APTC}\label{APTC}

$APTC$ captures several computational properties in the form of algebraic laws, and proves the soundness and completeness modulo truly concurrent bisimulation/rooted branching truly concurrent bisimulation equivalence. These computational properties are organized in a modular way by use of the concept of conservational extension, which include the following modules, note that, every algebra are composed of constants and operators, the constants are the computational objects, while operators capture the computational properties.

\begin{enumerate}
  \item \textbf{$BATC$ (Basic Algebras for True Concurrency)}. $BATC$ has sequential composition $\cdot$ and alternative composition $+$ to capture causality computation and conflict. The constants are ranged over $\mathbb{E}$, the set of atomic events. The algebraic laws on $\cdot$ and $+$ are sound and complete modulo truly concurrent bisimulation equivalences, such as pomset bisimulation $\sim_p$, step bisimulation $\sim_s$, history-preserving (hp-) bisimulation $\sim_{hp}$ and hereditary history-preserving (hhp-) bisimulation $\sim_{hhp}$.
  \item \textbf{$APTC$ (Algebra for Parallelism for True Concurrency)}. $APTC$ uses the whole parallel operator $\between$, the parallel operator $\parallel$ to model parallelism, and the communication merge $\mid$ to model causality (communication) among different parallel branches. Since a communication may be blocked, a new constant called deadlock $\delta$ is extended to $\mathbb{E}$, and also a new unary encapsulation operator $\partial_H$ is introduced to eliminate $\delta$, which may exist in the processes. And also a conflict elimination operator $\Theta$ to eliminate conflicts existing in different parallel branches. The algebraic laws on these operators are also sound and complete modulo truly concurrent bisimulation equivalences, such as pomset bisimulation $\sim_p$, step bisimulation $\sim_s$, history-preserving (hp-) bisimulation $\sim_{hp}$. Note that, these operators in a process except the parallel operator $\parallel$ can be eliminated by deductions on the process using axioms of $APTC$, and eventually be steadied by $\cdot$, $+$ and $\parallel$, this is also why bisimulations are called an \emph{truly concurrent} semantics.
  \item \textbf{Recursion}. To model infinite computation, recursion is introduced into $APTC$. In order to obtain a sound and complete theory, guarded recursion and linear recursion are needed. The corresponding axioms are $RSP$ (Recursive Specification Principle) and $RDP$ (Recursive Definition Principle), $RDP$ says the solutions of a recursive specification can represent the behaviors of the specification, while $RSP$ says that a guarded recursive specification has only one solution, they are sound with respect to $APTC$ with guarded recursion modulo truly concurrent bisimulation equivalences, such as pomset bisimulation $\sim_p$, step bisimulation $\sim_s$, history-preserving (hp-) bisimulation $\sim_{hp}$, and they are complete with respect to $APTC$ with linear recursion modulo truly concurrent bisimulation equivalence, such as pomset bisimulation $\sim_p$, step bisimulation $\sim_s$, history-preserving (hp-) bisimulation $\sim_{hp}$.
  \item \textbf{Abstraction}. To abstract away internal implementations from the external behaviors, a new constant $\tau$ called silent step is added to $\mathbb{E}$, and also a new unary abstraction operator $\tau_I$ is used to rename actions in $I$ into $\tau$ (the resulted $APTC$ with silent step and abstraction operator is called $APTC_{\tau}$). The recursive specification is adapted to guarded linear recursion to prevent infinite $\tau$-loops specifically. The axioms for $\tau$ and $\tau_I$ are sound modulo rooted branching truly concurrent bisimulation equivalences (a kind of weak truly concurrent bisimulation equivalence), such as rooted branching pomset bisimulation $\approx_p$, rooted branching step bisimulation $\approx_s$, rooted branching history-preserving (hp-) bisimulation $\approx_{hp}$. To eliminate infinite $\tau$-loops caused by $\tau_I$ and obtain the completeness, $CFAR$ (Cluster Fair Abstraction Rule) is used to prevent infinite $\tau$-loops in a constructible way.
\end{enumerate}

$APTC$ can be used to verify the correctness of system behaviors, by deduction on the description of the system using the axioms of $APTC$. Base on the modularity of $APTC$, it can be extended easily and elegantly. For more details, please refer to the manuscript of $APTC$ \cite{APTC}.

\subsection{Operational Semantics}\label{OS}

The semantics of $APTC$ is based on truly concurrent bisimulation/rooted branching truly concurrent bisimulation equivalences, and the modularity of $APTC$ relies on the concept of conservative extension, for the conveniences, we introduce some concepts and conclusions on them.

\begin{definition}[Prime event structure with silent event]\label{PES}
Let $\Lambda$ be a fixed set of labels, ranged over $a,b,c,\cdots$ and $\tau$. A ($\Lambda$-labelled) prime event structure with silent event $\tau$ is a tuple $\mathcal{E}=\langle \mathbb{E}, \leq, \sharp, \lambda\rangle$, where $\mathbb{E}$ is a denumerable set of events, including the silent event $\tau$. Let $\hat{\mathbb{E}}=\mathbb{E}\backslash\{\tau\}$, exactly excluding $\tau$, it is obvious that $\hat{\tau^*}=\epsilon$, where $\epsilon$ is the empty event. Let $\lambda:\mathbb{E}\rightarrow\Lambda$ be a labelling function and let $\lambda(\tau)=\tau$. And $\leq$, $\sharp$ are binary relations on $\mathbb{E}$, called causality and conflict respectively, such that:

\begin{enumerate}
  \item $\leq$ is a partial order and $\lceil e \rceil = \{e'\in \mathbb{E}|e'\leq e\}$ is finite for all $e\in \mathbb{E}$. It is easy to see that $e\leq\tau^*\leq e'=e\leq\tau\leq\cdots\leq\tau\leq e'$, then $e\leq e'$.
  \item $\sharp$ is irreflexive, symmetric and hereditary with respect to $\leq$, that is, for all $e,e',e''\in \mathbb{E}$, if $e\sharp e'\leq e''$, then $e\sharp e''$.
\end{enumerate}

Then, the concepts of consistency and concurrency can be drawn from the above definition:

\begin{enumerate}
  \item $e,e'\in \mathbb{E}$ are consistent, denoted as $e\frown e'$, if $\neg(e\sharp e')$. A subset $X\subseteq \mathbb{E}$ is called consistent, if $e\frown e'$ for all $e,e'\in X$.
  \item $e,e'\in \mathbb{E}$ are concurrent, denoted as $e\parallel e'$, if $\neg(e\leq e')$, $\neg(e'\leq e)$, and $\neg(e\sharp e')$.
\end{enumerate}
\end{definition}

\begin{definition}[Configuration]
Let $\mathcal{E}$ be a PES. A (finite) configuration in $\mathcal{E}$ is a (finite) consistent subset of events $C\subseteq \mathcal{E}$, closed with respect to causality (i.e. $\lceil C\rceil=C$). The set of finite configurations of $\mathcal{E}$ is denoted by $\mathcal{C}(\mathcal{E})$. We let $\hat{C}=C\backslash\{\tau\}$.
\end{definition}

A consistent subset of $X\subseteq \mathbb{E}$ of events can be seen as a pomset. Given $X, Y\subseteq \mathbb{E}$, $\hat{X}\sim \hat{Y}$ if $\hat{X}$ and $\hat{Y}$ are isomorphic as pomsets. In the following of the paper, we say $C_1\sim C_2$, we mean $\hat{C_1}\sim\hat{C_2}$.

\begin{definition}[Pomset transitions and step]
Let $\mathcal{E}$ be a PES and let $C\in\mathcal{C}(\mathcal{E})$, and $\emptyset\neq X\subseteq \mathbb{E}$, if $C\cap X=\emptyset$ and $C'=C\cup X\in\mathcal{C}(\mathcal{E})$, then $C\xrightarrow{X} C'$ is called a pomset transition from $C$ to $C'$. When the events in $X$ are pairwise concurrent, we say that $C\xrightarrow{X}C'$ is a step.
\end{definition}

\begin{definition}[Weak pomset transitions and weak step]
Let $\mathcal{E}$ be a PES and let $C\in\mathcal{C}(\mathcal{E})$, and $\emptyset\neq X\subseteq \hat{\mathbb{E}}$, if $C\cap X=\emptyset$ and $\hat{C'}=\hat{C}\cup X\in\mathcal{C}(\mathcal{E})$, then $C\xRightarrow{X} C'$ is called a weak pomset transition from $C$ to $C'$, where we define $\xRightarrow{e}\triangleq\xrightarrow{\tau^*}\xrightarrow{e}\xrightarrow{\tau^*}$. And $\xRightarrow{X}\triangleq\xrightarrow{\tau^*}\xrightarrow{e}\xrightarrow{\tau^*}$, for every $e\in X$. When the events in $X$ are pairwise concurrent, we say that $C\xRightarrow{X}C'$ is a weak step.
\end{definition}

We will also suppose that all the PESs in this paper are image finite, that is, for any PES $\mathcal{E}$ and $C\in \mathcal{C}(\mathcal{E})$ and $a\in \Lambda$, $\{e\in \mathbb{E}|C\xrightarrow{e} C'\wedge \lambda(e)=a\}$ and $\{e\in\hat{\mathbb{E}}|C\xRightarrow{e} C'\wedge \lambda(e)=a\}$ is finite.

\begin{definition}[Pomset, step bisimulation]\label{PSB}
Let $\mathcal{E}_1$, $\mathcal{E}_2$ be PESs. A pomset bisimulation is a relation $R\subseteq\mathcal{C}(\mathcal{E}_1)\times\mathcal{C}(\mathcal{E}_2)$, such that if $(C_1,C_2)\in R$, and $C_1\xrightarrow{X_1}C_1'$ then $C_2\xrightarrow{X_2}C_2'$, with $X_1\subseteq \mathbb{E}_1$, $X_2\subseteq \mathbb{E}_2$, $X_1\sim X_2$ and $(C_1',C_2')\in R$, and vice-versa. We say that $\mathcal{E}_1$, $\mathcal{E}_2$ are pomset bisimilar, written $\mathcal{E}_1\sim_p\mathcal{E}_2$, if there exists a pomset bisimulation $R$, such that $(\emptyset,\emptyset)\in R$. By replacing pomset transitions with steps, we can get the definition of step bisimulation. When PESs $\mathcal{E}_1$ and $\mathcal{E}_2$ are step bisimilar, we write $\mathcal{E}_1\sim_s\mathcal{E}_2$.
\end{definition}

\begin{definition}[Weak pomset, step bisimulation]\label{WPSB}
Let $\mathcal{E}_1$, $\mathcal{E}_2$ be PESs. A weak pomset bisimulation is a relation $R\subseteq\mathcal{C}(\mathcal{E}_1)\times\mathcal{C}(\mathcal{E}_2)$, such that if $(C_1,C_2)\in R$, and $C_1\xRightarrow{X_1}C_1'$ then $C_2\xRightarrow{X_2}C_2'$, with $X_1\subseteq \hat{\mathbb{E}_1}$, $X_2\subseteq \hat{\mathbb{E}_2}$, $X_1\sim X_2$ and $(C_1',C_2')\in R$, and vice-versa. We say that $\mathcal{E}_1$, $\mathcal{E}_2$ are weak pomset bisimilar, written $\mathcal{E}_1\approx_p\mathcal{E}_2$, if there exists a weak pomset bisimulation $R$, such that $(\emptyset,\emptyset)\in R$. By replacing weak pomset transitions with weak steps, we can get the definition of weak step bisimulation. When PESs $\mathcal{E}_1$ and $\mathcal{E}_2$ are weak step bisimilar, we write $\mathcal{E}_1\approx_s\mathcal{E}_2$.
\end{definition}

\begin{definition}[Posetal product]
Given two PESs $\mathcal{E}_1$, $\mathcal{E}_2$, the posetal product of their configurations, denoted $\mathcal{C}(\mathcal{E}_1)\overline{\times}\mathcal{C}(\mathcal{E}_2)$, is defined as

$$\{(C_1,f,C_2)|C_1\in\mathcal{C}(\mathcal{E}_1),C_2\in\mathcal{C}(\mathcal{E}_2),f:C_1\rightarrow C_2 \textrm{ isomorphism}\}.$$

A subset $R\subseteq\mathcal{C}(\mathcal{E}_1)\overline{\times}\mathcal{C}(\mathcal{E}_2)$ is called a posetal relation. We say that $R$ is downward closed when for any $(C_1,f,C_2),(C_1',f',C_2')\in \mathcal{C}(\mathcal{E}_1)\overline{\times}\mathcal{C}(\mathcal{E}_2)$, if $(C_1,f,C_2)\subseteq (C_1',f',C_2')$ pointwise and $(C_1',f',C_2')\in R$, then $(C_1,f,C_2)\in R$.

For $f:X_1\rightarrow X_2$, we define $f[x_1\mapsto x_2]:X_1\cup\{x_1\}\rightarrow X_2\cup\{x_2\}$, $z\in X_1\cup\{x_1\}$,(1)$f[x_1\mapsto x_2](z)=
x_2$,if $z=x_1$;(2)$f[x_1\mapsto x_2](z)=f(z)$, otherwise. Where $X_1\subseteq \mathbb{E}_1$, $X_2\subseteq \mathbb{E}_2$, $x_1\in \mathbb{E}_1$, $x_2\in \mathbb{E}_2$.
\end{definition}

\begin{definition}[Weakly posetal product]
Given two PESs $\mathcal{E}_1$, $\mathcal{E}_2$, the weakly posetal product of their configurations, denoted $\mathcal{C}(\mathcal{E}_1)\overline{\times}\mathcal{C}(\mathcal{E}_2)$, is defined as

$$\{(C_1,f,C_2)|C_1\in\mathcal{C}(\mathcal{E}_1),C_2\in\mathcal{C}(\mathcal{E}_2),f:\hat{C_1}\rightarrow \hat{C_2} \textrm{ isomorphism}\}.$$

A subset $R\subseteq\mathcal{C}(\mathcal{E}_1)\overline{\times}\mathcal{C}(\mathcal{E}_2)$ is called a weakly posetal relation. We say that $R$ is downward closed when for any $(C_1,f,C_2),(C_1',f,C_2')\in \mathcal{C}(\mathcal{E}_1)\overline{\times}\mathcal{C}(\mathcal{E}_2)$, if $(C_1,f,C_2)\subseteq (C_1',f',C_2')$ pointwise and $(C_1',f',C_2')\in R$, then $(C_1,f,C_2)\in R$.

For $f:X_1\rightarrow X_2$, we define $f[x_1\mapsto x_2]:X_1\cup\{x_1\}\rightarrow X_2\cup\{x_2\}$, $z\in X_1\cup\{x_1\}$,(1)$f[x_1\mapsto x_2](z)=
x_2$,if $z=x_1$;(2)$f[x_1\mapsto x_2](z)=f(z)$, otherwise. Where $X_1\subseteq \hat{\mathbb{E}_1}$, $X_2\subseteq \hat{\mathbb{E}_2}$, $x_1\in \hat{\mathbb{E}}_1$, $x_2\in \hat{\mathbb{E}}_2$. Also, we define $f(\tau^*)=f(\tau^*)$.
\end{definition}

\begin{definition}[(Hereditary) history-preserving bisimulation]\label{HHPB}
A history-preserving (hp-) bisimulation is a posetal relation $R\subseteq\mathcal{C}(\mathcal{E}_1)\overline{\times}\mathcal{C}(\mathcal{E}_2)$ such that if $(C_1,f,C_2)\in R$, and $C_1\xrightarrow{e_1} C_1'$, then $C_2\xrightarrow{e_2} C_2'$, with $(C_1',f[e_1\mapsto e_2],C_2')\in R$, and vice-versa. $\mathcal{E}_1,\mathcal{E}_2$ are history-preserving (hp-)bisimilar and are written $\mathcal{E}_1\sim_{hp}\mathcal{E}_2$ if there exists a hp-bisimulation $R$ such that $(\emptyset,\emptyset,\emptyset)\in R$.

A hereditary history-preserving (hhp-)bisimulation is a downward closed hp-bisimulation. $\mathcal{E}_1,\mathcal{E}_2$ are hereditary history-preserving (hhp-)bisimilar and are written $\mathcal{E}_1\sim_{hhp}\mathcal{E}_2$.
\end{definition}

\begin{definition}[Weak (hereditary) history-preserving bisimulation]\label{WHHPB}
A weak history-preserving (hp-) bisimulation is a weakly posetal relation $R\subseteq\mathcal{C}(\mathcal{E}_1)\overline{\times}\mathcal{C}(\mathcal{E}_2)$ such that if $(C_1,f,C_2)\in R$, and $C_1\xRightarrow{e_1} C_1'$, then $C_2\xRightarrow{e_2} C_2'$, with $(C_1',f[e_1\mapsto e_2],C_2')\in R$, and vice-versa. $\mathcal{E}_1,\mathcal{E}_2$ are weak history-preserving (hp-)bisimilar and are written $\mathcal{E}_1\approx_{hp}\mathcal{E}_2$ if there exists a hp-bisimulation $R$ such that $(\emptyset,\emptyset,\emptyset)\in R$.

A weakly hereditary history-preserving (hhp-)bisimulation is a downward closed weak hp-bisimulation. $\mathcal{E}_1,\mathcal{E}_2$ are weakly hereditary history-preserving (hhp-)bisimilar and are written $\mathcal{E}_1\approx_{hhp}\mathcal{E}_2$.
\end{definition}

\begin{definition}[Congruence]
Let $\Sigma$ be a signature. An equivalence relation $R$ on $\mathcal{T}(\Sigma)$ is a congruence if for each $f\in\Sigma$, if $s_i R t_i$ for $i\in\{1,\cdots,ar(f)\}$, then $f(s_1,\cdots,s_{ar(f)}) R f(t_1,\cdots,t_{ar(f)})$.
\end{definition}

\begin{definition}[Conservative extension]
Let $T_0$ and $T_1$ be TSSs (transition system specifications) over signatures $\Sigma_0$ and $\Sigma_1$, respectively. The TSS $T_0\oplus T_1$ is a conservative extension of $T_0$ if the LTSs (labeled transition systems) generated by $T_0$ and $T_0\oplus T_1$ contain exactly the same transitions $t\xrightarrow{a}t'$ and $tP$ with $t\in \mathcal{T}(\Sigma_0)$.
\end{definition}

\begin{definition}[Source-dependency]
The source-dependent variables in a transition rule of $\rho$ are defined inductively as follows: (1) all variables in the source of $\rho$ are source-dependent; (2) if $t\xrightarrow{a}t'$ is a premise of $\rho$ and all variables in $t$ are source-dependent, then all variables in $t'$ are source-dependent. A transition rule is source-dependent if all its variables are. A TSS is source-dependent if all its rules are.
\end{definition}

\begin{definition}[Freshness]
Let $T_0$ and $T_1$ be TSSs over signatures $\Sigma_0$ and $\Sigma_1$, respectively. A term in $\mathbb{T}(T_0\oplus T_1)$ is said to be fresh if it contains a function symbol from $\Sigma_1\setminus\Sigma_0$. Similarly, a transition label or predicate symbol in $T_1$ is fresh if it does not occur in $T_0$.
\end{definition}

\begin{theorem}[Conservative extension]\label{TCE}
Let $T_0$ and $T_1$ be TSSs over signatures $\Sigma_0$ and $\Sigma_1$, respectively, where $T_0$ and $T_0\oplus T_1$ are positive after reduction. Under the following conditions, $T_0\oplus T_1$ is a conservative extension of $T_0$. (1) $T_0$ is source-dependent. (2) For each $\rho\in T_1$, either the source of $\rho$ is fresh, or $\rho$ has a premise of the form $t\xrightarrow{a}t'$ or $tP$, where $t\in \mathbb{T}(\Sigma_0)$, all variables in $t$ occur in the source of $\rho$ and $t'$, $a$ or $P$ is fresh.
\end{theorem}

\subsection{Proof Techniques}\label{PT}

In this subsection, we introduce the concepts and conclusions about elimination, which is very important in the proof of completeness theorem.

\begin{definition}[Elimination property]
Let a process algebra with a defined set of basic terms as a subset of the set of closed terms over the process algebra. Then the process algebra has the elimination to basic terms property if for every closed term $s$ of the algebra, there exists a basic term $t$ of the algebra such that the algebra$\vdash s=t$.
\end{definition}

\begin{definition}[Strongly normalizing]
A term $s_0$ is called strongly normalizing if does not an infinite series of reductions beginning in $s_0$.
\end{definition}

\begin{definition}
We write $s>_{lpo} t$ if $s\rightarrow^+ t$ where $\rightarrow^+$ is the transitive closure of the reduction relation defined by the transition rules of a algebra.
\end{definition}

\begin{theorem}[Strong normalization]\label{SN}
Let a term rewriting (TRS) system with finitely many rewriting rules and let $>$ be a well-founded ordering on the signature of the corresponding algebra. If $s>_{lpo} t$ for each rewriting rule $s\rightarrow t$ in the TRS, then the term rewriting system is strongly normalizing.
\end{theorem}

\section{Forward-reverse Truly Concurrent Bisimulations}{\label{ftc}}

\begin{definition}[Forward-reverse (FR) pomset transitions and forward-reverse (FR) step]
Let $\mathcal{E}$ be a PES and let $C\in\mathcal{C}(\mathcal{E})$, $\emptyset\neq X\subseteq \mathbb{E}$, $\mathcal{K}\subseteq \mathbb{N}$, and $X[\mathcal{K}]$ denotes that for each $e\in X$, there is $e[m]\in X[\mathcal{K}]$ where $(m\in\mathcal{K})$, which is called the past of $e$, and we extend $\mathbb{E}$ to $\mathbb{E}\cup\tau\mathbb{E}[\mathcal{K}]$. If $C\cap X[\mathcal{K}]=\emptyset$ and $C'=C\cup X[\mathcal{K}], X\in\mathcal{C}(\mathcal{E})$, then $C\xrightarrow{X} C'$ is called a forward pomset transition from $C$ to $C'$, and $C'\xtworightarrow{X[\mathcal{K}]} C$ is called a reverse pomset transition from $C'$ to $C$. When the events in $X$ are pairwise concurrent, we say that $C\xrightarrow{X}C'$ is a forward step and $C'\xtworightarrow{X[\mathcal{K}]} C$ is a reverse step.
\end{definition}

\begin{definition}[Weak forward-reverse (FR) pomset transitions and weak forward-reverse (FR) step]
Let $\mathcal{E}$ be a PES and let $C\in\mathcal{C}(\mathcal{E})$, and $\emptyset\neq X\subseteq \hat{\mathbb{E}}$, $\mathcal{K}\subseteq \mathbb{N}$, and $X[\mathcal{K}]$ denotes that for each $e\in X$, there is $e[m]\in X[\mathcal{K}]$ where $(m\in\mathcal{K})$, which is called the past of $e$. If $C\cap X[\mathcal{K}]=\emptyset$ and $\hat{C'}=\hat{C}\cup X[\mathcal{K}], X\in\mathcal{C}(\mathcal{E})$, then $C\xRightarrow{X} C'$ is called a weak forward pomset transition from $C$ to $C'$, where we define $\xRightarrow{e}\triangleq\xrightarrow{\tau^*}\xrightarrow{e}\xrightarrow{\tau^*}$ and $\xRightarrow{X}\triangleq\xrightarrow{\tau^*}\xrightarrow{e}\xrightarrow{\tau^*}$, for every $e\in X$. And $C'\xTworightarrow{X[\mathcal{K}]} C$ is called a weak reverse pomset transition from $C'$ to $C$, where we define $\xTworightarrow{e[m]}\triangleq\xtworightarrow{\tau^*}\xtworightarrow{e[m]}\xtworightarrow{\tau^*}$, $\xTworightarrow{X[\mathcal{K}]}\triangleq\xtworightarrow{\tau^*}\xtworightarrow{e[m]} \xtworightarrow{\tau^*}$, for every $e\in X$ and $m\in\mathcal{K}$. When the events in $X$ are pairwise concurrent, we say that $C\xRightarrow{X}C'$ is a weak forward step and $C'\xTworightarrow{X[\mathcal{K}]} C$ is a weak reverse step.
\end{definition}

We will also suppose that all the PESs in this paper are image finite, that is, for any PES $\mathcal{E}$ and $C\in \mathcal{C}(\mathcal{E})$, and $a\in \Lambda$, $\{e\in \mathbb{E}|C\xrightarrow{e} C'\wedge \lambda(e)=a\}$ and $\{e\in\hat{\mathbb{E}}|C\xRightarrow{e} C'\wedge \lambda(e)=a\}$, and $a\in \Lambda$, $\{e\in \mathbb{E}|C'\xtworightarrow{e[m]} C\wedge \lambda(e)=a\}$ and $\{e\in\hat{\mathbb{E}}|C'\xTworightarrow{e[m]} C\wedge \lambda(e)=a\}$ are finite.

\begin{definition}[Forward-reverse (FR) pomset, step bisimulation]\label{FRPSB}
Let $\mathcal{E}_1$, $\mathcal{E}_2$ be PESs. An FR pomset bisimulation is a relation $R\subseteq\mathcal{C}(\mathcal{E}_1)\times\mathcal{C}(\mathcal{E}_2)$, such that (1) if $(C_1,C_2)\in R$, and $C_1\xrightarrow{X_1}C_1'$ then $C_2\xrightarrow{X_2}C_2'$, with $X_1\subseteq \mathbb{E}_1$, $X_2\subseteq \mathbb{E}_2$, $X_1\sim X_2$ and $(C_1',C_2')\in R$, and vice-versa; (2) if $(C_1',C_2')\in R$, and $C_1'\xtworightarrow{X_1[\mathcal{K}_1]}C_1$ then $C_2'\xtworightarrow{X_2[\mathcal{K}_2]}C_2$, with $X_1\subseteq \mathbb{E}_1$, $X_2\subseteq \mathbb{E}_2$, $\mathcal{K}_1,\mathcal{K}_2\subseteq\mathbb{N}$, $X_1\sim X_2$ and $(C_1,C_2)\in R$, and vice-versa. We say that $\mathcal{E}_1$, $\mathcal{E}_2$ are FR pomset bisimilar, written $\mathcal{E}_1\sim_p^{fr}\mathcal{E}_2$, if there exists an FR pomset bisimulation $R$, such that $(\emptyset,\emptyset)\in R$. By replacing FR pomset transitions with FR steps, we can get the definition of FR step bisimulation. When PESs $\mathcal{E}_1$ and $\mathcal{E}_2$ are FR step bisimilar, we write $\mathcal{E}_1\sim_s^{fr}\mathcal{E}_2$.
\end{definition}

\begin{definition}[Weak forward-reverse (FR) pomset, step bisimulation]\label{FRWPSB}
Let $\mathcal{E}_1$, $\mathcal{E}_2$ be PESs. A weak FR pomset bisimulation is a relation $R\subseteq\mathcal{C}(\mathcal{E}_1)\times\mathcal{C}(\mathcal{E}_2)$, such that (1) if $(C_1,C_2)\in R$, and $C_1\xRightarrow{X_1}C_1'$ then $C_2\xRightarrow{X_2}C_2'$, with $X_1\subseteq \hat{\mathbb{E}_1}$, $X_2\subseteq \hat{\mathbb{E}_2}$, $X_1\sim X_2$ and $(C_1',C_2')\in R$, and vice-versa; (2) if $(C_1',C_2')\in R$, and $C_1'\xTworightarrow{X_1[\mathcal{K}_1]}C_1$ then $C_2'\xTworightarrow{X_2[\mathcal{K}_2]}C_2$, with $X_1\subseteq \hat{\mathbb{E}_1}$, $X_2\subseteq \hat{\mathbb{E}_2}$, $\mathcal{K}_1,\mathcal{K}_2\subseteq\mathbb{N}$, $X_1\sim X_2$ and $(C_1,C_2)\in R$, and vice-versa. We say that $\mathcal{E}_1$, $\mathcal{E}_2$ are weak FR pomset bisimilar, written $\mathcal{E}_1\approx_p^{fr}\mathcal{E}_2$, if there exists a weak FR pomset bisimulation $R$, such that $(\emptyset,\emptyset)\in R$. By replacing weak FR pomset transitions with weak FR steps, we can get the definition of weak FR step bisimulation. When PESs $\mathcal{E}_1$ and $\mathcal{E}_2$ are weak FR step bisimilar, we write $\mathcal{E}_1\approx_s^{fr}\mathcal{E}_2$.
\end{definition}

\begin{definition}[Forward-reverse (FR) (hereditary) history-preserving bisimulation]\label{FRHHPB}
An FR history-preserving (hp-) bisimulation is a posetal relation $R\subseteq\mathcal{C}(\mathcal{E}_1)\overline{\times}\mathcal{C}(\mathcal{E}_2)$ such that (1) if $(C_1,f,C_2)\in R$, and $C_1\xrightarrow{e_1} C_1'$, then $C_2\xrightarrow{e_2} C_2'$, with $(C_1',f[e_1\mapsto e_2],C_2')\in R$, and vice-versa, (2) if $(C_1',f',C_2')\in R$, and $C_1'\xtworightarrow{e_1[m]} C_1$, then $C_2'\xtworightarrow{e_2[n]} C_2$, with $(C_1,f'[e_1[m]\mapsto e_2[n]],C_2)\in R$, and vice-versa. $\mathcal{E}_1,\mathcal{E}_2$ are FR history-preserving (hp-) bisimilar and are written $\mathcal{E}_1\sim_{hp}^{fr}\mathcal{E}_2$ if there exists an FR hp-bisimulation $R$ such that $(\emptyset,\emptyset,\emptyset)\in R$.

An FR hereditary history-preserving (hhp-)bisimulation is a downward closed FR hp-bisimulation. $\mathcal{E}_1,\mathcal{E}_2$ are FR hereditary history-preserving (hhp-)bisimilar and are written $\mathcal{E}_1\sim_{hhp}^{fr}\mathcal{E}_2$.
\end{definition}

\begin{definition}[Weak forward-reverse (FR) (hereditary) history-preserving bisimulation]\label{FRWHHPB}
A weak FR history-preserving (hp-) bisimulation is a weakly posetal relation $R\subseteq\mathcal{C}(\mathcal{E}_1)\overline{\times}\mathcal{C}(\mathcal{E}_2)$ such that (1) if $(C_1,f,C_2)\in R$, and $C_1\xRightarrow{e_1} C_1'$, then $C_2\xRightarrow{e_2} C_2'$, with $(C_1',f[e_1\mapsto e_2],C_2')\in R$, and vice-versa, (2) if $(C_1',f',C_2')\in R$, and $C_1'\xTworightarrow{e_1[m]} C_1$, then $C_2'\xTworightarrow{e_2[n]} C_2$, with $(C_1,f'[e_1[m]\mapsto e_2[n]],C_2)\in R$, and vice-versa. $\mathcal{E}_1,\mathcal{E}_2$ are weak FR history-preserving (hp-) bisimilar and are written $\mathcal{E}_1\approx_{hp}^{fr}\mathcal{E}_2$ if there exists a weak FR hp-bisimulation $R$ such that $(\emptyset,\emptyset,\emptyset)\in R$.

A weak FR hereditary history-preserving (hhp-) bisimulation is a downward closed weak FR hp-bisimulation. $\mathcal{E}_1,\mathcal{E}_2$ are weak FR hereditary history-preserving (hhp-) bisimilar and are written $\mathcal{E}_1\approx_{hhp}^{fr}\mathcal{E}_2$.
\end{definition}

\begin{definition}[Branching forward-reverse pomset, step bisimulation]\label{FRBPSB}
Assume a special termination predicate $\downarrow$, and let $\surd$ represent a state with $\surd\downarrow$. Let $\mathcal{E}_1$, $\mathcal{E}_2$ be PESs. A branching FR pomset bisimulation is a relation $R\subseteq\mathcal{C}(\mathcal{E}_1)\times\mathcal{C}(\mathcal{E}_2)$, such that:
 \begin{enumerate}
   \item if $(C_1,C_2)\in R$, and $C_1\xrightarrow{X}C_1'$ then
   \begin{itemize}
     \item either $X\equiv \tau^*$, and $(C_1',C_2)\in R$;
     \item or there is a sequence of (zero or more) $\tau$-transitions $C_2\xrightarrow{\tau^*} C_2^0$, such that $(C_1,C_2^0)\in R$ and $C_2^0\xRightarrow{X}C_2'$ with $(C_1',C_2')\in R$;
   \end{itemize}
   \item if $(C_1,C_2)\in R$, and $C_2\xrightarrow{X}C_2'$ then
   \begin{itemize}
     \item either $X\equiv \tau^*$, and $(C_1,C_2')\in R$;
     \item or there is a sequence of (zero or more) $\tau$-transitions $C_1\xrightarrow{\tau^*} C_1^0$, such that $(C_1^0,C_2)\in R$ and $C_1^0\xRightarrow{X}C_1'$ with $(C_1',C_2')\in R$;
   \end{itemize}
   \item if $(C_1,C_2)\in R$ and $C_1\downarrow$, then there is a sequence of (zero or more) $\tau$-transitions $C_2\xrightarrow{\tau^*}C_2^0$ such that $(C_1,C_2^0)\in R$ and $C_2^0\downarrow$;
   \item if $(C_1,C_2)\in R$ and $C_2\downarrow$, then there is a sequence of (zero or more) $\tau$-transitions $C_1\xrightarrow{\tau^*}C_1^0$ such that $(C_1^0,C_2)\in R$ and $C_1^0\downarrow$;
   \item if $(C_1',C_2')\in R$, and $C_1'\xtworightarrow{X[\mathcal{K}]}C_1$ then
   \begin{itemize}
     \item either $X[\mathcal{K}]\equiv \tau^*$, and $(C_1,C_2')\in R$;
     \item or there is a sequence of (zero or more) $\tau$-transitions $C_2'\xtworightarrow{\tau^*} C_2'^0$, such that $(C_1',C_2'^0)\in R$ and $C_2'^0\xTworightarrow{X[\mathcal{K}]}C_2$ with $(C_1,C_2)\in R$;
   \end{itemize}
   \item if $(C_1',C_2')\in R$, and $C_2'\xtworightarrow{X}C_2$ then
   \begin{itemize}
     \item either $X[\mathcal{K}]\equiv \tau^*$, and $(C_1',C_2)\in R$;
     \item or there is a sequence of (zero or more) $\tau$-transitions $C_1'\xtworightarrow{\tau^*} C_1'^0$, such that $(C_1'^0,C_2')\in R$ and $C_1'^0\xTworightarrow{X[\mathcal{K}]}C_1$ with $(C_1,C_2)\in R$;
   \end{itemize}
   \item if $(C_1',C_2')\in R$ and $C_1'\downarrow$, then there is a sequence of (zero or more) $\tau$-transitions $C_2'\xtworightarrow{\tau^*}C_2'^0$ such that $(C_1',C_2'^0)\in R$ and $C_2'^0\downarrow$;
   \item if $(C_1',C_2')\in R$ and $C_2'\downarrow$, then there is a sequence of (zero or more) $\tau$-transitions $C_1'\xtworightarrow{\tau^*}C_1'^0$ such that $(C_1'^0,C_2')\in R$ and $C_1'^0\downarrow$.
 \end{enumerate}

We say that $\mathcal{E}_1$, $\mathcal{E}_2$ are branching FR pomset bisimilar, written $\mathcal{E}_1\approx_{bp}^{fr}\mathcal{E}_2$, if there exists a branching FR pomset bisimulation $R$, such that $(\emptyset,\emptyset)\in R$.

By replacing FR pomset transitions with FR steps, we can get the definition of branching FR step bisimulation. When PESs $\mathcal{E}_1$ and $\mathcal{E}_2$ are branching FR step bisimilar, we write $\mathcal{E}_1\approx_{bs}^{fr}\mathcal{E}_2$.
\end{definition}

\begin{definition}[Rooted branching forward-reverse (FR) pomset, step bisimulation]\label{FRRBPSB}
Assume a special termination predicate $\downarrow$, and let $\surd$ represent a state with $\surd\downarrow$. Let $\mathcal{E}_1$, $\mathcal{E}_2$ be PESs. A rooted branching FR pomset bisimulation is a relation $R\subseteq\mathcal{C}(\mathcal{E}_1)\times\mathcal{C}(\mathcal{E}_2)$, such that:
 \begin{enumerate}
   \item if $(C_1,C_2)\in R$, and $C_1\xrightarrow{X}C_1'$ then $C_2\xrightarrow{X}C_2'$ with $C_1'\approx_{bp}C_2'$;
   \item if $(C_1,C_2)\in R$, and $C_2\xrightarrow{X}C_2'$ then $C_1\xrightarrow{X}C_1'$ with $C_1'\approx_{bp}C_2'$;
   \item if $(C_1',C_2')\in R$, and $C_1'\xtworightarrow{X[\mathcal{K}]}C_1$ then $C_2'\xtworightarrow{X[\mathcal{K}]}C_2$ with $C_1\approx_{bp}^{fr}C_2$;
   \item if $(C_1',C_2')\in R$, and $C_2'\xtworightarrow{X[\mathcal{K}]}C_2$ then $C_1'\xtworightarrow{X[\mathcal{K}]}C_1$ with $C_1\approx_{bp}^{fr}C_2$;
   \item if $(C_1,C_2)\in R$ and $C_1\downarrow$, then $C_2\downarrow$;
   \item if $(C_1,C_2)\in R$ and $C_2\downarrow$, then $C_1\downarrow$.
 \end{enumerate}

We say that $\mathcal{E}_1$, $\mathcal{E}_2$ are rooted branching FR pomset bisimilar, written $\mathcal{E}_1\approx_{rbp}^{fr}\mathcal{E}_2$, if there exists a rooted branching FR pomset bisimulation $R$, such that $(\emptyset,\emptyset)\in R$.

By replacing FR pomset transitions with FR steps, we can get the definition of rooted branching FR step bisimulation. When PESs $\mathcal{E}_1$ and $\mathcal{E}_2$ are rooted branching FR step bisimilar, we write $\mathcal{E}_1\approx_{rbs}^{fr}\mathcal{E}_2$.
\end{definition}

\begin{definition}[Branching forward-reverse (FR) (hereditary) history-preserving bisimulation]\label{FRBHHPB}
Assume a special termination predicate $\downarrow$, and let $\surd$ represent a state with $\surd\downarrow$. A branching FR history-preserving (hp-) bisimulation is a weakly posetal relation $R\subseteq\mathcal{C}(\mathcal{E}_1)\overline{\times}\mathcal{C}(\mathcal{E}_2)$ such that:

 \begin{enumerate}
   \item if $(C_1,f,C_2)\in R$, and $C_1\xrightarrow{e_1}C_1'$ then
   \begin{itemize}
     \item either $e_1\equiv \tau$, and $(C_1',f[e_1\mapsto \tau],C_2)\in R$;
     \item or there is a sequence of (zero or more) $\tau$-transitions $C_2\xrightarrow{\tau^*} C_2^0$, such that $(C_1,f,C_2^0)\in R$ and $C_2^0\xrightarrow{e_2}C_2'$ with $(C_1',f[e_1\mapsto e_2],C_2')\in R$;
   \end{itemize}
   \item if $(C_1,f,C_2)\in R$, and $C_2\xrightarrow{e_2}C_2'$ then
   \begin{itemize}
     \item either $e_2\equiv \tau$, and $(C_1,f[e_2\mapsto \tau],C_2')\in R$;
     \item or there is a sequence of (zero or more) $\tau$-transitions $C_1\xrightarrow{\tau^*} C_1^0$, such that $(C_1^0,f,C_2)\in R$ and $C_1^0\xrightarrow{e_1}C_1'$ with $(C_1',f[e_2\mapsto e_1],C_2')\in R$;
   \end{itemize}
   \item if $(C_1,f,C_2)\in R$ and $C_1\downarrow$, then there is a sequence of (zero or more) $\tau$-transitions $C_2\xrightarrow{\tau^*}C_2^0$ such that $(C_1,f,C_2^0)\in R$ and $C_2^0\downarrow$;
   \item if $(C_1,f,C_2)\in R$ and $C_2\downarrow$, then there is a sequence of (zero or more) $\tau$-transitions $C_1\xrightarrow{\tau^*}C_1^0$ such that $(C_1^0,f,C_2)\in R$ and $C_1^0\downarrow$;
   \item if $(C_1',f',C_2')\in R$, and $C_1'\xtworightarrow{e_1[m]}C_1$ then
   \begin{itemize}
     \item either $e_1[m]\equiv \tau$, and $(C_1,f'[e_1[m]\mapsto \tau],C_2')\in R$;
     \item or there is a sequence of (zero or more) $\tau$-transitions $C_2'\xtworightarrow{\tau^*} C_2'^0$, such that $(C_1',f',C_2'^0)\in R$ and $C_2'^0\xtworightarrow{e_2[n]}C_2$ with $(C_1,f'[e_1[m]\mapsto e_2[n]],C_2)\in R$;
   \end{itemize}
   \item if $(C_1',f',C_2')\in R$, and $C_2'\xtworightarrow{e_2[n]}C_2$ then
   \begin{itemize}
     \item either $e_2[n]\equiv \tau$, and $(C_1',f'[e_2[n]\mapsto \tau],C_2)\in R$;
     \item or there is a sequence of (zero or more) $\tau$-transitions $C_1'\xtworightarrow{\tau^*} C_1'^0$, such that $(C_1'^0,f',C_2')\in R$ and $C_1'^0\xtworightarrow{e_1[m]}C_1$ with $(C_1,f[e_2[n]\mapsto e_1[m]],C_2)\in R$;
   \end{itemize}
   \item if $(C_1',f',C_2')\in R$ and $C_1'\downarrow$, then there is a sequence of (zero or more) $\tau$-transitions $C_2'\xtworightarrow{\tau^*}C_2'^0$ such that $(C_1',f',C_2'^0)\in R$ and $C_2'^0\downarrow$;
   \item if $(C_1',f',C_2')\in R$ and $C_2'\downarrow$, then there is a sequence of (zero or more) $\tau$-transitions $C_1'\xtworightarrow{\tau^*}C_1'^0$ such that $(C_1'^0,f',C_2')\in R$ and $C_1'^0\downarrow$.
 \end{enumerate}

$\mathcal{E}_1,\mathcal{E}_2$ are branching FR history-preserving (hp-)bisimilar and are written $\mathcal{E}_1\approx_{bhp}^{fr}\mathcal{E}_2$ if there exists a branching FR hp-bisimulation $R$ such that $(\emptyset,\emptyset,\emptyset)\in R$.

A branching FR hereditary history-preserving (hhp-)bisimulation is a downward closed branching FR hp-bisimulation. $\mathcal{E}_1,\mathcal{E}_2$ are branching FR hereditary history-preserving (hhp-)bisimilar and are written $\mathcal{E}_1\approx_{bhhp}^{fr}\mathcal{E}_2$.
\end{definition}

\begin{definition}[Rooted branching forward-reverse (FR) (hereditary) history-preserving bisimulation]\label{FRRBHHPB}
Assume a special termination predicate $\downarrow$, and let $\surd$ represent a state with $\surd\downarrow$. A rooted branching FR history-preserving (hp-) bisimulation is a weakly posetal relation $R\subseteq\mathcal{C}(\mathcal{E}_1)\overline{\times}\mathcal{C}(\mathcal{E}_2)$ such that:

 \begin{enumerate}
   \item if $(C_1,f,C_2)\in R$, and $C_1\xrightarrow{e_1}C_1'$, then $C_2\xrightarrow{e_2}C_2'$ with $C_1'\approx_{bhp}C_2'$;
   \item if $(C_1,f,C_2)\in R$, and $C_2\xrightarrow{e_2}C_2'$, then $C_1\xrightarrow{e_1}C_1'$ with $C_1'\approx_{bhp}C_2'$;
   \item if $(C_1',f',C_2')\in R$, and $C_1'\xtworightarrow{e_1[m]}C_1$, then $C_2'\xtworightarrow{e_2[n]}C_2$ with $C_1\approx_{bhp}^{fr}C_2$;
   \item if $(C_1',f',C_2')\in R$, and $C_2'\xtworightarrow{e_2[n]}C_2$, then $C_1'\xtworightarrow{e_1[m]}C_1$ with $C_1\approx_{bhp}^{fr}C_2$;
   \item if $(C_1,f,C_2)\in R$ and $C_1\downarrow$, then $C_2\downarrow$;
   \item if $(C_1,f,C_2)\in R$ and $C_2\downarrow$, then $C_1\downarrow$.
 \end{enumerate}

$\mathcal{E}_1,\mathcal{E}_2$ are rooted branching FR history-preserving (hp-)bisimilar and are written $\mathcal{E}_1\approx_{rbhp}^{fr}\mathcal{E}_2$ if there exists a rooted branching FR hp-bisimulation $R$ such that $(\emptyset,\emptyset,\emptyset)\in R$.

A rooted branching FR hereditary history-preserving (hhp-)bisimulation is a downward closed rooted branching FR hp-bisimulation. $\mathcal{E}_1,\mathcal{E}_2$ are rooted branching FR hereditary history-preserving (hhp-)bisimilar and are written $\mathcal{E}_1\approx_{rbhhp}^{fr}\mathcal{E}_2$.
\end{definition}

\section{Basic Reversible Algebra for True Concurrency}{\label{bratc}}

In this section, we will discuss the algebraic laws for prime event structure $\mathcal{E}$, exactly for causality $\leq$ and conflict $\sharp$, with a reversible flavor. The resulted algebra is called Basic Reversible Algebra for True Concurrency, abbreviated $BRATC$.

\subsection{Axiom System of $BRATC$}

In the following, let $e_1, e_2, e_1', e_2'\in \mathbb{E}$, and let variables $x,y,z$ range over the set of terms for true concurrency, $p,q,s$ range over the set of closed terms, the the predicate $\textrm{Std(p)}$ represents that $p$ is a standard process containing no past events, the the predicate $\textrm{NStd(p)}$ represents that $p$ is a process full of past events. The set of axioms of $BRATC$ consists of the laws given in Table \ref{AxiomsForBRATC}.

\begin{center}
    \begin{table}
        \begin{tabular}{@{}ll@{}}
            \hline No. &Axiom\\
            $A1$ & $x+ y = y+ x$\\
            $A2$ & $(x+ y)+ z = x+ (y+ z)$\\
            $A3$ & $x+ x = x$\\
            $A4$ & $x\cdot( y + z) = x\cdot y + x\cdot z(\textrm{Std}(x), \textrm{Std}(y), \textrm{Std}(z))$ \\
            $RA4$ & $(x + y)\cdot z= x\cdot z + y\cdot z(\textrm{NStd}(x), \textrm{NStd}(y), \textrm{NStd}(z))$ \\
            $A5$ & $(x\cdot y)\cdot z = x\cdot(y\cdot z)$\\
        \end{tabular}
        \caption{Axioms of $BRATC$}
        \label{AxiomsForBRATC}
    \end{table}
\end{center}

Intuitively, the axiom $A1$ says that the binary operator $+$ satisfies commutative law. The axiom $A2$ says that $+$ satisfies associativity. $A3$ says that $+$ satisfies idempotency. The axiom $A4$ is the left distributivity of the binary operator $\cdot$ to $+$. The axiom $RA4$ is the right distributivity of the binary operator $\cdot$ to $+$. And $A5$ is the associativity of $\cdot$.

\subsection{Properties of $BRATC$}

\begin{definition}[Basic terms of $BRATC$]\label{BTBRATC}
The set of basic terms of $BRATC$, $\mathcal{B}(BRATC)$, is inductively defined as follows:
\begin{enumerate}
  \item $\mathbb{E}\subset\mathcal{B}(BRATC)$;
  \item if $e\in \mathbb{E}, t\in\mathcal{B}(BRATC)$ then $e\cdot t\in\mathcal{B}(BRATC)$;
  \item if $t,s\in\mathcal{B}(BRATC)$ then $t+ s\in\mathcal{B}(BRATC)$.
\end{enumerate}
\end{definition}

\begin{theorem}[Elimination theorem of $BRATC$]\label{ETBRATC}
Let $p$ be a closed $BRATC$ term. Then there is a basic $BRATC$ term $q$ such that $BRATC\vdash p=q$.
\end{theorem}

\begin{proof}
(1) Firstly, suppose that the following ordering on the signature of $BRATC$ is defined: $\cdot > +$ and the symbol $\cdot$ is given the lexicographical status for the first argument, then for each rewrite rule $p\rightarrow q$ (or $p\twoheadrightarrow q$) in Table \ref{TRSForBRATC} relation $p>_{lpo} q$ can easily be proved. We obtain that the term rewrite system shown in Table \ref{TRSForBRATC} is strongly normalizing, for it has finitely many rewriting rules, and $>$ is a well-founded ordering on the signature of $BRATC$, and if $s>_{lpo} t$, for each rewriting rule $s\rightarrow t$ is in Table \ref{TRSForBRATC} (see Theorem \ref{SN}).

\begin{center}
    \begin{table}
        \begin{tabular}{@{}ll@{}}
            \hline No. &Rewriting Rule\\
            $RRA3$ & $x+ x \rightarrow x$\\
            $RRA4$ & $x\cdot( y+ z) \rightarrow x\cdot y + x\cdot z$\\
            $RRRA4$ & $(x + y)\cdot z\rightarrow x\cdot z + y\cdot z$\\
            $RRA5$ & $(x\cdot y)\cdot z \rightarrow x\cdot(y\cdot z)$\\
        \end{tabular}
        \caption{Term rewrite system of $BRATC$}
        \label{TRSForBRATC}
    \end{table}
\end{center}

(2) Then we prove that the normal forms of closed $BRATC$ terms are basic $BRATC$ terms.

Suppose that $p$ is a normal form of some closed $BRATC$ term and suppose that $p$ is not a basic term. Let $p'$ denote the smallest sub-term of $p$ which is not a basic term. It implies that each sub-term of $p'$ is a basic term. Then we prove that $p$ is not a term in normal form. It is sufficient to induct on the structure of $p'$:

\begin{itemize}
  \item Case $p'\equiv e or e[m], e\in \mathbb{E}$. $p'$ is a basic term, which contradicts the assumption that $p'$ is not a basic term, so this case should not occur.
  \item Case $p'\equiv p_1\cdot p_2$. By induction on the structure of the basic term $p_1$:
      \begin{itemize}
        \item Subcase $p_1\in \mathbb{E}$. $p'$ would be a basic term, which contradicts the assumption that $p'$ is not a basic term;
        \item Subcase $p_1\equiv e\cdot p_1'$. $RRA5$ rewriting rule can be applied. So $p$ is not a normal form;
        \item Subcase $p_1\equiv p_1'\cdot e[m]$. $RRA5$ rewriting rule can be applied. So $p$ is not a normal form;
        \item Subcase $p_1\equiv p_1'+ p_1''$. $RRA4$ or $RRRA4$ rewriting rule can be applied. So $p$ is not a normal form.
      \end{itemize}
  \item Case $p'\equiv p_1+ p_2$. By induction on the structure of the basic terms both $p_1$ and $p_2$, all subcases will lead to that $p'$ would be a basic term, which contradicts the assumption that $p'$ is not a basic term.
\end{itemize}
\end{proof}

\subsection{Structured Operational Semantics of $BRATC$}

In this subsection, we will define a term-deduction system which gives the operational semantics of $BRATC$. We give the operational transition rules for operators $\cdot$ and $+$ as Table \ref{SETRForBRATC} and Table \ref{RSETRForBRATC} show. And the predicate $\xrightarrow{\alpha}\alpha[m]$ represents successful forward termination after execution of the action $\alpha$, the predicate $\xtworightarrow{\alpha[m]}\alpha$ represents successful reverse termination after execution of the event $\alpha[m]$.
\begin{center}
    \begin{table}
        $$\frac{}{e\xrightarrow{e}e[m]}$$

        $$\frac{x\xrightarrow{e}e[m] \quad e\notin y}{x+y\xrightarrow{e}e[m]+y}
        \quad\frac{x\xrightarrow{e}x' \quad e\notin y}{x+y\xrightarrow{e}x'+y}$$
        $$\frac{y\xrightarrow{e}e[m] \quad e\notin x}{x+y\xrightarrow{e}x+e[m]}
        \quad\frac{y\xrightarrow{e}y'\quad e\notin x}{x+y\xrightarrow{e}x+y'}$$

        $$\frac{x\xrightarrow{e}e[m]\quad y\xrightarrow{e}e[m]}{x+y\xrightarrow{e}e[m]+e[m]}
        \quad\frac{x\xrightarrow{e}x'\quad y\xrightarrow{e}e[m]}{x+y\xrightarrow{e}x'+e[m]}$$
        $$\frac{x\xrightarrow{e}e[m]\quad y\xrightarrow{e}y'}{x+y\xrightarrow{e}e[m]+y'}
        \quad\frac{x\xrightarrow{e}x'\quad y\xrightarrow{e}y'}{x+y\xrightarrow{e}x'+y'}$$

        $$\frac{x\xrightarrow{e}e[m]\quad\textrm{Std}(y)}{x\cdot y\xrightarrow{e} e[m]\cdot y} \quad\frac{x\xrightarrow{e}x' \quad \textrm{Std}(y)}{x\cdot y\xrightarrow{e}x'\cdot y}$$
        $$\frac{y\xrightarrow{e'}e'[n]\quad \textrm{NStd}(x)}{x\cdot y\xrightarrow{e'}x\cdot e'[n]} \quad\frac{y\xrightarrow{e'}y'\quad \textrm{NStd}(x)}{x\cdot y\xrightarrow{e'}x\cdot y'}$$
        \caption{Forward single event transition rules of $BRATC$}
        \label{SETRForBRATC}
    \end{table}
\end{center}

\begin{center}
    \begin{table}
        $$\frac{}{e[m]\xtworightarrow{e[m]}e}$$

        $$\frac{x\xtworightarrow{e[m]}e\quad e\notin y}{x+y\xtworightarrow{e[m]}e+y}
        \quad\frac{x\xtworightarrow{e[m]}x' \quad e\notin y}{x+y\xtworightarrow{e[m]}x'+y}$$
        $$\frac{y\xtworightarrow{e[m]}e \quad e\notin x}{x+y\xtworightarrow{e[m]}x+e}
        \quad\frac{y\xtworightarrow{e[m]}y' \quad e\notin x}{x+y\xtworightarrow{e[m]}x+y'}$$

        $$\frac{x\xtworightarrow{e[m]}e\quad y\xtworightarrow{e[m]}e}{x+y\xtworightarrow{e[m]}e+e}
        \quad\frac{x\xtworightarrow{e[m]}x'\quad y\xtworightarrow{e[m]}e}{x+y\xtworightarrow{e[m]}x'+e}$$
        $$\frac{x\xtworightarrow{e[m]}e\quad y\xtworightarrow{e[m]}y'}{x+y\xtworightarrow{e[m]}e+y'}
        \quad\frac{x\xtworightarrow{e[m]}x'\quad y\xtworightarrow{e[m]}y'}{x+y\xtworightarrow{e[m]}x'+y'}$$

        $$\frac{x\xtworightarrow{e[m]}e \quad \textrm{Std}(y)}{x\cdot y\xtworightarrow{e[m]} e\cdot y} \quad\frac{x\xtworightarrow{e[m]}x'\quad \textrm{Std}(y)}{x\cdot y\xtworightarrow{e[m]}x'\cdot y}$$
        $$\frac{y\xtworightarrow{e'[n]}e' \quad \textrm{NStd}(x)}{x\cdot y\xtworightarrow{e'[n]}x\cdot e'}\quad \frac{y\xtworightarrow{e'[n]}y' \quad \textrm{NStd}(x)}{x\cdot y\xtworightarrow{e'[n]}x\cdot y'}$$
        \caption{Reverse single event transition rules of $BRATC$}
        \label{RSETRForBRATC}
    \end{table}
\end{center}

The pomset transition rules are shown in Table \ref{PTRForBRATC} and Table \ref{RPTRForBRATC}, different to single event transition rules in Table \ref{SETRForBRATC} and Table \ref{RSETRForBRATC}, the pomset transition rules are labeled by pomsets, which are defined by causality $\cdot$ and conflict $+$.

\begin{center}
    \begin{table}
        $$\frac{}{X\xrightarrow{X}X[\mathcal{K}]}$$

        $$\frac{x\xrightarrow{X}X[\mathcal{K}] \quad X\nsubseteq y}{x+y\xrightarrow{X}X[\mathcal{K}]+y}
        \quad\frac{x\xrightarrow{X}x' \quad X\nsubseteq y}{x+y\xrightarrow{X}x'+y}$$
        $$\frac{y\xrightarrow{Y}Y[\mathcal{J}] \quad Y\nsubseteq x}{x+y\xrightarrow{Y}x+Y[\mathcal{J}]}
        \quad\frac{y\xrightarrow{Y}y'\quad Y\nsubseteq x}{x+y\xrightarrow{Y}x+y'}$$

        $$\frac{x\xrightarrow{X}X[\mathcal{K}]\quad y\xrightarrow{X}X[\mathcal{K}]}{x+y\xrightarrow{X}X[\mathcal{K}]+X[\mathcal{K}]}
        \quad\frac{x\xrightarrow{X}x'\quad y\xrightarrow{X}X[\mathcal{K}]}{x+y\xrightarrow{X}x'+X[\mathcal{K}]}$$
        $$\frac{x\xrightarrow{X}X[\mathcal{K}]\quad y\xrightarrow{X}y'}{x+y\xrightarrow{X}X[\mathcal{K}]+y'}
        \quad\frac{x\xrightarrow{X}x'\quad y\xrightarrow{X}y'}{x+y\xrightarrow{X}x'+y'}$$

        $$\frac{x\xrightarrow{X}X[\mathcal{K}]\quad\textrm{Std}(y)}{x\cdot y\xrightarrow{X} X[\mathcal{K}]\cdot y}(X\subseteq x) \quad\frac{x\xrightarrow{X}x' \quad \textrm{Std}(y)}{x\cdot y\xrightarrow{X}x'\cdot y}(X\subseteq x)$$
        $$\frac{y\xrightarrow{Y}Y[\mathcal{J}]\quad \textrm{NStd}(x)}{x\cdot y\xrightarrow{Y}x\cdot Y[\mathcal{J}]}(Y\subseteq y) \quad\frac{y\xrightarrow{Y}y'\quad \textrm{NStd}(x)}{x\cdot y\xrightarrow{Y}x\cdot y'}(Y\subseteq y)$$
        \caption{Forward pomset transition rules of $BRATC$}
        \label{PTRForBRATC}
    \end{table}
\end{center}

\begin{center}
    \begin{table}
        $$\frac{}{X[\mathcal{K}]\xtworightarrow{X[\mathcal{K}]}X}$$

        $$\frac{x\xtworightarrow{X[\mathcal{K}]}X\quad X\nsubseteq y}{x+y\xtworightarrow{X[\mathcal{K}]}X+y}
        \quad\frac{x\xtworightarrow{X[\mathcal{K}]}x' \quad X\nsubseteq y}{x+y\xtworightarrow{X[\mathcal{K}]}x'+y}$$
        $$\frac{y\xtworightarrow{Y[\mathcal{J}]}Y \quad Y\nsubseteq x}{x+y\xtworightarrow{Y[\mathcal{J}]}x+Y}
        \quad\frac{y\xtworightarrow{Y[\mathcal{J}]}y' \quad Y\nsubseteq x}{x+y\xtworightarrow{Y[\mathcal{J}]}x+y'}$$

        $$\frac{x\xtworightarrow{X[\mathcal{K}]}X\quad y\xtworightarrow{X[\mathcal{K}]}X}{x+y\xtworightarrow{X[\mathcal{K}]}X+X}
        \quad\frac{x\xtworightarrow{X[\mathcal{K}]}x'\quad y\xtworightarrow{X[\mathcal{K}]}X}{x+y\xtworightarrow{X[\mathcal{K}]}x'+X}$$
        $$\frac{x\xtworightarrow{X[\mathcal{K}]}X\quad y\xtworightarrow{X[\mathcal{K}]}y'}{x+y\xtworightarrow{X[\mathcal{K}]}X+y'}
        \quad\frac{x\xtworightarrow{X[\mathcal{K}]}x'\quad y\xtworightarrow{X[\mathcal{K}]}y'}{x+y\xtworightarrow{X[\mathcal{K}]}x'+y'}$$

        $$\frac{x\xtworightarrow{X[\mathcal{K}]}X \quad \textrm{Std}(y)}{x\cdot y\xtworightarrow{X[\mathcal{K}]} X\cdot y}(X\subseteq x) \quad\frac{x\xtworightarrow{X[\mathcal{K}]}x'\quad \textrm{Std}(y)}{x\cdot y\xtworightarrow{X[\mathcal{K}]}x'\cdot y}(X\subseteq x)$$
        $$\frac{y\xtworightarrow{Y[\mathcal{J}]}Y \quad \textrm{NStd}(x)}{x\cdot y\xtworightarrow{Y[\mathcal{J}]}x\cdot Y}(Y\subseteq y)\quad \frac{y\xtworightarrow{Y[\mathcal{J}]}y' \quad \textrm{NStd}(x)}{x\cdot y\xtworightarrow{Y[\mathcal{J}]}x\cdot y'}(Y\subseteq y)$$
        \caption{Reverse pomset transition rules of $BRATC$}
        \label{RPTRForBRATC}
    \end{table}
\end{center}

\begin{theorem}[Congruence of $BRATC$ with respect to FR pomset bisimulation equivalence]
FR pomset bisimulation equivalence $\sim_{p}^{fr}$ is a congruence with respect to $BRATC$.
\end{theorem}

\begin{proof}
It is easy to see that FR pomset bisimulation is an equivalent relation on $BRATC$ terms, we only need to prove that $\sim_{p}^{fr}$ is preserved by the operators $\cdot$ and $+$.

\begin{itemize}
  \item Causality operator $\cdot$. Let $x_1,x_2$ and $y_1,y_2$ be $BRATC$ processes, and $x_1\sim_{p}^{fr} y_1$, $x_2\sim_{p}^{fr} y_2$, it is sufficient to prove that $x_1\cdot x_2\sim_{p}^{fr} y_1\cdot y_2$.

      By the definition of FR pomset bisimulation $\sim_p^{fr}$ (Definition \ref{FRPSB}), $x_1\sim_p^{fr} y_1$ means that

      $$x_1\xrightarrow{X_1} x_1' \quad y_1\xrightarrow{Y_1} y_1'$$

      $$x_1\xtworightarrow{X_1[\mathcal{K}]} x_1'' \quad y_1\xtworightarrow{Y_1[\mathcal{J}]} y_1''$$

      with $X_1\subseteq x_1$, $Y_1\subseteq y_1$, $X_1\sim Y_1$, $x_1'\sim_p^{fr} y_1'$ and $x_1''\sim_p^{fr}y_1''$. The meaning of $x_2\sim_p^{fr} y_2$ is similar.

      By the FR pomset transition rules for causality operator $\cdot$ in Table \ref{PTRForBRATC} and Table \ref{RPTRForBRATC}, we can get

      $$x_1\cdot x_2\xrightarrow{X_1} X_1[\mathcal{K}]\cdot x_2 \quad y_1\cdot y_2\xrightarrow{Y_1} Y_1[\mathcal{J}]\cdot y_2$$

      $$x_1\cdot x_2\xtworightarrow{X_2[\mathcal{K}]} x_1\cdot X_2 \quad y_1\cdot y_2\xtworightarrow{Y_2[\mathcal{J}]} y_1\cdot Y_2$$

      with $X_1\subseteq x_1$, $Y_1\subseteq y_1$, $X_2\subseteq x_2$, $Y_2\subseteq y_2$, $X_1\sim Y_1$, $X_2\sim Y_2$, and the assumptions $X_1[\mathcal{K}]\cdot x_2\sim_p^{fr} Y_1[\mathcal{J}]\cdot y_2$ and $x_1\cdot X_2\sim_p^{fr} y_1\cdot Y_2$, so, we get $x_1\cdot x_2\sim_p^{fr} y_1\cdot y_2$, as desired.

      Or, we can get

      $$x_1\cdot x_2\xrightarrow{X_1} x_1'\cdot x_2 \quad y_1\cdot y_2\xrightarrow{Y_1} y_1'\cdot y_2$$

      $$x_1\cdot x_2\xtworightarrow{X_2[\mathcal{K}]} x_1\cdot x_2' \quad y_1\cdot y_2\xtworightarrow{Y_2[\mathcal{J}]} y_1\cdot y_2'$$

      with $X_1\subseteq x_1$, $Y_1\subseteq y_1$, $X_2\subseteq x_2$, $Y_2\subseteq y_2$, $X_1\sim Y_1$, $X_2\sim Y_2$, and the assumptions $x_1'\cdot x_2\sim_p^{fr} y_1'\cdot y_2$, $x_1\cdot x_2'\sim_p^{fr} y_1\cdot y_2'$, so, we get $x_1\cdot x_2\sim_p^{fr} y_1\cdot y_2$, as desired.
  \item Conflict operator $+$. Let $x_1, x_2$ and $y_1, y_2$ be $BRATC$ processes, and $x_1\sim_p^{fr} y_1$, $x_2\sim_p^{fr} y_2$, it is sufficient to prove that $x_1+ x_2 \sim_p^{fr} y_1+ y_2$. The meanings of $x_1\sim_p^{fr} y_1$ and $x_2\sim_p^{fr} y_2$ are the same as the above case, according to the definition of FR pomset bisimulation $\sim_p^{fr}$ in Definition \ref{FRPSB}.

      By the FR pomset transition rules for conflict operator $+$ in Table \ref{PTRForBRATC} and Table \ref{RPTRForBRATC}, we can get several cases:

      $$x_1+ x_2\xrightarrow{X_1} X_1[\mathcal{K}]+x_2 \quad y_1+ y_2\xrightarrow{Y_1} Y_1[\mathcal{J}]+y_2$$
      $$x_1+ x_2\xtworightarrow{X_1[\mathcal{K}]} X_1+x_2 \quad y_1+ y_2\xtworightarrow{Y_1[\mathcal{J}]} Y_1+y_2$$

      with $X_1\subseteq x_1$, $Y_1\subseteq y_1$, $X_1\sim Y_1$, and the assumptions $X_1[\mathcal{K}]+x_2\sim_p^{fr}Y_1[\mathcal{J}]+y_2$ and $X_1+x_2\sim_p^{fr}Y_1+y_2$, so, we get $x_1+ x_2\sim_p^{fr} y_1+ y_2$, as desired.

      Or, we can get

      $$x_1+ x_2\xrightarrow{X_1} x_1'+x_2 \quad y_1+ y_2\xrightarrow{Y_1} y_1'+y_2$$

      $$x_1+ x_2\xtworightarrow{X_1[\mathcal{K}]} x_1'+x_2 \quad y_1+ y_2\xtworightarrow{Y_1[\mathcal{J}]} y_1'+y_2$$

      with $X_1\subseteq x_1$, $Y_1\subseteq y_1$, $X_1\sim Y_1$, and $x_1'+x_2\sim_p^{fr} y_1'+y_2$, so, we get $x_1+ x_2\sim_p^{fr} y_1+ y_2$, as desired.

      Or, we can get

      $$x_1+ x_2\xrightarrow{X_2} x_1+X_2[\mathcal{K}] \quad y_1+ y_2\xrightarrow{Y_2} y_1+Y_2[\mathcal{J}]$$

      $$x_1+ x_2\xrightarrow{X_2[\mathcal{K}]} x_1+X_2 \quad y_1+ y_2\xrightarrow{Y_2[\mathcal{J}]} y_1+Y_2$$

      with $X_2\subseteq x_2$, $Y_2\subseteq y_2$, $X_2\sim Y_2$, and the assumptions $x_1+X_2[\mathcal{K}]\sim_p^{fr}y_1+Y_2[\mathcal{J}]$ and $x_1+X_2\sim_p^{fr}y_1+Y_2$, so, we get $x_1+ x_2\sim_p^{fr} y_1+ y_2$, as desired.

      Or, we can get

      $$x_1+ x_2\xrightarrow{X_2} x_1+x_2' \quad y_1+ y_2\xrightarrow{Y_2} y_1+y_2'$$

      $$x_1+ x_2\xtworightarrow{X_2[\mathcal{K}]} x_1+x_2' \quad y_1+ y_2\xtworightarrow{Y_2[\mathcal{J}]} y_1+y_2'$$

      with $X_2\subseteq x_2$, $Y_2\subseteq y_2$, $X_2\sim Y_2$, and the assumption $x_1+x_2'\sim_p^{fr}y_1+y_2'$, so, we get $x_1+ x_2\sim_p^{fr} y_1+ y_2$, as desired.

      Or, we can get

      $$x_1+ x_2\xrightarrow{X} x_1'+x_2' \quad y_1+ y_2\xrightarrow{Y} y_1'+y_2'$$

      $$x_1+ x_2\xtworightarrow{X[\mathcal{K}]} x_1'+x_2' \quad y_1+ y_2\xtworightarrow{Y[\mathcal{J}]} y_1'+y_2'$$

      with $X\subseteq x_1$, $Y\subseteq y_1$, $X\subseteq x_2$, $Y\subseteq y_2$, $X\sim Y$, and the assumption $x_1'+x_2'\sim_p^{fr}y_1'+y_2'$, so, we get $x_1+ x_2\sim_p^{fr} y_1+ y_2$, as desired.
\end{itemize}
\end{proof}

\begin{theorem}[Soundness of $BRATC$ modulo FR pomset bisimulation equivalence]\label{SBRATCPBE}
Let $x$ and $y$ be $BRATC$ terms. If $BRATC\vdash x=y$, then $x\sim_{p}^{fr} y$.
\end{theorem}

\begin{proof}
Since FR pomset bisimulation $\sim_p^{fr}$ is both an equivalent and a congruent relation, we only need to check if each axiom in Table \ref{AxiomsForBRATC} is sound modulo FR pomset bisimulation equivalence.

\begin{itemize}
  \item \textbf{Axiom $A1$}. Let $p,q$ be $BRATC$ processes, and $p+ q=q+ p$, it is sufficient to prove that $p+ q\sim_p^{fr} q+ p$. By the forward pomset transition rules for operator $+$ in Table \ref{PTRForBRATC}, we get

      $$\frac{p\xrightarrow{P}P[\mathcal{K}]}{p+ q\xrightarrow{P}P[\mathcal{K}]+q} (P\subseteq p,P\nsubseteq q) \quad \frac{p\xrightarrow{P}P[\mathcal{K}]}{q+ p\xrightarrow{P}q+P[\mathcal{K}]}(P\subseteq p,P\nsubseteq q)$$

      $$\frac{p\xrightarrow{P}p'}{p+ q\xrightarrow{P}p'+q}(P\subseteq p,P\nsubseteq q) \quad \frac{p\xrightarrow{P}p'}{q+ p\xrightarrow{P}q+p'}(P\subseteq p,P\nsubseteq q)$$

      $$\frac{q\xrightarrow{Q}Q[\mathcal{J}]}{p+ q\xrightarrow{Q}p+Q[\mathcal{J}]}(Q\subseteq q,Q\nsubseteq p) \quad \frac{q\xrightarrow{Q}Q[\mathcal{J}]}{q+ p\xrightarrow{Q}Q[\mathcal{J}]+p}(Q\subseteq q,Q\nsubseteq p)$$

      $$\frac{q\xrightarrow{Q}q'}{p+ q\xrightarrow{Q}p+q'}(Q\subseteq q,Q\nsubseteq p) \quad \frac{q\xrightarrow{Q}q'}{q+ p\xrightarrow{Q}q'+p}(Q\subseteq q, Q\nsubseteq p)$$

      $$\frac{p\xrightarrow{P}p'\quad q\xrightarrow{P}q'}{p+ q\xrightarrow{P}p'+q'}(P\subseteq p,P\subseteq q) \quad \frac{p\xrightarrow{P}p'\quad q\xrightarrow{P}q'}{q+ p\xrightarrow{P}q'+p'}(P\subseteq p,P\subseteq q)$$

      By the reverse pomset transition rules for operator $+$ in Table \ref{RPTRForBRATC}, we get

      $$\frac{p\xtworightarrow{P[\mathcal{K}]}P\quad P\nsubseteq q}{p+ q\xrightarrow{P[\mathcal{K}]}P+q} (P\subseteq p) \quad \frac{p\xtworightarrow{P[\mathcal{K}]}P\quad P\nsubseteq q}{q+ p\xtworightarrow{P[\mathcal{K}]}q+P}(P\subseteq p)$$

      $$\frac{p\xtworightarrow{P[\mathcal{K}]}p'\quad P\nsubseteq q}{p+ q\xtworightarrow{P[\mathcal{K}]}p'+q}(P\subseteq p) \quad \frac{p\xtworightarrow{P[\mathcal{K}]}p'\quad P\nsubseteq q}{q+ p\xtworightarrow{P[\mathcal{K}]}q+p'}(P\subseteq p)$$

      $$\frac{q\xtworightarrow{Q[\mathcal{J}]}Q\quad Q\nsubseteq p}{p+ q\xtworightarrow{Q[\mathcal{J}]}p+Q}(Q\subseteq q) \quad \frac{q\xtworightarrow{Q[\mathcal{J}]}Q\quad Q\nsubseteq p}{q+ p\xtworightarrow{Q[\mathcal{J}]}Q+p}(Q\subseteq q)$$

      $$\frac{q\xtworightarrow{Q[\mathcal{J}]}q'\quad Q\nsubseteq p}{p+ q\xtworightarrow{Q[\mathcal{J}]}p+q'}(Q\subseteq q) \quad \frac{q\xtworightarrow{Q[\mathcal{J}]}q'\quad Q\nsubseteq p}{q+ p\xtworightarrow{Q[\mathcal{J}]}q'+p}(Q\subseteq q)$$

      $$\frac{p\xtworightarrow{P[\mathcal{K}]}p'\quad q\xtworightarrow{P[\mathcal{K}]}q'}{p+ q\xtworightarrow{P[\mathcal{K}]}p'+q'}(P\subseteq p,P\subseteq q) \quad \frac{p\xtworightarrow{P[\mathcal{K}]}p'\quad q\xtworightarrow{P[\mathcal{K}]}q'}{q+ p\xtworightarrow{P[\mathcal{K}]}q'+p'}(P\subseteq p,P\subseteq q)$$

      With the assumptions $P[\mathcal{K}]+q\sim_p^{fr}q+P[\mathcal{K}]$, $P+q\sim_p^{fr}q+P$, $p+Q[\mathcal{J}]\sim_p^{fr}Q[\mathcal{J}]+p$, $p'+q\sim_p^{fr}q+p'$, $p+q'\sim_p^{fr}q'+p$ and $p'+q'\sim_p^{fr}q'+p'$ so, $p+ q\sim_p^{fr} q+ p$, as desired.
  \item \textbf{Axiom $A2$}. Let $p,q,s$ be $BRATC$ processes, and $(p+ q)+ s=p+ (q+ s)$, it is sufficient to prove that $(p+ q)+ s \sim_p^{fr} p+ (q+ s)$. By the forward pomset transition rules for operator $+$ in Table \ref{PTRForBRATC}, we get

      $$\frac{p\xrightarrow{P}P[\mathcal{K}]\quad P\nsubseteq q\quad P\nsubseteq s}{(p+ q)+ s\xrightarrow{P}(P[\mathcal{K}]+q)+s} (P\subseteq p) \quad \frac{p\xrightarrow{P}P[\mathcal{K}]\quad P\nsubseteq q\quad P\nsubseteq s}{p+ (q+ s)\xrightarrow{P}P[\mathcal{K}]+(q+s)}(P\subseteq p)$$

      $$\frac{p\xrightarrow{P}p'\quad P\nsubseteq q\quad P\nsubseteq s}{(p+ q)+ s\xrightarrow{P}(p'+q)+s}(P\subseteq p) \quad \frac{p\xrightarrow{P}p'\quad P\nsubseteq q\quad P\nsubseteq s}{p+ (q+ s)\xrightarrow{P}p'+(q+s)}(P\subseteq p)$$

      $$\frac{q\xrightarrow{Q}Q[\mathcal{J}]\quad Q\nsubseteq p\quad Q\nsubseteq s}{(p+ q)+ s\xrightarrow{Q}(p+Q[\mathcal{J}])+s}(Q\subseteq q) \quad \frac{q\xrightarrow{Q}Q[\mathcal{J}]\quad Q\nsubseteq p\quad Q\nsubseteq s}{p+ (q+ s)\xrightarrow{Q}p+(Q[\mathcal{J}]+s)}(Q\subseteq q)$$

      $$\frac{q\xrightarrow{Q}q'\quad Q\nsubseteq p\quad Q\nsubseteq s}{(p+ q)+ s\xrightarrow{Q}(p+q')+s}(Q\subseteq q) \quad \frac{q\xrightarrow{Q}q'\quad Q\nsubseteq p\quad Q\nsubseteq s}{p+ (q+ s)\xrightarrow{Q}p+(q'+s)}(Q\subseteq q)$$

      $$\frac{s\xrightarrow{S}S[\mathcal{I}]\quad S\nsubseteq p\quad S\nsubseteq q}{(p+ q)+ s\xrightarrow{S}(p+q)+S[\mathcal{I}]}(S\subseteq s) \quad \frac{s\xrightarrow{S}S[\mathcal{I}]\quad S\nsubseteq p\quad S\nsubseteq q}{p+ (q+ s)\xrightarrow{S}p+(q+S[\mathcal{I}])}(S\subseteq s)$$

      $$\frac{s\xrightarrow{S}s'\quad S\nsubseteq p\quad S\nsubseteq q}{(p+ q)+ s\xrightarrow{S}(p+q)+s'}(S\subseteq s) \quad \frac{s\xrightarrow{S}s'\quad S\nsubseteq p\quad S\nsubseteq q}{p+ (q+ s)\xrightarrow{S}p+(q+s')}(S\subseteq s)$$

      $$\frac{p\xrightarrow{P}p'\quad q\xrightarrow{P}q'\quad s\xrightarrow{P}s'}{(p+ q)+ s\xrightarrow{P}(p'+q')+s'}(P\subseteq p,P\subseteq q,P\subseteq s) \quad \frac{p\xrightarrow{P}p'\quad q\xrightarrow{P}q'\quad s\xrightarrow{P}s'}{p+ (q+ s)\xrightarrow{P}p'+(q'+s')}(P\subseteq p,P\subseteq q,P\subseteq s)$$

      By the reverse pomset transition rules for operator $+$ in Table \ref{RPTRForBRATC}, we get

      $$\frac{p\xtworightarrow{P[\mathcal{K}]}P\quad P\nsubseteq q\quad P\nsubseteq s}{(p+ q)+ s\xtworightarrow{P[\mathcal{K}]}(P+q)+s} (P\subseteq p) \quad \frac{p\xtworightarrow{P[\mathcal{K}]}P\quad P\nsubseteq q\quad P\nsubseteq s}{p+ (q+ s)\xtworightarrow{P[\mathcal{K}]}P+(q+s)}(P\subseteq p)$$

      $$\frac{p\xtworightarrow{P[\mathcal{K}]}p'\quad P\nsubseteq q\quad P\nsubseteq s}{(p+ q)+ s\xtworightarrow{P[\mathcal{K}]}(p'+q)+s}(P\subseteq p) \quad \frac{p\xtworightarrow{P[\mathcal{K}]}p'\quad P\nsubseteq q\quad P\nsubseteq s}{p+ (q+ s)\xtworightarrow{P[\mathcal{K}]}p'+(q+s)}(P\subseteq p)$$

      $$\frac{q\xtworightarrow{Q[\mathcal{J}]}Q\quad Q\nsubseteq p\quad Q\nsubseteq s}{(p+ q)+ s\xtworightarrow{Q[\mathcal{J}]}(p+Q)+s}(Q\subseteq q) \quad \frac{q\xtworightarrow{Q[\mathcal{J}]}Q\quad Q\nsubseteq p\quad Q\nsubseteq s}{p+ (q+ s)\xtworightarrow{Q[\mathcal{J}]}p+(Q+s)}(Q\subseteq q)$$

      $$\frac{q\xtworightarrow{Q[\mathcal{J}]}q'\quad Q\nsubseteq p\quad Q\nsubseteq s}{(p+ q)+ s\xrightarrow{Q[\mathcal{J}]}(p+q')+s}(Q\subseteq q) \quad \frac{q\xtworightarrow{Q[\mathcal{J}]}q'\quad Q\nsubseteq p\quad Q\nsubseteq s}{p+ (q+ s)\xtworightarrow{Q[\mathcal{J}]}p+(q'+s)}(Q\subseteq q)$$

      $$\frac{s\xtworightarrow{S[\mathcal{I}]}S\quad S\nsubseteq p\quad S\nsubseteq q}{(p+ q)+ s\xtworightarrow{S[\mathcal{I}]}(p+q)+S}(S\subseteq s) \quad \frac{s\xtworightarrow{S[\mathcal{I}]}S\quad S\nsubseteq p\quad S\nsubseteq q}{p+ (q+ s)\xtworightarrow{S[\mathcal{I}]}p+(q+S)}(S\subseteq s)$$

      $$\frac{s\xtworightarrow{S[\mathcal{I}]}s'\quad S\nsubseteq p\quad S\nsubseteq q}{(p+ q)+ s\xtworightarrow{S[\mathcal{I}]}(p+q)+s'}(S\subseteq s) \quad \frac{s\xtworightarrow{S[\mathcal{I}]}s'\quad S\nsubseteq p\quad S\nsubseteq q}{p+ (q+ s)\xtworightarrow{S[\mathcal{I}]}p+(q+s')}(S\subseteq s)$$

      $$\frac{p\xtworightarrow{P[\mathcal{K}]}p'\quad q\xtworightarrow{P[\mathcal{K}]}q'\quad s\xtworightarrow{P[\mathcal{K}]}s'}{(p+ q)+ s\xtworightarrow{P[\mathcal{K}]}(p'+q')+s'}(P\subseteq p,P\subseteq q,P\subseteq s) \quad \frac{p\xtworightarrow{P[\mathcal{K}]}p'\quad q\xtworightarrow{P[\mathcal{K}]}q'\quad s\xtworightarrow{P[\mathcal{K}]}s'}{p+ (q+ s)\xtworightarrow{P[\mathcal{K}]}p'+(q'+s')}(P\subseteq p,P\subseteq q,P\subseteq s)$$

      with the assumptions $(P[\mathcal{K}]+q)+s\sim_p^{fr}P[\mathcal{K}]+(q+s)$, $(P+q)+s\sim_p^{fr}P+(q+s)$, $(p+Q[\mathcal{J}])+s\sim_p^{fr}p+(Q[\mathcal{J}]+s)$, $(p+Q)+s\sim_p^{fr}p+(Q+s)$, $(p+q)+S[\mathcal{I}]\sim_p^{fr}p+(q+S[\mathcal{I}])$, $(p+q)+S\sim_p^{fr}p+(q+S)$, $(p'+q)+s\sim_p^{fr}p'+(q+s)$, $(p+q')+s\sim_p^{fr}p+(q'+s)$, $(p+q)+s'\sim_p^{fr}p+(q+s')$ and $(p'+q')+s'\sim_p^{fr}p'+(q'+s')$ so, $(p+ q)+ s\sim_p^{fr} p+ (q+ s)$, as desired.
  \item \textbf{Axiom $A3$}. Let $p$ be a $BRATC$ process, and $p+ p=p$, it is sufficient to prove that $p+ p\sim_p^{fr} p$. By the forward pomset transition rules for operator $+$ in Table \ref{PTRForBRATC}, we get

      $$\frac{p\xrightarrow{P}P[\mathcal{K}]}{p+ p\xrightarrow{P}P[\mathcal{K}]+P[\mathcal{K}]} (P\subseteq p) \quad \frac{p\xrightarrow{P}P[\mathcal{K}]}{p\xrightarrow{P}P[\mathcal{K}]}(P\subseteq p)$$

      $$\frac{p\xrightarrow{P}p'}{p+ p\xrightarrow{P}p'+p'}(P\subseteq p) \quad \frac{p\xrightarrow{P}p'}{p\xrightarrow{P}p'}(P\subseteq p)$$

      By the reverse pomset transition rules for operator $+$ in Table \ref{RPTRForBRATC}, we get

      $$\frac{p\xtworightarrow{P[\mathcal{K}]}P}{p+ p\xtworightarrow{P[\mathcal{K}]}P+P} (P\subseteq p) \quad \frac{p\xtworightarrow{P[\mathcal{K}]}P}{p\xtworightarrow{P[\mathcal{K}]}P}(P\subseteq p)$$

      $$\frac{p\xtworightarrow{P[\mathcal{K}]}p'}{p+ p\xtworightarrow{P[\mathcal{K}]}p'+p'}(P\subseteq p) \quad \frac{p\xtworightarrow{P[\mathcal{K}]}p'}{p\xtworightarrow{P[\mathcal{K}]}p'}(P\subseteq p)$$

      with the assumptions $P[\mathcal{K}]+P[\mathcal{K}]\sim_p^{fr}P[\mathcal{K}]$, $P+P\sim_p^{fr}P$ and $p'+p'\sim_p^{fr}p'$, so, $p+ p\sim_p^{fr} p$, as desired.
  \item \textbf{Axiom $A4$}. Let $p,q,s$ be $BRATC$ processes, and $p\cdot (q+ s)=p\cdot q + p\cdot s(\textrm{Std}(p), \textrm{Std}(q), \textrm{Std}(s))$, it is sufficient to prove that $p\cdot (q+ s)\sim_p^{fr}p\cdot q + p\cdot s$. By the pomset transition rules for operators $+$ and $\cdot$ in Table \ref{PTRForBRATC}, we get

      $$\frac{p\xrightarrow{P}P[\mathcal{K}]}{p\cdot (q+ s)\xrightarrow{P}P[\mathcal{K}]\cdot(q+s)} (P\subseteq p) \quad \frac{p\xrightarrow{P}P[\mathcal{K}]}{p\cdot q + p\cdot s\xrightarrow{P}P[\mathcal{K}]\cdot q +P[\mathcal{K}]\cdot s}(P\subseteq p)$$

      $$\frac{p\xrightarrow{P}p'}{p\cdot (q+ s)\xrightarrow{P}p'\cdot (q+s)}(P\subseteq p) \quad \frac{p\xrightarrow{P}p'}{p\cdot q + p\cdot s\xrightarrow{P}p'\cdot q+p'\cdot s}(P\subseteq p)$$

      By the reverse transition rules for operators $+$ and $\cdot$ in Table \ref{RPTRForBRATC}, there are no transitions.

      with the assumptions $P[\mathcal{K}]\cdot(q+s)\sim_p^{fr}P[\mathcal{K}]\cdot q + P[\mathcal{K}]\cdot s$, $p'\cdot(q+s)\sim_p^{fr}p'\cdot q + p'\cdot s$, so, $p\cdot (q+ s)\sim_p^{fr}p\cdot q + p\cdot s(\textrm{Std}(p), \textrm{Std}(q), \textrm{Std}(s))$, as desired.
  \item \textbf{Axiom $RA4$}. Let $p,q,s$ be $BRATC$ processes, and $ (q+ s)\cdot p=q\cdot p + s\cdot p(\textrm{NStd}(p), \textrm{NStd}(q), \textrm{NStd}(s))$, it is sufficient to prove that $(q+ s)\cdot p \sim_p^{fr}q\cdot p + s\cdot p$. By the pomset transition rules for operators $+$ and $\cdot$ in Table \ref{PTRForBRATC}, there are no transitions.

      By the reverse transition rules for operators $+$ and $\cdot$ in Table \ref{RPTRForBRATC}, we get

      $$\frac{p\xtworightarrow{P[\mathcal{K}]}P}{(q+ s)\cdot p \xtworightarrow{P[\mathcal{K}]}(q+s)\cdot P} (P\subseteq p) \quad \frac{p\xtworightarrow{P[\mathcal{K}]}P}{q\cdot p + s\cdot p\xtworightarrow{P[\mathcal{K}]}q\cdot P +s\cdot P}(P\subseteq p)$$

      $$\frac{p\xtworightarrow{P[\mathcal{K}]}p'}{(q+ s)\cdot p \xtworightarrow{P[\mathcal{K}]}(q+s)\cdot p'}(P\subseteq p) \quad \frac{p\xtworightarrow{P[\mathcal{K}]}p'}{q\cdot p + s\cdot p\xtworightarrow{P[\mathcal{K}]}q\cdot p'+s\cdot p'}(P\subseteq p)$$

      with the assumptions $(q+s)\cdot P\sim_p^{fr}q\cdot P + s\cdot P$, $(q+s)\cdot p'\sim_p^{fr}q\cdot p' + s\cdot p'$, so, $(q+ s)\cdot p \sim_p^{fr}q\cdot p + s\cdot p(\textrm{NStd}(p), \textrm{NStd}(q), \textrm{NStd}(s))$, as desired.
  \item \textbf{Axiom $A5$}. Let $p,q,s$ be $BRATC$ processes, and $(p\cdot q)\cdot s=p\cdot (q\cdot s)$, it is sufficient to prove that $(p\cdot q)\cdot s \sim_p^{fr} p\cdot (q\cdot s)$. By the forward pomset transition rules for operator $\cdot$ in Table \ref{PTRForBRATC}, we get

      $$\frac{p\xrightarrow{P}P[\mathcal{K}]}{(p\cdot q)\cdot s\xrightarrow{P}(P[\mathcal{K}]\cdot q)\cdot s} (P\subseteq p) \quad \frac{p\xrightarrow{P}P[\mathcal{K}]}{p\cdot (q\cdot s)\xrightarrow{P}P[\mathcal{K}]\cdot(q\cdot s)}(P\subseteq p)$$

      $$\frac{p\xrightarrow{P}p'}{(p\cdot q)\cdot s\xrightarrow{P}(p'\cdot q)\cdot s}(P\subseteq p) \quad \frac{p\xrightarrow{P}p'}{p\cdot (q\cdot s)\xrightarrow{P}p'\cdot (q\cdot s)}(P\subseteq p)$$

      By the reverse pomset transition rules for operator $\cdot$ in Table \ref{RPTRForBRATC}, we get

      $$\frac{s\xtworightarrow{S[\mathcal{I}]}S}{(p\cdot q)\cdot s\xtworightarrow{S[\mathcal{I}]}(p\cdot q)\cdot S} (S\subseteq s) \quad \frac{s\xtworightarrow{S[\mathcal{I}]}S}{p\cdot (q\cdot s)\xtworightarrow{S[\mathcal{I}]}p\cdot(q\cdot S)}(S\subseteq s)$$

      $$\frac{s\xtworightarrow{S[\mathcal{I}]}s'}{(p\cdot q)\cdot s\xtworightarrow{S[\mathcal{I}]}(p\cdot q)\cdot s'}(S\subseteq s) \quad \frac{s\xtworightarrow{S[\mathcal{I}]}s'}{p\cdot (q\cdot s)\xtworightarrow{S[\mathcal{I}]}p\cdot (q\cdot s')}(S\subseteq s)$$

      With assumptions $(P[\mathcal{K}]\cdot q)\cdot s\sim_p^{fr}P[\mathcal{K}]\cdot(q\cdot s)$, $(p'\cdot q)\cdot s\sim_p^{fr}p'\cdot(q\cdot s)$, $(p\cdot q)\cdot S\sim_p^{fr}p\cdot(q\cdot S)$, $(p\cdot q)\cdot s'\sim_p^{fr}p\cdot(q\cdot s')$, so, $(p\cdot q)\cdot s\sim_p^{fr} p\cdot (q\cdot s)$, as desired.
\end{itemize}
\end{proof}

\begin{proposition}[About Completeness of $BRATC$ modulo FR truly concurrent bisimulation equivalence]\label{CBRATCPBE}
Let $p$ and $q$ be closed $BRATC$ terms, if $p\sim_{p}^{fr} q$ then there may be $p\neq q$.
\end{proposition}

\begin{proof}
Firstly, by the elimination theorem of $BRATC$, we know that for each closed $BRATC$ term $p$, there exists a closed basic $BRATC$ term $p'$, such that $BRATC\vdash p=p'$, so, we only need to consider closed basic $BRATC$ terms.

The basic terms (see Definition \ref{BTBRATC}) modulo associativity and commutativity (AC) of conflict $+$ (defined by axioms $A1$ and $A2$ in Table \ref{AxiomsForBRATC}), and this equivalence is denoted by $=_{AC}$. Then, each equivalence class $s$ modulo AC of $+$ has the following normal form

$$s_1+\cdots+ s_k$$

with each $s_i$ either an atomic event or of the form $t_1\cdot t_2$, and each $s_i$ is called the summand of $s$.

Now, we try to prove that for normal forms $n$ and $n'$, if $n\sim_{p}^{fr} n'$ then $n=_{AC}n'$. It is sufficient to induct on the sizes of $n$ and $n'$.

Consider a summand $e_1\cdot e_2\cdot e_3$ of $n$. Then $n\xrightarrow{e_1}\xrightarrow{e_2}\xrightarrow{e_3}e_1[1]\cdot e_2[2]\cdot e_3[3]$, $n'$ should also have $n'\xrightarrow{e_1}\xrightarrow{e_2}\xrightarrow{e_3}n''$, but maybe $n''=e_1[1]\cdot e_2[2]\cdot e_3[3]$, or maybe $n''=e_1[1]\cdot e_3[3]+e_2[2]\cdot e_4$ according to the transition rules of $+$. Note that in the reversible version of $APTC$, the choice $+$ is different to that alternative composition $+$ in $APTC$. Though we define in $+$, if one branch forward or reverse executes successfully, then $+$ forward or reverse executes successfully, the above situation still stands.

That is, we cannot get $n=_{AC} n'$. So, we cannot give the completeness of $BRATC$ modulo FR pomset bisimulation equivalence. Similarly, we cannot give the completeness of $BRATC$ modulo any FR truly concurrent bisimulation equivalence. And in section \ref{raptc}, since $BRATC$ is an embedding of $RAPTC$, so we also cannot give the completeness of $RAPTC$ modulo any FR truly concurrent bisimulation equivalence. And more, in section \ref{abs}, since $RAPTC_{\tau}$ is a conservative extension of $RAPTC$, so we also cannot give the completeness of $RAPTC_{\tau}$ modulo any weakly FR truly concurrent bisimulation equivalence.
\end{proof}

The FR step transition rules are defined in Table \ref{STRForBRATC} and Table \ref{RSTRForBRATC}, different to FR pomset transition rules, the FR step transition rules are labeled by steps, in which every event is pairwise concurrent.

\begin{center}
    \begin{table}
        $$\frac{}{X\xrightarrow{X}X[\mathcal{K}]}(\forall e_1, e_2\in X \textrm{ are pairwise concurrent.})$$

        $$\frac{x\xrightarrow{X}X[\mathcal{K}] \quad X\nsubseteq y}{x+y\xrightarrow{X}X[\mathcal{K}]+y}(\forall e_1, e_2\in X \textrm{ are pairwise concurrent.})$$
        $$\frac{x\xrightarrow{X}x' \quad X\nsubseteq y}{x+y\xrightarrow{X}x'+y}(\forall e_1, e_2\in X \textrm{ are pairwise concurrent.})$$
        $$\frac{y\xrightarrow{Y}Y[\mathcal{J}] \quad Y\nsubseteq x}{x+y\xrightarrow{Y}x+Y[\mathcal{J}]}(\forall e_1, e_2\in Y \textrm{ are pairwise concurrent.})$$
        $$\frac{y\xrightarrow{Y}y'\quad Y\nsubseteq x}{x+y\xrightarrow{Y}x+y'}(\forall e_1, e_2\in Y \textrm{ are pairwise concurrent.})$$

        $$\frac{x\xrightarrow{X}X[\mathcal{K}]\quad y\xrightarrow{X}X[\mathcal{K}]}{x+y\xrightarrow{X}X[\mathcal{K}]+X[\mathcal{K}]}(\forall e_1, e_2\in X \textrm{ are pairwise concurrent.})$$
        $$\frac{x\xrightarrow{X}x'\quad y\xrightarrow{X}X[\mathcal{K}]}{x+y\xrightarrow{X}x'+X[\mathcal{K}]}(\forall e_1, e_2\in X \textrm{ are pairwise concurrent.})$$
        $$\frac{x\xrightarrow{X}X[\mathcal{K}]\quad y\xrightarrow{X}y'}{x+y\xrightarrow{X}X[\mathcal{K}]+y'}(\forall e_1, e_2\in X \textrm{ are pairwise concurrent.})$$
        $$\frac{x\xrightarrow{X}x'\quad y\xrightarrow{X}y'}{x+y\xrightarrow{X}x'+y'}(\forall e_1, e_2\in X \textrm{ are pairwise concurrent.})$$

        $$\frac{x\xrightarrow{X}X[\mathcal{K}]\quad\textrm{Std}(y)}{x\cdot y\xrightarrow{X} X[\mathcal{K}]\cdot y}(X\subseteq x,\forall e_1, e_2\in X \textrm{ are pairwise concurrent.})$$
        $$\frac{x\xrightarrow{X}x' \quad \textrm{Std}(y)}{x\cdot y\xrightarrow{X}x'\cdot y}(X\subseteq x,\forall e_1, e_2\in X \textrm{ are pairwise concurrent.})$$
        $$\frac{y\xrightarrow{Y}Y[\mathcal{J}]\quad \textrm{NStd}(x)}{x\cdot y\xrightarrow{Y}x\cdot Y[\mathcal{J}]}(Y\subseteq y,\forall e_1, e_2\in Y \textrm{ are pairwise concurrent.})$$
        $$\frac{y\xrightarrow{Y}y'\quad \textrm{NStd}(x)}{x\cdot y\xrightarrow{Y}x\cdot y'}(Y\subseteq y,\forall e_1, e_2\in Y \textrm{ are pairwise concurrent.})$$
        \caption{Forward step transition rules of $BRATC$}
        \label{STRForBRATC}
    \end{table}
\end{center}

\begin{center}
    \begin{table}
        $$\frac{}{X[\mathcal{K}]\xtworightarrow{X[\mathcal{K}]}X}(\forall e_1, e_2\in X \textrm{ are pairwise concurrent.})$$

        $$\frac{x\xtworightarrow{X[\mathcal{K}]}X\quad X\nsubseteq y}{x+y\xtworightarrow{X[\mathcal{K}]}X+y}(\forall e_1, e_2\in X \textrm{ are pairwise concurrent.})$$
        $$\frac{x\xtworightarrow{X[\mathcal{K}]}x' \quad X\nsubseteq y}{x+y\xtworightarrow{X[\mathcal{K}]}x'+y}(\forall e_1, e_2\in X \textrm{ are pairwise concurrent.})$$
        $$\frac{y\xtworightarrow{Y[\mathcal{J}]}Y \quad Y\nsubseteq x}{x+y\xtworightarrow{Y[\mathcal{J}]}x+Y}(\forall e_1, e_2\in Y \textrm{ are pairwise concurrent.})$$
        $$\frac{y\xtworightarrow{Y[\mathcal{J}]}y' \quad Y\nsubseteq x}{x+y\xtworightarrow{Y[\mathcal{J}]}x+y'}(\forall e_1, e_2\in Y \textrm{ are pairwise concurrent.})$$

        $$\frac{x\xtworightarrow{X[\mathcal{K}]}X\quad y\xtworightarrow{X[\mathcal{K}]}X}{x+y\xtworightarrow{X[\mathcal{K}]}X+X}(\forall e_1, e_2\in X \textrm{ are pairwise concurrent.})$$
        $$\frac{x\xtworightarrow{X[\mathcal{K}]}x'\quad y\xtworightarrow{X[\mathcal{K}]}X}{x+y\xtworightarrow{X[\mathcal{K}]}x'+X}(\forall e_1, e_2\in X \textrm{ are pairwise concurrent.})$$
        $$\frac{x\xtworightarrow{X[\mathcal{K}]}X\quad y\xtworightarrow{X[\mathcal{K}]}y'}{x+y\xtworightarrow{X[\mathcal{K}]}X+y'}(\forall e_1, e_2\in X \textrm{ are pairwise concurrent.})$$
        $$\frac{x\xtworightarrow{X[\mathcal{K}]}x'\quad y\xtworightarrow{X[\mathcal{K}]}y'}{x+y\xtworightarrow{X[\mathcal{K}]}x'+y'}(\forall e_1, e_2\in X \textrm{ are pairwise concurrent.})$$

        $$\frac{x\xtworightarrow{X[\mathcal{K}]}X \quad \textrm{Std}(y)}{x\cdot y\xtworightarrow{X[\mathcal{K}]} X\cdot y}(X\subseteq x,\forall e_1, e_2\in X \textrm{ are pairwise concurrent.})$$
        $$\frac{x\xtworightarrow{X[\mathcal{K}]}x'\quad \textrm{Std}(y)}{x\cdot y\xtworightarrow{X[\mathcal{K}]}x'\cdot y}(X\subseteq x,\forall e_1, e_2\in X \textrm{ are pairwise concurrent.})$$
        $$\frac{y\xtworightarrow{Y[\mathcal{J}]}Y \quad \textrm{NStd}(x)}{x\cdot y\xtworightarrow{Y[\mathcal{J}]}x\cdot Y}(Y\subseteq y,\forall e_1, e_2\in Y \textrm{ are pairwise concurrent.})$$
        $$\frac{y\xtworightarrow{Y[\mathcal{J}]}y' \quad \textrm{NStd}(x)}{x\cdot y\xtworightarrow{Y[\mathcal{J}]}x\cdot y'}(Y\subseteq y,\forall e_1, e_2\in Y \textrm{ are pairwise concurrent.})$$
        \caption{Reverse step transition rules of $BRATC$}
        \label{RSTRForBRATC}
    \end{table}
\end{center}

\begin{theorem}[Congruence of $BRATC$ with respect to FR step bisimulation equivalence]
FR step bisimulation equivalence $\sim_s^{fr}$ is a congruence with respect to $BRATC$.
\end{theorem}

\begin{proof}
It is easy to see that FR step bisimulation is an equivalent relation on $BRATC$ terms, we only need to prove that $\sim_{s}^{fr}$ is preserved by the operators $\cdot$ and $+$.

\begin{itemize}
  \item Causality operator $\cdot$. Let $x_1,x_2$ and $y_1,y_2$ be $BRATC$ processes, and $x_1\sim_{s}^{fr} y_1$, $x_2\sim_{s}^{fr} y_2$, it is sufficient to prove that $x_1\cdot x_2\sim_{s}^{fr} y_1\cdot y_2$.

      By the definition of FR step bisimulation $\sim_s^{fr}$ (Definition \ref{FRPSB}), $x_1\sim_s^{fr} y_1$ means that

      $$x_1\xrightarrow{X_1} x_1' \quad y_1\xrightarrow{Y_1} y_1'$$

      $$x_1\xtworightarrow{X_1[\mathcal{K}]} x_1'' \quad y_1\xtworightarrow{Y_1[\mathcal{J}]} y_1''$$

      with $X_1\subseteq x_1$, $\forall e_1, e_2\in X_1 \textrm{ are pairwise concurrent}$, $Y_1\subseteq y_1$, $\forall e_1, e_2\in Y_1 \textrm{ are pairwise concurrent}$, $X_1\sim Y_1$, $x_1'\sim_s^{fr} y_1'$ and $x_1''\sim_s^{fr}y_1''$. The meaning of $x_2\sim_s^{fr} y_2$ is similar.

      By the FR step transition rules for causality operator $\cdot$ in Table \ref{PTRForBRATC} and Table \ref{RPTRForBRATC}, we can get

      $$x_1\cdot x_2\xrightarrow{X_1} X_1[\mathcal{K}]\cdot x_2 \quad y_1\cdot y_2\xrightarrow{Y_1} Y_1[\mathcal{J}]\cdot y_2$$

      $$x_1\cdot x_2\xtworightarrow{X_2[\mathcal{K}]} x_1\cdot X_2 \quad y_1\cdot y_2\xtworightarrow{Y_2[\mathcal{J}]} y_1\cdot Y_2$$

      with $X_1\subseteq x_1$, $\forall e_1, e_2\in X_1 \textrm{ are pairwise concurrent}$, $Y_1\subseteq y_1$, $\forall e_1, e_2\in Y_1 \textrm{ are pairwise concurrent}$, $X_2\subseteq x_2$, $\forall e_1, e_2\in X_2 \textrm{ are pairwise concurrent}$, $Y_2\subseteq y_2$, $\forall e_1, e_2\in Y_2 \textrm{ are pairwise concurrent}$, $X_1\sim Y_1$, $X_2\sim Y_2$, and the assumptions $X_1[\mathcal{K}]\cdot x_2\sim_s^{fr} Y_1[\mathcal{J}]\cdot y_2$ and $x_1\cdot X_2\sim_s^{fr} y_1\cdot Y_2$, so, we get $x_1\cdot x_2\sim_s^{fr} y_1\cdot y_2$, as desired.

      Or, we can get

      $$x_1\cdot x_2\xrightarrow{X_1} x_1'\cdot x_2 \quad y_1\cdot y_2\xrightarrow{Y_1} y_1'\cdot y_2$$

      $$x_1\cdot x_2\xtworightarrow{X_2[\mathcal{K}]} x_1\cdot x_2' \quad y_1\cdot y_2\xtworightarrow{Y_2[\mathcal{J}]} y_1\cdot y_2'$$

      with $X_1\subseteq x_1$, $\forall e_1, e_2\in X_1 \textrm{ are pairwise concurrent}$, $Y_1\subseteq y_1$, $\forall e_1, e_2\in Y_1 \textrm{ are pairwise concurrent}$, $X_2\subseteq x_2$, $\forall e_1, e_2\in X_2 \textrm{ are pairwise concurrent}$, $Y_2\subseteq y_2$, $\forall e_1, e_2\in Y_2 \textrm{ are pairwise concurrent}$, $X_1\sim Y_1$, $X_2\sim Y_2$, and the assumptions $x_1'\cdot x_2\sim_s^{fr} y_1'\cdot y_2$, $x_1\cdot x_2'\sim_s^{fr} y_1\cdot y_2'$, so, we get $x_1\cdot x_2\sim_s^{fr} y_1\cdot y_2$, as desired.
  \item Conflict operator $+$. Let $x_1, x_2$ and $y_1, y_2$ be $BRATC$ processes, and $x_1\sim_s^{fr} y_1$, $x_2\sim_s^{fr} y_2$, it is sufficient to prove that $x_1+ x_2 \sim_s^{fr} y_1+ y_2$. The meanings of $x_1\sim_s^{fr} y_1$ and $x_2\sim_s^{fr} y_2$ are the same as the above case, according to the definition of FR step bisimulation $\sim_s^{fr}$ in Definition \ref{FRPSB}.

      By the FR step transition rules for conflict operator $+$ in Table \ref{PTRForBRATC} and Table \ref{RPTRForBRATC}, we can get several cases:

      $$x_1+ x_2\xrightarrow{X_1} X_1[\mathcal{K}]+x_2 \quad y_1+ y_2\xrightarrow{Y_1} Y_1[\mathcal{J}]+y_2$$
      $$x_1+ x_2\xtworightarrow{X_1[\mathcal{K}]} X_1+x_2 \quad y_1+ y_2\xtworightarrow{Y_1[\mathcal{J}]} Y_1+y_2$$

      with $X_1\subseteq x_1$, $\forall e_1, e_2\in X_1 \textrm{ are pairwise concurrent}$, $Y_1\subseteq y_1$, $\forall e_1, e_2\in Y_1 \textrm{ are pairwise concurrent}$, $X_1\sim Y_1$, and the assumptions $X_1[\mathcal{K}]+x_2\sim_s^{fr}Y_1[\mathcal{J}]+y_2$ and $X_1+x_2\sim_s^{fr}Y_1+y_2$, so, we get $x_1+ x_2\sim_s^{fr} y_1+ y_2$, as desired.

      Or, we can get

      $$x_1+ x_2\xrightarrow{X_1} x_1'+x_2 \quad y_1+ y_2\xrightarrow{Y_1} y_1'+y_2$$

      $$x_1+ x_2\xtworightarrow{X_1[\mathcal{K}]} x_1'+x_2 \quad y_1+ y_2\xtworightarrow{Y_1[\mathcal{J}]} y_1'+y_2$$

      with $X_1\subseteq x_1$, $\forall e_1, e_2\in X_1 \textrm{ are pairwise concurrent}$, $Y_1\subseteq y_1$, $\forall e_1, e_2\in Y_1 \textrm{ are pairwise concurrent}$, $X_1\sim Y_1$, and $x_1'+x_2\sim_s^{fr} y_1'+y_2$, so, we get $x_1+ x_2\sim_s^{fr} y_1+ y_2$, as desired.

      Or, we can get

      $$x_1+ x_2\xrightarrow{X_2} x_1+X_2[\mathcal{K}] \quad y_1+ y_2\xrightarrow{Y_2} y_1+Y_2[\mathcal{J}]$$

      $$x_1+ x_2\xrightarrow{X_2[\mathcal{K}]} x_1+X_2 \quad y_1+ y_2\xrightarrow{Y_2[\mathcal{J}]} y_1+Y_2$$

      with $X_2\subseteq x_2$, $\forall e_1, e_2\in X_2 \textrm{ are pairwise concurrent}$, $Y_2\subseteq y_2$, $\forall e_1, e_2\in Y_2 \textrm{ are pairwise concurrent}$, $X_2\sim Y_2$, and the assumptions $x_1+X_2[\mathcal{K}]\sim_s^{fr}y_1+Y_2[\mathcal{J}]$ and $x_1+X_2\sim_s^{fr}y_1+Y_2$, so, we get $x_1+ x_2\sim_s^{fr} y_1+ y_2$, as desired.

      Or, we can get

      $$x_1+ x_2\xrightarrow{X_2} x_1+x_2' \quad y_1+ y_2\xrightarrow{Y_2} y_1+y_2'$$

      $$x_1+ x_2\xtworightarrow{X_2[\mathcal{K}]} x_1+x_2' \quad y_1+ y_2\xtworightarrow{Y_2[\mathcal{J}]} y_1+y_2'$$

      with $X_2\subseteq x_2$, $\forall e_1, e_2\in X_2 \textrm{ are pairwise concurrent}$, $Y_2\subseteq y_2$, $X_2\sim Y_2$, $\forall e_1, e_2\in Y_2 \textrm{ are pairwise concurrent}$, and the assumption $x_1+x_2'\sim_s^{fr}y_1+y_2'$, so, we get $x_1+ x_2\sim_s^{fr} y_1+ y_2$, as desired.

      Or, we can get

      $$x_1+ x_2\xrightarrow{X} x_1'+x_2' \quad y_1+ y_2\xrightarrow{Y} y_1'+y_2'$$

      $$x_1+ x_2\xtworightarrow{X[\mathcal{K}]} x_1'+x_2' \quad y_1+ y_2\xtworightarrow{Y[\mathcal{J}]} y_1'+y_2'$$

      with $X\subseteq x_1$, $Y\subseteq y_1$, $X\subseteq x_2$, $Y\subseteq y_2$, $\forall e_1, e_2\in X \textrm{ are pairwise concurrent}$, $\forall e_1, e_2\in Y \textrm{ are pairwise concurrent}$, $X\sim Y$, and the assumption $x_1'+x_2'\sim_s^{fr}y_1'+y_2'$, so, we get $x_1+ x_2\sim_s^{fr} y_1+ y_2$, as desired.
\end{itemize}
\end{proof}

\begin{theorem}[Soundness of $BRATC$ modulo FR step bisimulation equivalence]\label{SBRATCSBE}
Let $x$ and $y$ be $BRATC$ terms. If $BRATC\vdash x=y$, then $x\sim_{s}^{fr} y$.
\end{theorem}

\begin{proof}
Since FR step bisimulation $\sim_s^{fr}$ is both an equivalent and a congruent relation, we only need to check if each axiom in Table \ref{AxiomsForBRATC} is sound modulo FR step bisimulation equivalence.

\begin{itemize}
  \item \textbf{Axiom $A1$}. Let $p,q$ be $BRATC$ processes, and $p+ q=q+ p$, it is sufficient to prove that $p+ q\sim_s^{fr} q+ p$. By the forward step transition rules for operator $+$ in Table \ref{STRForBRATC}, we get

      $$\frac{p\xrightarrow{P}P[\mathcal{K}]}{p+ q\xrightarrow{P}P[\mathcal{K}]+q} (P\subseteq p,P\nsubseteq q,\forall e_1, e_2\in P \textrm{ are pairwise concurrent.})$$
      $$\frac{p\xrightarrow{P}P[\mathcal{K}]}{q+ p\xrightarrow{P}q+P[\mathcal{K}]}(P\subseteq p,P\nsubseteq q,\forall e_1, e_2\in P \textrm{ are pairwise concurrent.})$$

      $$\frac{p\xrightarrow{P}p'}{p+ q\xrightarrow{P}p'+q}(P\subseteq p,P\nsubseteq q,\forall e_1, e_2\in P \textrm{ are pairwise concurrent.})$$
      $$\frac{p\xrightarrow{P}p'}{q+ p\xrightarrow{P}q+p'}(P\subseteq p,P\nsubseteq q,\forall e_1, e_2\in P \textrm{ are pairwise concurrent.})$$

      $$\frac{q\xrightarrow{Q}Q[\mathcal{J}]}{p+ q\xrightarrow{Q}p+Q[\mathcal{J}]}(Q\subseteq q,Q\nsubseteq p,\forall e_1, e_2\in Q \textrm{ are pairwise concurrent.})$$
      $$\frac{q\xrightarrow{Q}Q[\mathcal{J}]}{q+ p\xrightarrow{Q}Q[\mathcal{J}]+p}(Q\subseteq q,Q\nsubseteq p,\forall e_1, e_2\in Q \textrm{ are pairwise concurrent.})$$

      $$\frac{q\xrightarrow{Q}q'}{p+ q\xrightarrow{Q}p+q'}(Q\subseteq q,Q\nsubseteq p,\forall e_1, e_2\in Q \textrm{ are pairwise concurrent.})$$
      $$\frac{q\xrightarrow{Q}q'}{q+ p\xrightarrow{Q}q'+p}(Q\subseteq q, Q\nsubseteq p,\forall e_1, e_2\in Q \textrm{ are pairwise concurrent.})$$

      $$\frac{p\xrightarrow{P}p'\quad q\xrightarrow{P}q'}{p+ q\xrightarrow{P}p'+q'}(P\subseteq p,P\subseteq q,\forall e_1, e_2\in P \textrm{ are pairwise concurrent.})$$
      $$\frac{p\xrightarrow{P}p'\quad q\xrightarrow{P}q'}{q+ p\xrightarrow{P}q'+p'}(P\subseteq p,P\subseteq q,\forall e_1, e_2\in P \textrm{ are pairwise concurrent.})$$

      By the reverse step transition rules for operator $+$ in Table \ref{RSTRForBRATC}, we get

      $$\frac{p\xtworightarrow{P[\mathcal{K}]}P\quad P\nsubseteq q}{p+ q\xrightarrow{P[\mathcal{K}]}P+q} (P\subseteq p,\forall e_1, e_2\in P \textrm{ are pairwise concurrent.})$$
      $$\frac{p\xtworightarrow{P[\mathcal{K}]}P\quad P\nsubseteq q}{q+ p\xtworightarrow{P[\mathcal{K}]}q+P}(P\subseteq p,\forall e_1, e_2\in P \textrm{ are pairwise concurrent.})$$

      $$\frac{p\xtworightarrow{P[\mathcal{K}]}p'\quad P\nsubseteq q}{p+ q\xtworightarrow{P[\mathcal{K}]}p'+q}(P\subseteq p,\forall e_1, e_2\in P \textrm{ are pairwise concurrent.})$$
      $$\frac{p\xtworightarrow{P[\mathcal{K}]}p'\quad P\nsubseteq q}{q+ p\xtworightarrow{P[\mathcal{K}]}q+p'}(P\subseteq p,\forall e_1, e_2\in P \textrm{ are pairwise concurrent.})$$

      $$\frac{q\xtworightarrow{Q[\mathcal{J}]}Q\quad Q\nsubseteq p}{p+ q\xtworightarrow{Q[\mathcal{J}]}p+Q}(Q\subseteq q,\forall e_1, e_2\in Q \textrm{ are pairwise concurrent.})$$
      $$\frac{q\xtworightarrow{Q[\mathcal{J}]}Q\quad Q\nsubseteq p}{q+ p\xtworightarrow{Q[\mathcal{J}]}Q+p}(Q\subseteq q,\forall e_1, e_2\in Q \textrm{ are pairwise concurrent.})$$

      $$\frac{q\xtworightarrow{Q[\mathcal{J}]}q'\quad Q\nsubseteq p}{p+ q\xtworightarrow{Q[\mathcal{J}]}p+q'}(Q\subseteq q,\forall e_1, e_2\in Q \textrm{ are pairwise concurrent.})$$
      $$\frac{q\xtworightarrow{Q[\mathcal{J}]}q'\quad Q\nsubseteq p}{q+ p\xtworightarrow{Q[\mathcal{J}]}q'+p}(Q\subseteq q,\forall e_1, e_2\in Q \textrm{ are pairwise concurrent.})$$

      $$\frac{p\xtworightarrow{P[\mathcal{K}]}p'\quad q\xtworightarrow{P[\mathcal{K}]}q'}{p+ q\xtworightarrow{P[\mathcal{K}]}p'+q'}(P\subseteq p,P\subseteq q,\forall e_1, e_2\in P \textrm{ are pairwise concurrent.})$$
      $$\frac{p\xtworightarrow{P[\mathcal{K}]}p'\quad q\xtworightarrow{P[\mathcal{K}]}q'}{q+ p\xtworightarrow{P[\mathcal{K}]}q'+p'}(P\subseteq p,P\subseteq q,\forall e_1, e_2\in P \textrm{ are pairwise concurrent.})$$

      With the assumptions $P[\mathcal{K}]+q\sim_s^{fr}q+P[\mathcal{K}]$, $P+q\sim_s^{fr}q+P$, $p+Q[\mathcal{J}]\sim_s^{fr}Q[\mathcal{J}]+p$, $p'+q\sim_s^{fr}q+p'$, $p+q'\sim_s^{fr}q'+p$ and $p'+q'\sim_s^{fr}q'+p'$ so, $p+ q\sim_s^{fr} q+ p$, as desired.
  \item \textbf{Axiom $A2$}. Let $p,q,s$ be $BRATC$ processes, and $(p+ q)+ s=p+ (q+ s)$, it is sufficient to prove that $(p+ q)+ s \sim_s^{fr} p+ (q+ s)$. By the forward step transition rules for operator $+$ in Table \ref{STRForBRATC}, we get

      $$\frac{p\xrightarrow{P}P[\mathcal{K}]\quad P\nsubseteq q\quad P\nsubseteq s}{(p+ q)+ s\xrightarrow{P}(P[\mathcal{K}]+q)+s} (P\subseteq p,\forall e_1, e_2\in P \textrm{ are pairwise concurrent.})$$
      $$\frac{p\xrightarrow{P}P[\mathcal{K}]\quad P\nsubseteq q\quad P\nsubseteq s}{p+ (q+ s)\xrightarrow{P}P[\mathcal{K}]+(q+s)}(P\subseteq p,\forall e_1, e_2\in P \textrm{ are pairwise concurrent.})$$

      $$\frac{p\xrightarrow{P}p'\quad P\nsubseteq q\quad P\nsubseteq s}{(p+ q)+ s\xrightarrow{P}(p'+q)+s}(P\subseteq p,\forall e_1, e_2\in P \textrm{ are pairwise concurrent.})$$
      $$\frac{p\xrightarrow{P}p'\quad P\nsubseteq q\quad P\nsubseteq s}{p+ (q+ s)\xrightarrow{P}p'+(q+s)}(P\subseteq p,\forall e_1, e_2\in P \textrm{ are pairwise concurrent.})$$

      $$\frac{q\xrightarrow{Q}Q[\mathcal{J}]\quad Q\nsubseteq p\quad Q\nsubseteq s}{(p+ q)+ s\xrightarrow{Q}(p+Q[\mathcal{J}])+s}(Q\subseteq q,\forall e_1, e_2\in Q \textrm{ are pairwise concurrent.})$$
      $$\frac{q\xrightarrow{Q}Q[\mathcal{J}]\quad Q\nsubseteq p\quad Q\nsubseteq s}{p+ (q+ s)\xrightarrow{Q}p+(Q[\mathcal{J}]+s)}(Q\subseteq q,\forall e_1, e_2\in Q \textrm{ are pairwise concurrent.})$$

      $$\frac{q\xrightarrow{Q}q'\quad Q\nsubseteq p\quad Q\nsubseteq s}{(p+ q)+ s\xrightarrow{Q}(p+q')+s}(Q\subseteq q,\forall e_1, e_2\in Q \textrm{ are pairwise concurrent.})$$
      $$\frac{q\xrightarrow{Q}q'\quad Q\nsubseteq p\quad Q\nsubseteq s}{p+ (q+ s)\xrightarrow{Q}p+(q'+s)}(Q\subseteq q,\forall e_1, e_2\in Q \textrm{ are pairwise concurrent.})$$

      $$\frac{s\xrightarrow{S}S[\mathcal{I}]\quad S\nsubseteq p\quad S\nsubseteq q}{(p+ q)+ s\xrightarrow{S}(p+q)+S[\mathcal{I}]}(S\subseteq s,\forall e_1, e_2\in S \textrm{ are pairwise concurrent.})$$
      $$\frac{s\xrightarrow{S}S[\mathcal{I}]\quad S\nsubseteq p\quad S\nsubseteq q}{p+ (q+ s)\xrightarrow{S}p+(q+S[\mathcal{I}])}(S\subseteq s,\forall e_1, e_2\in S \textrm{ are pairwise concurrent.})$$

      $$\frac{s\xrightarrow{S}s'\quad S\nsubseteq p\quad S\nsubseteq q}{(p+ q)+ s\xrightarrow{S}(p+q)+s'}(S\subseteq s,\forall e_1, e_2\in S \textrm{ are pairwise concurrent.})$$
      $$\frac{s\xrightarrow{S}s'\quad S\nsubseteq p\quad S\nsubseteq q}{p+ (q+ s)\xrightarrow{S}p+(q+s')}(S\subseteq s,\forall e_1, e_2\in S \textrm{ are pairwise concurrent.})$$

      $$\frac{p\xrightarrow{P}p'\quad q\xrightarrow{P}q'\quad s\xrightarrow{P}s'}{(p+ q)+ s\xrightarrow{P}(p'+q')+s'}(P\subseteq p,P\subseteq q,P\subseteq s,\forall e_1, e_2\in P \textrm{ are pairwise concurrent.})$$
      $$\frac{p\xrightarrow{P}p'\quad q\xrightarrow{P}q'\quad s\xrightarrow{P}s'}{p+ (q+ s)\xrightarrow{P}p'+(q'+s')}(P\subseteq p,P\subseteq q,P\subseteq s,\forall e_1, e_2\in P \textrm{ are pairwise concurrent.})$$

      By the reverse step transition rules for operator $+$ in Table \ref{RSTRForBRATC}, we get

      $$\frac{p\xtworightarrow{P[\mathcal{K}]}P\quad P\nsubseteq q\quad P\nsubseteq s}{(p+ q)+ s\xtworightarrow{P[\mathcal{K}]}(P+q)+s} (P\subseteq p,\forall e_1, e_2\in P \textrm{ are pairwise concurrent.})$$
      $$\frac{p\xtworightarrow{P[\mathcal{K}]}P\quad P\nsubseteq q\quad P\nsubseteq s}{p+ (q+ s)\xtworightarrow{P[\mathcal{K}]}P+(q+s)}(P\subseteq p,\forall e_1, e_2\in P \textrm{ are pairwise concurrent.})$$

      $$\frac{p\xtworightarrow{P[\mathcal{K}]}p'\quad P\nsubseteq q\quad P\nsubseteq s}{(p+ q)+ s\xtworightarrow{P[\mathcal{K}]}(p'+q)+s}(P\subseteq p,\forall e_1, e_2\in P \textrm{ are pairwise concurrent.})$$
      $$\frac{p\xtworightarrow{P[\mathcal{K}]}p'\quad P\nsubseteq q\quad P\nsubseteq s}{p+ (q+ s)\xtworightarrow{P[\mathcal{K}]}p'+(q+s)}(P\subseteq p,\forall e_1, e_2\in P \textrm{ are pairwise concurrent.})$$

      $$\frac{q\xtworightarrow{Q[\mathcal{J}]}Q\quad Q\nsubseteq p\quad Q\nsubseteq s}{(p+ q)+ s\xtworightarrow{Q[\mathcal{J}]}(p+Q)+s}(Q\subseteq q,\forall e_1, e_2\in Q \textrm{ are pairwise concurrent.})$$
      $$\frac{q\xtworightarrow{Q[\mathcal{J}]}Q\quad Q\nsubseteq p\quad Q\nsubseteq s}{p+ (q+ s)\xtworightarrow{Q[\mathcal{J}]}p+(Q+s)}(Q\subseteq q,\forall e_1, e_2\in Q \textrm{ are pairwise concurrent.})$$

      $$\frac{q\xtworightarrow{Q[\mathcal{J}]}q'\quad Q\nsubseteq p\quad Q\nsubseteq s}{(p+ q)+ s\xrightarrow{Q[\mathcal{J}]}(p+q')+s}(Q\subseteq q,\forall e_1, e_2\in Q \textrm{ are pairwise concurrent.})$$
      $$\frac{q\xtworightarrow{Q[\mathcal{J}]}q'\quad Q\nsubseteq p\quad Q\nsubseteq s}{p+ (q+ s)\xtworightarrow{Q[\mathcal{J}]}p+(q'+s)}(Q\subseteq q,\forall e_1, e_2\in Q \textrm{ are pairwise concurrent.})$$

      $$\frac{s\xtworightarrow{S[\mathcal{I}]}S\quad S\nsubseteq p\quad S\nsubseteq q}{(p+ q)+ s\xtworightarrow{S[\mathcal{I}]}(p+q)+S}(S\subseteq s,\forall e_1, e_2\in S \textrm{ are pairwise concurrent.})$$
      $$\frac{s\xtworightarrow{S[\mathcal{I}]}S\quad S\nsubseteq p\quad S\nsubseteq q}{p+ (q+ s)\xtworightarrow{S[\mathcal{I}]}p+(q+S)}(S\subseteq s,\forall e_1, e_2\in S \textrm{ are pairwise concurrent.})$$

      $$\frac{s\xtworightarrow{S[\mathcal{I}]}s'\quad S\nsubseteq p\quad S\nsubseteq q}{(p+ q)+ s\xtworightarrow{S[\mathcal{I}]}(p+q)+s'}(S\subseteq s,\forall e_1, e_2\in S \textrm{ are pairwise concurrent.})$$
      $$\frac{s\xtworightarrow{S[\mathcal{I}]}s'\quad S\nsubseteq p\quad S\nsubseteq q}{p+ (q+ s)\xtworightarrow{S[\mathcal{I}]}p+(q+s')}(S\subseteq s,\forall e_1, e_2\in S \textrm{ are pairwise concurrent.})$$

      $$\frac{p\xtworightarrow{P[\mathcal{K}]}p'\quad q\xtworightarrow{P[\mathcal{K}]}q'\quad s\xtworightarrow{P[\mathcal{K}]}s'}{(p+ q)+ s\xtworightarrow{P[\mathcal{K}]}(p'+q')+s'}(P\subseteq p,P\subseteq q,P\subseteq s,\forall e_1, e_2\in P \textrm{ are pairwise concurrent.})$$
      $$\frac{p\xtworightarrow{P[\mathcal{K}]}p'\quad q\xtworightarrow{P[\mathcal{K}]}q'\quad s\xtworightarrow{P[\mathcal{K}]}s'}{p+ (q+ s)\xtworightarrow{P[\mathcal{K}]}p'+(q'+s')}(P\subseteq p,P\subseteq q,P\subseteq s,\forall e_1, e_2\in P \textrm{ are pairwise concurrent.})$$

      with the assumptions $(P[\mathcal{K}]+q)+s\sim_s^{fr}P[\mathcal{K}]+(q+s)$, $(P+q)+s\sim_s^{fr}P+(q+s)$, $(p+Q[\mathcal{J}])+s\sim_s^{fr}p+(Q[\mathcal{J}]+s)$, $(p+Q)+s\sim_s^{fr}p+(Q+s)$, $(p+q)+S[\mathcal{I}]\sim_s^{fr}p+(q+S[\mathcal{I}])$, $(p+q)+S\sim_s^{fr}p+(q+S)$, $(p'+q)+s\sim_s^{fr}p'+(q+s)$, $(p+q')+s\sim_s^{fr}p+(q'+s)$, $(p+q)+s'\sim_s^{fr}p+(q+s')$ and $(p'+q')+s'\sim_s^{fr}p'+(q'+s')$ so, $(p+ q)+ s\sim_s^{fr} p+ (q+ s)$, as desired.
  \item \textbf{Axiom $A3$}. Let $p$ be a $BRATC$ process, and $p+ p=p$, it is sufficient to prove that $p+ p\sim_s^{fr} p$. By the forward step transition rules for operator $+$ in Table \ref{STRForBRATC}, we get

      $$\frac{p\xrightarrow{P}P[\mathcal{K}]}{p+ p\xrightarrow{P}P[\mathcal{K}]+P[\mathcal{K}]} (P\subseteq p,\forall e_1, e_2\in P \textrm{ are pairwise concurrent.})$$
      $$\frac{p\xrightarrow{P}P[\mathcal{K}]}{p\xrightarrow{P}P[\mathcal{K}]}(P\subseteq p,\forall e_1, e_2\in P \textrm{ are pairwise concurrent.})$$

      $$\frac{p\xrightarrow{P}p'}{p+ p\xrightarrow{P}p'+p'}(P\subseteq p,\forall e_1, e_2\in P \textrm{ are pairwise concurrent.})$$
      $$\frac{p\xrightarrow{P}p'}{p\xrightarrow{P}p'}(P\subseteq p,\forall e_1, e_2\in P \textrm{ are pairwise concurrent.})$$

      By the reverse step transition rules for operator $+$ in Table \ref{RSTRForBRATC}, we get

      $$\frac{p\xtworightarrow{P[\mathcal{K}]}P}{p+ p\xtworightarrow{P[\mathcal{K}]}P+P} (P\subseteq p,\forall e_1, e_2\in P \textrm{ are pairwise concurrent.})$$
      $$\frac{p\xtworightarrow{P[\mathcal{K}]}P}{p\xtworightarrow{P[\mathcal{K}]}P}(P\subseteq p,\forall e_1, e_2\in P \textrm{ are pairwise concurrent.})$$

      $$\frac{p\xtworightarrow{P[\mathcal{K}]}p'}{p+ p\xtworightarrow{P[\mathcal{K}]}p'+p'}(P\subseteq p,\forall e_1, e_2\in P \textrm{ are pairwise concurrent.})$$
      $$\frac{p\xtworightarrow{P[\mathcal{K}]}p'}{p\xtworightarrow{P[\mathcal{K}]}p'}(P\subseteq p,\forall e_1, e_2\in P \textrm{ are pairwise concurrent.})$$

      with the assumptions $P[\mathcal{K}]+P[\mathcal{K}]\sim_s^{fr}P[\mathcal{K}]$, $P+P\sim_s^{fr}P$ and $p'+p'\sim_s^{fr}p'$, so, $p+ p\sim_s^{fr} p$, as desired.
  \item \textbf{Axiom $A4$}. Let $p,q,s$ be $BRATC$ processes, and $p\cdot (q+ s)=p\cdot q + p\cdot s(\textrm{Std}(p), \textrm{Std}(q), \textrm{Std}(s))$, it is sufficient to prove that $p\cdot (q+ s)\sim_p^{fr}p\cdot q + p\cdot s$. By the pomset transition rules for operators $+$ and $\cdot$ in Table \ref{PTRForBRATC}, we get

      $$\frac{p\xrightarrow{P}P[\mathcal{K}]}{p\cdot (q+ s)\xrightarrow{P}P[\mathcal{K}]\cdot(q+s)} (P\subseteq p,\forall e_1, e_2\in P \textrm{ are pairwise concurrent.})$$
      $$\frac{p\xrightarrow{P}P[\mathcal{K}]}{p\cdot q + p\cdot s\xrightarrow{P}P[\mathcal{K}]\cdot q +P[\mathcal{K}]\cdot s}(P\subseteq p,\forall e_1, e_2\in P \textrm{ are pairwise concurrent.})$$

      $$\frac{p\xrightarrow{P}p'}{p\cdot (q+ s)\xrightarrow{P}p'\cdot (q+s)}(P\subseteq p,\forall e_1, e_2\in P \textrm{ are pairwise concurrent.})$$
      $$\frac{p\xrightarrow{P}p'}{p\cdot q + p\cdot s\xrightarrow{P}p'\cdot q+p'\cdot s}(P\subseteq p,\forall e_1, e_2\in P \textrm{ are pairwise concurrent.})$$

      By the reverse transition rules for operators $+$ and $\cdot$ in Table \ref{RPTRForBRATC}, there are no transitions.

      with the assumptions $P[\mathcal{K}]\cdot(q+s)\sim_p^{fr}P[\mathcal{K}]\cdot q + P[\mathcal{K}]\cdot s$, $p'\cdot(q+s)\sim_p^{fr}p'\cdot q + p'\cdot s$, so, $p\cdot (q+ s)\sim_p^{fr}p\cdot q + p\cdot s(\textrm{Std}(p), \textrm{Std}(q), \textrm{Std}(s))$, as desired.
  \item \textbf{Axiom $RA4$}. Let $p,q,s$ be $BRATC$ processes, and $ (q+ s)\cdot p=q\cdot p + s\cdot p(\textrm{NStd}(p), \textrm{NStd}(q), \textrm{NStd}(s))$, it is sufficient to prove that $(q+ s)\cdot p \sim_p^{fr}q\cdot p + s\cdot p$. By the pomset transition rules for operators $+$ and $\cdot$ in Table \ref{PTRForBRATC}, there are no transitions.

      By the reverse transition rules for operators $+$ and $\cdot$ in Table \ref{RPTRForBRATC}, we get

      $$\frac{p\xtworightarrow{P[\mathcal{K}]}P}{(q+ s)\cdot p \xtworightarrow{P[\mathcal{K}]}(q+s)\cdot P} (P\subseteq p,\forall e_1, e_2\in P \textrm{ are pairwise concurrent.})$$
      $$\frac{p\xtworightarrow{P[\mathcal{K}]}P}{q\cdot p + s\cdot p\xtworightarrow{P[\mathcal{K}]}q\cdot P +s\cdot P}(P\subseteq p,\forall e_1, e_2\in P \textrm{ are pairwise concurrent.})$$

      $$\frac{p\xtworightarrow{P[\mathcal{K}]}p'}{(q+ s)\cdot p \xtworightarrow{P[\mathcal{K}]}(q+s)\cdot p'}(P\subseteq p,\forall e_1, e_2\in P \textrm{ are pairwise concurrent.})$$
      $$\frac{p\xtworightarrow{P[\mathcal{K}]}p'}{q\cdot p + s\cdot p\xtworightarrow{P[\mathcal{K}]}q\cdot p'+s\cdot p'}(P\subseteq p,\forall e_1, e_2\in P \textrm{ are pairwise concurrent.})$$

      with the assumptions $(q+s)\cdot P\sim_p^{fr}q\cdot P + s\cdot P$, $(q+s)\cdot p'\sim_p^{fr}q\cdot p' + s\cdot p'$, so, $(q+ s)\cdot p \sim_p^{fr}q\cdot p + s\cdot p(\textrm{NStd}(p), \textrm{NStd}(q), \textrm{NStd}(s))$, as desired.
  \item \textbf{Axiom $A5$}. Let $p,q,s$ be $BRATC$ processes, and $(p\cdot q)\cdot s=p\cdot (q\cdot s)$, it is sufficient to prove that $(p\cdot q)\cdot s \sim_s^{fr} p\cdot (q\cdot s)$. By the forward step transition rules for operator $\cdot$ in Table \ref{STRForBRATC}, we get

      $$\frac{p\xrightarrow{P}P[\mathcal{K}]}{(p\cdot q)\cdot s\xrightarrow{P}(P[\mathcal{K}]\cdot q)\cdot s} (P\subseteq p,\forall e_1, e_2\in P \textrm{ are pairwise concurrent.})$$
      $$\frac{p\xrightarrow{P}P[\mathcal{K}]}{p\cdot (q\cdot s)\xrightarrow{P}P[\mathcal{K}]\cdot(q\cdot s)}(P\subseteq p,\forall e_1, e_2\in P \textrm{ are pairwise concurrent.})$$

      $$\frac{p\xrightarrow{P}p'}{(p\cdot q)\cdot s\xrightarrow{P}(p'\cdot q)\cdot s}(P\subseteq p,\forall e_1, e_2\in P \textrm{ are pairwise concurrent.})$$
      $$\frac{p\xrightarrow{P}p'}{p\cdot (q\cdot s)\xrightarrow{P}p'\cdot (q\cdot s)}(P\subseteq p,\forall e_1, e_2\in P \textrm{ are pairwise concurrent.})$$

      By the reverse step transition rules for operator $\cdot$ in Table \ref{RSTRForBRATC}, we get

      $$\frac{s\xtworightarrow{S[\mathcal{I}]}S}{(p\cdot q)\cdot s\xtworightarrow{S[\mathcal{I}]}(p\cdot q)\cdot S} (S\subseteq s,\forall e_1, e_2\in S \textrm{ are pairwise concurrent.})$$
      $$\frac{s\xtworightarrow{S[\mathcal{I}]}S}{p\cdot (q\cdot s)\xtworightarrow{S[\mathcal{I}]}p\cdot(q\cdot S)}(S\subseteq s,\forall e_1, e_2\in S \textrm{ are pairwise concurrent.})$$

      $$\frac{s\xtworightarrow{S[\mathcal{I}]}s'}{(p\cdot q)\cdot s\xtworightarrow{S[\mathcal{I}]}(p\cdot q)\cdot s'}(S\subseteq s,\forall e_1, e_2\in S \textrm{ are pairwise concurrent.})$$
      $$\frac{s\xtworightarrow{S[\mathcal{I}]}s'}{p\cdot (q\cdot s)\xtworightarrow{S[\mathcal{I}]}p\cdot (q\cdot s')}(S\subseteq s,\forall e_1, e_2\in S \textrm{ are pairwise concurrent.})$$

      With assumptions $(P[\mathcal{K}]\cdot q)\cdot s\sim_s^{fr}P[\mathcal{K}]\cdot(q\cdot s)$, $(p'\cdot q)\cdot s\sim_s^{fr}p'\cdot(q\cdot s)$, $(p\cdot q)\cdot S\sim_s^{fr}p\cdot(q\cdot S)$, $(p\cdot q)\cdot s'\sim_s^{fr}p\cdot(q\cdot s')$, so, $(p\cdot q)\cdot s\sim_s^{fr} p\cdot (q\cdot s)$, as desired.
\end{itemize}
\end{proof}

The transition rules for FR (hereditary) hp-bisimulation of $BRATC$ are the same as single event transition rules in Table \ref{SETRForBRATC}.

\begin{theorem}[Congruence of $BRATC$ with respect to FR hp-bisimulation equivalence]
FR hp-bisimulation equivalence $\sim_{hp}^{fr}$ is a congruence with respect to $BRATC$.
\end{theorem}

\begin{proof}
It is easy to see that FR history-preserving bisimulation is an equivalent relation on $BRATC$ terms, we only need to prove that $\sim_{hp}^{fr}$ is preserved by the operators $\cdot$ and $+$.

\begin{itemize}
  \item Causality operator $\cdot$. Let $x_1,x_2$ and $y_1,y_2$ be $BRATC$ processes, and $x_1\sim_{hp}^{fr} y_1$, $x_2\sim_{hp}^{fr} y_2$, it is sufficient to prove that $x_1\cdot x_2\sim_{hp}^{fr} y_1\cdot y_2$.

      By the definition of FR hp-bisimulation $\sim_{hp}^{fr}$ (Definition \ref{FRHHPB}), $x_1\sim_{hp}^{fr} y_1$ means that there is a posetal relation $(C(x_1),f,C(y_1))\in\sim_{hp}^{fr}$, and

      $$x_1\xrightarrow{e_1} x_1' \quad y_1\xrightarrow{e_2} y_1'$$

      $$x_1\xtworightarrow{e_1[m]} x_1' \quad y_1\xtworightarrow{e_2[m]} y_1'$$

      with $(C(x_1'),f[e_1\mapsto e_2],C(y_1'))\in\sim_{hp}^{fr}$. The meaning of $x_2\sim_{hp}^{fr} y_2$ is similar.

      By the FR hp-transition rules for causality operator $\cdot$ in Table \ref{SETRForBRATC} and Table \ref{RSETRForBRATC}, we can get

      $$x_1\cdot x_2\xrightarrow{e_1} e_1[m]\cdot x_2 \quad y_1\cdot y_2\xrightarrow{e_2} e_2[n]\cdot y_2$$

      $$x_1\cdot x_2\xtworightarrow{e_1'[m} x_1\cdot e_1' \quad y_1\cdot y_2\xtworightarrow{e_2'[n} y_1\cdot e_2'$$

      with the assumptions $(C(e_1[m]\cdot x_2),f[e_1\mapsto e_2],C(e_2[n]\cdot y_2))\in\sim_{hp}^{fr}$ and $(C(x_1\cdot e_1'),f[e_1'\mapsto e_2'],C(y_1\cdot e_2'))\in\sim_{hp}^{fr}$, so, we get $x_1\cdot x_2\sim_{hp}^{fr} y_1\cdot y_2$, as desired.

      Or, we can get

      $$x_1\cdot x_2\xrightarrow{e_1} x_1'\cdot x_2 \quad y_1\cdot y_2\xrightarrow{e_2} y_1'\cdot y_2$$

      $$x_1\cdot x_2\xtworightarrow{e1'[m]} x_1\cdot x_2' \quad y_1\cdot y_2\xtworightarrow{e_2'[n]} y_1\cdot y_2'$$

      with the assumptions $(C(x_1'\cdot x_2),f[e_1\mapsto e_2],C(y_1'\cdot y_2))\in\sim_{hp}^{fr}$ and $(C(x_1\cdot x_2'),f[e_1'\mapsto e_2'],C(y_1\cdot y_2'))\in\sim_{hp}^{fr}$, so, we get $x_1\cdot x_2\sim_{hp}^{fr} y_1\cdot y_2$, as desired.

  \item Conflict operator $+$. Let $x_1, x_2$ and $y_1, y_2$ be $BRATC$ processes, and $x_1\sim_{hp}^{fr} y_1$, $x_2\sim_{hp}^{fr} y_2$, it is sufficient to prove that $x_1+ x_2 \sim_{hp}^{fr} y_1+ y_2$. The meanings of $x_1\sim_{hp}^{fr} y_1$ and $x_2\sim_{hp}^{fr} y_2$ are the same as the above case, according to the definition of FR hp-bisimulation $\sim_{hp}^{fr}$ in Definition \ref{FRHHPB}.

      By the FR hp-transition rules for conflict operator $+$ in Table \ref{SETRForBRATC} and Table \ref{RSETRForBRATC}, we can get several cases:

      $$x_1+ x_2\xrightarrow{e_1} e_1[m]+x_2 \quad y_1+ y_2\xrightarrow{e_2} e_2[n]+y_2$$
      $$x_1+ x_2\xtworightarrow{e_1[m]} e_1+x_2 \quad y_1+ y_2\xtworightarrow{e_2[n]} e_2+y_2$$

      with the assumptions $(C(e_1[m]+ x_2),f[e_1\mapsto e_2],C(e_2[n]+ y_2))\in\sim_{hp}^{fr}$ and $(C(e_1+ x_2),f[e_1\mapsto e_2],C(e_2+ y_2))\in\sim_{hp}^{fr}$, so, we get $x_1+ x_2\sim_{hp}^{fr} y_1+ y_2$, as desired.

      Or, we can get

      $$x_1+ x_2\xrightarrow{e_1} x_1'+x_2 \quad y_1+ y_2\xrightarrow{e_2} y_1'+y_2$$

      $$x_1+ x_2\xtworightarrow{e_1[m]} x_1'+x_2 \quad y_1+ y_2\xtworightarrow{e_2[n]} y_1'+y_2$$

      with the assumptions $(C(x_1'+ x_2),f[e_1\mapsto e_2],C(y_1'+ y_2))\in\sim_{hp}^{fr}$, so, we get $x_1+ x_2\sim_{hp}^{fr} y_1+ y_2$, as desired.

      Or, we can get

      $$x_1+ x_2\xrightarrow{e_1'} x_1+e_1'[m] \quad y_1+ y_2\xrightarrow{e_2'} y_1+e_2'[n]$$

      $$x_1+ x_2\xrightarrow{e_1'[m]} x_1+e_1' \quad y_1+ y_2\xrightarrow{e_2'[n]} y_1+e_2'$$

      with the assumptions $(C(x_1 +e_1'[m]),f[e_1'\mapsto e_2'],C(y_1+e_2'[n]))\in\sim_{hp}^{fr}$ and $(C(x_1+e_1'),f[e_1'\mapsto e_2'],C(y_1+e_2'))\in\sim_{hp}^{fr}$, so, we get $x_1+ x_2\sim_{hp}^{fr} y_1+ y_2$, as desired.

      Or, we can get

      $$x_1+ x_2\xrightarrow{e_1'} x_1+x_2' \quad y_1+ y_2\xrightarrow{e_2'} y_1+y_2'$$

      $$x_1+ x_2\xtworightarrow{e_1'[m]} x_1+x_2' \quad y_1+ y_2\xtworightarrow{e_2'[n]} y_1+y_2'$$

      with the assumptions $(C(x_1+ x_2'),f[e_1'\mapsto e_2'],C(y_1+ y_2'))\in\sim_{hp}^{fr}$, so, we get $x_1+ x_2\sim_{hp}^{fr} y_1+ y_2$, as desired.

      Or, we can get

      $$x_1+ x_2\xrightarrow{e_1} x_1'+x_2' \quad y_1+ y_2\xrightarrow{e_2} y_1'+y_2'$$

      $$x_1+ x_2\xtworightarrow{e_1[m]} x_1'+x_2' \quad y_1+ y_2\xtworightarrow{e_2[n]} y_1'+y_2'$$

      with the assumptions $(C(x_1'+ x_2'),f[e_1\mapsto e_2],C(y_1'+ y_2'))\in\sim_{hp}^{fr}$, so, we get $x_1+ x_2\sim_{hp}^{fr} y_1+ y_2$, as desired.
\end{itemize}
\end{proof}

\begin{theorem}[Soundness of $BRATC$ modulo FR hp-bisimulation equivalence]\label{SBRATCHPBE}
Let $x$ and $y$ be $BRATC$ terms. If $BRATC\vdash x=y$, then $x\sim_{hp}^{fr} y$.
\end{theorem}

\begin{proof}
Since FR hp-bisimulation $\sim_{hp}^{fr}$ is both an equivalent and a congruent relation, we only need to check if each axiom in Table \ref{AxiomsForBRATC} is sound modulo FR hp-bisimulation equivalence.

\begin{itemize}
  \item \textbf{Axiom $A1$}. Let $p,q$ be $BRATC$ processes, and $p+ q=q+ p$, it is sufficient to prove that $p+ q\sim_{hp}^{fr} q+ p$. By the forward hp-transition rules for operator $+$ in Table \ref{SETRForBRATC}, we get

      $$\frac{p\xrightarrow{e_1}e_1[m]}{p+ q\xrightarrow{e_1}e_1[m]+q} (e_1\in p,e_1\notin q) \quad \frac{p\xrightarrow{e_1}e_1[m]}{q+ p\xrightarrow{e_1}q+e_1[m]}(e_1\in p,e_1\notin q)$$

      $$\frac{p\xrightarrow{e_1}p'}{p+ q\xrightarrow{e_1}p'+q}(e_1\in p,e_1\notin q) \quad \frac{p\xrightarrow{e_1}p'}{q+ p\xrightarrow{e_1}q+p'}(e_1\in p,e_1\notin q)$$

      $$\frac{q\xrightarrow{e_2}e_2[n]}{p+ q\xrightarrow{e_2}p+e_2[n]}(e_2\in q,e_2\notin p) \quad \frac{q\xrightarrow{e_2}e_2[n]}{q+ p\xrightarrow{e_2}e_2[n]+p}(e_2\in q,e_2\notin p)$$

      $$\frac{q\xrightarrow{e_2}q'}{p+ q\xrightarrow{e_2}p+q'}(e_2\in q,e_2\notin p) \quad \frac{q\xrightarrow{e_2}q'}{q+ p\xrightarrow{e_2}q'+p}(e_2\in q, e_2\notin p)$$

      $$\frac{p\xrightarrow{e_1}p'\quad q\xrightarrow{e_1}q'}{p+ q\xrightarrow{e_1}p'+q'}(e_1\in p,e_1\in q) \quad \frac{p\xrightarrow{e_1}p'\quad q\xrightarrow{e_1}q'}{q+ p\xrightarrow{e_1}q'+p'}(e_1\in p,e_1\in q)$$

      By the reverse hp-transition rules for operator $+$ in Table \ref{RSETRForBRATC}, we get

      $$\frac{p\xtworightarrow{e_1[m]}e_1\quad e_1\notin q}{p+ q\xrightarrow{e_1[m]}e_1+q} (e_1\in p) \quad \frac{p\xtworightarrow{e_1[m]}e_1\quad e_1\notin q}{q+ p\xtworightarrow{e_1[m]}q+e_1}(e_1\in p)$$

      $$\frac{p\xtworightarrow{e_1[m]}p'\quad e_1\notin q}{p+ q\xtworightarrow{e_1[m]}p'+q}(e_1\in p) \quad \frac{p\xtworightarrow{e_1[m]}p'\quad e_1\notin q}{q+ p\xtworightarrow{e_1[m]}q+p'}(e_1\in p)$$

      $$\frac{q\xtworightarrow{e_2[n]}e_2\quad e_2\notin p}{p+ q\xtworightarrow{e_2[n]}p+e_2}(e_2\in q) \quad \frac{q\xtworightarrow{e_2[n]}e_2\quad e_2\notin p}{q+ p\xtworightarrow{e_2[n]}e_2+p}(e_2\in q)$$

      $$\frac{q\xtworightarrow{e_2[n]}q'\quad e_2\notin p}{p+ q\xtworightarrow{e_2[n]}p+q'}(e_2\in q) \quad \frac{q\xtworightarrow{e_2[n]}q'\quad e_2\notin p}{q+ p\xtworightarrow{e_2[n]}q'+p}(e_2\in q)$$

      $$\frac{p\xtworightarrow{e_1[m]}p'\quad q\xtworightarrow{e_1[m]}q'}{p+ q\xtworightarrow{e_1[m]}p'+q'}(e_1\in p,e_1\in q) \quad \frac{p\xtworightarrow{e_1[m]}p'\quad q\xtworightarrow{e_1[m]}q'}{q+ p\xtworightarrow{e_1[m]}q'+p'}(e_1\in p,e_1\in q)$$

      With the assumptions $(C(e_1[m]+q),f[e_1\mapsto e_1],C(q+e_1[m]))\in\sim_{hp}^{fr}$, $C((e_1+q),f[e_1\mapsto e_1],C(q+e_1))\in\sim_{hp}^{fr}$, $(C(p+e_2[n]),f[e_2\mapsto e_2],C(e_2[n]+p))\in\sim_{hp}^{fr}$, $(C(p'+q),f[e_1\mapsto e_1],C(q+p'))\in\sim_{hp}^{fr}$, $(C(p+q'),f[e_2\mapsto e_2],C(q'+p))\in\sim_{hp}^{fr}$ and $(C(p'+q'),f[e_1\mapsto e_1],C(q'+p'))\in\sim_{hp}^{fr}$ so, $p+ q\sim_{hp}^{fr} q+ p$, as desired.
  \item \textbf{Axiom $A2$}. Let $p,q,s$ be $BRATC$ processes, and $(p+ q)+ s=p+ (q+ s)$, it is sufficient to prove that $(p+ q)+ s \sim_{hp}^{fr} p+ (q+ s)$. By the forward hp- transition rules for operator $+$ in Table \ref{SETRForBRATC}, we get

      $$\frac{p\xrightarrow{e_1}e_1[m]\quad e_1\notin q\quad e_1\notin s}{(p+ q)+ s\xrightarrow{e_1}(e_1[m]+q)+s} (e_1\in p) \quad \frac{p\xrightarrow{e_1}e_1[m]\quad e_1\notin q\quad e_1\notin s}{p+ (q+ s)\xrightarrow{e_1}e_1[m]+(q+s)}(e_1\in p)$$

      $$\frac{p\xrightarrow{e_1}p'\quad e_1\notin q\quad e_1\notin s}{(p+ q)+ s\xrightarrow{e_1}(p'+q)+s}(e_1\in p) \quad \frac{p\xrightarrow{e_1}p'\quad e_1\notin q\quad e_1\notin s}{p+ (q+ s)\xrightarrow{e_1}p'+(q+s)}(e_1\in p)$$

      $$\frac{q\xrightarrow{e_2}e_2[n]\quad e_2\notin p\quad e_2\notin s}{(p+ q)+ s\xrightarrow{e_2}(p+e_2[n])+s}(e_2\in q) \quad \frac{q\xrightarrow{e_2}e_2[n]\quad e_2\notin p\quad e_2\notin s}{p+ (q+ s)\xrightarrow{e_2}e_1+(e_2[n]+s)}(e_2\in q)$$

      $$\frac{q\xrightarrow{e_2}q'\quad e_2\notin p\quad e_2\notin s}{(p+ q)+ s\xrightarrow{e_2}(p+q')+s}(e_2\in q) \quad \frac{q\xrightarrow{e_2}q'\quad e_2\notin p\quad e_2\notin s}{p+ (q+ s)\xrightarrow{e_2}p+(q'+s)}(e_2\in q)$$

      $$\frac{s\xrightarrow{e_3}e_3[l]\quad e_3\notin p\quad e_3\notin q}{(p+ q)+ s\xrightarrow{e_3}(p+q)+e_3[l]}(e_3\in s) \quad \frac{s\xrightarrow{e_3}e_3[l]\quad e_3\notin p\quad e_3\notin q}{p+ (q+ s)\xrightarrow{e_3}p+(q+e_3[l])}(e_3\in s)$$

      $$\frac{s\xrightarrow{e_3}s'\quad e_3\notin p\quad e_3\notin q}{(p+ q)+ s\xrightarrow{e_3}(p+q)+s'}(e_3\in s) \quad \frac{s\xrightarrow{e_3}s'\quad e_3\notin p\quad e_3\notin q}{p+ (q+ s)\xrightarrow{e_3}p+(q+s')}(e_3\in s)$$

      $$\frac{p\xrightarrow{e_1}p'\quad q\xrightarrow{e_1}q'\quad s\xrightarrow{e_1}s'}{(p+ q)+ s\xrightarrow{e_1}(p'+q')+s'}(e_1\in p,e_1\in q,e_1\in s) \quad \frac{p\xrightarrow{e_1}p'\quad q\xrightarrow{e_1}q'\quad s\xrightarrow{e_1}s'}{p+ (q+ s)\xrightarrow{e_1}p'+(q'+s')}(e_1\in p,e_1\in q,e_1\in s)$$

      By the reverse hp-transition rules for operator $+$ in Table \ref{RSETRForBRATC}, we get

      $$\frac{p\xtworightarrow{e_1[m]}e_1\quad e_1\notin q\quad e_1\notin s}{(p+ q)+ s\xtworightarrow{e_1[m]}(e_1+q)+s} (e_1\in p) \quad \frac{p\xtworightarrow{e_1[m]}e_1\quad e_1\notin q\quad e_1\notin s}{p+ (q+ s)\xtworightarrow{e_1[m]}e_1+(q+s)}(e_1\in p)$$

      $$\frac{p\xtworightarrow{e_1[m]}p'\quad e_1\notin q\quad e_1\notin s}{(p+ q)+ s\xtworightarrow{e_1[m]}(p'+q)+s}(e_1\in p) \quad \frac{p\xtworightarrow{e_1[m]}p'\quad e_1\notin q\quad e_1\notin s}{p+ (q+ s)\xtworightarrow{e_1[m]}p'+(q+s)}(e_1\in p)$$

      $$\frac{q\xtworightarrow{e_2[n]}e_2\quad e_2\notin p\quad e_2\notin s}{(p+ q)+ s\xtworightarrow{e_2[n]}(p+e_2)+s}(e_2\in q) \quad \frac{q\xtworightarrow{e_2[n]}e_2\quad e_2\notin p\quad e_2\notin s}{p+ (q+ s)\xtworightarrow{e_2[n]}p+(e_2+s)}(e_2\in q)$$

      $$\frac{q\xtworightarrow{e_2[n]}q'\quad e_2\notin p\quad e_2\notin s}{(p+ q)+ s\xrightarrow{e_2[n]}(p+q')+s}(e_2\in q) \quad \frac{q\xtworightarrow{e_2[n]}q'\quad e_2\notin p\quad e_2\notin s}{p+ (q+ s)\xtworightarrow{e_2[n]}p+(q'+s)}(e_2\in q)$$

      $$\frac{s\xtworightarrow{e_3[l]}e_3\quad e_3\notin p\quad e_3\notin q}{(p+ q)+ s\xtworightarrow{e_3[l]}(p+q)+e_3}(e_3\in s) \quad \frac{s\xtworightarrow{e_3[l]}e_3\quad e_3\notin p\quad e_3\notin q}{p+ (q+ s)\xtworightarrow{e_3[l]}p+(q+e_3)}(e_3\in s)$$

      $$\frac{s\xtworightarrow{e_3[l]}s'\quad e_3\notin p\quad e_3\notin q}{(p+ q)+ s\xtworightarrow{e_3[l]}(p+q)+s'}(e_3\in s) \quad \frac{s\xtworightarrow{e_3[l]}s'\quad e_3\notin p\quad e_3\notin q}{p+ (q+ s)\xtworightarrow{e_3[l]}p+(q+s')}(e_3\in s)$$

      $$\frac{p\xtworightarrow{e_1[m]}p'\quad q\xtworightarrow{e_1[m]}q'\quad s\xtworightarrow{e_1[m]}s'}{(p+ q)+ s\xtworightarrow{e_1[m]}(p'+q')+s'}(e_1\in p,e_1\in q,e_1\in s) \quad \frac{p\xtworightarrow{e_1[m]}p'\quad q\xtworightarrow{e_1[m]}q'\quad s\xtworightarrow{e_1[m]}s'}{p+ (q+ s)\xtworightarrow{e_1[m]}p'+(q'+s')}(e_1\in p,e_1\in q,e_1\in s)$$

      with the assumptions $(C((e_1[m]+q)+s),f[e_1\mapsto e_1],C(e_1[m]+(q+s)))\in\sim_{hp}^{fr}$, $(C((e_1+q)+s),f[e_1\mapsto e_1],C(e_1+(q+s)))\in\sim_{hp}^{fr}$, $(C((p+e_2[n])+s),f[e_2\mapsto e_2],C(p+(e_2[n]+s)))\in\sim_{hp}^{fr}$, $(C((p+e_2)+s),f[e_2\mapsto e_2],C(p+(e_2+s)))\in\sim_{hp}^{fr}$, $(C((p+q)+e_3[l]),f[e_3\mapsto e_3],C(p+(q+e_3[l])))\in \sim_{hp}^{fr}$, $(C((p+q)+e_3\sim_{hp}^{fr}p+(q+e_3)$, $(p'+q)+s\sim_{hp}^{fr}p'+(q+s)$, $(C((p+q')+s),f[e_2\mapsto e_2],C(p+(q'+s)))\in\sim_{hp}^{fr}$, $(C((p+q)+s'),f[e_3\mapsto e_3],C(p+(q+s')))\in\sim_{hp}^{fr}$ and $(C((p'+q')+s'),f[e_1\mapsto e_1],C(p'+(q'+s')))\in\sim_{hp}^{fr}$ so, $(p+ q)+ s\sim_{hp}^{fr} p+ (q+ s)$, as desired.
  \item \textbf{Axiom $A3$}. Let $p$ be a $BRATC$ process, and $p+ p=p$, it is sufficient to prove that $p+ p\sim_{hp}^{fr} p$. By the forward hp-transition rules for operator $+$ in Table \ref{SETRForBRATC}, we get

      $$\frac{p\xrightarrow{e_1}e_1[m]}{p+ p\xrightarrow{e_1}e_1[m]+e_1[m]} (e_1\in p) \quad \frac{p\xrightarrow{e_1}e_1[m]}{p\xrightarrow{e_1}e_1[m]}(e_1\in p)$$

      $$\frac{p\xrightarrow{e_1}p'}{p+ p\xrightarrow{e_1}p'+p'}(e_1\in p) \quad \frac{p\xrightarrow{e_1}p'}{p\xrightarrow{e_1}p'}(e_1\in p)$$

      By the reverse hp-transition rules for operator $+$ in Table \ref{RSETRForBRATC}, we get

      $$\frac{p\xtworightarrow{e_1[m]}e_1}{p+ p\xtworightarrow{e_1[m]}e_1+e_1} (e_1\in p) \quad \frac{p\xtworightarrow{e_1[m]}e_1}{p\xtworightarrow{e_1[m]}e_1}(e_1\in p)$$

      $$\frac{p\xtworightarrow{e_1[m]}p'}{p+ p\xtworightarrow{e_1[m]}p'+p'}(e_1\in p) \quad \frac{p\xtworightarrow{e_1[m]}p'}{p\xtworightarrow{e_1[m]}p'}(e_1\in p)$$

      with the assumptions $(C(e_1[m]+e_1[m]),f[e_1\mapsto e_1],C(e_1[m]))\in\sim_{hp}^{fr}$, $(C(e_1+e_1),f[e_1\mapsto e_1],C(e_1))\in\sim_{hp}^{fr}$ and $(C(p'+p'),f[e_1\mapsto e_1],C(p'))\in\sim_{hp}^{fr}$, so, $p+ p\sim_{hp}^{fr} p$, as desired.
  \item \textbf{Axiom $A4$}. Let $p,q,s$ be $BRATC$ processes, and $p\cdot (q+ s)=p\cdot q + p\cdot s(\textrm{Std}(p), \textrm{Std}(q), \textrm{Std}(s))$, it is sufficient to prove that $p\cdot (q+ s)\sim_{hp}^{fr}p\cdot q + p\cdot s$. By the hp-transition rules for operators $+$ and $\cdot$ in Table \ref{SETRForBRATC}, we get

      $$\frac{p\xrightarrow{e_1}e_1[m]}{p\cdot (q+ s)\xrightarrow{e_1}e_1[m]\cdot(q+s)} (e_1\in p) \quad \frac{p\xrightarrow{e_1}e_1[m]}{p\cdot q + p\cdot s\xrightarrow{e_1}e_1[m]\cdot q +e_1[m]\cdot s}(e_1\in p)$$

      $$\frac{p\xrightarrow{e_1}p'}{p\cdot (q+ s)\xrightarrow{e_1}p'\cdot (q+s)}(e_1\in p) \quad \frac{p\xrightarrow{e_1}p'}{p\cdot q + p\cdot s\xrightarrow{e_1}p'\cdot q+p'\cdot s}(e_1\in p)$$

      By the reverse transition rules for operators $+$ and $\cdot$ in Table \ref{RSETRForBRATC}, there are no transition rules.

      with the assumptions $(C(e_1[m]\cdot(q+s)),f[e_1\mapsto e_1],C(e_1[m]\cdot q + e_1[m]\cdot s))\in\sim_{hp}^{fr}$, $(C(p'\cdot(q+s)),f[e_1\mapsto e_1],C(p'\cdot q + p'\cdot s))\in\sim_{hp}^{fr}$, so, $p\cdot (q+ s)\sim_{hp}^{fr}p\cdot q + p\cdot s(\textrm{Std}(p), \textrm{Std}(q), \textrm{Std}(s))$, as desired.
  \item \textbf{Axiom $RA4$}. Let $p,q,s$ be $BRATC$ processes, and $(q+ s)\cdot p =q\cdot p + s\cdot p(\textrm{NStd}(p), \textrm{NStd}(q), \textrm{NStd}(s))$, it is sufficient to prove that $(q+ s)\cdot p \sim_{hp}^{fr}q\cdot p + s\cdot p$. By the hp-transition rules for operators $+$ and $\cdot$ in Table \ref{SETRForBRATC}, there are no transition rules.

      By the reverse transition rules for operators $+$ and $\cdot$ in Table \ref{RSETRForBRATC}, we get

      $$\frac{p\xtworightarrow{e_1[m]}e_1}{(q+ s)\cdot p\xtworightarrow{e_1[m]}(q+s)\cdot e_1} (e_1\in p) \quad \frac{p\xtworightarrow{e_1[m]}e_1}{q\cdot p + s\cdot p\xtworightarrow{e_1[m]}q\cdot e_1 +s\cdot e_1 }(e_1\in p)$$

      $$\frac{p\xtworightarrow{e_1[m]}p'}{(q+ s)\cdot p \xtworightarrow{e_1[m]}(q+s\cdot p' )}(e_1\in p) \quad \frac{p\xtworightarrow{e_1[m]}p'}{q\cdot p + s\cdot p\xtworightarrow{e_1[m]}q\cdot p'+s\cdot p'}(e_1\in p)$$

      with the assumptions $(C((q+s)\cdot e_1),f[e_1\mapsto e_1],C( q\cdot e_1 + s\cdot e_1 ))\in\sim_{hp}^{fr}$, $(C((q+s)\cdot p'),f[e_1\mapsto e_1],C(q\cdot p' + s\cdot p'))\in\sim_{hp}^{fr}$, so, $ (q+ s)\cdot p\sim_{hp}^{fr}q\cdot p + s\cdot p(\textrm{NStd}(p), \textrm{NStd}(q), \textrm{NStd}(s))$, as desired.
  \item \textbf{Axiom $A5$}. Let $p,q,s$ be $BRATC$ processes, and $(p\cdot q)\cdot s=p\cdot (q\cdot s)$, it is sufficient to prove that $(p\cdot q)\cdot s \sim_{hp}^{fr} p\cdot (q\cdot s)$. By the forward hp-transition rules for operator $\cdot$ in Table \ref{SETRForBRATC}, we get

      $$\frac{p\xrightarrow{e_1}e_1[m]}{(p\cdot q)\cdot s\xrightarrow{e_1}(e_1[m]\cdot q)\cdot s} (e_1\in p) \quad \frac{p\xrightarrow{e_1}e_1[m]}{p\cdot (q\cdot s)\xrightarrow{e_1}e_1[m]\cdot(q\cdot s)}(e_1\in p)$$

      $$\frac{p\xrightarrow{e_1}p'}{(p\cdot q)\cdot s\xrightarrow{e_1}(p'\cdot q)\cdot s}(e_1\in p) \quad \frac{p\xrightarrow{e_1}p'}{p\cdot (q\cdot s)\xrightarrow{e_1}p'\cdot (q\cdot s)}(e_1\in p)$$

      By the reverse hp-transition rules for operator $\cdot$ in Table \ref{RSETRForBRATC}, we get

      $$\frac{s\xtworightarrow{e_3[l]}e_3}{(p\cdot q)\cdot s\xtworightarrow{e_3[l]}(p\cdot q)\cdot e_3} (e_3\in s) \quad \frac{s\xtworightarrow{e_3[l]}e_3}{p\cdot (q\cdot s)\xtworightarrow{e_3[l]}p\cdot(q\cdot e_3)}(e_3\in s)$$

      $$\frac{s\xtworightarrow{e_3[l]}s'}{(p\cdot q)\cdot s\xtworightarrow{e_3[l]}(p\cdot q)\cdot s'}(e_3\in s) \quad \frac{s\xtworightarrow{e_3[l]}s'}{p\cdot (q\cdot s)\xtworightarrow{e_3[l]}p\cdot (q\cdot s')}(e_3\in s)$$

      With assumptions $(C((e_1[m]\cdot q)\cdot s),f[e_1\mapsto e_1],C(e_1[m]\cdot(q\cdot s)))\in\sim_{hp}^{fr}$, $(C((p'\cdot q)\cdot s),f[e_1\mapsto e_1],C(p'\cdot(q\cdot s)))\in\sim_{hp}^{fr}$, $(C((p\cdot q)\cdot e_3),f[e_3\mapsto e_3],C(p\cdot(q\cdot e_3)))\in\sim_{hp}^{fr}$, $(C((p\cdot q)\cdot s'),f[e_3\mapsto e_3],C(p\cdot(q\cdot s')))\in\sim_{hp}^{fr}$, so, $(p\cdot q)\cdot s\sim_{hp}^{fr} p\cdot (q\cdot s)$, as desired.
\end{itemize}
\end{proof}

\begin{theorem}[Congruence of $BRATC$ with respect to FR hhp-bisimulation equivalence]
FR hhp-bisimulation equivalence $\sim_{hhp}^{fr}$ is a congruence with respect to $BRATC$.
\end{theorem}

\begin{proof}
It is easy to see that FR hhp-bisimulation is an equivalent relation on $BRATC$ terms, we only need to prove that $\sim_{hhp}^{fr}$ is preserved by the operators $\cdot$ and $+$.

\begin{itemize}
  \item Causality operator $\cdot$. Let $x_1,x_2$ and $y_1,y_2$ be $BRATC$ processes, and $x_1\sim_{hhp}^{fr} y_1$, $x_2\sim_{hhp}^{fr} y_2$, it is sufficient to prove that $x_1\cdot x_2\sim_{hhp}^{fr} y_1\cdot y_2$.

      By the definition of FR hhp-bisimulation $\sim_{hhp}^{fr}$ (Definition \ref{FRHHPB}), $x_1\sim_{hhp}^{fr} y_1$ means that there is a posetal relation $(C(x_1),f,C(y_1))\in\sim_{hhp}^{fr}$, and

      $$x_1\xrightarrow{e_1} x_1' \quad y_1\xrightarrow{e_2} y_1'$$

      $$x_1\xtworightarrow{e_1[m]} x_1' \quad y_1\xtworightarrow{e_2[m]} y_1'$$

      with $(C(x_1'),f[e_1\mapsto e_2],C(y_1'))\in\sim_{hhp}^{fr}$. The meaning of $x_2\sim_{hhp}^{fr} y_2$ is similar.

      By the FR hhp-transition rules for causality operator $\cdot$ in Table \ref{SETRForBRATC} and Table \ref{RSETRForBRATC}, we can get

      $$x_1\cdot x_2\xrightarrow{e_1} e_1[m]\cdot x_2 \quad y_1\cdot y_2\xrightarrow{e_2} e_2[n]\cdot y_2$$

      $$x_1\cdot x_2\xtworightarrow{e_1'[m} x_1\cdot e_1' \quad y_1\cdot y_2\xtworightarrow{e_2'[n} y_1\cdot e_2'$$

      with the assumptions $(C(e_1[m]\cdot x_2),f[e_1\mapsto e_2],C(e_2[n]\cdot y_2))\in\sim_{hhp}^{fr}$ and $(C(x_1\cdot e_1'),f[e_1'\mapsto e_2'],C(y_1\cdot e_2'))\in\sim_{hhp}^{fr}$, so, we get $x_1\cdot x_2\sim_{hhp}^{fr} y_1\cdot y_2$, as desired.

      Or, we can get

      $$x_1\cdot x_2\xrightarrow{e_1} x_1'\cdot x_2 \quad y_1\cdot y_2\xrightarrow{e_2} y_1'\cdot y_2$$

      $$x_1\cdot x_2\xtworightarrow{e1'[m]} x_1\cdot x_2' \quad y_1\cdot y_2\xtworightarrow{e_2'[n]} y_1\cdot y_2'$$

      with the assumptions $(C(x_1'\cdot x_2),f[e_1\mapsto e_2],C(y_1'\cdot y_2))\in\sim_{hhp}^{fr}$ and $(C(x_1\cdot x_2'),f[e_1'\mapsto e_2'],C(y_1\cdot y_2'))\in\sim_{hhp}^{fr}$, so, we get $x_1\cdot x_2\sim_{hhp}^{fr} y_1\cdot y_2$, as desired.

  \item Conflict operator $+$. Let $x_1, x_2$ and $y_1, y_2$ be $BRATC$ processes, and $x_1\sim_{hhp}^{fr} y_1$, $x_2\sim_{hhp}^{fr} y_2$, it is sufficient to prove that $x_1+ x_2 \sim_{hhp}^{fr} y_1+ y_2$. The meanings of $x_1\sim_{hhp}^{fr} y_1$ and $x_2\sim_{hhp}^{fr} y_2$ are the same as the above case, according to the definition of FR hhp-bisimulation $\sim_{hp}^{fr}$ in Definition \ref{FRHHPB}.

      By the FR hhp-transition rules for conflict operator $+$ in Table \ref{SETRForBRATC} and Table \ref{RSETRForBRATC}, we can get several cases:

      $$x_1+ x_2\xrightarrow{e_1} e_1[m]+x_2 \quad y_1+ y_2\xrightarrow{e_2} e_2[n]+y_2$$
      $$x_1+ x_2\xtworightarrow{e_1[m]} e_1+x_2 \quad y_1+ y_2\xtworightarrow{e_2[n]} e_2+y_2$$

      with the assumptions $(C(e_1[m]+ x_2),f[e_1\mapsto e_2],C(e_2[n]+ y_2))\in\sim_{hhp}^{fr}$ and $(C(e_1+ x_2),f[e_1\mapsto e_2],C(e_2+ y_2))\in\sim_{hhp}^{fr}$, so, we get $x_1+ x_2\sim_{hhp}^{fr} y_1+ y_2$, as desired.

      Or, we can get

      $$x_1+ x_2\xrightarrow{e_1} x_1'+x_2 \quad y_1+ y_2\xrightarrow{e_2} y_1'+y_2$$

      $$x_1+ x_2\xtworightarrow{e_1[m]} x_1'+x_2 \quad y_1+ y_2\xtworightarrow{e_2[n]} y_1'+y_2$$

      with the assumptions $(C(x_1'+ x_2),f[e_1\mapsto e_2],C(y_1'+ y_2))\in\sim_{hhp}^{fr}$, so, we get $x_1+ x_2\sim_{hhp}^{fr} y_1+ y_2$, as desired.

      Or, we can get

      $$x_1+ x_2\xrightarrow{e_1'} x_1+e_1'[m] \quad y_1+ y_2\xrightarrow{e_2'} y_1+e_2'[n]$$

      $$x_1+ x_2\xrightarrow{e_1'[m]} x_1+e_1' \quad y_1+ y_2\xrightarrow{e_2'[n]} y_1+e_2'$$

      with the assumptions $(C(x_1 +e_1'[m]),f[e_1'\mapsto e_2'],C(y_1+e_2'[n]))\in\sim_{hhp}^{fr}$ and $(C(x_1+e_1'),f[e_1'\mapsto e_2'],C(y_1+e_2'))\in\sim_{hhp}^{fr}$, so, we get $x_1+ x_2\sim_{hhp}^{fr} y_1+ y_2$, as desired.

      Or, we can get

      $$x_1+ x_2\xrightarrow{e_1'} x_1+x_2' \quad y_1+ y_2\xrightarrow{e_2'} y_1+y_2'$$

      $$x_1+ x_2\xtworightarrow{e_1'[m]} x_1+x_2' \quad y_1+ y_2\xtworightarrow{e_2'[n]} y_1+y_2'$$

      with the assumptions $(C(x_1+ x_2'),f[e_1'\mapsto e_2'],C(y_1+ y_2'))\in\sim_{hhp}^{fr}$, so, we get $x_1+ x_2\sim_{hhp}^{fr} y_1+ y_2$, as desired.

      Or, we can get

      $$x_1+ x_2\xrightarrow{e_1} x_1'+x_2' \quad y_1+ y_2\xrightarrow{e_2} y_1'+y_2'$$

      $$x_1+ x_2\xtworightarrow{e_1[m]} x_1'+x_2' \quad y_1+ y_2\xtworightarrow{e_2[n]} y_1'+y_2'$$

      with the assumptions $(C(x_1'+ x_2'),f[e_1\mapsto e_2],C(y_1'+ y_2'))\in\sim_{hhp}^{fr}$, so, we get $x_1+ x_2\sim_{hhp}^{fr} y_1+ y_2$, as desired.
\end{itemize}
\end{proof}

\begin{theorem}[Soundness of $BRATC$ modulo FR hhp-bisimulation equivalence]\label{SBRATCHHPBE}
Let $x$ and $y$ be $BRATC$ terms. If $BRATC\vdash x=y$, then $x\sim_{hhp}^{fr} y$.
\end{theorem}

\begin{proof}
Since FR hhp-bisimulation $\sim_{hhp}^{fr}$ is both an equivalent and a congruent relation, we only need to check if each axiom in Table \ref{AxiomsForBRATC} is sound modulo FR hhp-bisimulation equivalence.

\begin{itemize}
  \item \textbf{Axiom $A1$}. Let $p,q$ be $BRATC$ processes, and $p+ q=q+ p$, it is sufficient to prove that $p+ q\sim_{hhp}^{fr} q+ p$. By the forward hhp-transition rules for operator $+$ in Table \ref{SETRForBRATC}, we get

      $$\frac{p\xrightarrow{e_1}e_1[m]}{p+ q\xrightarrow{e_1}e_1[m]+q} (e_1\in p,e_1\notin q) \quad \frac{p\xrightarrow{e_1}e_1[m]}{q+ p\xrightarrow{e_1}q+e_1[m]}(e_1\in p,e_1\notin q)$$

      $$\frac{p\xrightarrow{e_1}p'}{p+ q\xrightarrow{e_1}p'+q}(e_1\in p,e_1\notin q) \quad \frac{p\xrightarrow{e_1}p'}{q+ p\xrightarrow{e_1}q+p'}(e_1\in p,e_1\notin q)$$

      $$\frac{q\xrightarrow{e_2}e_2[n]}{p+ q\xrightarrow{e_2}p+e_2[n]}(e_2\in q,e_2\notin p) \quad \frac{q\xrightarrow{e_2}e_2[n]}{q+ p\xrightarrow{e_2}e_2[n]+p}(e_2\in q,e_2\notin p)$$

      $$\frac{q\xrightarrow{e_2}q'}{p+ q\xrightarrow{e_2}p+q'}(e_2\in q,e_2\notin p) \quad \frac{q\xrightarrow{e_2}q'}{q+ p\xrightarrow{e_2}q'+p}(e_2\in q, e_2\notin p)$$

      $$\frac{p\xrightarrow{e_1}p'\quad q\xrightarrow{e_1}q'}{p+ q\xrightarrow{e_1}p'+q'}(e_1\in p,e_1\in q) \quad \frac{p\xrightarrow{e_1}p'\quad q\xrightarrow{e_1}q'}{q+ p\xrightarrow{e_1}q'+p'}(e_1\in p,e_1\in q)$$

      By the reverse hhp-transition rules for operator $+$ in Table \ref{RSETRForBRATC}, we get

      $$\frac{p\xtworightarrow{e_1[m]}e_1\quad e_1\notin q}{p+ q\xrightarrow{e_1[m]}e_1+q} (e_1\in p) \quad \frac{p\xtworightarrow{e_1[m]}e_1\quad e_1\notin q}{q+ p\xtworightarrow{e_1[m]}q+e_1}(e_1\in p)$$

      $$\frac{p\xtworightarrow{e_1[m]}p'\quad e_1\notin q}{p+ q\xtworightarrow{e_1[m]}p'+q}(e_1\in p) \quad \frac{p\xtworightarrow{e_1[m]}p'\quad e_1\notin q}{q+ p\xtworightarrow{e_1[m]}q+p'}(e_1\in p)$$

      $$\frac{q\xtworightarrow{e_2[n]}e_2\quad e_2\notin p}{p+ q\xtworightarrow{e_2[n]}p+e_2}(e_2\in q) \quad \frac{q\xtworightarrow{e_2[n]}e_2\quad e_2\notin p}{q+ p\xtworightarrow{e_2[n]}e_2+p}(e_2\in q)$$

      $$\frac{q\xtworightarrow{e_2[n]}q'\quad e_2\notin p}{p+ q\xtworightarrow{e_2[n]}p+q'}(e_2\in q) \quad \frac{q\xtworightarrow{e_2[n]}q'\quad e_2\notin p}{q+ p\xtworightarrow{e_2[n]}q'+p}(e_2\in q)$$

      $$\frac{p\xtworightarrow{e_1[m]}p'\quad q\xtworightarrow{e_1[m]}q'}{p+ q\xtworightarrow{e_1[m]}p'+q'}(e_1\in p,e_1\in q) \quad \frac{p\xtworightarrow{e_1[m]}p'\quad q\xtworightarrow{e_1[m]}q'}{q+ p\xtworightarrow{e_1[m]}q'+p'}(e_1\in p,e_1\in q)$$

      With the assumptions $(C(e_1[m]+q),f[e_1\mapsto e_1],C(q+e_1[m]))\in\sim_{hhp}^{fr}$, $C((e_1+q),f[e_1\mapsto e_1],C(q+e_1))\in\sim_{hhp}^{fr}$, $(C(p+e_2[n]),f[e_2\mapsto e_2],C(e_2[n]+p))\in\sim_{hhp}^{fr}$, $(C(p'+q),f[e_1\mapsto e_1],C(q+p'))\in\sim_{hhp}^{fr}$, $(C(p+q'),f[e_2\mapsto e_2],C(q'+p))\in\sim_{hhp}^{fr}$ and $(C(p'+q'),f[e_1\mapsto e_1],C(q'+p'))\in\sim_{hhp}^{fr}$ so, $p+ q\sim_{hhp}^{fr} q+ p$, as desired.
  \item \textbf{Axiom $A2$}. Let $p,q,s$ be $BRATC$ processes, and $(p+ q)+ s=p+ (q+ s)$, it is sufficient to prove that $(p+ q)+ s \sim_{hhp}^{fr} p+ (q+ s)$. By the forward hhp- transition rules for operator $+$ in Table \ref{SETRForBRATC}, we get

      $$\frac{p\xrightarrow{e_1}e_1[m]\quad e_1\notin q\quad e_1\notin s}{(p+ q)+ s\xrightarrow{e_1}(e_1[m]+q)+s} (e_1\in p) \quad \frac{p\xrightarrow{e_1}e_1[m]\quad e_1\notin q\quad e_1\notin s}{p+ (q+ s)\xrightarrow{e_1}e_1[m]+(q+s)}(e_1\in p)$$

      $$\frac{p\xrightarrow{e_1}p'\quad e_1\notin q\quad e_1\notin s}{(p+ q)+ s\xrightarrow{e_1}(p'+q)+s}(e_1\in p) \quad \frac{p\xrightarrow{e_1}p'\quad e_1\notin q\quad e_1\notin s}{p+ (q+ s)\xrightarrow{e_1}p'+(q+s)}(e_1\in p)$$

      $$\frac{q\xrightarrow{e_2}e_2[n]\quad e_2\notin p\quad e_2\notin s}{(p+ q)+ s\xrightarrow{e_2}(p+e_2[n])+s}(e_2\in q) \quad \frac{q\xrightarrow{e_2}e_2[n]\quad e_2\notin p\quad e_2\notin s}{p+ (q+ s)\xrightarrow{e_2}e_1+(e_2[n]+s)}(e_2\in q)$$

      $$\frac{q\xrightarrow{e_2}q'\quad e_2\notin p\quad e_2\notin s}{(p+ q)+ s\xrightarrow{e_2}(p+q')+s}(e_2\in q) \quad \frac{q\xrightarrow{e_2}q'\quad e_2\notin p\quad e_2\notin s}{p+ (q+ s)\xrightarrow{e_2}p+(q'+s)}(e_2\in q)$$

      $$\frac{s\xrightarrow{e_3}e_3[l]\quad e_3\notin p\quad e_3\notin q}{(p+ q)+ s\xrightarrow{e_3}(p+q)+e_3[l]}(e_3\in s) \quad \frac{s\xrightarrow{e_3}e_3[l]\quad e_3\notin p\quad e_3\notin q}{p+ (q+ s)\xrightarrow{e_3}p+(q+e_3[l])}(e_3\in s)$$

      $$\frac{s\xrightarrow{e_3}s'\quad e_3\notin p\quad e_3\notin q}{(p+ q)+ s\xrightarrow{e_3}(p+q)+s'}(e_3\in s) \quad \frac{s\xrightarrow{e_3}s'\quad e_3\notin p\quad e_3\notin q}{p+ (q+ s)\xrightarrow{e_3}p+(q+s')}(e_3\in s)$$

      $$\frac{p\xrightarrow{e_1}p'\quad q\xrightarrow{e_1}q'\quad s\xrightarrow{e_1}s'}{(p+ q)+ s\xrightarrow{e_1}(p'+q')+s'}(e_1\in p,e_1\in q,e_1\in s) \quad \frac{p\xrightarrow{e_1}p'\quad q\xrightarrow{e_1}q'\quad s\xrightarrow{e_1}s'}{p+ (q+ s)\xrightarrow{e_1}p'+(q'+s')}(e_1\in p,e_1\in q,e_1\in s)$$

      By the reverse hhp-transition rules for operator $+$ in Table \ref{RSETRForBRATC}, we get

      $$\frac{p\xtworightarrow{e_1[m]}e_1\quad e_1\notin q\quad e_1\notin s}{(p+ q)+ s\xtworightarrow{e_1[m]}(e_1+q)+s} (e_1\in p) \quad \frac{p\xtworightarrow{e_1[m]}e_1\quad e_1\notin q\quad e_1\notin s}{p+ (q+ s)\xtworightarrow{e_1[m]}e_1+(q+s)}(e_1\in p)$$

      $$\frac{p\xtworightarrow{e_1[m]}p'\quad e_1\notin q\quad e_1\notin s}{(p+ q)+ s\xtworightarrow{e_1[m]}(p'+q)+s}(e_1\in p) \quad \frac{p\xtworightarrow{e_1[m]}p'\quad e_1\notin q\quad e_1\notin s}{p+ (q+ s)\xtworightarrow{e_1[m]}p'+(q+s)}(e_1\in p)$$

      $$\frac{q\xtworightarrow{e_2[n]}e_2\quad e_2\notin p\quad e_2\notin s}{(p+ q)+ s\xtworightarrow{e_2[n]}(p+e_2)+s}(e_2\in q) \quad \frac{q\xtworightarrow{e_2[n]}e_2\quad e_2\notin p\quad e_2\notin s}{p+ (q+ s)\xtworightarrow{e_2[n]}p+(e_2+s)}(e_2\in q)$$

      $$\frac{q\xtworightarrow{e_2[n]}q'\quad e_2\notin p\quad e_2\notin s}{(p+ q)+ s\xrightarrow{e_2[n]}(p+q')+s}(e_2\in q) \quad \frac{q\xtworightarrow{e_2[n]}q'\quad e_2\notin p\quad e_2\notin s}{p+ (q+ s)\xtworightarrow{e_2[n]}p+(q'+s)}(e_2\in q)$$

      $$\frac{s\xtworightarrow{e_3[l]}e_3\quad e_3\notin p\quad e_3\notin q}{(p+ q)+ s\xtworightarrow{e_3[l]}(p+q)+e_3}(e_3\in s) \quad \frac{s\xtworightarrow{e_3[l]}e_3\quad e_3\notin p\quad e_3\notin q}{p+ (q+ s)\xtworightarrow{e_3[l]}p+(q+e_3)}(e_3\in s)$$

      $$\frac{s\xtworightarrow{e_3[l]}s'\quad e_3\notin p\quad e_3\notin q}{(p+ q)+ s\xtworightarrow{e_3[l]}(p+q)+s'}(e_3\in s) \quad \frac{s\xtworightarrow{e_3[l]}s'\quad e_3\notin p\quad e_3\notin q}{p+ (q+ s)\xtworightarrow{e_3[l]}p+(q+s')}(e_3\in s)$$

      $$\frac{p\xtworightarrow{e_1[m]}p'\quad q\xtworightarrow{e_1[m]}q'\quad s\xtworightarrow{e_1[m]}s'}{(p+ q)+ s\xtworightarrow{e_1[m]}(p'+q')+s'}(e_1\in p,e_1\in q,e_1\in s) \quad \frac{p\xtworightarrow{e_1[m]}p'\quad q\xtworightarrow{e_1[m]}q'\quad s\xtworightarrow{e_1[m]}s'}{p+ (q+ s)\xtworightarrow{e_1[m]}p'+(q'+s')}(e_1\in p,e_1\in q,e_1\in s)$$

      with the assumptions $(C((e_1[m]+q)+s),f[e_1\mapsto e_1],C(e_1[m]+(q+s)))\in\sim_{hhp}^{fr}$, $(C((e_1+q)+s),f[e_1\mapsto e_1],C(e_1+(q+s)))\in\sim_{hhp}^{fr}$, $(C((p+e_2[n])+s),f[e_2\mapsto e_2],C(p+(e_2[n]+s)))\in\sim_{hhp}^{fr}$, $(C((p+e_2)+s),f[e_2\mapsto e_2],C(p+(e_2+s)))\in\sim_{hhp}^{fr}$, $(C((p+q)+e_3[l]),f[e_3\mapsto e_3],C(p+(q+e_3[l])))\in \sim_{hhp}^{fr}$, $(C((p+q)+e_3\sim_{hhp}^{fr}p+(q+e_3)$, $(p'+q)+s\sim_{hhp}^{fr}p'+(q+s)$, $(C((p+q')+s),f[e_2\mapsto e_2],C(p+(q'+s)))\in\sim_{hhp}^{fr}$, $(C((p+q)+s'),f[e_3\mapsto e_3],C(p+(q+s')))\in\sim_{hhp}^{fr}$ and $(C((p'+q')+s'),f[e_1\mapsto e_1],C(p'+(q'+s')))\in\sim_{hhp}^{fr}$ so, $(p+ q)+ s\sim_{hhp}^{fr} p+ (q+ s)$, as desired.
  \item \textbf{Axiom $A3$}. Let $p$ be a $BRATC$ process, and $p+ p=p$, it is sufficient to prove that $p+ p\sim_{hhp}^{fr} p$. By the forward hhp-transition rules for operator $+$ in Table \ref{SETRForBRATC}, we get

      $$\frac{p\xrightarrow{e_1}e_1[m]}{p+ p\xrightarrow{e_1}e_1[m]+e_1[m]} (e_1\in p) \quad \frac{p\xrightarrow{e_1}e_1[m]}{p\xrightarrow{e_1}e_1[m]}(e_1\in p)$$

      $$\frac{p\xrightarrow{e_1}p'}{p+ p\xrightarrow{e_1}p'+p'}(e_1\in p) \quad \frac{p\xrightarrow{e_1}p'}{p\xrightarrow{e_1}p'}(e_1\in p)$$

      By the reverse hhp-transition rules for operator $+$ in Table \ref{RSETRForBRATC}, we get

      $$\frac{p\xtworightarrow{e_1[m]}e_1}{p+ p\xtworightarrow{e_1[m]}e_1+e_1} (e_1\in p) \quad \frac{p\xtworightarrow{e_1[m]}e_1}{p\xtworightarrow{e_1[m]}e_1}(e_1\in p)$$

      $$\frac{p\xtworightarrow{e_1[m]}p'}{p+ p\xtworightarrow{e_1[m]}p'+p'}(e_1\in p) \quad \frac{p\xtworightarrow{e_1[m]}p'}{p\xtworightarrow{e_1[m]}p'}(e_1\in p)$$

      with the assumptions $(C(e_1[m]+e_1[m]),f[e_1\mapsto e_1],C(e_1[m]))\in\sim_{hhp}^{fr}$, $(C(e_1+e_1),f[e_1\mapsto e_1],C(e_1))\in\sim_{hhp}^{fr}$ and $(C(p'+p'),f[e_1\mapsto e_1],C(p'))\in\sim_{hhp}^{fr}$, so, $p+ p\sim_{hhp}^{fr} p$, as desired.
  \item \textbf{Axiom $A4$}. Let $p,q,s$ be $BRATC$ processes, and $p\cdot (q+ s)=p\cdot q + p\cdot s(\textrm{Std}(p), \textrm{Std}(q), \textrm{Std}(s))$, it is sufficient to prove that $p\cdot (q+ s)\sim_{hhp}^{fr}p\cdot q + p\cdot s$. By the hhp-transition rules for operators $+$ and $\cdot$ in Table \ref{SETRForBRATC}, we get

      $$\frac{p\xrightarrow{e_1}e_1[m]}{p\cdot (q+ s)\xrightarrow{e_1}e_1[m]\cdot(q+s)} (e_1\in p) \quad \frac{p\xrightarrow{e_1}e_1[m]}{p\cdot q + p\cdot s\xrightarrow{e_1}e_1[m]\cdot q +e_1[m]\cdot s}(e_1\in p)$$

      $$\frac{p\xrightarrow{e_1}p'}{p\cdot (q+ s)\xrightarrow{e_1}p'\cdot (q+s)}(e_1\in p) \quad \frac{p\xrightarrow{e_1}p'}{p\cdot q + p\cdot s\xrightarrow{e_1}p'\cdot q+p'\cdot s}(e_1\in p)$$

      By the reverse transition rules for operators $+$ and $\cdot$ in Table \ref{RSETRForBRATC}, there are no transition rules.

      with the assumptions $(C(e_1[m]\cdot(q+s)),f[e_1\mapsto e_1],C(e_1[m]\cdot q + e_1[m]\cdot s))\in\sim_{hhp}^{fr}$, $(C(p'\cdot(q+s)),f[e_1\mapsto e_1],C(p'\cdot q + p'\cdot s))\in\sim_{hhp}^{fr}$, so, $p\cdot (q+ s)\sim_{hhp}^{fr}p\cdot q + p\cdot s(\textrm{Std}(p), \textrm{Std}(q), \textrm{Std}(s))$, as desired.
  \item \textbf{Axiom $RA4$}. Let $p,q,s$ be $BRATC$ processes, and $(q+ s)\cdot p =q\cdot p + s\cdot p(\textrm{NStd}(p), \textrm{NStd}(q), \textrm{NStd}(s))$, it is sufficient to prove that $(q+ s)\cdot p \sim_{hhp}^{fr}q\cdot p + s\cdot p$. By the hhp-transition rules for operators $+$ and $\cdot$ in Table \ref{SETRForBRATC}, there are no transition rules.

      By the reverse transition rules for operators $+$ and $\cdot$ in Table \ref{RSETRForBRATC}, we get

      $$\frac{p\xtworightarrow{e_1[m]}e_1}{(q+ s)\cdot p\xtworightarrow{e_1[m]}(q+s)\cdot e_1} (e_1\in p) \quad \frac{p\xtworightarrow{e_1[m]}e_1}{q\cdot p + s\cdot p\xtworightarrow{e_1[m]}q\cdot e_1 +s\cdot e_1 }(e_1\in p)$$

      $$\frac{p\xtworightarrow{e_1[m]}p'}{(q+ s)\cdot p \xtworightarrow{e_1[m]}(q+s\cdot p' )}(e_1\in p) \quad \frac{p\xtworightarrow{e_1[m]}p'}{q\cdot p + s\cdot p\xtworightarrow{e_1[m]}q\cdot p'+s\cdot p'}(e_1\in p)$$

      with the assumptions $(C((q+s)\cdot e_1),f[e_1\mapsto e_1],C( q\cdot e_1 + s\cdot e_1 ))\in\sim_{hhp}^{fr}$, $(C((q+s)\cdot p'),f[e_1\mapsto e_1],C(q\cdot p' + s\cdot p'))\in\sim_{hhp}^{fr}$, so, $ (q+ s)\cdot p\sim_{hhp}^{fr}q\cdot p + s\cdot p(\textrm{NStd}(p), \textrm{NStd}(q), \textrm{NStd}(s))$, as desired.
  \item \textbf{Axiom $A5$}. Let $p,q,s$ be $BRATC$ processes, and $(p\cdot q)\cdot s=p\cdot (q\cdot s)$, it is sufficient to prove that $(p\cdot q)\cdot s \sim_{hhp}^{fr} p\cdot (q\cdot s)$. By the forward hhp-transition rules for operator $\cdot$ in Table \ref{SETRForBRATC}, we get

      $$\frac{p\xrightarrow{e_1}e_1[m]}{(p\cdot q)\cdot s\xrightarrow{e_1}(e_1[m]\cdot q)\cdot s} (e_1\in p) \quad \frac{p\xrightarrow{e_1}e_1[m]}{p\cdot (q\cdot s)\xrightarrow{e_1}e_1[m]\cdot(q\cdot s)}(e_1\in p)$$

      $$\frac{p\xrightarrow{e_1}p'}{(p\cdot q)\cdot s\xrightarrow{e_1}(p'\cdot q)\cdot s}(e_1\in p) \quad \frac{p\xrightarrow{e_1}p'}{p\cdot (q\cdot s)\xrightarrow{e_1}p'\cdot (q\cdot s)}(e_1\in p)$$

      By the reverse hhp-transition rules for operator $\cdot$ in Table \ref{RSETRForBRATC}, we get

      $$\frac{s\xtworightarrow{e_3[l]}e_3}{(p\cdot q)\cdot s\xtworightarrow{e_3[l]}(p\cdot q)\cdot e_3} (e_3\in s) \quad \frac{s\xtworightarrow{e_3[l]}e_3}{p\cdot (q\cdot s)\xtworightarrow{e_3[l]}p\cdot(q\cdot e_3)}(e_3\in s)$$

      $$\frac{s\xtworightarrow{e_3[l]}s'}{(p\cdot q)\cdot s\xtworightarrow{e_3[l]}(p\cdot q)\cdot s'}(e_3\in s) \quad \frac{s\xtworightarrow{e_3[l]}s'}{p\cdot (q\cdot s)\xtworightarrow{e_3[l]}p\cdot (q\cdot s')}(e_3\in s)$$

      With assumptions $(C((e_1[m]\cdot q)\cdot s),f[e_1\mapsto e_1],C(e_1[m]\cdot(q\cdot s)))\in\sim_{hhp}^{fr}$, $(C((p'\cdot q)\cdot s),f[e_1\mapsto e_1],C(p'\cdot(q\cdot s)))\in\sim_{hhp}^{fr}$, $(C((p\cdot q)\cdot e_3),f[e_3\mapsto e_3],C(p\cdot(q\cdot e_3)))\in\sim_{hhp}^{fr}$, $(C((p\cdot q)\cdot s'),f[e_3\mapsto e_3],C(p\cdot(q\cdot s')))\in\sim_{hhp}^{fr}$, so, $(p\cdot q)\cdot s\sim_{hhp}^{fr} p\cdot (q\cdot s)$, as desired.
\end{itemize}
\end{proof}

\section{Reversible Algebra for Parallelism in True Concurrency}\label{raptc}

In this section, we will discuss reversible parallelism in true concurrency. The resulted algebra is called Reversible Algebra for Parallelism in True Concurrency, abbreviated $RAPTC$.

\subsection{Parallelism}

The forward transition rules for parallelism $\parallel$ are shown in Table \ref{TRForParallel}, and the reverse transition rules for $\parallel$ are shown in Table \ref{RTRForParallel}.

\begin{center}
    \begin{table}
        $$\frac{x\xrightarrow{e_1}e_1[m]\quad y\xrightarrow{e_2}e_2[m]}{x\parallel y\xrightarrow{\{e_1,e_2\}}e_1[m]\parallel e_2[m]} \quad\frac{x\xrightarrow{e_1}x'\quad y\xrightarrow{e_2}e_2[m]}{x\parallel y\xrightarrow{\{e_1,e_2\}}x'\parallel e_2[m]}$$
        $$\frac{x\xrightarrow{e_1}e_1[m]\quad y\xrightarrow{e_2}y'}{x\parallel y\xrightarrow{\{e_1,e_2\}}e_1[m]\parallel y'} \quad\frac{x\xrightarrow{e_1}x'\quad y\xrightarrow{e_2}y'}{x\parallel y\xrightarrow{\{e_1,e_2\}}x'\between y'}$$
        \caption{Forward transition rules of parallel operator $\parallel$}
        \label{TRForParallel}
    \end{table}
\end{center}

\begin{center}
    \begin{table}
        $$\frac{x\xtworightarrow{e_1[m]}e_1\quad y\xtworightarrow{e_2[m]}e_2}{x\parallel y\xtworightarrow{\{e_1[m],e_2[m]\}}e_1\parallel e_2} \quad\frac{x\xtworightarrow{e_1[m]}x'\quad y\xtworightarrow{e_2[m]}e_2}{x\parallel y\xtworightarrow{\{e_1[m],e_2[m]\}}x'\parallel e_2}$$
        $$\frac{x\xtworightarrow{e_1[m]}e_1\quad y\xtworightarrow{e_2[m]}y'}{x\parallel y\xtworightarrow{\{e_1[m],e_2[m]\}}e_1\parallel y'} \quad\frac{x\xtworightarrow{e_1[m]}x'\quad y\xtworightarrow{e_2[m]}y'}{x\parallel y\xtworightarrow{\{e_1[m],e_2[m]\}}x'\between y'}$$
        \caption{Reverse transition rules of parallel operator $\parallel$}
        \label{RTRForParallel}
    \end{table}
\end{center}

The forward and reverse transition rules of communication $\mid$ are shown in Table \ref{TRForCommunication} and Table \ref{RTRForCommunication}.

\begin{center}
    \begin{table}
        $$\frac{x\xrightarrow{e_1}e_1[m]\quad y\xrightarrow{e_2}e_2[m]}{x\mid y\xrightarrow{\gamma(e_1,e_2)}\gamma(e_1,e_2)[m]} \quad\frac{x\xrightarrow{e_1}x'\quad y\xrightarrow{e_2}e_2[m]}{x\mid y\xrightarrow{\gamma(e_1,e_2)}\gamma(e_1,e_2)[m]\cdot x'}$$
        $$\frac{x\xrightarrow{e_1}e_1[m]\quad y\xrightarrow{e_2}y'}{x\mid y\xrightarrow{\gamma(e_1,e_2)}\gamma(e_1,e_2)[m]\cdot y'} \quad\frac{x\xrightarrow{e_1}x'\quad y\xrightarrow{e_2}y'}{x\mid y\xrightarrow{\gamma(e_1,e_2)}\gamma(e_1,e_2)[m]\cdot x'\between y'}$$
        \caption{Forward transition rules of communication operator $\mid$}
        \label{TRForCommunication}
    \end{table}
\end{center}

\begin{center}
    \begin{table}
        $$\frac{x\xtworightarrow{e_1[m]}e_1\quad y\xtworightarrow{e_2[m]}e_2}{x\mid y\xtworightarrow{\gamma(e_1,e_2)[m]}\gamma(e_1,e_2)} \quad\frac{x\xtworightarrow{e_1[m]}x'\quad y\xtworightarrow{e_2[m]}e_2}{x\mid y\xtworightarrow{\gamma(e_1,e_2)[m]}\gamma(e_1,e_2)\cdot x'}$$
        $$\frac{x\xtworightarrow{e_1[m]}e_1\quad y\xtworightarrow{e_2[m]}y'}{x\mid y\xtworightarrow{\gamma(e_1,e_2)[m]}\gamma(e_1,e_2)\cdot y'} \quad\frac{x\xtworightarrow{e_1[m]}x'\quad y\xtworightarrow{e_2[m]}y'}{x\mid y\xtworightarrow{\gamma(e_1,e_2)[m]}\gamma(e_1,e_2)\cdot x'\between y'}$$
        \caption{Reverse transition rules of communication operator $\mid$}
        \label{RTRForCommunication}
    \end{table}
\end{center}

The conflict elimination is also captured by two auxiliary operators, the unary conflict elimination operator $\Theta$ and the binary unless operator $\triangleleft$. The forward and reverse transition rules for $\Theta$ and $\triangleleft$ are expressed by ten transition rules in Table \ref{TRForConflict} and Table \ref{RTRForConflict}.

\begin{center}
    \begin{table}
        $$\frac{x\xrightarrow{e_1}e_1[m]\quad (\sharp(e_1,e_2))}{\Theta(x)\xrightarrow{e_1}e_1[m]} \quad\frac{x\xrightarrow{e_2}e_2[n]\quad (\sharp(e_1,e_2))}{\Theta(x)\xrightarrow{e_2}e_2[n]}$$
        $$\frac{x\xrightarrow{e_1}x'\quad (\sharp(e_1,e_2))}{\Theta(x)\xrightarrow{e_1}\Theta(x')} \quad\frac{x\xrightarrow{e_2}x'\quad (\sharp(e_1,e_2))}{\Theta(x)\xrightarrow{e_2}\Theta(x')}$$
        $$\frac{x\xrightarrow{e_1}e_1[m] \quad y\nrightarrow^{e_2}\quad (\sharp(e_1,e_2))}{x\triangleleft y\xrightarrow{\tau}\surd}
        \quad\frac{x\xrightarrow{e_1}x' \quad y\nrightarrow^{e_2}\quad (\sharp(e_1,e_2))}{x\triangleleft y\xrightarrow{\tau}x'}$$
        $$\frac{x\xrightarrow{e_1}e_1[m] \quad y\nrightarrow^{e_3}\quad (\sharp(e_1,e_2),e_2\leq e_3)}{x\triangleleft y\xrightarrow{e_1}e_1[m]}
        \quad\frac{x\xrightarrow{e_1}x' \quad y\nrightarrow^{e_3}\quad (\sharp(e_1,e_2),e_2\leq e_3)}{x\triangleleft y\xrightarrow{e_1}x'}$$
        $$\frac{x\xrightarrow{e_3}e_3[l] \quad y\nrightarrow^{e_2}\quad (\sharp(e_1,e_2),e_1\leq e_3)}{x\triangleleft y\xrightarrow{\tau}\surd}
        \quad\frac{x\xrightarrow{e_3}x' \quad y\nrightarrow^{e_2}\quad (\sharp(e_1,e_2),e_1\leq e_3)}{x\triangleleft y\xrightarrow{\tau}x'}$$
        \caption{Forward transition rules of conflict elimination}
        \label{TRForConflict}
    \end{table}
\end{center}

\begin{center}
    \begin{table}
        $$\frac{x\xtworightarrow{e_1[m]}e_1\quad (\sharp(e_1,e_2))}{\Theta(x)\xtworightarrow{e_1[m]}e_1} \quad\frac{x\xtworightarrow{e_2[n]}e_2\quad (\sharp(e_1,e_2))}{\Theta(x)\xtworightarrow{e_2[n]}e_2}$$
        $$\frac{x\xtworightarrow{e_1[m]}x'\quad (\sharp(e_1,e_2))}{\Theta(x)\xtworightarrow{e_1[m]}\Theta(x')} \quad\frac{x\xtworightarrow{e_2[n]}x'\quad (\sharp(e_1,e_2))}{\Theta(x)\xtworightarrow{e_2[n]}\Theta(x')}$$
        $$\frac{x\xtworightarrow{e_1[m]}e_1 \quad y\xntworightarrow{e_2[n]}\quad (\sharp(e_1,e_2))}{x\triangleleft y\xtworightarrow{\tau}\surd}
        \quad\frac{x\xtworightarrow{e_1[m]}x' \quad y\xntworightarrow{e_2[n]}\quad (\sharp(e_1,e_2))}{x\triangleleft y\xtworightarrow{\tau}x'}$$
        $$\frac{x\xtworightarrow{e_1[m]}e_1 \quad y\xntworightarrow{e_3[l]}\quad (\sharp(e_1,e_2),e_2\geq e_3)}{x\triangleleft y\xtworightarrow{e_1[m]}e_1}
        \quad\frac{x\xtworightarrow{e_1[m]}x' \quad y\xntworightarrow{e_3[l]}\quad (\sharp(e_1,e_2),e_2\geq e_3)}{x\triangleleft y\xtworightarrow{e_1[m]}x'}$$
        $$\frac{x\xtworightarrow{e_3[l]}e_3 \quad y\xntworightarrow{e_2[n]}\quad (\sharp(e_1,e_2),e_1\geq e_3)}{x\triangleleft y\xtworightarrow{\tau}\surd}
        \quad\frac{x\xtworightarrow{e_3[l]}x' \quad y\xntworightarrow{e_2[n]}\quad (\sharp(e_1,e_2),e_1\geq e_3)}{x\triangleleft y\xtworightarrow{\tau}x'}$$
        \caption{Reverse transition rules of conflict elimination}
        \label{RTRForConflict}
    \end{table}
\end{center}

\begin{theorem}[Congruence theorem of $RAPTC$]
FR truly concurrent bisimulation equivalences $\sim_{p}^{fr}$, $\sim_s^{fr}$, $\sim_{hp}^{fr}$ and $\sim_{hhp}^{fr}$ are all congruences with respect to $RAPTC$.
\end{theorem}

\begin{proof}
(1) Case FR pomset bisimulation equivalence $\sim_p^{fr}$.

\begin{itemize}
  \item Case parallel operator $\parallel$. Let $x_1,x_2$ and $y_1,y_2$ be $RAPTC$ processes, and $x_1\sim_{p}^{fr} y_1$, $x_2\sim_{p}^{fr} y_2$, it is sufficient to prove that $x_1\parallel x_2\sim_{p}^{fr} y_1\parallel y_2$.

      By the definition of FR pomset bisimulation $\sim_p^{fr}$ (Definition \ref{FRPSB}), $x_1\sim_p^{fr} y_1$ means that

      $$x_1\xrightarrow{X_1} x_1' \quad y_1\xrightarrow{Y_1} y_1'$$

      with $X_1\subseteq x_1$, $Y_1\subseteq y_1$, $X_1\sim Y_1$ and $x_1'\sim_p^{fr} y_1'$. The meaning of $x_2\sim_p^{fr} y_2$ is similar.

      By the forward transition rules for parallel operator $\parallel$ in Table \ref{TRForParallel}, we can get

      $$x_1\parallel x_2\xrightarrow{\{X_1,X_2\}} X_1[\mathcal{K}]\parallel X_2[\mathcal{K}] \quad y_1\parallel y_2\xrightarrow{\{Y_1,Y_2\}} Y_1[\mathcal{J}]\parallel Y_2[\mathcal{J}]$$

      $$x_1\parallel x_2\xtworightarrow{\{X_1[\mathcal{K}],X_2[\mathcal{K}]\}} X_1\parallel X_2 \quad y_1\parallel y_2\xtworightarrow{\{Y_1[\mathcal{J}],Y_2[\mathcal{J}]\}} Y_1\parallel Y_2$$

      with $X_1\subseteq x_1$, $Y_1\subseteq y_1$, $X_2\subseteq x_2$, $Y_2\subseteq y_2$, $X_1\sim Y_1$ and $X_2\sim Y_2$, and the assumptions $X_1[\mathcal{K}\parallel X_2[\mathcal{K}]]\sim_p^{fr}Y_1[\mathcal{J}]\parallel Y_2[\mathcal{J}]$ and $X_1\parallel X_2\sim_p^{fr}Y_1\parallel Y_2$, so, we get $x_1\parallel x_2\sim_p^{fr} y_1\parallel y_2$, as desired.

      Or, we can get

      $$x_1\parallel x_2\xrightarrow{\{X_1,X_2\}} x_1'\parallel X_2[\mathcal{K}] \quad y_1\parallel y_2\xrightarrow{\{Y_1,Y_2\}} y_1'\parallel Y_2[\mathcal{J}]$$

      $$x_1\parallel x_2\xtworightarrow{\{X_1[\mathcal{K}],X_2[\mathcal{K}]\}} x_1'\parallel X_2 \quad y_1\parallel y_2\xtworightarrow{\{Y_1[\mathcal{J}],Y_2[\mathcal{J}]\}} y_1'\parallel Y_2$$

      with $X_1\subseteq x_1$, $Y_1\subseteq y_1$, $X_2\subseteq x_2$, $Y_2\subseteq y_2$, $X_1\sim Y_1$, $X_2\sim Y_2$, and the assumptions $x_1'\parallel X_2[\mathcal{K}]]\sim_p^{fr}y_1'\parallel Y_2[\mathcal{J}]$ and $x_1'\parallel X_2\sim_p^{fr}y_1'\parallel Y_2$ so, we get $x_1\parallel x_2\sim_p^{fr} y_1\parallel y_2$, as desired.

      Or, we can get

      $$x_1\parallel x_2\xrightarrow{\{X_1,X_2\}}X_1[\mathcal{K}]\parallel x_2' \quad y_1\parallel y_2\xrightarrow{\{Y_1,Y_2\}}Y_1[\mathcal{J}]\parallel y_2'$$

      $$x_1\parallel x_2\xtworightarrow{\{X_1[\mathcal{K}],X_2[\mathcal{K}]\}}X_1\parallel x_2' \quad y_1\parallel y_2\xtworightarrow{\{Y_1[\mathcal{J}],Y_2[\mathcal{J}]\}}Y_1\parallel y_2'$$

      with $X_1\subseteq x_1$, $Y_1\subseteq y_1$, $X_2\subseteq x_2$, $Y_2\subseteq y_2$, $X_1\sim Y_1$, $X_2\sim Y_2$, and the assumptions $X_1[\mathcal{K}\parallel x_2'\sim_p^{fr}Y_1[\mathcal{J}]\parallel y_2'$ and $X_1\parallel x_2'\sim_p^{fr}Y_1\parallel y_2'$, so, we get $x_1\parallel x_2\sim_p^{fr} y_1\parallel y_2$, as desired.

      Or, we can get

      $$x_1\parallel x_2\xrightarrow{\{X_1,X_2\}} x_1'\between x_2' \quad y_1\parallel y_2\xrightarrow{\{Y_1,Y_2\}} y_1'\between y_2'$$

      $$x_1\parallel x_2\xtworightarrow{\{X_1[\mathcal{K}],X_2[\mathcal{K}]\}} x_1'\between x_2' \quad y_1\parallel y_2\xtworightarrow{\{Y_1[\mathcal{J}],Y_2[\mathcal{J}]\}} y_1'\between y_2'$$

      with $X_1\subseteq x_1$, $Y_1\subseteq y_1$, $X_2\subseteq x_2$, $Y_2\subseteq y_2$, $X_1\sim Y_1$, $X_2\sim Y_2$, and the assumption $x_1'\between x_2'\sim_p^{fr}y_1'\between y_2'$, so, we get $x_1\parallel x_2\sim_p^{fr} y_1\parallel y_2$, as desired.

  \item Case communication operator $\mid$. It can be proved similarly to the case of parallel operator $\parallel$, we omit it. Note that, a communication is defined between two single communicating events.

  \item Case conflict elimination operator $\Theta$. It can be proved similarly to the above cases, we omit it. Note that the conflict elimination operator $\Theta$ is a unary operator.

  \item Case unless operator $\triangleleft$. It can be proved similarly to the case of parallel operator $\parallel$, we omit it. Note that, a conflict relation is defined between two single events.

\end{itemize}

(2) The cases of FR step bisimulation $\sim_s^{fr}$, FR hp-bisimulation $\sim_{hp}^{fr}$ and FR hhp-bisimulation $\sim_{hhp}^{fr}$ can be proven similarly, we omit them.
\end{proof}

\subsection{Axiom System of Parallelism}

\begin{definition}[Basic terms of $RAPTC$]\label{BTAPTC}
The set of basic terms of $RAPTC$, $\mathcal{B}(RAPTC)$, is inductively defined as follows:
\begin{enumerate}
  \item $\mathbb{E}\subset\mathcal{B}(RAPTC)$;
  \item if $e\in \mathbb{E}, t\in\mathcal{B}(RAPTC)$ then $e\cdot t\in\mathcal{B}(RAPTC)$;
  \item if $t,s\in\mathcal{B}(RAPTC)$ then $t+ s\in\mathcal{B}(RAPTC)$;
  \item if $t,s\in\mathcal{B}(RAPTC)$ then $t\parallel s\in\mathcal{B}(RAPTC)$.
\end{enumerate}
\end{definition}

We design the axioms of parallelism in Table \ref{AxiomsForParallelism}, including algebraic laws for parallel operator $\parallel$, communication operator $\mid$, conflict elimination operator $\Theta$ and unless operator $\triangleleft$, and also the whole parallel operator $\between$. Since the communication between two communicating events in different parallel branches may cause deadlock (a state of inactivity), which is caused by mismatch of two communicating events or the imperfectness of the communication channel. We introduce a new constant $\delta$ to denote the deadlock, and let the atomic event $e\in \mathbb{E}\cup\{\delta\}$.

\begin{center}
    \begin{table}
        \begin{tabular}{@{}ll@{}}
            \hline No. &Axiom\\
            $A6$ & $x+ \delta = x$\\
            $A7$ & $\delta\cdot x =\delta$\\
            $P1$ & $x\between y = x\parallel y + x\mid y$\\
            $P2$ & $x\parallel y = y \parallel x$\\
            $P3$ & $(x\parallel y)\parallel z = x\parallel (y\parallel z)$\\
            $P4$ & $e_1\parallel (e_2\cdot y) = (e_1\parallel e_2)\cdot y$\\
            $RP4$ & $e_1[m]\parallel (y\cdot e_2[m]) = y\cdot(e_1[m]\parallel e_2[m])$\\
            $P5$ & $(e_1\cdot x)\parallel e_2 = (e_1\parallel e_2)\cdot x$\\
            $RP5$ & $(x\cdot e_1[m])\parallel e_2[m] = x\cdot(e_1[m]\parallel e_2[m])$\\
            $P6$ & $(e_1\cdot x)\parallel (e_2\cdot y) = (e_1\parallel e_2)\cdot (x\between y)$\\
            $RP6$ & $(x\cdot e_1[m])\parallel (y\cdot e_2[m]) = (x\between y)\cdot(e_1[m]\parallel e_2[m])$\\
            $P7$ & $(x+ y)\parallel z = (x\parallel z)+ (y\parallel z)$\\
            $P8$ & $x\parallel (y+ z) = (x\parallel y)+ (x\parallel z)$\\
            $P9$ & $\delta\parallel x = \delta$\\
            $P10$ & $x\parallel \delta = \delta$\\
            $C11$ & $e_1\mid e_2 = \gamma(e_1,e_2)$\\
            $RC11$ & $e_1[m]\mid e_2[m] = \gamma(e_1,e_2)[m]$\\
            $C12$ & $e_1\mid (e_2\cdot y) = \gamma(e_1,e_2)\cdot y$\\
            $RC12$ & $e_1[m]\mid (y \cdot e_2[m]) =y\cdot \gamma(e_1,e_2)[m]$\\
            $C13$ & $(e_1\cdot x)\mid e_2 = \gamma(e_1,e_2)\cdot x$\\
            $RC13$ & $(x \cdot e_1[m])\mid e_2[m] =x\cdot \gamma(e_1,e_2)[m]$\\
            $C14$ & $(e_1\cdot x)\mid (e_2\cdot y) = \gamma(e_1,e_2)\cdot (x\between y)$\\
            $RC14$ & $(x \cdot e_1[m])\mid (y \cdot e_2[m]) =(x\between y)\cdot \gamma(e_1,e_2)[m]$\\
            $C15$ & $(x+ y)\mid z = (x\mid z) + (y\mid z)$\\
            $C16$ & $x\mid (y+ z) = (x\mid y)+ (x\mid z)$\\
            $C17$ & $\delta\mid x = \delta$\\
            $C18$ & $x\mid\delta = \delta$\\
            $CE19$ & $\Theta(e) = e$\\
            $RCE19$ & $\Theta(e[m]) = e[m]$\\
            $CE20$ & $\Theta(\delta) = \delta$\\
            $CE21$ & $\Theta(x+ y) = \Theta(x)\triangleleft y + \Theta(y)\triangleleft x$\\
            $CE22$ & $\Theta(x\cdot y)=\Theta(x)\cdot\Theta(y)$\\
            $CE23$ & $\Theta(x\parallel y) = ((\Theta(x)\triangleleft y)\parallel y)+ ((\Theta(y)\triangleleft x)\parallel x)$\\
            $CE24$ & $\Theta(x\mid y) = ((\Theta(x)\triangleleft y)\mid y)+ ((\Theta(y)\triangleleft x)\mid x)$\\
            $U25$ & $(\sharp(e_1,e_2))\quad e_1\triangleleft e_2 = \tau$\\
            $RU25$ & $(\sharp(e_1[m],e_2[n]))\quad e_1[m]\triangleleft e_2[n] = \tau$\\
            $U26$ & $(\sharp(e_1,e_2),e_2\leq e_3)\quad e_1\triangleleft e_3 = e_1$\\
            $RU26$ & $(\sharp(e_1[m],e_2[n]),e_2[n]\geq e_3[l])\quad e_1[m]\triangleleft e_3[l] = e_1[m]$\\
            $U27$ & $(\sharp(e_1,e_2),e_2\leq e_3)\quad e3\triangleleft e_1 = \tau$\\
            $RU27$ & $(\sharp(e_1[m],e_2[n]),e_2[n]\geq e_3[l])\quad e3[l]\triangleleft e_1[m] = \tau$\\
            $U28$ & $e\triangleleft \delta = e$\\
            $U29$ & $\delta \triangleleft e = \delta$\\
            $U30$ & $(x+ y)\triangleleft z = (x\triangleleft z)+ (y\triangleleft z)$\\
            $U31$ & $(x\cdot y)\triangleleft z = (x\triangleleft z)\cdot (y\triangleleft z)$\\
            $U32$ & $(x\parallel y)\triangleleft z = (x\triangleleft z)\parallel (y\triangleleft z)$\\
            $U33$ & $(x\mid y)\triangleleft z = (x\triangleleft z)\mid (y\triangleleft z)$\\
            $U34$ & $x\triangleleft (y+ z) = (x\triangleleft y)\triangleleft z$\\
            $U35$ & $x\triangleleft (y\cdot z)=(x\triangleleft y)\triangleleft z$\\
            $U36$ & $x\triangleleft (y\parallel z) = (x\triangleleft y)\triangleleft z$\\
            $U37$ & $x\triangleleft (y\mid z) = (x\triangleleft y)\triangleleft z$\\
        \end{tabular}
        \caption{Axioms of parallelism}
        \label{AxiomsForParallelism}
    \end{table}
\end{center}

Based on the definition of basic terms for $RAPTC$ (see Definition \ref{BTAPTC}) and axioms of parallelism (see Table \ref{AxiomsForParallelism}), we can prove the elimination theorem of parallelism.

\begin{theorem}[Elimination theorem of FR parallelism]\label{ETParallelism}
Let $p$ be a closed $RAPTC$ term. Then there is a basic $RAPTC$ term $q$ such that $RAPTC\vdash p=q$.
\end{theorem}

\begin{proof}
(1) Firstly, suppose that the following ordering on the signature of $RAPTC$ is defined: $\parallel > \cdot > +$ and the symbol $\parallel$ is given the lexicographical status for the first argument, then for each rewrite rule $p\rightarrow q$ in Table \ref{TRSForRAPTC} relation $p>_{lpo} q$ can easily be proved. We obtain that the term rewrite system shown in Table \ref{TRSForRAPTC} is strongly normalizing, for it has finitely many rewriting rules, and $>$ is a well-founded ordering on the signature of $RAPTC$, and if $s>_{lpo} t$, for each rewriting rule $s\rightarrow t$ is in Table \ref{TRSForRAPTC} (see Theorem \ref{SN}).

\begin{center}
    \begin{table}
        \begin{tabular}{@{}ll@{}}
            \hline No. &Rewriting Rule\\
            $RRA6$ & $x+ \delta \rightarrow x$\\
            $RRA7$ & $\delta\cdot x \rightarrow\delta$\\
            $RRP1$ & $x\between y \rightarrow x\parallel y + x\mid y$\\
            $RRP2$ & $x\parallel y \rightarrow y \parallel x$\\
            $RRP3$ & $(x\parallel y)\parallel z \rightarrow x\parallel (y\parallel z)$\\
            $RRP4$ & $e_1\parallel (e_2\cdot y) \rightarrow (e_1\parallel e_2)\cdot y$\\
            $RRRP4$ & $e_1[m]\parallel (y\cdot e_2[m]) \rightarrow y\cdot(e_1[m]\parallel e_2[m])$\\
            $RRP5$ & $(e_1\cdot x)\parallel e_2 \rightarrow (e_1\parallel e_2)\cdot x$\\
            $RRRP5$ & $(x\cdot e_1[m])\parallel e_2[m] \rightarrow x\cdot(e_1[m]\parallel e_2[m])$\\
            $RRP6$ & $(e_1\cdot x)\parallel (e_2\cdot y) \rightarrow (e_1\parallel e_2)\cdot (x\between y)$\\
            $RRRP6$ & $(x\cdot e_1[m])\parallel (y\cdot e_2[m]) \rightarrow (x\between y)\cdot(e_1[m]\parallel e_2[m])$\\
            $RRP7$ & $(x+ y)\parallel z \rightarrow (x\parallel z)+ (y\parallel z)$\\
            $RRP8$ & $x\parallel (y+ z) \rightarrow (x\parallel y)+ (x\parallel z)$\\
            $RRP9$ & $\delta\parallel x \rightarrow \delta$\\
            $RRP10$ & $x\parallel \delta \rightarrow \delta$\\
            $RRC11$ & $e_1\mid e_2 \rightarrow \gamma(e_1,e_2)$\\
            $RRRC11$ & $e_1[m]\mid e_2[m] \rightarrow \gamma(e_1,e_2)[m]$\\
            $RRC12$ & $e_1\mid (e_2\cdot y) \rightarrow \gamma(e_1,e_2)\cdot y$\\
            $RRRC12$ & $e_1[m]\mid (y \cdot e_2[m]) \rightarrow y\cdot \gamma(e_1,e_2)[m]$\\
            $RRC13$ & $(e_1\cdot x)\mid e_2 \rightarrow \gamma(e_1,e_2)\cdot x$\\
            $RRRC13$ & $(x \cdot e_1[m])\mid e_2[m] \rightarrow x\cdot \gamma(e_1,e_2)[m]$\\
            $RRC14$ & $(e_1\cdot x)\mid (e_2\cdot y) \rightarrow \gamma(e_1,e_2)\cdot (x\between y)$\\
            $RRRC14$ & $(x \cdot e_1[m])\mid (y \cdot e_2[m]) \rightarrow(x\between y)\cdot \gamma(e_1,e_2)[m]$\\
            $RRC15$ & $(x+ y)\mid z \rightarrow (x\mid z) + (y\mid z)$\\
            $RRC16$ & $x\mid (y+ z) \rightarrow (x\mid y)+ (x\mid z)$\\
            $RRC17$ & $\delta\mid x \rightarrow \delta$\\
            $RRC18$ & $x\mid\delta \rightarrow \delta$\\
            $RRCE19$ & $\Theta(e) \rightarrow e$\\
            $RRRCE19$ & $\Theta(e[m]) \rightarrow e[m]$\\
            $RRCE20$ & $\Theta(\delta) \rightarrow \delta$\\
            $RRCE21$ & $\Theta(x+ y) \rightarrow \Theta(x)\triangleleft y + \Theta(y)\triangleleft x$\\
            $RRCE22$ & $\Theta(x\cdot y)\rightarrow\Theta(x)\cdot\Theta(y)$\\
            $RRCE23$ & $\Theta(x\parallel y) \rightarrow ((\Theta(x)\triangleleft y)\parallel y)+ ((\Theta(y)\triangleleft x)\parallel x)$\\
            $RRCE24$ & $\Theta(x\mid y) \rightarrow ((\Theta(x)\triangleleft y)\mid y)+ ((\Theta(y)\triangleleft x)\mid x)$\\
            $RRU25$ & $(\sharp(e_1,e_2))\quad e_1\triangleleft e_2 \rightarrow \tau$\\
            $RRRU25$ & $(\sharp(e_1[m],e_2[n]))\quad e_1[m]\triangleleft e_2[n] \rightarrow \tau$\\
            $RRU26$ & $(\sharp(e_1,e_2),e_2\leq e_3)\quad e_1\triangleleft e_3 \rightarrow e_1$\\
            $RRRU26$ & $(\sharp(e_1[m],e_2[n]),e_2[n]\geq e_3[l])\quad e_1[m]\triangleleft e_3[l] \rightarrow e_1[m]$\\
            $RRU27$ & $(\sharp(e_1,e_2),e_2\leq e_3)\quad e3\triangleleft e_1 \rightarrow \tau$\\
            $RRRU27$ & $(\sharp(e_1[m],e_2[n]),e_2[n]\geq e_3[l])\quad e3[l]\triangleleft e_1[m] \rightarrow \tau$\\
            $RRU28$ & $e\triangleleft \delta \rightarrow e$\\
            $RRU29$ & $\delta \triangleleft e \rightarrow \delta$\\
            $RRU30$ & $(x+ y)\triangleleft z \rightarrow (x\triangleleft z)+ (y\triangleleft z)$\\
            $RRU31$ & $(x\cdot y)\triangleleft z \rightarrow (x\triangleleft z)\cdot (y\triangleleft z)$\\
            $RRU32$ & $(x\parallel y)\triangleleft z \rightarrow (x\triangleleft z)\parallel (y\triangleleft z)$\\
            $RRU33$ & $(x\mid y)\triangleleft z \rightarrow (x\triangleleft z)\mid (y\triangleleft z)$\\
            $RRU34$ & $x\triangleleft (y+ z) \rightarrow (x\triangleleft y)\triangleleft z$\\
            $RRU35$ & $x\triangleleft (y\cdot z)\rightarrow(x\triangleleft y)\triangleleft z$\\
            $RRU36$ & $x\triangleleft (y\parallel z) \rightarrow (x\triangleleft y)\triangleleft z$\\
            $RRU37$ & $x\triangleleft (y\mid z) \rightarrow (x\triangleleft y)\triangleleft z$\\
        \end{tabular}
        \caption{Term rewrite system of $RAPTC$}
        \label{TRSForRAPTC}
    \end{table}
\end{center}

(2) Then we prove that the normal forms of closed $RAPTC$ terms are basic $RAPTC$ terms.

Suppose that $p$ is a normal form of some closed $RAPTC$ term and suppose that $p$ is not a basic $RAPTC$ term. Let $p'$ denote the smallest sub-term of $p$ which is not a basic $RAPTC$ term. It implies that each sub-term of $p'$ is a basic $RAPTC$ term. Then we prove that $p$ is not a term in normal form. It is sufficient to induct on the structure of $p'$:

\begin{itemize}
  \item Case $p'\equiv e$ or $e[m], e\in \mathbb{E}$. $p'$ is a basic $RAPTC$ term, which contradicts the assumption that $p'$ is not a basic $RAPTC$ term, so this case should not occur.
  \item Case $p'\equiv p_1\cdot p_2$. By induction on the structure of the basic $RAPTC$ term $p_1$:
      \begin{itemize}
        \item Subcase $p_1\in \mathbb{E}$. $p'$ would be a basic $RAPTC$ term, which contradicts the assumption that $p'$ is not a basic $RAPTC$ term;
        \item Subcase $p_1\equiv e\cdot p_1'$. $RRA5$ rewriting rule in Table \ref{TRSForBRATC} can be applied. So $p$ is not a normal form;
        \item Subcase $p_1\equiv p_1'\cdot e[m]$. $RRA5$ rewriting rule in Table \ref{TRSForBRATC} can be applied. So $p$ is not a normal form;
        \item Subcase $p_1\equiv p_1'+ p_1''$. $RRA4$ rewriting rule in Table \ref{TRSForBRATC} can be applied. So $p$ is not a normal form;
        \item Subcase $p_1\equiv p_1'\parallel p_1''$. $p'$ would be a basic $RAPTC$ term, which contradicts the assumption that $p'$ is not a basic $RAPTC$ term;
        \item Subcase $p_1\equiv p_1'\mid p_1''$. $RRC11$ and $RRRC11$ rewrite rule in Table \ref{TRSForRAPTC} can be applied. So $p$ is not a normal form;
        \item Subcase $p_1\equiv \Theta(p_1')$. $RRCE19$, $RRRCE19$ and $RRCE20$ rewrite rules in Table \ref{TRSForRAPTC} can be applied. So $p$ is not a normal form.
      \end{itemize}
  \item Case $p'\equiv p_1+ p_2$. By induction on the structure of the basic $RAPTC$ terms both $p_1$ and $p_2$, all subcases will lead to that $p'$ would be a basic $RAPTC$ term, which contradicts the assumption that $p'$ is not a basic $RAPTC$ term.
  \item Case $p'\equiv p_1\parallel p_2$. By induction on the structure of the basic $RAPTC$ terms both $p_1$ and $p_2$, all subcases will lead to that $p'$ would be a basic $RAPTC$ term, which contradicts the assumption that $p'$ is not a basic $RAPTC$ term.
  \item Case $p'\equiv p_1\mid p_2$. By induction on the structure of the basic $RAPTC$ terms both $p_1$ and $p_2$, all subcases will lead to that $p'$ would be a basic $RAPTC$ term, which contradicts the assumption that $p'$ is not a basic $RAPTC$ term.
  \item Case $p'\equiv \Theta(p_1)$. By induction on the structure of the basic $RAPTC$ term $p_1$, $RRCE19-RRCE24$ rewrite rules in Table \ref{TRSForRAPTC} can be applied. So $p$ is not a normal form.
  \item Case $p'\equiv p_1\triangleleft p_2$. By induction on the structure of the basic $RAPTC$ terms both $p_1$ and $p_2$, all subcases will lead to that $p'$ would be a basic $RAPTC$ term, which contradicts the assumption that $p'$ is not a basic $RAPTC$ term.
\end{itemize}
\end{proof}

\subsection{Structured Operational Semantics of Parallelism}

\begin{theorem}[Generalization of the reversible algebra for parallelism with respect to $BRATC$]
The algebra for parallelism is a generalization of $BRATC$.
\end{theorem}

\begin{proof}
It follows from the following two facts (see Theorem \ref{TCE}).

\begin{enumerate}
  \item The transition rules of $BRATC$ in section \ref{bratc} are all source-dependent;
  \item The sources of the transition rules for the algebra for parallelism contain an occurrence of $\between$, or $\parallel$, or $\mid$, or $\Theta$, or $\triangleleft$;
  \item The transition rules of $RAPTC$ are all source-dependent.
\end{enumerate}

So, the reversible algebra for parallelism is a generalization of $BRATC$, as desired.
\end{proof}

\begin{theorem}[Soundness of parallelism modulo FR step bisimulation equivalence]\label{SPSBE}
Let $x$ and $y$ be $RAPTC$ terms. If $RAPTC\vdash x=y$, then $x\sim_{s}^{fr} y$.
\end{theorem}

\begin{proof}
Since FR step bisimulation $\sim_s^{fr}$ is both an equivalent and a congruent relation with respect to the operators $\between$, $\parallel$, $\mid$, $\Theta$ and $\triangleleft$, we only need to check if each axiom in Table \ref{AxiomsForParallelism} is sound modulo FR step bisimulation equivalence.

Though transition rules in Table \ref{TRForParallel}, \ref{TRForCommunication}, and \ref{TRForConflict}, \ref{RTRForParallel}, \ref{RTRForCommunication}, and \ref{RTRForConflict} are defined in the flavor of single event, they can be modified into a step (a set of events within which each event is pairwise concurrent), we omit them. If we treat a single event as a step containing just one event, the proof of this soundness theorem does not exist any problem, so we use this way and still use the transition rules in Table \ref{TRForParallel}, \ref{TRForCommunication}, and \ref{TRForConflict}, \ref{RTRForParallel}, \ref{RTRForCommunication}, and \ref{RTRForConflict}.

We omit the defining axioms, and the trivial axioms related to $\delta$, in the following, we only prove the soundness of the non-trivial axioms, including axioms $P2-P8$, $C12-C16$, $CE21-CE24$ and $U30-U37$.

\begin{itemize}
  \item \textbf{Axiom $P2$}. Let $p,q$ be $RAPTC$ processes, and $p\parallel q=q\parallel p$, it is sufficient to prove that $p\parallel q\sim_s^{fr} q\parallel p$. By the forward transition rules for operator $\parallel$ in Table \ref{TRForParallel}, we get

      $$\frac{p\xrightarrow{e_1}e_1[m]\quad q\xrightarrow{e_2}e_2[m]}{p\parallel q\xrightarrow{\{e_1,e_2\}}e_1[m]\parallel e_2[m]}
      \quad\frac{p\xrightarrow{e_1}e_1[m]\quad q\xrightarrow{e_2}e_2[m]}{q\parallel p\xrightarrow{\{e_1,e_2\}}e_2[m]\parallel e_1[m]}$$

      $$\frac{p\xrightarrow{e_1}p'\quad q\xrightarrow{e_2}e_2[m]}{p\parallel q\xrightarrow{\{e_1,e_2\}}p'\parallel e_2[m]}
      \quad\frac{p\xrightarrow{e_1}p'\quad q\xrightarrow{e_2}e_2[m]}{q\parallel p\xrightarrow{\{e_1,e_2\}}e_2[m]\parallel p'}$$

      $$\frac{p\xrightarrow{e_1}e_1[m]\quad q\xrightarrow{e_2}q'}{p\parallel q\xrightarrow{\{e_1,e_2\}}e_1[m]\parallel q'}
      \quad\frac{p\xrightarrow{e_1}e_1[m]\quad q\xrightarrow{e_2}q'}{q\parallel p\xrightarrow{\{e_1,e_2\}}q'\parallel e_1[m]}$$

      $$\frac{p\xrightarrow{e_1}p'\quad q\xrightarrow{e_2}q'}{p\parallel q\xrightarrow{\{e_1,e_2\}}p'\between q'}
      \quad\frac{p\xrightarrow{e_1}p'\quad q\xrightarrow{e_2}q'}{q\parallel p\xrightarrow{\{e_1,e_2\}}q'\between p'}$$

      By the reverse transition rules for operator $\parallel$ in Table \ref{RTRForParallel}, we get

      $$\frac{p\xtworightarrow{e_1[m]}e_1\quad q\xtworightarrow{e_2[m]}e_2}{p\parallel q\xtworightarrow{\{e_1[m],e_2[m]\}}e_1\parallel e_2}
      \quad\frac{p\xtworightarrow{e_1[m]}e_1\quad q\xtworightarrow{e_2[m]}e_2}{q\parallel p\xtworightarrow{\{e_1[m],e_2[m]\}}e_2\parallel e_1}$$

      $$\frac{p\xtworightarrow{e_1[m]}p'\quad q\xtworightarrow{e_2[m]}e_2}{p\parallel q\xtworightarrow{\{e_1[m],e_2[m]\}}p'\parallel e_2}
      \quad\frac{p\xtworightarrow{e_1[m]}p'\quad q\xtworightarrow{e_2[m]}e_2}{q\parallel p\xtworightarrow{\{e_1[m],e_2[m]\}}e_2\parallel p'}$$

      $$\frac{p\xtworightarrow{e_1[m]}e_1\quad q\xtworightarrow{e_2[m]}q'}{p\parallel q\xtworightarrow{\{e_1[m],e_2[m]\}}e_1\parallel q'}
      \quad\frac{p\xtworightarrow{e_1[m]}e_1\quad q\xtworightarrow{e_2[m]}q'}{q\parallel p\xtworightarrow{\{e_1[m],e_2[m]\}}q'\parallel e_1}$$

      $$\frac{p\xtworightarrow{e_1[m]}p'\quad q\xtworightarrow{e_2[m]}q'}{p\parallel q\xtworightarrow{\{e_1[m],e_2[m]\}}p'\between q'}
      \quad\frac{p\xtworightarrow{e_1[m]}p'\quad q\xtworightarrow{e_2[m]}q'}{q\parallel p\xtworightarrow{\{e_1[m],e_2[m]\}}q'\between p'}$$

      So, with the assumption $e_1[m]\parallel e_2[m]\sim_s^{fr}e_2[m]\parallel e_1[m]$, $e_1\parallel e_2\sim_s^{fr}e_2\parallel e_1$, $p'\parallel e_2[m]\sim_s^{fr}e_2[m]\parallel p'$, $p'\parallel e_2\sim_s^{fr} e_2\parallel p'$, $e_1[m]\parallel q'\sim_s^{fr}q'\parallel e_1[m]$, $e_1\parallel q'\sim_s^{fr}q'\parallel e_1$, $p'\between q' \sim_s^{fr} q'\between p'$, $p\parallel q\sim_s^{fr} q\parallel p$, as desired.
  \item \textbf{Axiom $P3$}. Let $p,q,r$ be $RAPTC$ processes, and $(p\parallel q)\parallel r=p\parallel(q\parallel r)$, it is sufficient to prove that $(p\parallel q)\parallel r\sim_s^{fr} p\parallel(q\parallel r)$. By the forward transition rules for operator $\parallel$ in Table \ref{TRForParallel}, we get

      $$\frac{p\xrightarrow{e_1}e_1[m]\quad q\xrightarrow{e_2}e_2[m]\quad r\xrightarrow{e_3}e_3[m]}{(p\parallel q)\parallel r\xrightarrow{\{e_1,e_2,e_3\}}(e_1[m]\parallel e_2[m])\parallel e_3[m]}
      \quad\frac{p\xrightarrow{e_1}e_1[m]\quad q\xrightarrow{e_2}e_2[m]\quad r\xrightarrow{e_3}e_3[m]}{p\parallel(q\parallel r)\xrightarrow{\{e_1,e_2,e_3\}}e_1[m]\parallel(e_2[m]\parallel e_3[m])}$$

      $$\frac{p\xrightarrow{e_1}p'\quad q\xrightarrow{e_2}e_2[m]\quad r\xrightarrow{e_3}e_3[m]}{(p\parallel q)\parallel r\xrightarrow{\{e_1,e_2,e_3\}}(p'\parallel e_2[m])\parallel e_3[m]}
      \quad\frac{p\xrightarrow{e_1}p'\quad q\xrightarrow{e_2}e_2[m]\quad r\xrightarrow{e_3}e_3[m]}{p\parallel(q\parallel r)\xrightarrow{\{e_1,e_2,e_3\}}p'\parallel(e_2[m]\parallel e_3[m])}$$

      There are also two other cases, we omit them.

      $$\frac{p\xrightarrow{e_1}p'\quad q\xrightarrow{e_2}q'\quad r\xrightarrow{e_3}e_3[m]}{(p\parallel q)\parallel r\xrightarrow{\{e_1,e_2,e_3\}}(p'\between q')\between e_3[m]}
      \quad\frac{p\xrightarrow{e_1}p'\quad q\xrightarrow{e_2}q'\quad r\xrightarrow{e_3}e_3[m]}{p\parallel(q\parallel r)\xrightarrow{\{e_1,e_2,e_3\}}p'\between (q'\between e_3[m])}$$

      There are also other cases, we also omit them.

      $$\frac{p\xrightarrow{e_1}p'\quad q\xrightarrow{e_2}q'\quad r\xrightarrow{e_3} r'}{(p\parallel q)\parallel r'\xrightarrow{\{e_1,e_2,e_3\}}(p'\between q')\between r'}
      \quad\frac{p\xrightarrow{e_1}p'\quad q\xrightarrow{e_2}q'\quad r\xrightarrow{e_3} r'}{p\parallel (q\parallel r)\xrightarrow{\{e_1,e_2,e_3\}}p'\between(q'\between r')}$$

      By the reverse transition rules for operator $\parallel$ in Table \ref{RTRForParallel}, we get

      $$\frac{p\xtworightarrow{e_1[m]}e_1\quad q\xtworightarrow{e_2[m]}e_2\quad r\xtworightarrow{e_3[m]}e_3}{(p\parallel q)\parallel r\xtworightarrow{\{e_1[m],e_2[m],e_3[m]\}}(e_1\parallel e_2)\parallel e_3}
      \quad\frac{p\xtworightarrow{e_1[m]}e_1\quad q\xtworightarrow{e_2[m]}e_2\quad r\xtworightarrow{e_3[m]}e_3}{p\parallel(q\parallel r)\xtworightarrow{\{e_1[m],e_2[m],e_3[m]\}}e_1\parallel(e_2\parallel e_3)}$$

      $$\frac{p\xtworightarrow{e_1[m]}p'\quad q\xtworightarrow{e_2[m]}e_2\quad r\xtworightarrow{e_3[m]}e_3}{(p\parallel q)\parallel r\xtworightarrow{\{e_1[m],e_2[m],e_3[m]\}}(p'\parallel e_2)\parallel e_3}
      \quad\frac{p\xtworightarrow{e_1[m]}p'\quad q\xtworightarrow{e_2[m]}e_2\quad r\xtworightarrow{e_3[m]}e_3}{p\parallel(q\parallel r)\xtworightarrow{\{e_1[m],e_2[m],e_3[m]\}}p'\parallel(e_2\parallel e_3)}$$

      There are also two other cases, we omit them.

      $$\frac{p\xtworightarrow{e_1[m]}p'\quad q\xtworightarrow{e_2[m]}q'\quad r\xtworightarrow{e_3[m]}e_3}{(p\parallel q)\parallel r\xtworightarrow{\{e_1[m],e_2[m],e_3[m]\}}(p'\between q')\between e_3}
      \quad\frac{p\xtworightarrow{e_1[m]}p'\quad q\xtworightarrow{e_2[m]}q'\quad r\xtworightarrow{e_3[m]}e_3}{p\parallel(q\parallel r)\xtworightarrow{\{e_1[m],e_2[m],e_3[m]\}}p'\between (q'\between e_3)}$$

      There are also other cases, we also omit them.

      $$\frac{p\xtworightarrow{e_1[m]}p'\quad q\xtworightarrow{e_2[m]}q'\quad r\xtworightarrow{e_3[m]} r'}{(p\parallel q)\parallel r'\xtworightarrow{\{e_1[m],e_2[m],e_3[m]\}}(p'\between q')\between r'}
      \quad\frac{p\xtworightarrow{e_1[m]}p'\quad q\xtworightarrow{e_2[m]}q'\quad r\xtworightarrow{e_3[m]} r'}{p\parallel (q\parallel r)\xtworightarrow{\{e_1[m],e_2[m],e_3[m]\}}p'\between(q'\between r')}$$

      So, with the assumption $(e_1[m]\parallel e_2[m])\parallel e_3[m]\sim_s^{fr}e_1[m]\parallel(e_2[m]\parallel e_3[m])$, $(e_1\parallel e_2)\parallel e_3\sim_s^{fr}e_1\parallel(e_2\parallel e_3)$, $(p'\parallel e_2[m])\parallel e_3[m]\sim_s^{fr}p'\parallel(e_2[m]\parallel e_3[m])$, $(p'\parallel e_2)\parallel e_3\sim_s^{fr}p'\parallel(e_2\parallel e_3)$, $(p'\between q')\between e_3[m]\sim_s^{fr}p'\between (q'\between e_3[m])$, $(p'\between q')\between e_3\sim_s^{fr}p'\between(q'\between e_3)$, $(p'\between q')\between r'\sim_s^{fr}p'\between(q'\between r')$, $(p\parallel q)\parallel r\sim_s^{fr} p\parallel (q\parallel r)$, as desired.
  \item \textbf{Axiom $P4$}. Let $q$ be an $RAPTC$ process, and $e_1\parallel (e_2\cdot q)=(e_1\parallel e_2)\cdot q$, it is sufficient to prove that $e_1\parallel(e_2\cdot q)\sim_s^{fr} (e_1\parallel e_2)\cdot q$. By the forward transition rules for operator $\parallel$ in Table \ref{TRForParallel}, we get

      $$\frac{e_1\xrightarrow{e_1}e_1[m]\quad e_2\xrightarrow{e_2}e_2[m]}{e_1\parallel (e_2\cdot q)\xrightarrow{\{e_1,e_2\}}e_1[m]\parallel (e_2[m]\cdot q)}$$

      $$\frac{e_1\xrightarrow{e_1}e_1[m]\quad e_2\xrightarrow{e_2}e_2[m]}{(e_1\parallel e_2)\cdot q\xrightarrow{\{e_1,e_2\}}(e_1[m]\parallel e_2[m])\cdot q}$$

      By the reverse transition rules for operator $\parallel$ in Table \ref{RTRForParallel}, there are no transitions.

      So, with the assumption $e_1[m]\parallel (e_2[m]\cdot q)\sim_s^{fr}(e_1[m]\parallel e_2[m])\cdot q$, $e_1\parallel (e_2\cdot q)\sim_s^{fr} (e_1\parallel e_2)\cdot q$, as desired.
  \item \textbf{Axiom $RP4$}. Let $q$ be an $RAPTC$ process, and $e_1[m]\parallel (q\cdot e_2[m])=q\cdot (e_1[m]\parallel e_2[m])$, it is sufficient to prove that $e_1[m]\parallel (q\cdot e_2[m])\sim_s^{fr}q\cdot (e_1[m]\parallel e_2[m])$. By the forward transition rules for operator $\parallel$ in Table \ref{TRForParallel}, there are no transitions.

      By the reverse transition rules for operator $\parallel$ in Table \ref{RTRForParallel}, we get

      $$\frac{e_1[m]\xtworightarrow{e_1[m]}e_1\quad e_2[m] \xtworightarrow{e_2[m]}e_2}{e_1[m]\parallel (q\cdot e_2[m])\xtworightarrow{\{e_1[m],e_2[m]\}}e_1\parallel (q\cdot e_2)}$$

      $$\frac{e_1[m]\xtworightarrow{e_1[m]}e_1\quad e_2[m]\xtworightarrow{e_2[m]}e_2}{q\cdot (e_1[m]\parallel e_2[m])\xtworightarrow{\{e_1[m],e_2[m]\}}q\cdot (e_1\parallel e_2)}$$

      So, with the assumption $e_1\parallel (q\cdot e_2)\sim_s^{fr}q\cdot (e_1\parallel e_2)$, $e_1[m]\parallel (q\cdot e_2[m])\sim_s^{fr}q\cdot (e_1[m]\parallel e_2[m])$, as desired.
  \item \textbf{Axiom $P5$}. Let $p$ be an $RAPTC$ process, and $(e_1\cdot p)\parallel e_2=(e_1\parallel e_2)\cdot p$, it is sufficient to prove that $(e_1\cdot p)\parallel e_2\sim_s^{fr} (e_1\parallel e_2)\cdot p$. By the forward transition rules for operator $\parallel$ in Table \ref{TRForParallel}, we get

      $$\frac{e_1\xrightarrow{e_1}e_1[m]\quad e_2\xrightarrow{e_2}e_2[m]}{(e_1\cdot p)\parallel e_2\xrightarrow{\{e_1,e_2\}}(e_1[m]\cdot p)\parallel e_2[m]}$$

      $$\frac{e_1\xrightarrow{e_1}e_1[m]\quad e_2\xrightarrow{e_2}e_2[m]}{(e_1\parallel e_2)\cdot p\xrightarrow{\{e_1,e_2\}}(e_1[m]\parallel e_2[m])\cdot p}$$

      By the reverse transition rules for operator $\parallel$ in Table \ref{RTRForParallel}, there are no transitions.

      So, with the assumption $(e_1[m]\cdot p)\parallel e_2[m]\sim_s^{fr}(e_1[m]\parallel e_2[m])\cdot p$, $(e_1\cdot p)\parallel e_2\sim_s^{fr} (e_1\parallel e_2)\cdot p$, as desired.
  \item \textbf{Axiom $RP5$}. Let $p$ be an $RAPTC$ process, and $(p \cdot e_1[m])\parallel e_2[m]=p\cdot (e_1[m]\parallel e_2[m])$, it is sufficient to prove that $(p \cdot e_1[m])\parallel e_2[m]\sim_s^{fr}p\cdot (e_1[m]\parallel e_2[m])$. By the forward transition rules for operator $\parallel$ in Table \ref{TRForParallel}, there are no transitions.

      By the reverse transition rules for operator $\parallel$ in Table \ref{RTRForParallel}, we get

      $$\frac{e_1[m]\xtworightarrow{e_1[m]}e_1\quad e_2[m]\xtworightarrow{e_2[m]}e_2}{(p \cdot e_1[m])\parallel e_2[m]\xtworightarrow{\{e_1[m],e_2[m]\}}(p \cdot e_1)\parallel e_2}$$

      $$\frac{e_1[m]\xtworightarrow{e_1[m]}e_1\quad e_2[m]\xtworightarrow{e_2[m]}e_2}{p\cdot (e_1[m]\parallel e_2[m])\xtworightarrow{\{e_1[m],e_2[m]\}}p\cdot (e_1\parallel e_2)}$$

      So, with the assumption $(p \cdot e_1)\parallel e_2\sim_s^{fr}p\cdot (e_1\parallel e_2)$, $(p \cdot e_1[m])\parallel e_2[m]\sim_s^{fr}p\cdot (e_1[m]\parallel e_2[m])$, as desired.
  \item \textbf{Axiom $P6$}. Let $p,q$ be $RAPTC$ processes, and $(e_1\cdot p)\parallel (e_2\cdot q)=(e_1\parallel e_2)\cdot (p\between q)$, it is sufficient to prove that $(e_1\cdot p)\parallel(e_2\cdot q)\sim_s^{fr} (e_1\parallel e_2)\cdot (p\between q)$. By the forward transition rules for operator $\parallel$ in Table \ref{TRForParallel}, we get

      $$\frac{e_1\xrightarrow{e_1}e_1[m]\quad e_2\xrightarrow{e_2}e_2[m]}{(e_1\cdot p)\parallel (e_2\cdot q)\xrightarrow{\{e_1,e_2\}}(e_1[m]\cdot p)\parallel (e_2[m]\cdot q)}$$

      $$\frac{e_1\xrightarrow{e_1}e_1[m]\quad e_2\xrightarrow{e_2}e_2[m]}{(e_1\parallel e_2)\cdot (p\between q)\xrightarrow{\{e_1,e_2\}}(e_1[m]\parallel e_2[m])\cdot (p\between q)}$$

      By the reverse transition rules for operator $\parallel$ in Table \ref{RTRForParallel}, there are no transitions.

      So, with the assumption $(e_1[m]\cdot p)\parallel (e_1[m]\cdot q)\sim_s^{fr}(e_1[m]\parallel e_2[m])\cdot (p\between q)$, $(e_1\cdot p)\parallel (e_2\cdot q)\sim_s^{fr} (e_1\parallel e_2)\cdot (p\between q)$, as desired.
 \item \textbf{Axiom $RP6$}. Let $p,q$ be $RAPTC$ processes, and $(p\cdot e_1[m])\parallel (q\cdot e_2[m])=(p\between q)\cdot(e_1[m]\parallel e_2[m])$, it is sufficient to prove that $(p\cdot e_1[m])\parallel (q\cdot e_2[m])\sim_s^{fr}(p\between q)\cdot(e_1[m]\parallel e_2[m])$. By the forward transition rules for operator $\parallel$ in Table \ref{TRForParallel}, there are no transitions.

      By the reverse transition rules for operator $\parallel$ in Table \ref{RTRForParallel}, we get

      $$\frac{e_1[m]\xtworightarrow{e_1[m]}e_1\quad e_2[m]\xtworightarrow{e_2[m]}e_2}{(p\cdot e_1[m])\parallel (q\cdot e_2[m])\xtworightarrow{\{e_1[m],e_2[m]\}}(p\cdot e_1)\parallel (q\cdot e_2)}$$

      $$\frac{e_1\xtworightarrow{e_1[m]}e_1\quad e_2\xtworightarrow{e_2[m]}e_2}{(p\between q)\cdot(e_1[m]\parallel e_2[m])\xtworightarrow{\{e_1[m],e_2[m]\}}(p\between q)\cdot(e_1\parallel e_2)}$$

      So, with the assumption $(p\cdot e_1)\parallel (q\cdot e_2)\sim_s^{fr}(p\between q)\cdot(e_1\parallel e_2)$, $(p\cdot e_1[m])\parallel (q\cdot e_2[m])\sim_s^{fr}(p\between q)\cdot(e_1[m]\parallel e_2[m])$, as desired.
  \item \textbf{Axiom $P7$}. Let $p,q,r$ be $RAPTC$ processes, and $(p+ q)\parallel r = (p\parallel r) + (q\parallel r)$, it is sufficient to prove that $(p+ q)\parallel r \sim_s^{fr} (p\parallel r) + (q\parallel r)$. There are several cases, we will not enumerate all. By the forward transition rules for operators $+$ and $\parallel$ in Table \ref{SETRForBRATC} and \ref{TRForParallel}, we get

      $$\frac{p\xrightarrow{e_1}e_1[m]\quad q\xrightarrow{e_1}e_1[m]\quad r\xrightarrow{e_2}e_2[m]}{(p+ q)\parallel r\xrightarrow{\{e_1,e_2\}}(e_1[m]+e_1[m])\parallel e_2[m]}
      \quad \frac{p\xrightarrow{e_1}e_1[m]\quad q\xrightarrow{e_1}e_1[m]\quad r\xrightarrow{e_2}e_2[m]}{(p\parallel r)+ (q\parallel r)\xrightarrow{\{e_1,e_2\}}(e_1[m]\parallel e_2[m])+(e_1[m]\parallel e_2[m])}$$

      $$\frac{p\xrightarrow{e_1}p'\quad q\xrightarrow{e_1}q'\quad r\xrightarrow{e_2}r'}{(p+ q)\parallel r\xrightarrow{\{e_1,e_2\}}(p'+q')\parallel r'}
      \quad \frac{p\xrightarrow{e_1}p'\quad q\xrightarrow{e_1}q'\quad r\xrightarrow{e_2}r'}{(p\parallel r)+ (q\parallel r)\xrightarrow{\{e_1,e_2\}}(p'\parallel r')+(q'\parallel r')}$$

      By the reverse transition rules for operators $+$ and $\parallel$ in Table \ref{RSETRForBRATC} and \ref{RTRForParallel}, we get

      $$\frac{p\xtworightarrow{e_1[m]}e_1\quad q\xtworightarrow{e_1[m]}e_1\quad r\xtworightarrow{e_2[m]}e_2}{(p+ q)\parallel r\xtworightarrow{\{e_1[m],e_2[m]\}}(e_1+e_1)\parallel e_2}
      \quad \frac{p\xtworightarrow{e_1[m]}e_1\quad q\xtworightarrow{e_1[m]}e_1\quad r\xtworightarrow{e_2[m]}e_2}{(p\parallel r)+ (q\parallel r)\xtworightarrow{\{e_1[m],e_2[m]\}}(e_1\parallel e_2)+(e_1\parallel e_2)}$$

      $$\frac{p\xtworightarrow{e_1[m]}p'\quad q\xtworightarrow{e_1[m]}q'\quad r\xtworightarrow{e_2[m]}r'}{(p+ q)\parallel r\xtworightarrow{\{e_1[m],e_2[m]\}}(p'+q')\parallel r'}
      \quad \frac{p\xtworightarrow{e_1[m]}p'\quad q\xtworightarrow{e_1[m]}q'\quad r\xtworightarrow{e_2[m]}r'}{(p\parallel r)+ (q\parallel r)\xtworightarrow{\{e_1[m],e_2[m]\}}(p'\parallel r')+(q'\parallel r')}$$

      So, with the assumptions $(e_1[m]+ e_1[m])\parallel e_2[m]\sim_s^{fr} (e_1[m]\parallel e_2[m])+ (e_1[m]\parallel e_2[m])$, $(e_1+ e_1)\parallel e_2\sim_s^{fr} (e_1\parallel e_2)+ (e_1\parallel e_2)$, $(p'+ q')\parallel r'\sim_s^{fr} (p'\parallel r')+ (q'\parallel r')$, $(p+ q)\parallel r\sim_s^{fr} (p\parallel r)+ (q\parallel r)$, as desired.
  \item \textbf{Axiom $P8$}. Let $p,q,r$ be $RAPTC$ processes, and $p\parallel(q+ r) = (p\parallel q) + (p\parallel r)$, it is sufficient to prove that $p\parallel(q+ r) \sim_s^{fr} (p\parallel q) + (p\parallel r)$. There are several cases, we will not enumerate all. By the forward transition rules for operators $+$ and $\parallel$ in Table \ref{SETRForBRATC} and \ref{TRForParallel}, we get

      $$\frac{p\xrightarrow{e_1}e_1[m]\quad q\xrightarrow{e_2}e_2[m]\quad r\xrightarrow{e_2}e_2[m]}{p\parallel (q+ r)\xrightarrow{\{e_1,e_2\}}e_1[m]\parallel(e_2[m]+e_2[m])}
      \quad \frac{p\xrightarrow{e_1}e_1[m]\quad q\xrightarrow{e_2}e_2[m]\quad r\xrightarrow{e_2}e_2[m]}{(p\parallel q)+ (p\parallel r)\xrightarrow{\{e_1,e_2\}}(e_1[m]\parallel e_2[m])+(e_1[m]+e_2[m])}$$

      $$\frac{p\xrightarrow{e_1}p'\quad q\xrightarrow{e_2}q'\quad r\xrightarrow{e_2}r'}{p\parallel (q+ r)\xrightarrow{\{e_1,e_2\}}p'\parallel(q'+r')}
      \quad \frac{p\xrightarrow{e_1}p'\quad q\xrightarrow{e_2}q'\quad r\xrightarrow{e_2}r'}{(p\parallel q)+ (p\parallel r)\xrightarrow{\{e_1,e_2\}}(p'\parallel q')+(p'+r')}$$

      By the reverse transition rules for operators $+$ and $\parallel$ in Table \ref{RSETRForBRATC} and \ref{RTRForParallel}, we get

      $$\frac{p\xtworightarrow{e_1[m]}e_1\quad q\xtworightarrow{e_2[m]}e_2\quad r\xtworightarrow{e_2[m]}e_2}{p\parallel (q+ r)\xtworightarrow{\{e_1[m],e_2[m]\}}e_1\parallel(e_2+e_2)}
      \quad \frac{p\xtworightarrow{e_1[m]}e_1\quad q\xtworightarrow{e_2[m]}e_2\quad r\xtworightarrow{e_2[m]}e_2}{(p\parallel q)+ (p\parallel r)\xtworightarrow{\{e_1[m],e_2[m]\}}(e_1\parallel e_2)+(e_1+e_2)}$$

      $$\frac{p\xtworightarrow{e_1[m]}p'\quad q\xtworightarrow{e_2[m]}q'\quad r\xtworightarrow{e_2[m]}r'}{p\parallel (q+ r)\xtworightarrow{\{e_1[m],e_2[m]\}}p'\parallel(q'+r')}
      \quad \frac{p\xtworightarrow{e_1[m]}p'\quad q\xtworightarrow{e_2[m]}q'\quad r\xtworightarrow{e_2[m]}r'}{(p\parallel q)+ (p\parallel r)\xtworightarrow{\{e_1[m],e_2[m]\}}(p'\parallel q')+(p'+r')}$$

      So, with the assumptions $e_1[m]\parallel(e_2[m]+ e_2[m]) \sim_s^{fr} (e_1[m]\parallel e_2[m]) + (e_1[m]\parallel e_2[m])$, $e_1\parallel(e_2+ e_2) \sim_s^{fr} (e_1\parallel e_2) + (e_1\parallel e_2)$, $p'\parallel(q'+ r') \sim_s^{fr} (p'\parallel q') + (p'\parallel r')$, $p\parallel(q+ r) \sim_s^{fr} (p\parallel q) + (p\parallel r)$, as desired.
  \item \textbf{Axiom $C12$}. Let $q$ be an $RAPTC$ process, and $e_1\mid (e_2\cdot q)=\gamma(e_1,e_2)\cdot q$, it is sufficient to prove that $e_1\mid(e_2\cdot q)\sim_s^{fr} \gamma(e_1,e_2)\cdot q$. By the forward transition rules for operator $\mid$ in Table \ref{TRForCommunication}, we get

      $$\frac{e_1\xrightarrow{e_1}e_1[m]\quad e_2\xrightarrow{e_2}e_2[m]}{e_1\mid (e_2\cdot q)\xrightarrow{\gamma(e_1,e_2)}e_1[m]\mid(e_2[m]\cdot q)}$$

      $$\frac{e_1\xrightarrow{e_1}e_1[m]\quad e_2\xrightarrow{e_2}e_2[m]}{\gamma(e_1,e_2)\cdot q\xrightarrow{\gamma(e_1,e_2)}\gamma(e_1,e_2)[m]\cdot q}$$

      By the reverse transition rules for operator $\mid$ in Table \ref{RTRForCommunication}, there are no transitions.

      So, with the assumptions $e_1[m]\mid (e_2[m]\cdot q)\sim_s^{fr} \gamma(e_1,e_2)[m]\cdot q$, $e_1\mid (e_2\cdot q)\sim_s^{fr} \gamma(e_1,e_2)\cdot q$, as desired.
  \item \textbf{Axiom $RC12$}. Let $q$ be an $RAPTC$ process, and $e_1[m]\mid (q\cdot e_2[m])=q\cdot \gamma(e_1,e_2)[m]$, it is sufficient to prove that $e_1[m]\mid (q\cdot e_2[m])\sim_s^{fr}q\cdot \gamma(e_1,e_2)[m]$. By the forward transition rules for operator $\mid$ in Table \ref{TRForCommunication}, there are no transitions.

      By the reverse transition rules for operator $\mid$ in Table \ref{RTRForCommunication}, we get

      $$\frac{e_1\xtworightarrow{e_1[m]}e_1\quad e_2\xtworightarrow{e_2[m]}e_2}{e_1[m]\mid (q\cdot e_2[m])\xtworightarrow{\gamma(e_1,e_2)[m]}e_1\mid (q\cdot e_2)}$$

      $$\frac{e_1\xtworightarrow{e_1[m]}e_1\quad e_2\xtworightarrow{e_2[m]}e_2}{q\cdot \gamma(e_1,e_2)[m]\xtworightarrow{\gamma(e_1,e_2)[m]}q\cdot \gamma(e_1,e_2)}$$

      So, with the assumptions $e_1\mid (q\cdot e_2)\sim_s^{fr}q\cdot \gamma(e_1,e_2)$, $e_1[m]\mid (q\cdot e_2[m])\sim_s^{fr}q\cdot \gamma(e_1,e_2)[m]$, as desired.
  \item \textbf{Axiom $C13$}. Let $p$ be an $RAPTC$ process, and $(e_1\cdot p)\mid e_2=\gamma(e_1,e_2)\cdot p$, it is sufficient to prove that $(e_1\cdot p)\mid e_2\sim_s^{fr} \gamma(e_1,e_2)\cdot p$. By forward the transition rules for operator $\mid$ in Table \ref{TRForCommunication}, we get

      $$\frac{e_1\xrightarrow{e_1}e_1[m]\quad e_2\xrightarrow{e_2}e_2[m]}{(e_1\cdot p)\mid e_2\xrightarrow{\gamma(e_1,e_2)}(e_1[m]\cdot p)\mid e_2[m]}$$

      $$\frac{e_1\xrightarrow{e_1}e_1[m]\quad e_2\xrightarrow{e_2}e_2[m]}{\gamma(e_1,e_2)\cdot p\xrightarrow{\gamma(e_1,e_2)}\gamma(e_1,e_2)[m]\cdot p}$$

      By reverse the transition rules for operator $\mid$ in Table \ref{RTRForCommunication}, there are no transitions.

      So, with the assumption $(e_1[m]\cdot p)\mid e_2[m]\sim_s^{fr} \gamma(e_1,e_2)[m]\cdot p$, $(e_1\cdot p)\mid e_2\sim_s^{fr} \gamma(e_1,e_2)\cdot p$, as desired.
  \item \textbf{Axiom $RC13$}. Let $p$ be an $RAPTC$ process, and $(p\cdot e_1[m])\mid e_2[m]=p\cdot \gamma(e_1,e_2)[m]$, it is sufficient to prove that $(p\cdot e_1[m])\mid e_2[m]\sim_s^{fr}p\cdot \gamma(e_1,e_2)[m]$. By forward the transition rules for operator $\mid$ in Table \ref{TRForCommunication}, there are no transitions.

      By reverse the transition rules for operator $\mid$ in Table \ref{RTRForCommunication}, we get

      $$\frac{e_1\xtworightarrow{e_1[m]}e_1\quad e_2\xtworightarrow{e_2[m]}e_2}{(p\cdot e_1[m])\mid e_2[m]\xtworightarrow{\gamma(e_1,e_2)[m]}(p\cdot e_1)\mid e_2}$$

      $$\frac{e_1\xtworightarrow{e_1[m]}e_1\quad e_2\xtworightarrow{e_2[m]}e_2}{p\cdot \gamma(e_1,e_2)[m]\xtworightarrow{\gamma(e_1,e_2)[m]}p\cdot \gamma(e_1,e_2)}$$

      So, with the assumption $(p\cdot e_1)\mid e_2\sim_s^{fr}p\cdot \gamma(e_1,e_2)$, $(p\cdot e_1[m])\mid e_2[m]\sim_s^{fr}p\cdot \gamma(e_1,e_2)[m]$, as desired.
  \item \textbf{Axiom $C14$}. Let $p,q$ be $RAPTC$ processes, and $(e_1\cdot p)\mid (e_2\cdot q)=\gamma(e_1,e_2)\cdot (p\between q)$, it is sufficient to prove that $(e_1\cdot p)\mid(e_2\cdot q)\sim_s^{fr} \gamma(e_1,e_2)\cdot (p\between q)$. By the forward transition rules for operator $\mid$ in Table \ref{TRForCommunication}, we get

      $$\frac{e_1\xrightarrow{e_1}e_1[m]\quad e_2\xrightarrow{e_2}e_2[m]}{(e_1\cdot p)\mid (e_2\cdot q)\xrightarrow{\gamma(e_1,e_2)}(e_1[m]\cdot p)\between (e_2[m]\cdot q)}$$

      $$\frac{e_1\xrightarrow{e_1}e_1[m]\quad e_2\xrightarrow{e_2}e_2[m]}{\gamma(e_1,e_2)\cdot (p\between q)\xrightarrow{\gamma(e_1,e_2)}\gamma(e_1,e_2)[m]\cdot (p\between q)}$$

      By the reverse transition rules for operator $\mid$ in Table \ref{RTRForCommunication}, there are no transitions.

      So, with the assumption $(e_1[m]\cdot p)\mid (e_2[m]\cdot q)\sim_s^{fr} \gamma(e_1,e_2)[m]\cdot (p\between q)$, $(e_1\cdot p)\mid (e_2\cdot q)\sim_s^{fr} \gamma(e_1,e_2)\cdot (p\between q)$, as desired.
  \item \textbf{Axiom $RC14$}. Let $p,q$ be $RAPTC$ processes, and $(p\cdot e_1[m])\mid (q\cdot e_2[m])=(p\between q)\cdot \gamma(e_1,e_2)[m]$, it is sufficient to prove that $(p\cdot e_1[m])\mid (q\cdot e_2[m])\sim_s^{fr}(p\between q)\cdot \gamma(e_1,e_2)[m]$. By the forward transition rules for operator $\mid$ in Table \ref{TRForCommunication}, there are no transitions.

      By the reverse transition rules for operator $\mid$ in Table \ref{RTRForCommunication}, we get

      $$\frac{e_1\xtworightarrow{e_1[m]}e_1\quad e_2\xtworightarrow{e_2[m]}e_2}{(p\cdot e_1[m])\mid (q\cdot e_2[m])\xtworightarrow{\gamma(e_1,e_2)[m]}(p\cdot e_1)\mid (q\cdot e_2)}$$

      $$\frac{e_1\xtworightarrow{e_1[m]}e_1\quad e_2\xtworightarrow{e_2[m]}e_2}{(p\between q)\cdot \gamma(e_1,e_2)[m]\xtworightarrow{\gamma(e_1,e_2)[m]}(p\between q)\cdot \gamma(e_1,e_2)}$$

      So, with the assumption $(p\cdot e_1)\mid (q\cdot e_2)\sim_s^{fr}(p\between q)\cdot \gamma(e_1,e_2)$, $(p\cdot e_1[m])\mid (q\cdot e_2[m])\sim_s^{fr}(p\between q)\cdot \gamma(e_1,e_2)[m]$, as desired.
  \item \textbf{Axiom $C15$}. Let $p,q,r$ be $RAPTC$ processes, and $(p+ q)\mid r = (p\mid r) + (q\mid r)$, it is sufficient to prove that $(p+ q)\mid r \sim_s^{fr} (p\mid r) + (q\mid r)$. There are several cases, we will not enumerate all. By the forward transition rules for operators $+$ and $\mid$ in Table \ref{SETRForBRATC} and \ref{TRForCommunication}, we get

      $$\frac{p\xrightarrow{e_1}e_1[m]\quad q\xrightarrow{e_1}e_1[m]\quad r\xrightarrow{e_2}e_2[m]}{(p+ q)\mid r\xrightarrow{\gamma(e_1,e_2)}(e_1[m]+e_1[m])\mid e_2[m]}
      \quad \frac{p\xrightarrow{e_1}e_1[m]\quad q\xrightarrow{e_1}e_1[m]\quad r\xrightarrow{e_2}e_2[m]}{(p\mid r)+ (q\mid r)\xrightarrow{\gamma(e_1,e_2)}(e_1[m]\mid e_2[m])+(e_1[m]\mid e_2[m])}$$

      $$\frac{p\xrightarrow{e_1}p'\quad q\xrightarrow{e_1}q'\quad r\xrightarrow{e_2}r'}{(p+ q)\mid r\xrightarrow{\gamma(e_1,e_2)}(p'+q')\mid r'}
      \quad \frac{p\xrightarrow{e_1}p'\quad q\xrightarrow{e_1}q'\quad r\xrightarrow{e_2}r'}{(p\mid r)+ (q\mid r)\xrightarrow{\gamma(e_1,e_2)}(p'\mid r')+(q'\mid r')}$$

      By the reverse transition rules for operators $+$ and $\mid$ in Table \ref{RSETRForBRATC} and \ref{RTRForCommunication}, we get

      $$\frac{p\xtworightarrow{e_1[m]}e_1\quad q\xtworightarrow{e_1[m]}e_1\quad r\xtworightarrow{e_2[m]}e_2}{(p+ q)\mid r\xtworightarrow{\gamma(e_1,e_2)[m]}(e_1+e_1)\mid e_2}
      \quad \frac{p\xtworightarrow{e_1[m]}e_1\quad q\xtworightarrow{e_1[m]}e_1\quad r\xtworightarrow{e_2[m]}e_2}{(p\mid r)+ (q\mid r)\xtworightarrow{\gamma(e_1,e_2)[m]}(e_1\mid e_2)+(e_1\mid e_2)}$$

      $$\frac{p\xtworightarrow{e_1[m]}p'\quad q\xtworightarrow{e_1[m]}q'\quad r\xtworightarrow{e_2[m]}r'}{(p+ q)\mid r\xtworightarrow{\gamma(e_1,e_2)[m]}(p'+q')\mid r'}
      \quad \frac{p\xtworightarrow{e_1[m]}p'\quad q\xtworightarrow{e_1[m]}q'\quad r\xtworightarrow{e_2[m]}r'}{(p\mid r)+ (q\mid r)\xtworightarrow{\gamma(e_1,e_2)[m]}(p'\mid r')+(q'\mid r')}$$

      So, with the assumptions $(e_1[m]+ e_1[m])\mid e_2[m]\sim_s^{fr} (e_1[m]\mid e_2[m])+ (e_1[m]\mid e_2[m])$, $(e_1+ e_1)\mid e_2\sim_s^{fr} (e_1\mid e_2)+ (e_1\mid e_2)$, $(p'+ q')\mid r'\sim_s^{fr} (p'\mid r')+ (q'\mid r')$, $(p+ q)\mid r\sim_s^{fr} (p\mid r)+ (q\mid r)$, as desired.
  \item \textbf{Axiom $C16$}. Let $p,q,r$ be $RAPTC$ processes, and $p\mid(q+ r) = (p\mid q) + (p\mid r)$, it is sufficient to prove that $p\mid(q+ r) \sim_s^{fr} (p\mid q) + (p\mid r)$. There are several cases, we will not enumerate all. By the forward transition rules for operators $+$ and $\mid$ in Table \ref{SETRForBRATC} and \ref{TRForCommunication}, we get

      $$\frac{p\xrightarrow{e_1}e_1[m]\quad q\xrightarrow{e_2}e_2[m]\quad r\xrightarrow{e_2}e_2[m]}{p\mid (q+ r)\xrightarrow{\gamma(e_1,e_2)}e_1[m]\parallel (e_2[m]+e_2[m])}
      \quad \frac{p\xrightarrow{e_1}e_1[m]\quad q\xrightarrow{e_2}e_2[m]\quad r\xrightarrow{e_2}e_2[m]}{(p\mid q)+ (p\mid r)\xrightarrow{\gamma(e_1,e_2)}(e_1[m]\mid e_2[m])+(e_1[m]\mid e_2[m])}$$

      $$\frac{p\xrightarrow{e_1}p'\quad q\xrightarrow{e_2}q'\quad r\xrightarrow{e_2}r'}{p\mid (q+ r)\xrightarrow{\gamma(e_1,e_2)}p'\parallel (q'+r')}
      \quad \frac{p\xrightarrow{e_1}p'\quad q\xrightarrow{e_2}q'\quad r\xrightarrow{e_2}r'}{(p\mid q)+ (p\mid r)\xrightarrow{\gamma(e_1,e_2)}(p'\mid q')+(p'\mid r')}$$

      By the reverse transition rules for operators $+$ and $\mid$ in Table \ref{RSETRForBRATC} and \ref{RTRForCommunication}, we get

      $$\frac{p\xtworightarrow{e_1[m]}e_1\quad q\xtworightarrow{e_2[m]}e_2\quad r\xtworightarrow{e_2[m]}e_2}{p\mid (q+ r)\xtworightarrow{\gamma(e_1,e_2)[m]}e_1\parallel (e_2+e_2)}
      \quad \frac{p\xtworightarrow{e_1[m]}e_1\quad q\xtworightarrow{e_2[m]}e_2\quad r\xtworightarrow{e_2[m]}e_2}{(p\mid q)+ (p\mid r)\xtworightarrow{\gamma(e_1,e_2)[m]}(e_1\mid e_2)+(e_1\mid e_2)}$$

      $$\frac{p\xtworightarrow{e_1[m]}p'\quad q\xtworightarrow{e_2[m]}q'\quad r\xtworightarrow{e_2[m]}r'}{p\mid (q+ r)\xtworightarrow{\gamma(e_1,e_2)[m]}p'\parallel (q'+r')}
      \quad \frac{p\xtworightarrow{e_1[m]}p'\quad q\xtworightarrow{e_2[m]}q'\quad r\xtworightarrow{e_2[m]}r'}{(p\mid q)+ (p\mid r)\xtworightarrow{\gamma(e_1,e_2)[m]}(p'\mid q')+(p'\mid r')}$$

      So, with the assumptions $e_1[m]\mid(e_2[m]+ e_2[m]) \sim_s^{fr} (e_1[m]\mid e_2[m]) + (e_1[m]\mid e_2[m])$, $e_1\mid(e_2+ e_2) \sim_s^{fr} (e_1\mid e_2) + (e_1\mid e_2)$, $p'\mid(q'+ r') \sim_s^{fr} (p'\mid q') + (p'\mid r')$, $p\mid(q+ r) \sim_s^{fr} (p\mid q) + (p\mid r)$, as desired.
  \item \textbf{Axiom $CE21$}. Let $p,q$ be $RAPTC$ processes, and $\Theta(p+ q)=\Theta(p)\triangleleft q + \Theta(q)\triangleleft p$, it is sufficient to prove that $\Theta(p+ q) \sim_s^{fr} \Theta(p)\triangleleft q + \Theta(q)\triangleleft p$. By the forward transition rules for operators $+$ in Table \ref{SETRForBRATC}, and $\Theta$ and $\triangleleft$ in Table \ref{TRForConflict}, we get

      $$\frac{p\xrightarrow{e_1}e_1[m] (\sharp(e_1,e_2))}{\Theta(p+ q)\xrightarrow{e_1}\Theta(e_1[m]+q)}
      \quad\frac{p\xrightarrow{e_1}e_1[m] (\sharp(e_1,e_2))}{\Theta(p)\triangleleft q + \Theta(q)\triangleleft p\xrightarrow{e_1}\Theta(e_1[m])\triangleleft q + \Theta(q)\triangleleft e_1[m]}$$

      $$\frac{q\xrightarrow{e_2}e_2[m] (\sharp(e_1,e_2))}{\Theta(p+ q)\xrightarrow{e_2}\Theta(p+e_2[m])}
      \quad\frac{q\xrightarrow{e_2}e_2[m] (\sharp(e_1,e_2))}{\Theta(p)\triangleleft q + \Theta(q)\triangleleft p\xrightarrow{e_2}\Theta(p)\triangleleft e_2[m] + \Theta(e_2[m])\triangleleft p}$$

      $$\frac{p\xrightarrow{e_1}p' (\sharp(e_1,e_2))}{\Theta(p+ q)\xrightarrow{e_1}\Theta(p'+q)}
      \quad\frac{p\xrightarrow{e_1}p' (\sharp(e_1,e_2))}{\Theta(p)\triangleleft q + \Theta(q)\triangleleft p\xrightarrow{e_1}\Theta(p')\triangleleft q + \Theta(q)\triangleleft p'}$$

      $$\frac{q\xrightarrow{e_2}q' (\sharp(e_1,e_2))}{\Theta(p+ q)\xrightarrow{e_2}\Theta(p+q')}
      \quad\frac{q\xrightarrow{e_2}q' (\sharp(e_1,e_2))}{\Theta(p)\triangleleft q + \Theta(q)\triangleleft p\xrightarrow{e_2}\Theta(p)\triangleleft q' + \Theta(q')\triangleleft p}$$

      By the reverse transition rules for operators $+$ in Table \ref{RSETRForBRATC}, and $\Theta$ and $\triangleleft$ in Table \ref{RTRForConflict}, we get

      $$\frac{p\xtworightarrow{e_1[m]}e_1 (\sharp(e_1,e_2))}{\Theta(p+ q)\xtworightarrow{e_1[m]}\Theta(e_1+q)}
      \quad\frac{p\xtworightarrow{e_1[m]}e_1 (\sharp(e_1,e_2))}{\Theta(p)\triangleleft q + \Theta(q)\triangleleft p\xtworightarrow{e_1[m]}\Theta(e_1)\triangleleft q + \Theta(q)\triangleleft e_1}$$

      $$\frac{q\xtworightarrow{e_2[m]}e_2 (\sharp(e_1,e_2))}{\Theta(p+ q)\xtworightarrow{e_2[m]}\Theta(p+e_2)}
      \quad\frac{q\xtworightarrow{e_2[m]}e_2 (\sharp(e_1,e_2))}{\Theta(p)\triangleleft q + \Theta(q)\triangleleft p\xtworightarrow{e_2[m]}\Theta(p)\triangleleft e_2 + \Theta(e_2)\triangleleft p}$$

      $$\frac{p\xtworightarrow{e_1[m]}p' (\sharp(e_1,e_2))}{\Theta(p+ q)\xtworightarrow{e_1[m]}\Theta(p'+q)}
      \quad\frac{p\xtworightarrow{e_1[m]}p' (\sharp(e_1,e_2))}{\Theta(p)\triangleleft q + \Theta(q)\triangleleft p\xtworightarrow{e_1[m]}\Theta(p')\triangleleft q + \Theta(q)\triangleleft p'}$$

      $$\frac{q\xtworightarrow{e_2[m]}q' (\sharp(e_1,e_2))}{\Theta(p+ q)\xtworightarrow{e_2[m]}\Theta(p+q')}
      \quad\frac{q\xtworightarrow{e_2[m]}q' (\sharp(e_1,e_2))}{\Theta(p)\triangleleft q + \Theta(q)\triangleleft p\xtworightarrow{e_2[m]}\Theta(p)\triangleleft q' + \Theta(q')\triangleleft p}$$

      So, with the assumptions $\Theta(e_1[m]+ q) \sim_s^{fr} \Theta(e_1[m])\triangleleft q + \Theta(q)\triangleleft e_1[m]$, $\Theta(e_1+ q) \sim_s^{fr} \Theta(e_1)\triangleleft q + \Theta(q)\triangleleft e_1$, $\Theta(p+ e_2[m]) \sim_s^{fr} \Theta(p)\triangleleft e_2[m] + \Theta(e_2[m])\triangleleft p$, $\Theta(p+ e_2) \sim_s^{fr} \Theta(p)\triangleleft e_2 + \Theta(e_2)\triangleleft p$, $\Theta(p+ q') \sim_s^{fr} \Theta(p)\triangleleft q' + \Theta(q')\triangleleft p$, $\Theta(p'+ q) \sim_s^{fr} \Theta(p')\triangleleft q + \Theta(q)\triangleleft p'$, $\Theta(p+ q) \sim_s^{fr} \Theta(p)\triangleleft q + \Theta(q)\triangleleft p$, as desired.
  \item \textbf{Axiom $CE22$}. Let $p,q$ be $RAPTC$ processes, and $\Theta(p\cdot q)=\Theta(p)\cdot \Theta(q)$, it is sufficient to prove that $\Theta(p\cdot q) \sim_s^{fr} \Theta(p)\cdot \Theta(q)$. There several cases, we will not enumerate all. By the forward transition rules for operators $\cdot$ in Table \ref{SETRForBRATC}, and $\Theta$ in Table \ref{TRForConflict}, we get

      $$\frac{p\xrightarrow{e_1}e_1[m]}{\Theta(p\cdot q)\xrightarrow{e_1}\Theta(e_1[m]\cdot q)}
      \quad\frac{p\xrightarrow{e_1}e_1[m]}{\Theta(p)\cdot\Theta(q)\xrightarrow{e_1}\Theta(e_1[m])\cdot\Theta(q)}$$

      $$\frac{p\xrightarrow{e_1}p'}{\Theta(p\cdot q)\xrightarrow{e_1}\Theta(p'\cdot q)}
      \quad\frac{p\xrightarrow{e_1}p'}{\Theta(p)\cdot\Theta(q)\xrightarrow{e_1}\Theta(p')\cdot\Theta(q)}$$

      By the reverse transition rules for operators $\cdot$ in Table \ref{RSETRForBRATC}, and $\Theta$ in Table \ref{RTRForConflict}, we get

      $$\frac{q\xtworightarrow{e_2[m]}e_1}{\Theta(p\cdot q)\xtworightarrow{e_2[m]}\Theta(p\cdot e_2)}
      \quad\frac{q\xtworightarrow{e_2[m]}e_2}{\Theta(p)\cdot\Theta(q)\xtworightarrow{e_2[m]}\Theta(p)\cdot\Theta(e_2)}$$

      $$\frac{q\xtworightarrow{e_2[m]}q'}{\Theta(p\cdot q)\xtworightarrow{e_2[m]}\Theta(p\cdot q')}
      \quad\frac{q\xtworightarrow{e_2[m]}q'}{\Theta(p)\cdot\Theta(q)\xtworightarrow{e_2[m]}\Theta(p)\cdot\Theta(q')}$$

      So, with the assumption $\Theta(e_1[m]\cdot q) \sim_s^{fr} \Theta(e_1[m])\cdot \Theta(q)$, $\Theta(p\cdot e_2) \sim_s^{fr} \Theta(p)\cdot \Theta(e_2)$, $\Theta(p'\cdot q) \sim_s^{fr} \Theta(p')\cdot \Theta(q)$, $\Theta(p\cdot q') \sim_s^{fr} \Theta(p)\cdot \Theta(q')$, $\Theta(p\cdot q) \sim_s^{fr} \Theta(p)\cdot \Theta(q)$, as desired.
  \item \textbf{Axiom $CE23$}. Let $p,q$ be $RAPTC$ processes, and $\Theta(p\parallel q)=((\Theta(p)\triangleleft q)\parallel q) + ((\Theta(q)\triangleleft p)\parallel p)$, it is sufficient to prove that $\Theta(p\parallel q) \sim_s^{fr} ((\Theta(p)\triangleleft q)\parallel q) + ((\Theta(q)\triangleleft p)\parallel p)$. By the forward transition rules for operators $+$ in Table \ref{SETRForBRATC}, and $\Theta$ and $\triangleleft$ in Table \ref{TRForConflict}, and $\parallel$ in Table \ref{TRForParallel} we get

      $$\frac{p\xrightarrow{e_1}e_1[m] \quad q\xrightarrow{e_2}e_2[m]}{\Theta(p\parallel q)\xrightarrow{\{e_1,e_2\}}\Theta(e_1[m]\parallel e_2[m])}$$
      $$\frac{p\xrightarrow{e_1}e_1[m] \quad q\xrightarrow{e_2}e_2[m]}{((\Theta(p)\triangleleft q)\parallel q) + ((\Theta(q)\triangleleft p)\parallel p)\xrightarrow{\{e_1,e_2\}}((\Theta(e_1[m])\triangleleft e_2[m])\parallel e_2[m]) + ((\Theta(e_2[m])\triangleleft e_1[m])\parallel e_1[m])}$$

      $$\frac{p\xrightarrow{e_1}p' \quad q\xrightarrow{e_2}e_2[m]}{\Theta(p\parallel q)\xrightarrow{\{e_1,e_2\}}\Theta(p'\parallel e_2[m])}$$
      $$\frac{p\xrightarrow{e_1}p' \quad q\xrightarrow{e_2}e_2[m]}{((\Theta(p)\triangleleft q)\parallel q) + ((\Theta(q)\triangleleft p)\parallel p)\xrightarrow{\{e_1,e_2\}}((\Theta(p)\triangleleft e_2[m])\parallel e_2[m]) + ((\Theta(e_2[m])\triangleleft p)\parallel p)}$$

      $$\frac{p\xrightarrow{e_1}e_1[m] \quad q\xrightarrow{e_2}q'}{\Theta(p\parallel q)\xrightarrow{\{e_1,e_2\}}\Theta(e_1[m]\parallel q')}$$
      $$\frac{p\xrightarrow{e_1}e_1[m] \quad q\xrightarrow{e_2}q'}{((\Theta(p)\triangleleft q)\parallel q) + ((\Theta(q)\triangleleft p)\parallel p)\xrightarrow{\{e_1,e_2\}}((\Theta(e_1[m])\triangleleft q)\parallel q) + ((\Theta(q)\triangleleft e_1[m])\parallel e_1[m])}$$

      $$\frac{p\xrightarrow{e_1}p' \quad q\xrightarrow{e_2}q'}{\Theta(p\parallel q)\xrightarrow{\{e_1,e_2\}}\Theta(p'\between q')}$$
      $$\frac{p\xrightarrow{e_1}p' \quad q\xrightarrow{e_2}q'}{((\Theta(p)\triangleleft q)\parallel q) + ((\Theta(q)\triangleleft p)\parallel p)\xrightarrow{\{e_1,e_2\}}((\Theta(p')\triangleleft q')\between q') + ((\Theta(q')\triangleleft p')\between p')}$$

      By the reverse transition rules for operators $+$ in Table \ref{RSETRForBRATC}, and $\Theta$ and $\triangleleft$ in Table \ref{RTRForConflict}, and $\parallel$ in Table \ref{RTRForParallel} we get

      $$\frac{p\xtworightarrow{e_1[m]}e_1 \quad q\xtworightarrow{e_2[m]}e_2}{\Theta(p\parallel q)\xtworightarrow{\{e_1[m],e_2[m]\}}\Theta(e_1\parallel e_2)}$$
      $$\frac{p\xtworightarrow{e_1[m]}e_1 \quad q\xtworightarrow{e_2[m]}e_2}{((\Theta(p)\triangleleft q)\parallel q) + ((\Theta(q)\triangleleft p)\parallel p)\xtworightarrow{\{e_1[m],e_2[m]\}}((\Theta(e_1)\triangleleft e_2)\parallel e_2) + ((\Theta(e_2)\triangleleft e_1)\parallel e_1)}$$

      $$\frac{p\xtworightarrow{e_1[m]}p' \quad q\xtworightarrow{e_2[m]}e_2}{\Theta(p\parallel q)\xtworightarrow{\{e_1[m],e_2[m]\}}\Theta(p'\parallel e_2)}$$
      $$\frac{p\xtworightarrow{e_1[m]}p' \quad q\xtworightarrow{e_2[m]}e_2}{((\Theta(p)\triangleleft q)\parallel q) + ((\Theta(q)\triangleleft p)\parallel p)\xtworightarrow{\{e_1[m],e_2[m]\}}((\Theta(p)\triangleleft e_2)\parallel e_2) + ((\Theta(e_2)\triangleleft p)\parallel p)}$$

      $$\frac{p\xtworightarrow{e_1[m]}e_1 \quad q\xtworightarrow{e_2[m]}q'}{\Theta(p\parallel q)\xtworightarrow{\{e_1[m],e_2[m]\}}\Theta(e_1\parallel q')}$$
      $$\frac{p\xtworightarrow{e_1[m]}e_1 \quad q\xtworightarrow{e_2[m]}q'}{((\Theta(p)\triangleleft q)\parallel q) + ((\Theta(q)\triangleleft p)\parallel p)\xtworightarrow{\{e_1[m],e_2[m]\}}((\Theta(e_1)\triangleleft q)\parallel q) + ((\Theta(q)\triangleleft e_1)\parallel e_1)}$$

      $$\frac{p\xtworightarrow{e_1[m]}p' \quad q\xtworightarrow{e_2[m]}q'}{\Theta(p\parallel q)\xtworightarrow{\{e_1[m],e_2[m]\}}\Theta(p'\between q')}$$
      $$\frac{p\xtworightarrow{e_1[m]}p' \quad q\xtworightarrow{e_2[m]}q'}{((\Theta(p)\triangleleft q)\parallel q) + ((\Theta(q)\triangleleft p)\parallel p)\xtworightarrow{\{e_1[m],e_2[m]\}}((\Theta(p')\triangleleft q')\between q') + ((\Theta(q')\triangleleft p')\between p')}$$

      So, with the assumptions $\Theta(e_1[m]\parallel e_2[m]) \sim_s^{fr} ((\Theta(e_1[m])\triangleleft e_2[m])\parallel e_2[m]) + ((\Theta(e_2[m])\triangleleft e_1[m])\parallel e_1[m])$, $\Theta(e_1\parallel e_2) \sim_s^{fr} ((\Theta(e_1)\triangleleft e_2)\parallel e_2) + ((\Theta(e_2)\triangleleft e_1)\parallel e_1)$, $\Theta(e_1[m]\parallel q') \sim_s^{fr} ((\Theta(e_1[m])\triangleleft q')\parallel q') + ((\Theta(q')\triangleleft e_1[m])\parallel e_1[m])$, $\Theta(e_1\parallel q') \sim_s^{fr} ((\Theta(e_1)\triangleleft q')\parallel q') + ((\Theta(q')\triangleleft e_1)\parallel e_1)$, $\Theta(p'\parallel e_2[m]) \sim_s^{fr} ((\Theta(p')\triangleleft e_2[m])\parallel e_2[m]) + ((\Theta(e_2[m])\triangleleft p')\parallel p')$, $\Theta(p'\parallel e_2) \sim_s^{fr} ((\Theta(p')\triangleleft e_2)\parallel e_2) + ((\Theta(e_2)\triangleleft p')\parallel p')$, $\Theta(p'\between q')\sim_s^{fr}((\Theta(p')\triangleleft q')\between q') + ((\Theta(q')\triangleleft p')\between p')$, $\Theta(p\parallel q) \sim_s^{fr} ((\Theta(p)\triangleleft q)\parallel q) + ((\Theta(q)\triangleleft p)\parallel p)$, as desired.
  \item \textbf{Axiom $CE24$}. Let $p,q$ be $RAPTC$ processes, and $\Theta(p\mid q)=((\Theta(p)\triangleleft q)\mid q) + ((\Theta(q)\triangleleft p)\mid p)$, it is sufficient to prove that $\Theta(p\mid q) \sim_s^{fr} ((\Theta(p)\triangleleft q)\mid q) + ((\Theta(q)\triangleleft p)\mid p)$. By the forward transition rules for operators $+$ in Table \ref{SETRForBRATC}, and $\Theta$ and $\triangleleft$ in Table \ref{TRForConflict}, and $\mid$ in Table \ref{TRForCommunication} we get

      $$\frac{p\xrightarrow{e_1}e_1[m] \quad q\xrightarrow{e_2}e_2[m]}{\Theta(p\mid q)\xrightarrow{\gamma(e_1,e_2)}\Theta(e_1[m]\mid e_2[m])}$$
      $$\frac{p\xrightarrow{e_1}e_1[m] \quad q\xrightarrow{e_2}e_2[m]}{((\Theta(p)\triangleleft q)\mid q) + ((\Theta(q)\triangleleft p)\mid p)\xrightarrow{\gamma(e_1,e_2)}((\Theta(e_1[m])\triangleleft e_2[m])\mid e_2[m]) + ((\Theta(e_2[m])\triangleleft e_1[m])\mid e_1[m]}$$

      $$\frac{p\xrightarrow{e_1}p' \quad q\xrightarrow{e_2}e_2[m]}{\Theta(p\mid q)\xrightarrow{\gamma(e_1,e_2)}\Theta(p'\mid e_2[m])}$$
      $$\frac{p\xrightarrow{e_1}p' \quad q\xrightarrow{e_2}e_2[m]}{((\Theta(p)\triangleleft q)\mid q) + ((\Theta(q)\triangleleft p)\mid p)\xrightarrow{\gamma(e_1,e_2)}((\Theta(p')\triangleleft e_2[m])\mid e_2[m]) + ((\Theta(e_2[m])\triangleleft p')\mid p'}$$

      $$\frac{p\xrightarrow{e_1}e_1[m] \quad q\xrightarrow{e_2}q'}{\Theta(p\mid q)\xrightarrow{\gamma(e_1,e_2)}\Theta(e_1[m]\mid q')}$$
      $$\frac{p\xrightarrow{e_1}e_1[m] \quad q\xrightarrow{e_2}q'}{((\Theta(p)\triangleleft q)\mid q) + ((\Theta(q)\triangleleft p)\mid p)\xrightarrow{\gamma(e_1,e_2)}((\Theta(e_1[m])\triangleleft q')\mid q') + ((\Theta(q')\triangleleft e_1[m])\mid e_1[m]}$$

      $$\frac{p\xrightarrow{e_1}p' \quad q\xrightarrow{e_2}q'}{\Theta(p\mid q)\xrightarrow{\gamma(e_1,e_2)}\Theta(p'\between q')}$$
      $$\frac{p\xrightarrow{e_1}p' \quad q\xrightarrow{e_2}q'}{((\Theta(p)\triangleleft q)\mid q) + ((\Theta(q)\triangleleft p)\mid p)\xrightarrow{\gamma(e_1,e_2)}((\Theta(p')\triangleleft q')\between q') + ((\Theta(q')\triangleleft p')\between p')}$$

      By the reverse transition rules for operators $+$ in Table \ref{RSETRForBRATC}, and $\Theta$ and $\triangleleft$ in Table \ref{RTRForConflict}, and $\mid$ in Table \ref{RTRForCommunication} we get

      $$\frac{p\xtworightarrow{e_1[m]}e_1 \quad q\xtworightarrow{e_2[m]}e_2}{\Theta(p\mid q)\xtworightarrow{\gamma(e_1,e_2)[m]}\Theta(e_1\mid e_2)}$$
      $$\frac{p\xtworightarrow{e_1[m]}e_1 \quad q\xtworightarrow{e_2[m]}e_2}{((\Theta(p)\triangleleft q)\mid q) + ((\Theta(q)\triangleleft p)\mid p)\xtworightarrow{\gamma(e_1,e_2)[m]}((\Theta(e_1)\triangleleft e_2)\mid e_2) + ((\Theta(e_2)\triangleleft e_1)\mid e_1}$$

      $$\frac{p\xtworightarrow{e_1[m]}p' \quad q\xtworightarrow{e_2[m]}e_2[m]}{\Theta(p\mid q)\xtworightarrow{\gamma(e_1,e_2)[m]}\Theta(p'\mid e_2)}$$
      $$\frac{p\xtworightarrow{e_1[m]}p' \quad q\xtworightarrow{e_2[m]}e_2}{((\Theta(p)\triangleleft q)\mid q) + ((\Theta(q)\triangleleft p)\mid p)\xtworightarrow{\gamma(e_1,e_2)[m]}((\Theta(p')\triangleleft e_2)\mid e_2) + ((\Theta(e_2)\triangleleft p')\mid p'}$$

      $$\frac{p\xtworightarrow{e_1[m]}e_1 \quad q\xtworightarrow{e_2[m]}q'}{\Theta(p\mid q)\xtworightarrow{\gamma(e_1,e_2)[m]}\Theta(e_1\mid q')}$$
      $$\frac{p\xtworightarrow{e_1[m]}e_1 \quad q\xtworightarrow{e_2[m]}q'}{((\Theta(p)\triangleleft q)\mid q) + ((\Theta(q)\triangleleft p)\mid p)\xtworightarrow{\gamma(e_1,e_2)[m]}((\Theta(e_1)\triangleleft q')\mid q') + ((\Theta(q')\triangleleft e_1)\mid e_1}$$

      $$\frac{p\xtworightarrow{e_1[m]}p' \quad q\xtworightarrow{e_2[m]}q'}{\Theta(p\mid q)\xtworightarrow{\gamma(e_1,e_2)[m]}\Theta(p'\between q')}$$
      $$\frac{p\xtworightarrow{e_1[m]}p' \quad q\xtworightarrow{e_2[m]}q'}{((\Theta(p)\triangleleft q)\mid q) + ((\Theta(q)\triangleleft p)\mid p)\xtworightarrow{\gamma(e_1,e_2)[m]}((\Theta(p')\triangleleft q')\between q') + ((\Theta(q')\triangleleft p')\between p')}$$

      So, with the assumptions $\Theta(e_1[m]\mid e_2[m]) \sim_s^{fr} ((\Theta(e_1[m])\triangleleft e_2[m])\mid e_2[m]) + ((\Theta(e_2[m])\triangleleft e_1[m])\mid e_1[m])$, $\Theta(e_1\mid e_2) \sim_s^{fr} ((\Theta(e_1)\triangleleft e_2)\mid e_2) + ((\Theta(e_2)\triangleleft e_1)\mid e_1)$, $\Theta(e_1[m]\mid q') \sim_s^{fr} ((\Theta(e_1[m])\triangleleft q')\mid q') + ((\Theta(q')\triangleleft e_1[m])\mid e_1[m])$, $\Theta(e_1\mid q') \sim_s^{fr} ((\Theta(e_1)\triangleleft q')\mid q') + ((\Theta(q')\triangleleft e_1)\mid e_1)$, $\Theta(p'\mid e_2[m]) \sim_s^{fr} ((\Theta(p')\triangleleft e_2[m])\mid e_2[m]) + ((\Theta(e_2[m])\triangleleft p')\mid p')$, $\Theta(p'\mid e_2) \sim_s^{fr} ((\Theta(p')\triangleleft e_2)\mid e_2) + ((\Theta(e_2)\triangleleft p')\mid p')$, $\Theta(p'\between q')\sim_s^{fr}((\Theta(p')\triangleleft q')\between q') + ((\Theta(q')\triangleleft p')\between p')$, $\Theta(p\mid q) \sim_s^{fr} ((\Theta(p)\triangleleft q)\mid q) + ((\Theta(q)\triangleleft p)\mid p)$, as desired.
  \item \textbf{Axiom $U30$}. Let $p,q,r$ be $RAPTC$ processes, and $(p+ q)\triangleleft r = (p\triangleleft r) + (q\triangleleft r)$, it is sufficient to prove that $(p+ q)\triangleleft r \sim_s^{fr} (p\triangleleft r) + (q\triangleleft r)$. By the forward transition rules for operators $+$ and $\triangleleft$ in Table \ref{SETRForBRATC} and \ref{TRForConflict}, we get

      $$\frac{p\xrightarrow{e_1}e_1[m]\quad e_1\notin q}{(p+ q)\triangleleft r\xrightarrow{e_1}(e_1[m]+q)\triangleleft r}
      \quad \frac{p\xrightarrow{e_1}e_1[m]\quad e_2\notin q}{(p\triangleleft r)+ (q\triangleleft r)\xrightarrow{e_1}(e_1[m]\triangleleft r)+ (q\triangleleft r)}$$

      $$\frac{q\xrightarrow{e_2}e_2[m]\quad e_2\notin p}{(p+ q)\triangleleft r\xrightarrow{e_2}(p+e_2[m])\triangleleft r}
      \quad \frac{q\xrightarrow{e_2}e_2[m]\quad e_2\notin p}{(p\triangleleft r)+ (q\triangleleft r)\xrightarrow{e_2}(p\triangleleft r)+ (e_2[m]\triangleleft r)}$$

      $$\frac{p\xrightarrow{e_1}p'\quad e_1\notin q}{(p+ q)\triangleleft r\xrightarrow{e_1}(p'+q)\triangleleft r}
      \quad \frac{p\xrightarrow{e_1}p'\quad e_1\notin q}{(p\triangleleft r)+ (q\triangleleft r)\xrightarrow{e_1}(p'\triangleleft r)+(q\triangleleft r)}$$

      $$\frac{q\xrightarrow{e_2}q'\quad e_2\notin p}{(p+ q)\triangleleft r\xrightarrow{e_2}(p+q')\triangleleft r}
      \quad \frac{q\xrightarrow{e_2}q'\quad e_2\notin p}{(p\triangleleft r)+ (q\triangleleft r)\xrightarrow{e_2}(p\triangleleft r)+(q'\triangleleft r)}$$

      $$\frac{p\xrightarrow{e_1}p'\quad q\xrightarrow{e_1}q'}{(p+ q)\triangleleft r\xrightarrow{e_1}(p'+q')\triangleleft r}
      \quad \frac{p\xrightarrow{e_1}p'\quad q\xrightarrow{e_1}q'}{(p\triangleleft r)+ (q\triangleleft r)\xrightarrow{e_1}(p'\triangleleft r)+(q'\triangleleft r)}$$

      By the reverse transition rules for operators $+$ and $\triangleleft$ in Table \ref{RSETRForBRATC} and \ref{RTRForConflict}, we get

      $$\frac{p\xtworightarrow{e_1[m]}e_1\quad e_1\notin q}{(p+ q)\triangleleft r\xtworightarrow{e_1[m]}(e_1+q)\triangleleft r}
      \quad \frac{p\xtworightarrow{e_1[m]}e_1\quad e_2\notin q}{(p\triangleleft r)+ (q\triangleleft r)\xtworightarrow{e_1[m]}(e_1\triangleleft r)+ (q\triangleleft r)}$$

      $$\frac{q\xtworightarrow{e_2}e_2[m]\quad e_2\notin p}{(p+ q)\triangleleft r\xtworightarrow{e_2[m]}(p+e_2)\triangleleft r}
      \quad \frac{q\xtworightarrow{e_2[m]}e_2\quad e_2\notin p}{(p\triangleleft r)+ (q\triangleleft r)\xtworightarrow{e_2[m]}(p\triangleleft r)+ (e_2\triangleleft r)}$$

      $$\frac{p\xtworightarrow{e_1[m]}p'\quad e_1\notin q}{(p+ q)\triangleleft r\xtworightarrow{e_1[m]}(p'+q)\triangleleft r}
      \quad \frac{p\xtworightarrow{e_1[m]}p'\quad e_1\notin q}{(p\triangleleft r)+ (q\triangleleft r)\xtworightarrow{e_1[m]}(p'\triangleleft r)+(q\triangleleft r)}$$

      $$\frac{q\xtworightarrow{e_2[m]}q'\quad e_2\notin p}{(p+ q)\triangleleft r\xtworightarrow{e_2[m]}(p+q')\triangleleft r}
      \quad \frac{q\xtworightarrow{e_2[m]}q'\quad e_2\notin p}{(p\triangleleft r)+ (q\triangleleft r)\xtworightarrow{e_2[m]}(p\triangleleft r)+(q'\triangleleft r)}$$

      $$\frac{p\xtworightarrow{e_1[m]}p'\quad q\xtworightarrow{e_1[m]}q'}{(p+ q)\triangleleft r\xtworightarrow{e_1[m]}(p'+q')\triangleleft r}
      \quad \frac{p\xtworightarrow{e_1[m]}p'\quad q\xtworightarrow{e_1[m]}q'}{(p\triangleleft r)+ (q\triangleleft r)\xtworightarrow{e_1[m]}(p'\triangleleft r)+(q'\triangleleft r)}$$

      So, with the assumptions $(e_1[m]+ q)\triangleleft r\sim_s^{fr} (e_1[m]\triangleleft r)+ (q\triangleleft r)$, $(e_1+ q)\triangleleft r\sim_s^{fr} (e_1\triangleleft r)+ (q\triangleleft r)$, $(p+ e_2[m])\triangleleft r\sim_s^{fr} (p\triangleleft r)+ (e_2[m]\triangleleft r)$, $(p'+ q)\triangleleft r\sim_s^{fr} (p'\triangleleft r)+ (q\triangleleft r)$, $(p+ q')\triangleleft r\sim_s^{fr} (p\triangleleft r)+ (q'\triangleleft r)$, $(p'+ q')\triangleleft r\sim_s^{fr} (p'\triangleleft r)+ (q'\triangleleft r)$, $(p+ q)\triangleleft r\sim_s^{fr} (p\triangleleft r)+ (q\triangleleft r)$, as desired.
  \item \textbf{Axiom $U31$}. Let $p,q,r$ be $RAPTC$ processes, and $(p\cdot q)\triangleleft r = (p\triangleleft r) \cdot (q\triangleleft r)$, it is sufficient to prove that $(p\cdot q)\triangleleft r \sim_s^{fr} (p\triangleleft r) \cdot (q\triangleleft r)$. By the forward transition rules for operators $\cdot$ and $\triangleleft$ in Table \ref{SETRForBRATC} and \ref{TRForConflict}, we get

      $$\frac{p\xrightarrow{e_1}e_1[m]}{(p\cdot q)\triangleleft r\xrightarrow{e_1}(e_1[m]\cdot q)\triangleleft r}
      \quad \frac{p\xrightarrow{e_1}e_1[m]}{(p\triangleleft r)\cdot (q\triangleleft r)\xrightarrow{e_1}(e_1[m]\triangleleft r)\cdot (q\triangleleft r)}$$

      $$\frac{p\xrightarrow{e_1}p'}{(p\cdot q)\triangleleft r\xrightarrow{e_1}(p'\cdot q)\triangleleft r}
      \quad \frac{p\xrightarrow{e_1}p'}{(p\triangleleft r)\cdot (q\triangleleft r)\xrightarrow{e_1}(p'\triangleleft r)\cdot (q\triangleleft r)}$$

      By the reverse transition rules for operators $\cdot$ and $\triangleleft$ in Table \ref{RSETRForBRATC} and \ref{RTRForConflict}, we get

      $$\frac{q\xtworightarrow{e_2[m]}e_2}{(p\cdot q)\triangleleft r\xtworightarrow{e_2[m]}(p\cdot e_2)\triangleleft r}
      \quad \frac{q\xtworightarrow{e_2[m]}e_2}{(p\triangleleft r)\cdot (q\triangleleft r)\xtworightarrow{e_2[m]}(p\triangleleft r)\cdot (e_2\triangleleft r)}$$

      $$\frac{q\xtworightarrow{e_2[m]}q'}{(p\cdot q)\triangleleft r\xtworightarrow{e_2[m]}(p\cdot q')\triangleleft r}
      \quad \frac{q\xtworightarrow{e_2[m]}q'}{(p\triangleleft r)\cdot (q\triangleleft r)\xtworightarrow{e_2[m]}(p\triangleleft r)\cdot (q'\triangleleft r)}$$

      With the assumptions $(e_1[m]\cdot q)\triangleleft r \sim_s^{fr} (e_1[m]\triangleleft r) \cdot (q\triangleleft r)$, $(e_1\cdot q)\triangleleft r \sim_s^{fr} (e_1\triangleleft r) \cdot (q\triangleleft r)$, $(p'\cdot q)\triangleleft r = (p'\triangleleft r) \cdot (q\triangleleft r)$, $(p\cdot e_2[m])\triangleleft r\sim_s^{fr} (p\triangleleft r)\cdot (e_2[m]\triangleleft r)$, $(p\cdot e_2)\triangleleft r\sim_s^{fr} (p\triangleleft r)\cdot (e_2\triangleleft r)$, $(p\cdot q')\triangleleft r = (p\triangleleft r) \cdot (q'\triangleleft r)$, so, $(p\cdot q)\triangleleft r\sim_s^{fr} (p\triangleleft r)\cdot (q\triangleleft r)$, as desired.
  \item \textbf{Axiom $U32$}. Let $p,q,r$ be $RAPTC$ processes, and $(p\parallel q)\triangleleft r = (p\triangleleft r) \parallel (q\triangleleft r)$, it is sufficient to prove that $(p\parallel q)\triangleleft r \sim_s^{fr} (p\triangleleft r) \parallel (q\triangleleft r)$. By the forward transition rules for operators $\parallel$ and $\triangleleft$ in Table \ref{TRForParallel} and \ref{TRForConflict}, we get

      $$\frac{p\xrightarrow{e_1}e_1[m]\quad q\xrightarrow{e_2}e_2[m]}{(p\parallel q)\triangleleft r\xrightarrow{\{e_1,e_2\}}(e_1[m]\parallel e_2[m])\triangleleft r}
      \quad \frac{p\xrightarrow{e_1}e_1[m]\quad q\xrightarrow{e_2}e_2[m]}{(p\triangleleft r)\parallel (q\triangleleft r)\xrightarrow{\{e_1,e_2\}}(e_1[m]\triangleleft r)\parallel (e_2[m]\triangleleft r)}$$

      $$\frac{p\xrightarrow{e_1}p'\quad q\xrightarrow{e_2}e_2[m]}{(p\parallel q)\triangleleft r\xrightarrow{\{e_1,e_2\}}(p'\parallel e_2[m])\triangleleft r}
      \quad \frac{p\xrightarrow{e_1}p'\quad q\xrightarrow{e_2}e_2[m]}{(p\triangleleft r)\parallel (q\triangleleft r)\xrightarrow{\{e_1,e_2\}}(p'\triangleleft r)\parallel (e_2[m]\triangleleft r)}$$

      $$\frac{p\xrightarrow{e_1}e_1[m]\quad q\xrightarrow{e_2}q'}{(p\parallel q)\triangleleft r\xrightarrow{\{e_1,e_2\}}(e_1[m]\parallel q')\triangleleft r}
      \quad \frac{p\xrightarrow{e_1}e_1[m]\quad q\xrightarrow{e_2}q'}{(p\triangleleft r)\parallel (q\triangleleft r)\xrightarrow{\{e_1,e_2\}}(e_1[m]\triangleleft r)\parallel(q'\triangleleft r)}$$

      $$\frac{p\xrightarrow{e_1}p'\quad q\xrightarrow{e_2}q'}{(p\parallel q)\triangleleft r\xrightarrow{\{e_1,e_2\}}(p'\between q')\triangleleft r}
      \quad \frac{p\xrightarrow{e_1}p'\quad q\xrightarrow{e_2}q'}{(p\triangleleft r)\parallel (q\triangleleft r)\xrightarrow{\{e_1,e_2\}}(p'\triangleleft r)\between (q'\triangleleft r)}$$

      By the reverse transition rules for operators $\parallel$ and $\triangleleft$ in Table \ref{RTRForParallel} and \ref{RTRForConflict}, we get

      $$\frac{p\xtworightarrow{e_1[m]}e_1\quad q\xtworightarrow{e_2[m]}e_2}{(p\parallel q)\triangleleft r\xtworightarrow{\{e_1[m],e_2[m]\}}(e_1\parallel e_2)\triangleleft r}
      \quad \frac{p\xtworightarrow{e_1[m]}e_1\quad q\xtworightarrow{e_2[m]}e_2}{(p\triangleleft r)\parallel (q\triangleleft r)\xtworightarrow{\{e_1[m],e_2[m]\}}(e_1\triangleleft r)\parallel (e_2\triangleleft r)}$$

      $$\frac{p\xtworightarrow{e_1[m]}p'\quad q\xtworightarrow{e_2[m]}e_2}{(p\parallel q)\triangleleft r\xtworightarrow{\{e_1[m],e_2[m]\}}(p'\parallel e_2)\triangleleft r}
      \quad \frac{p\xtworightarrow{e_1[m]}p'\quad q\xtworightarrow{e_2[m]}e_2}{(p\triangleleft r)\parallel (q\triangleleft r)\xtworightarrow{\{e_1[m],e_2[m]\}}(p'\triangleleft r)\parallel (e_2\triangleleft r)}$$

      $$\frac{p\xtworightarrow{e_1[m]}e_1\quad q\xtworightarrow{e_2[m]}q'}{(p\parallel q)\triangleleft r\xtworightarrow{\{e_1[m],e_2[m]\}}(e_1\parallel q')\triangleleft r}
      \quad \frac{p\xtworightarrow{e_1[m]}e_1\quad q\xtworightarrow{e_2[m]}q'}{(p\triangleleft r)\parallel (q\triangleleft r)\xtworightarrow{\{e_1[m],e_2[m]\}}(e_1\triangleleft r)\parallel(q'\triangleleft r)}$$

      $$\frac{p\xtworightarrow{e_1[m]}p'\quad q\xtworightarrow{e_2[m]}q'}{(p\parallel q)\triangleleft r\xtworightarrow{\{e_1[m],e_2[m]\}}(p'\between q')\triangleleft r}
      \quad \frac{p\xtworightarrow{e_1[m]}p'\quad q\xtworightarrow{e_2[m]}q'}{(p\triangleleft r)\parallel (q\triangleleft r)\xtworightarrow{\{e_1[m],e_2[m]\}}(p'\triangleleft r)\between (q'\triangleleft r)}$$

      With the assumptions $(e_1[m]\parallel e_2[m])\triangleleft r\sim_s^{fr} (e_1[m]\triangleleft r)\parallel (e_2[m]\triangleleft r)$, $(e_1\parallel e_2)\triangleleft r\sim_s^{fr} (e_1\triangleleft r)\parallel (e_2\triangleleft r)$, $(e_1[m]\parallel q')\triangleleft r\sim_s^{fr} (e_1[m]\triangleleft r)\parallel (q'\triangleleft r)$, $(e_1\parallel q')\triangleleft r\sim_s^{fr} (e_1\triangleleft r)\parallel (q'\triangleleft r)$, $(p'\parallel e_2[m])\triangleleft r\sim_s^{fr} (p'\triangleleft r)\parallel (e_2[m]\triangleleft r)$, $(p'\parallel e_2)\triangleleft r\sim_s^{fr} (p'\triangleleft r)\parallel (e_2\triangleleft r)$, $(p'\between q')\triangleleft r = (p'\triangleleft r) \between (q'\triangleleft r)$, so, $(p\parallel q)\triangleleft r\sim_s^{fr} (p\triangleleft r)\parallel (q\triangleleft r)$, as desired.
  \item \textbf{Axiom $U33$}. Let $p,q,r$ be $RAPTC$ processes, and $(p\mid q)\triangleleft r = (p\triangleleft r) \mid (q\triangleleft r)$, it is sufficient to prove that $(p\mid q)\triangleleft r \sim_s^{fr} (p\triangleleft r) \mid (q\triangleleft r)$. By the forward transition rules for operators $\mid$ and $\triangleleft$ in Table \ref{TRForCommunication} and \ref{TRForConflict}, we get

      $$\frac{p\xrightarrow{e_1}e_1[m]\quad q\xrightarrow{e_2}e_2[m]}{(p\mid q)\triangleleft r\xrightarrow{\gamma(e_1,e_2)}(e_1[m]\mid e_2[m])\triangleleft r}
      \quad \frac{p\xrightarrow{e_1}e_1[m]\quad q\xrightarrow{e_2}e_2[m]}{(p\triangleleft r)\mid (q\triangleleft r)\xrightarrow{\gamma(e_1,e_2)}(e_1[m]\triangleleft r)\mid (e_2[m]\triangleleft r)}$$

      $$\frac{p\xrightarrow{e_1}p'\quad q\xrightarrow{e_2}e_2[m]}{(p\mid q)\triangleleft r\xrightarrow{\gamma(e_1,e_2)}(p'\mid e_2[m])\triangleleft r}
      \quad \frac{p\xrightarrow{e_1}p'\quad q\xrightarrow{e_2}e_2[m]}{(p\triangleleft r)\mid (q\triangleleft r)\xrightarrow{\gamma(e_1,e_2)}(p'\triangleleft r)\mid(e_2[m]\triangleleft r)}$$

      $$\frac{p\xrightarrow{e_1}e_1[m]\quad q\xrightarrow{e_2}q'}{(p\mid q)\triangleleft r\xrightarrow{\gamma(e_1,e_2)}(e_1[m]\mid q')\triangleleft r}
      \quad \frac{p\xrightarrow{e_1}e_1[m]\quad q\xrightarrow{e_2}q'}{(p\triangleleft r)\mid (q\triangleleft r)\xrightarrow{\gamma(e_1,e_2)}(e_1[m]\triangleleft r)\mid(q'\triangleleft r)}$$

      $$\frac{p\xrightarrow{e_1}p'\quad q\xrightarrow{e_2}q'}{(p\mid q)\triangleleft r\xrightarrow{\gamma(e_1,e_2)}(p'\between q')\triangleleft r}
      \quad \frac{p\xrightarrow{e_1}p'\quad q\xrightarrow{e_2}q'}{(p\triangleleft r)\mid (q\triangleleft r)\xrightarrow{\gamma(e_1,e_2)}(p'\triangleleft r)\between (q'\triangleleft r)}$$

      By the reverse transition rules for operators $\mid$ and $\triangleleft$ in Table \ref{RTRForCommunication} and \ref{RTRForConflict}, we get

      $$\frac{p\xtworightarrow{e_1[m]}e_1\quad q\xtworightarrow{e_2[m]}e_2}{(p\mid q)\triangleleft r\xtworightarrow{\gamma(e_1,e_2)[m]}(e_1\mid e_2)\triangleleft r}
      \quad \frac{p\xtworightarrow{e_1[m]}e_1\quad q\xtworightarrow{e_2[m]}e_2}{(p\triangleleft r)\mid (q\triangleleft r)\xtworightarrow{\gamma(e_1,e_2)[m]}(e_1\triangleleft r)\mid (e_2\triangleleft r)}$$

      $$\frac{p\xtworightarrow{e_1[m]}p'\quad q\xtworightarrow{e_2[m]}e_2}{(p\mid q)\triangleleft r\xtworightarrow{\gamma(e_1,e_2)[m]}(p'\mid e_2)\triangleleft r}
      \quad \frac{p\xtworightarrow{e_1[m]}p'\quad q\xtworightarrow{e_2[m]}e_2}{(p\triangleleft r)\mid (q\triangleleft r)\xtworightarrow{\gamma(e_1,e_2)[m]}(p'\triangleleft r)\mid(e_2\triangleleft r)}$$

      $$\frac{p\xtworightarrow{e_1[m]}e_1\quad q\xtworightarrow{e_2[m]}q'}{(p\mid q)\triangleleft r\xtworightarrow{\gamma(e_1,e_2)[m]}(e_1\mid q')\triangleleft r}
      \quad \frac{p\xtworightarrow{e_1[m]}e_1\quad q\xtworightarrow{e_2[m]}q'}{(p\triangleleft r)\mid (q\triangleleft r)\xtworightarrow{\gamma(e_1,e_2)[m]}(e_1\triangleleft r)\mid(q'\triangleleft r)}$$

      $$\frac{p\xtworightarrow{e_1[m]}p'\quad q\xtworightarrow{e_2[m]}q'}{(p\mid q)\triangleleft r\xtworightarrow{\gamma(e_1,e_2)[m]}(p'\between q')\triangleleft r}
      \quad \frac{p\xtworightarrow{e_1[m]}p'\quad q\xtworightarrow{e_2[m]}q'}{(p\triangleleft r)\mid (q\triangleleft r)\xtworightarrow{\gamma(e_1,e_2)[m]}(p'\triangleleft r)\between (q'\triangleleft r)}$$

      With the assumptions $(e_1[m]\mid e_2[m])\triangleleft r\sim_s^{fr} (e_1[m]\triangleleft r)\mid (e_2[m]\triangleleft r)$, $(e_1[m]\mid q')\triangleleft r\sim_s^{fr} (e_1[m]\triangleleft r)\mid (q'\triangleleft r)$, $(e_1\mid q')\triangleleft r\sim_s^{fr} (e_1\triangleleft r)\mid (q'\triangleleft r)$, $(p'\mid e_2[m])\triangleleft r\sim_s^{fr} (p'\triangleleft r)\mid (e_2[m]\triangleleft r)$, $(p'\mid e_2)\triangleleft r\sim_s^{fr} (p'\triangleleft r)\mid (e_2\triangleleft r)$, $(p'\between q')\triangleleft r = (p'\triangleleft r) \between (q'\triangleleft r)$, so, $(p\mid q)\triangleleft r\sim_s^{fr} (p\triangleleft r)\mid (q\triangleleft r)$, as desired.
  \item \textbf{Axiom $U34$}. Let $p,q,r$ be $RAPTC$ processes, and $p\triangleleft (q+ r) = (p\triangleleft q)\triangleleft r$, it is sufficient to prove that $p\triangleleft (q+ r) \sim_s^{fr} (p\triangleleft q)\triangleleft r$. By the forward transition rules for operators $+$ and $\triangleleft$ in Table \ref{SETRForBRATC} and \ref{TRForConflict}, we get

      $$\frac{p\xrightarrow{e_1}e_1[m]}{p\triangleleft (q+ r)\xrightarrow{e_1}e_1[m]\triangleleft (q+ r)}
      \quad \frac{p\xrightarrow{e_1}e_1[m]}{(p\triangleleft q)\triangleleft r\xrightarrow{e_1}(e_1[m]\triangleleft q)\triangleleft r}$$

      $$\frac{p\xrightarrow{e_1}p'}{p\triangleleft (q+ r)\xrightarrow{e_1}p'\triangleleft (q+ r)}
      \quad \frac{p\xrightarrow{e_1}p'}{(p\triangleleft q)\triangleleft r\xrightarrow{e_1}(p'\triangleleft q)\triangleleft r}$$

      By the reverse transition rules for operators $+$ and $\triangleleft$ in Table \ref{RSETRForBRATC} and \ref{RTRForConflict}, we get

      $$\frac{p\xtworightarrow{e_1[m]}e_1}{p\triangleleft (q+ r)\xtworightarrow{e_1[m]}e_1\triangleleft (q+ r)}
      \quad \frac{p\xtworightarrow{e_1[m]}e_1}{(p\triangleleft q)\triangleleft r\xtworightarrow{e_1[m]}(e_1\triangleleft q)\triangleleft r}$$

      $$\frac{p\xtworightarrow{e_1[m]}p'}{p\triangleleft (q+ r)\xtworightarrow{e_1[m]}p'\triangleleft (q+ r)}
      \quad \frac{p\xtworightarrow{e_1[m]}p'}{(p\triangleleft q)\triangleleft r\xtworightarrow{e_1[m]}(p'\triangleleft q)\triangleleft r}$$

      With the assumptions $e_1[m]\triangleleft (q+ r) \sim_s^{fr} (e_1[m]\triangleleft q)\triangleleft r$, $e_1\triangleleft (q+ r) \sim_s^{fr} (e_1\triangleleft q)\triangleleft r$, $p'\triangleleft (q+ r) \sim_s^{fr} (p'\triangleleft q)\triangleleft r$, so, $p\triangleleft (q+ r) \sim_s^{fr} (p\triangleleft q)\triangleleft r$, as desired.
  \item \textbf{Axiom $U35$}. Let $p,q,r$ be $RAPTC$ processes, and $p\triangleleft (q\cdot r) = (p\triangleleft q)\triangleleft r$, it is sufficient to prove that $p\triangleleft (q\cdot r) \sim_s^{fr} (p\triangleleft q)\triangleleft r$. By the forward transition rules for operators $\cdot$ and $\triangleleft$ in Table \ref{SETRForBRATC} and \ref{TRForConflict}, we get

      $$\frac{p\xrightarrow{e_1}e_1[m]}{p\triangleleft (q\cdot r)\xrightarrow{e_1}e_1[m]\triangleleft (q\cdot r)}
      \quad \frac{p\xrightarrow{e_1}e_1[m]}{(p\triangleleft q)\triangleleft r\xrightarrow{e_1}(e_1[m]\triangleleft q)\triangleleft r}$$

      $$\frac{p\xrightarrow{e_1}p'}{p\triangleleft (q\cdot r)\xrightarrow{e_1}p'\triangleleft (q\cdot r)}
      \quad \frac{p\xrightarrow{e_1}p'}{(p\triangleleft q)\triangleleft r\xrightarrow{e_1}(p'\triangleleft q)\triangleleft r}$$

      By the reverse transition rules for operators $\cdot$ and $\triangleleft$ in Table \ref{RSETRForBRATC} and \ref{RTRForConflict}, we get

      $$\frac{p\xtworightarrow{e_1[m]}e_1}{p\triangleleft (q\cdot r)\xtworightarrow{e_1[m]}e_1\triangleleft (q\cdot r)}
      \quad \frac{p\xtworightarrow{e_1[m]}e_1}{(p\triangleleft q)\triangleleft r\xtworightarrow{e_1[m]}(e_1\triangleleft q)\triangleleft r}$$

      $$\frac{p\xtworightarrow{e_1[m]}p'}{p\triangleleft (q\cdot r)\xtworightarrow{e_1[m]}p'\triangleleft (q\cdot r)}
      \quad \frac{p\xtworightarrow{e_1[m]}p'}{(p\triangleleft q)\triangleleft r\xtworightarrow{e_1[m]}(p'\triangleleft q)\triangleleft r}$$

      With the assumptions $e_1[m]\triangleleft (q\cdot r) \sim_s^{fr} (e_1[m]\triangleleft q)\triangleleft r$, $e_1\triangleleft (q\cdot r) \sim_s^{fr} (e_1\triangleleft q)\triangleleft r$, $p'\triangleleft (q\cdot r) \sim_s^{fr} (p'\triangleleft q)\triangleleft r$, so, $p\triangleleft (q\cdot r) \sim_s^{fr} (p\triangleleft q)\triangleleft r$, as desired.
  \item \textbf{Axiom $U36$}. Let $p,q,r$ be $RAPTC$ processes, and $p\triangleleft (q\parallel r) = (p\triangleleft q)\triangleleft r$, it is sufficient to prove that $p\triangleleft (q\parallel r) \sim_s^{fr} (p\triangleleft q)\triangleleft r$. By the forward transition rules for operators $\parallel$ and $\triangleleft$ in Table \ref{TRForParallel} and \ref{TRForConflict}, we get

      $$\frac{p\xrightarrow{e_1}e_1[m]}{p\triangleleft (q\parallel r)\xrightarrow{e_1}e_1[m]\triangleleft (q\parallel r)}
      \quad \frac{p\xrightarrow{e_1}e_1[m]}{(p\triangleleft q)\triangleleft r\xrightarrow{e_1}(e_1[m]\triangleleft q)\triangleleft r}$$

      $$\frac{p\xrightarrow{e_1}p'}{p\triangleleft (q\parallel r)\xrightarrow{e_1}p'\triangleleft (q\parallel r)}
      \quad \frac{p\xrightarrow{e_1}p'}{(p\triangleleft q)\triangleleft r\xrightarrow{e_1}(p'\triangleleft q)\triangleleft r}$$

      By the reverse transition rules for operators $\parallel$ and $\triangleleft$ in Table \ref{RTRForParallel} and \ref{RTRForConflict}, we get

      $$\frac{p\xtworightarrow{e_1[m]}e_1}{p\triangleleft (q\parallel r)\xtworightarrow{e_1[m]}e_1\triangleleft (q\parallel r)}
      \quad \frac{p\xtworightarrow{e_1[m]}e_1}{(p\triangleleft q)\triangleleft r\xtworightarrow{e_1[m]}(e_1\triangleleft q)\triangleleft r}$$

      $$\frac{p\xtworightarrow{e_1[m]}p'}{p\triangleleft (q\parallel r)\xtworightarrow{e_1[m]}p'\triangleleft (q\parallel r)}
      \quad \frac{p\xtworightarrow{e_1[m]}p'}{(p\triangleleft q)\triangleleft r\xtworightarrow{e_1[m]}(p'\triangleleft q)\triangleleft r}$$

      With the assumptions $e_1[m]\triangleleft (q\parallel r) \sim_s^{fr} (e_1[m]\triangleleft q)\triangleleft r$, $e_1\triangleleft (q\parallel r) \sim_s^{fr} (e_1\triangleleft q)\triangleleft r$, $p'\triangleleft (q\parallel r) \sim_s^{fr} (p'\triangleleft q)\triangleleft r$, so, $p\triangleleft (q\parallel r) \sim_s^{fr} (p\triangleleft q)\triangleleft r$, as desired.
  \item \textbf{Axiom $U37$}. Let $p,q,r$ be $RAPTC$ processes, and $p\triangleleft (q\mid r) = (p\triangleleft q)\triangleleft r$, it is sufficient to prove that $p\triangleleft (q\mid r) \sim_s^{fr} (p\triangleleft q)\triangleleft r$. By the forward transition rules for operators $\mid$ and $\triangleleft$ in Table \ref{TRForCommunication} and \ref{TRForConflict}, we get

      $$\frac{p\xrightarrow{e_1}e_1[m]}{p\triangleleft (q\mid r)\xrightarrow{e_1}e_1[m]\triangleleft (q\mid r)}
      \quad \frac{p\xrightarrow{e_1}e_1[m]}{(p\triangleleft q)\triangleleft r\xrightarrow{e_1}e_1[m]\triangleleft (q\mid r)}$$

      $$\frac{p\xrightarrow{e_1}p'}{p\triangleleft (q\mid r)\xrightarrow{e_1}p'\triangleleft (q\mid r)}
      \quad \frac{p\xrightarrow{e_1}p'}{(p\triangleleft q)\triangleleft r\xrightarrow{e_1}(p'\triangleleft q)\triangleleft r}$$

      By the reverse transition rules for operators $\mid$ and $\triangleleft$ in Table \ref{RTRForCommunication} and \ref{RTRForConflict}, we get

      $$\frac{p\xtworightarrow{e_1[m]}e_1}{p\triangleleft (q\mid r)\xtworightarrow{e_1[m]}e_1\triangleleft (q\mid r)}
      \quad \frac{p\xtworightarrow{e_1[m]}e_1}{(p\triangleleft q)\triangleleft r\xrightarrow{e_1[m]}e_1\triangleleft (q\mid r)}$$

      $$\frac{p\xtworightarrow{e_1[m]}p'}{p\triangleleft (q\mid r)\xtworightarrow{e_1[m]}p'\triangleleft (q\mid r)}
      \quad \frac{p\xtworightarrow{e_1[m]}p'}{(p\triangleleft q)\triangleleft r\xtworightarrow{e_1[m]}(p'\triangleleft q)\triangleleft r}$$

      With the assumptions $e_1[m]\triangleleft (q\mid r) \sim_s^{fr} (e_1[m]\triangleleft q)\triangleleft r$, $e_1\triangleleft (q\mid r) \sim_s^{fr} (e_1\triangleleft q)\triangleleft r$, $p'\triangleleft (q\mid r) \sim_s^{fr} (p'\triangleleft q)\triangleleft r$, so, $p\triangleleft (q\mid r) \sim_s^{fr} (p\triangleleft q)\triangleleft r$, as desired.
\end{itemize}
\end{proof}

\begin{theorem}[Soundness of parallelism modulo FR pomset bisimulation equivalence]\label{SPPBE}
Let $x$ and $y$ be $RAPTC$ terms. If $RAPTC\vdash x=y$, then $x\sim_{p}^{fr} y$.
\end{theorem}

\begin{proof}
Since FR pomset bisimulation $\sim_{p}^{fr}$ is both an equivalent and a congruent relation with respect to the operators $\between$, $\parallel$, $\mid$, $\Theta$ and $\triangleleft$, we only need to check if each axiom in Table \ref{AxiomsForParallelism} is sound modulo FR pomset bisimulation equivalence.

From the definition of FR pomset bisimulation (see Definition \ref{FRPSB}), we know that FR pomset bisimulation is defined by pomset transitions, which are labeled by pomsets. In a pomset transition, the events in the pomset are either within causality relations (defined by $\cdot$) or in concurrency (implicitly defined by $\cdot$ and $+$, and explicitly defined by $\between$), of course, they are pairwise consistent (without conflicts). In Theorem \ref{SPSBE}, we have already proven the case that all events are pairwise concurrent, so, we only need to prove the case of events in causality. Without loss of generality, we take a pomset of $P=\{e_1,e_2:e_1\cdot e_2\}$. Then the pomset transition labeled by the above $P$ is just composed of one single event transition labeled by $e_1$ succeeded by another single event transition labeled by $e_2$, that is, $\xrightarrow{P}=\xrightarrow{e_1}\xrightarrow{e_2}$ or $\xrightarrow{P}=\xtworightarrow{e_2[n]}\xtworightarrow{e_1[m]}$.

Similarly to the proof of soundness of parallelism modulo FR step bisimulation equivalence (see Theorem \ref{SPSBE}), we can prove that each axiom in Table \ref{AxiomsForParallelism} is sound modulo FR pomset bisimulation equivalence, we omit them.
\end{proof}

\begin{theorem}[Soundness of parallelism modulo FR hp-bisimulation equivalence]\label{SPHPBE}
Let $x$ and $y$ be $RAPTC$ terms. If $RAPTC\vdash x=y$, then $x\sim_{hp}^{fr} y$.
\end{theorem}

\begin{proof}
Since FR hp-bisimulation $\sim_{hp}^{fr}$ is both an equivalent and a congruent relation with respect to the operators $\between$, $\parallel$, $\mid$, $\Theta$ and $\triangleleft$, we only need to check if each axiom in Table \ref{AxiomsForParallelism} is sound modulo FR hp-bisimulation equivalence.

From the definition of FR hp-bisimulation (see Definition \ref{FRHHPB}), we know that FR hp-bisimulation is defined on the posetal product $(C_1,f,C_2),f:C_1\rightarrow C_2\textrm{ isomorphism}$. Two process terms $s$ related to $C_1$ and $t$ related to $C_2$, and $f:C_1\rightarrow C_2\textrm{ isomorphism}$. Initially, $(C_1,f,C_2)=(\emptyset,\emptyset,\emptyset)$, and $(\emptyset,\emptyset,\emptyset)\in\sim_{hp}^{fr}$. When $s\xrightarrow{e}s'$ ($C_1\xrightarrow{e}C_1'$), there will be $t\xrightarrow{e}t'$ ($C_2\xrightarrow{e}C_2'$), and we define $f'=f[e\mapsto e]$. And when $s\xtworightarrow{e[m]}s'$ ($C_1\xtworightarrow{e[m]}C_1'$), there will be $t\xtworightarrow{e[m]}t'$ ($C_2\xtworightarrow{e[m]}C_2'$), and we define $f'=f[e[m]\mapsto e[m]]$. Then, if $(C_1,f,C_2)\in\sim_{hp}^{fr}$, then $(C_1',f',C_2')\in\sim_{hp}^{fr}$.

Similarly to the proof of soundness of parallelism modulo FR pomset bisimulation equivalence (see Theorem \ref{SPPBE}), we can prove that each axiom in Table \ref{AxiomsForParallelism} is sound modulo FR hp-bisimulation equivalence, we just need additionally to check the above conditions on FR hp-bisimulation, we omit them.
\end{proof}

\subsection{Encapsulation}

The mismatch of two communicating events in different parallel branches can cause deadlock, so the deadlocks in the concurrent processes should be eliminated. Like $APTC$ \cite{APTC}, we also introduce the unary encapsulation operator $\partial_H$ for set $H$ of atomic events, which renames all atomic events in $H$ into $\delta$. The whole algebra including parallelism for true concurrency in the above subsections, deadlock $\delta$ and encapsulation operator $\partial_H$, is called Reversible Algebra for Parallelism in True Concurrency, abbreviated $RAPTC$.

The forward transition rules of encapsulation operator $\partial_H$ are shown in Table \ref{TRForEncapsulation}, and the reverse transition rules of encapsulation operator $\partial_H$ are shown in Table \ref{RTRForEncapsulation}.

\begin{center}
    \begin{table}
        $$\frac{x\xrightarrow{e}e[m]}{\partial_H(x)\xrightarrow{e}\partial_H(e[m])}\quad (e\notin H)\quad\frac{x\xrightarrow{e}x'}{\partial_H(x)\xrightarrow{e}\partial_H(x')}\quad(e\notin H)$$
        \caption{Forward transition rules of encapsulation operator $\partial_H$}
        \label{TRForEncapsulation}
    \end{table}
\end{center}

\begin{center}
    \begin{table}
        $$\frac{x\xtworightarrow{e[m]}e}{\partial_H(x)\xtworightarrow{e[m]}e}\quad (e\notin H)\quad\quad\frac{x\xtworightarrow{e}x'}{\partial_H(x)\xtworightarrow{e}\partial_H(x')}\quad(e\notin H)$$
        \caption{Reverse transition rules of encapsulation operator $\partial_H$}
        \label{RTRForEncapsulation}
    \end{table}
\end{center}

Based on the transition rules for encapsulation operator $\partial_H$ in Table \ref{TRForEncapsulation} and Table \ref{RTRForEncapsulation}, we design the axioms as Table \ref{AxiomsForEncapsulation} shows.

\begin{center}
    \begin{table}
        \begin{tabular}{@{}ll@{}}
            \hline No. &Axiom\\
            $D1$ & $e\notin H\quad\partial_H(e) = e$\\
            $RD1$ & $e\notin H\quad\partial_H(e[m]) = e[m]$\\
            $D2$ & $e\in H\quad \partial_H(e) = \delta$\\
            $RD2$ & $e\in H\quad \partial_H(e[m]) = \delta$\\
            $D3$ & $\partial_H(\delta) = \delta$\\
            $D4$ & $\partial_H(x+ y) = \partial_H(x)+\partial_H(y)$\\
            $D5$ & $\partial_H(x\cdot y) = \partial_H(x)\cdot\partial_H(y)$\\
            $D6$ & $\partial_H(x\parallel y) = \partial_H(x)\parallel\partial_H(y)$\\
        \end{tabular}
        \caption{Axioms of encapsulation operator}
        \label{AxiomsForEncapsulation}
    \end{table}
\end{center}

\begin{theorem}[Conservativity of $RAPTC$ with respect to the reversible algebra for parallelism]
$RAPTC$ is a conservative extension of the reversible algebra for parallelism.
\end{theorem}

\begin{proof}
It follows from the following two facts (see Theorem \ref{TCE}).

\begin{enumerate}
  \item The transition rules of the reversible algebra for parallelism in the above subsections are all source-dependent;
  \item The sources of the transition rules for the encapsulation operator contain an occurrence of $\partial_H$.
\end{enumerate}

So, $RAPTC$ is a conservative extension of the reversible algebra for parallelism, as desired.
\end{proof}

\begin{theorem}[Congruence theorem of encapsulation operator $\partial_H$]
FR truly concurrent bisimulation equivalences $\sim_{p}^{fr}$, $\sim_s^{fr}$, $\sim_{hp}^{fr}$ and $\sim_{hhp}^{fr}$ are all congruences with respect to encapsulation operator $\partial_H$.
\end{theorem}

\begin{proof}
(1) Case FR pomset bisimulation equivalence $\sim_p^{fr}$.

Let $x$ and $y$ be $RAPTC$ processes, and $x\sim_{p}^{fr} y$, it is sufficient to prove that $\partial_H(x)\sim_{p}^{fr} \partial_H(y)$.

By the definition of FR pomset bisimulation $\sim_p^{fr}$ (Definition \ref{FRPSB}), $x\sim_p^{fr} y$ means that

$$x\xrightarrow{X} x' \quad y\xrightarrow{Y} y'$$

$$x\xtworightarrow{X[\mathcal{K}]} x' \quad y\xtworightarrow{Y[\mathcal{J}]} y'$$

with $X\subseteq x$, $Y\subseteq y$, $X\sim Y$ and $x'\sim_p^{fr} y'$.

By the FR pomset transition rules for encapsulation operator $\partial_H$ in Table \ref{TRForEncapsulation} and Table \ref{RTRForEncapsulation}, we can get

$$\partial_H(x)\xrightarrow{X} \partial_H(X[\mathcal{K}]) (X\nsubseteq H) \quad \partial_H(y)\xrightarrow{Y} \partial_H(Y[\mathcal{J}]) (Y\nsubseteq H)$$

$$\partial_H(x)\xtworightarrow{X[\mathcal{K}]} \partial_H(X) (X\nsubseteq H) \quad \partial_H(y)\xtworightarrow{Y[\mathcal{J}]} \partial_H(Y) (Y\nsubseteq H)$$

with $X\subseteq x$, $Y\subseteq y$, and $X\sim Y$, and the assumptions $\partial_H(X[\mathcal{K}])\sim_p^{fr} \partial_H(Y[\mathcal{J}])$, $\partial_H(X)\sim_p^{fr} \partial_H(Y)$ so, we get $\partial_H(x)\sim_p^{fr} \partial_H(y)$, as desired.

Or, we can get

$$\partial_H(x)\xrightarrow{X} \partial_H(x') (X\nsubseteq H) \quad \partial_H(y)\xrightarrow{Y} \partial_H(y') (Y\nsubseteq H)$$

$$\partial_H(x)\xtworightarrow{X} \partial_H(x') (X\nsubseteq H) \quad \partial_H(y)\xtworightarrow{Y} \partial_H(y') (Y\nsubseteq H)$$

with $X\subseteq x$, $Y\subseteq y$, $X\sim Y$, $x'\sim_p^{fr} y'$ and the assumption $\partial_H(x')\sim_p^{fr}\partial_H(y')$, so, we get $\partial_H(x)\sim_p^{fr} \partial_H(y)$, as desired.

(2) The cases of FR step bisimulation $\sim_s^{fr}$, FR hp-bisimulation $\sim_{hp}^{fr}$ and FR hhp-bisimulation $\sim_{hhp}^{fr}$ can be proven similarly, we omit them.
\end{proof}

\begin{theorem}[Elimination theorem of $RAPTC$]\label{ETEncapsulation}
Let $p$ be a closed $RAPTC$ term including the encapsulation operator $\partial_H$. Then there is a basic $RAPTC$ term $q$ such that $RAPTC\vdash p=q$.
\end{theorem}

\begin{proof}
(1) Firstly, suppose that the following ordering on the signature of $RAPTC$ is defined: $\parallel > \cdot > +$ and the symbol $\parallel$ is given the lexicographical status for the first argument, then for each rewrite rule $p\rightarrow q$ in Table \ref{TRSForEncapsulation} relation $p>_{lpo} q$ can easily be proved. We obtain that the term rewrite system shown in Table \ref{TRSForEncapsulation} is strongly normalizing, for it has finitely many rewriting rules, and $>$ is a well-founded ordering on the signature of $RAPTC$, and if $s>_{lpo} t$, for each rewriting rule $s\rightarrow t$ is in Table \ref{TRSForEncapsulation} (see Theorem \ref{SN}).

\begin{center}
    \begin{table}
        \begin{tabular}{@{}ll@{}}
            \hline No. &Rewriting Rule\\
            $RRD1$ & $e\notin H\quad\partial_H(e) \rightarrow e$\\
            $RRRD1$ & $e\notin H\quad\partial_H(e[m]) \rightarrow e[m]$\\
            $RRD2$ & $e\in H\quad \partial_H(e) \rightarrow \delta$\\
            $RRRD2$ & $e\in H\quad \partial_H(e[m]) \rightarrow \delta$\\
            $RRD3$ & $\partial_H(\delta) \rightarrow \delta$\\
            $RRD4$ & $\partial_H(x+ y) \rightarrow \partial_H(x)+\partial_H(y)$\\
            $RRD5$ & $\partial_H(x\cdot y) \rightarrow \partial_H(x)\cdot\partial_H(y)$\\
            $RRD6$ & $\partial_H(x\parallel y) \rightarrow \partial_H(x)\parallel\partial_H(y)$\\
        \end{tabular}
        \caption{Term rewrite system of encapsulation operator $\partial_H$}
        \label{TRSForEncapsulation}
    \end{table}
\end{center}

(2) Then we prove that the normal forms of closed $RAPTC$ terms including encapsulation operator $\partial_H$ are basic $RAPTC$ terms.

Suppose that $p$ is a normal form of some closed $RAPTC$ term and suppose that $p$ is not a basic $RAPTC$ term. Let $p'$ denote the smallest sub-term of $p$ which is not a basic $RAPTC$ term. It implies that each sub-term of $p'$ is a basic $RAPTC$ term. Then we prove that $p$ is not a term in normal form. It is sufficient to induct on the structure of $p'$, following from Theorem \ref{ETParallelism}, we only prove the new case $p'\equiv \partial_H(p_1)$:

\begin{itemize}
  \item Case $p_1\equiv e$. The transition rules $RRD1$ or $RRD2$ can be applied, so $p$ is not a normal form;
  \item Case $p_1\equiv e[m]$. The transition rules $RRRD1$ or $RRRD2$ can be applied, so $p$ is not a normal form;
  \item Case $p_1\equiv \delta$. The transition rules $RRD3$ can be applied, so $p$ is not a normal form;
  \item Case $p_1\equiv p_1'+ p_1''$. The transition rules $RRD4$ can be applied, so $p$ is not a normal form;
  \item Case $p_1\equiv p_1'\cdot p_1''$. The transition rules $RRD5$ can be applied, so $p$ is not a normal form;
  \item Case $p_1\equiv p_1'\parallel p_1''$. The transition rules $RRD6$ can be applied, so $p$ is not a normal form.
\end{itemize}
\end{proof}

\begin{theorem}[Soundness of $RAPTC$ modulo FR step bisimulation equivalence]\label{SAPTCSBE}
Let $x$ and $y$ be $RAPTC$ terms including encapsulation operator $\partial_H$. If $RAPTC\vdash x=y$, then $x\sim_{s}^{fr} y$.
\end{theorem}

\begin{proof}
Since FR step bisimulation $\sim_s^{fr}$ is both an equivalent and a congruent relation with respect to the operator $\partial_H$, we only need to check if each axiom in Table \ref{AxiomsForEncapsulation} is sound modulo FR step bisimulation equivalence.

Though transition rules in Table \ref{TRForEncapsulation} and Table \ref{RTRForEncapsulation} are defined in the flavor of single event, they can be modified into a step (a set of events within which each event is pairwise concurrent), we omit them. If we treat a single event as a step containing just one event, the proof of this soundness theorem does not exist any problem, so we use this way and still use the transition rules in Table \ref{TRForEncapsulation} and Table \ref{RTRForEncapsulation}.

We omit the defining axioms, including axioms $D1-D3$, and we only prove the soundness of the non-trivial axioms, including axioms $D4-D6$.

\begin{itemize}
  \item \textbf{Axiom $D4$}. Let $p,q$ be $RAPTC$ processes, and $\partial_H(p+ q)=\partial_H(p)+\partial_H(q)$, it is sufficient to prove that $\partial_H(p+ q)\sim_s^{fr} \partial_H(p)+\partial_H(q)$. By the forward transition rules for operator $+$ in Table \ref{SETRForBRATC} and $\partial_H$ in Table \ref{TRForEncapsulation}, we get

      $$\frac{p\xrightarrow{e_1}e_1[m]\quad e_1\notin q\quad(e_1\notin H)}{\partial_H(p+ q)\xrightarrow{e_1}\partial_H(e_1[m]+q)}
      \quad\frac{p\xrightarrow{e_1}e_1[m]\quad e_1\notin q\quad(e_1\notin H)}{\partial_H(p)+ \partial_H(q)\xrightarrow{e_1}\partial_H(e_1[m])+\partial_H(q)}$$

      $$\frac{q\xrightarrow{e_2}e_2[m]\quad e_2\notin p\quad(e_2\notin H)}{\partial_H(p+ q)\xrightarrow{e_2}\partial_H(p+e_2[m])}
      \quad\frac{q\xrightarrow{e_2}e_2[m]\quad e_2\notin p\quad(e_2\notin H)}{\partial_H(p)+ \partial_H(q)\xrightarrow{e_2}\partial_H(p)+\partial_H(e_2[m])}$$

      $$\frac{p\xrightarrow{e_1}p'\quad e_1\notin q\quad(e_1\notin H)}{\partial_H(p+ q)\xrightarrow{e_1}\partial_H(p'+q)}
      \quad\frac{p\xrightarrow{e_1}p'\quad e_1\notin q\quad(e_1\notin H)}{\partial_H(p)+ \partial_H(q)\xrightarrow{e_1}\partial_H(p')+\partial(q)}$$

      $$\frac{q\xrightarrow{e_2}q'\quad e_2\notin p\quad(e_2\notin H)}{\partial_H(p+ q)\xrightarrow{e_2}\partial_H(p+q')}
      \quad\frac{q\xrightarrow{e_2}q'\quad e_2\notin p\quad(e_2\notin H)}{\partial_H(p)+ \partial_H(q)\xrightarrow{e_2}\partial_H(p)+\partial_H(q')}$$

      $$\frac{q\xrightarrow{e_2}q'\quad e_2\notin p\quad(e_2\notin H)}{\partial_H(p+ q)\xrightarrow{e_2}\partial_H(p+q')}
      \quad\frac{q\xrightarrow{e_2}q'\quad e_2\notin p\quad(e_2\notin H)}{\partial_H(p)+ \partial_H(q)\xrightarrow{e_2}\partial_H(p)+\partial_H(q')}$$

      $$\frac{p\xrightarrow{e_1}p'\quad q\xrightarrow{e_1}q'\quad(e_1\notin H)}{\partial_H(p+ q)\xrightarrow{e_1}\partial_H(p'+q')}
      \quad\frac{p\xrightarrow{e_1}p'\quad q\xrightarrow{e_1}q' \quad(e_1\notin H)}{\partial_H(p)+ \partial_H(q)\xrightarrow{e_1}\partial_H(p')+\partial(q')}$$

      By the reverse transition rules for operator $+$ in Table \ref{RSETRForBRATC} and $\partial_H$ in Table \ref{RTRForEncapsulation}, we get

      $$\frac{p\xtworightarrow{e_1[m]}e_1\quad e_1\notin q\quad(e_1\notin H)}{\partial_H(p+ q)\xtworightarrow{e_1[m]}\partial_H(e_1+q)}
      \quad\frac{p\xtworightarrow{e_1[m]}e_1\quad e_1\notin q\quad(e_1\notin H)}{\partial_H(p)+ \partial_H(q)\xtworightarrow{e_1[m]}\partial_H(e_1)+\partial_H(q)}$$

      $$\frac{q\xtworightarrow{e_2[m]}e_2\quad e_2\notin p\quad(e_2\notin H)}{\partial_H(p+ q)\xtworightarrow{e_2[m]}\partial_H(p+e_2)}
      \quad\frac{q\xtworightarrow{e_2[m]}e_2\quad e_2\notin p\quad(e_2\notin H)}{\partial_H(p)+ \partial_H(q)\xtworightarrow{e_2[m]}\partial_H(p)+\partial_H(e_2)}$$

      $$\frac{p\xtworightarrow{e_1[m]}p'\quad e_1\notin q\quad(e_1\notin H)}{\partial_H(p+ q)\xtworightarrow{e_1[m]}\partial_H(p'+q)}
      \quad\frac{p\xtworightarrow{e_1[m]}p'\quad e_1\notin q\quad(e_1\notin H)}{\partial_H(p)+ \partial_H(q)\xtworightarrow{e_1[m]}\partial_H(p')+\partial(q)}$$

      $$\frac{q\xtworightarrow{e_2[m]}q'\quad e_2\notin p\quad(e_2\notin H)}{\partial_H(p+ q)\xtworightarrow{e_2[m]}\partial_H(p+q')}
      \quad\frac{q\xtworightarrow{e_2[m]}q'\quad e_2\notin p\quad(e_2\notin H)}{\partial_H(p)+ \partial_H(q)\xtworightarrow{e_2[m]}\partial_H(p)+\partial_H(q')}$$

      $$\frac{q\xtworightarrow{e_2[m]}q'\quad e_2\notin p\quad(e_2\notin H)}{\partial_H(p+ q)\xtworightarrow{e_2[m]}\partial_H(p+q')}
      \quad\frac{q\xtworightarrow{e_2[m]}q'\quad e_2\notin p\quad(e_2\notin H)}{\partial_H(p)+ \partial_H(q)\xtworightarrow{e_2[m]}\partial_H(p)+\partial_H(q')}$$

      $$\frac{p\xtworightarrow{e_1[m]}p'\quad q\xtworightarrow{e_1[m]}q'\quad(e_1\notin H)}{\partial_H(p+ q)\xtworightarrow{e_1[m]}\partial_H(p'+q')}
      \quad\frac{p\xtworightarrow{e_1[m]}p'\quad q\xtworightarrow{e_1[m]}q' \quad(e_1\notin H)}{\partial_H(p)+ \partial_H(q)\xtworightarrow{e_1[m]}\partial_H(p')+\partial(q')}$$

      So, with the assumptions $\partial_H(e_1[m]+ q)\sim_s^{fr} \partial_H(e_1[m])+\partial_H(q)$, $\partial_H(e_1+ q)\sim_s^{fr} \partial_H(e_1)+\partial_H(q)$, $\partial_H(p+ e_2[m])\sim_s^{fr} \partial_H(p)+\partial_H(e_2[m])$, $\partial_H(p+ e_2)\sim_s^{fr} \partial_H(p)+\partial_H(e_2)$, $\partial_H(p'+ q)\sim_s^{fr} \partial_H(p')+\partial_H(q)$, $\partial_H(p+ q')\sim_s^{fr} \partial_H(p)+\partial_H(q')$, $\partial_H(p'+ q')\sim_s^{fr} \partial_H(p')+\partial_H(q')$, $\partial_H(p+ q)\sim_s^{fr} \partial_H(p)+\partial_H(q)$, as desired.
  \item \textbf{Axiom $D5$}. Let $p,q$ be $RAPTC$ processes, and $\partial_H(p\cdot q)=\partial_H(p)\cdot\partial_H(q)$, it is sufficient to prove that $\partial_H(p\cdot q)\sim_s^{fr} \partial_H(p)\cdot\partial_H(q)$. By the forward transition rules for operator $\cdot$ in Table \ref{SETRForBRATC} and $\partial_H$ in Table \ref{TRForEncapsulation}, we get

      $$\frac{p\xrightarrow{e_1}e_1[m]\quad(e_1\notin H)}{\partial_H(p\cdot q)\xrightarrow{e_1}\partial_H(e_1[m]\cdot q)}
      \quad\frac{p\xrightarrow{e_1}e_1[m]\quad(e_1\notin H)}{\partial_H(p)\cdot \partial_H(q)\xrightarrow{e_1}\partial_H(e_1[m])\cdot\partial_H(q)}$$

      $$\frac{p\xrightarrow{e_1}p'\quad(e_1\notin H)}{\partial_H(p\cdot q)\xrightarrow{e_1}\partial_H(p'\cdot q)}
      \quad\frac{p\xrightarrow{e_1}p'\quad(e_1\notin H)}{\partial_H(p)\cdot \partial_H(q)\xrightarrow{e_1}\partial_H(p')\cdot\partial_H(q)}$$

      By the reverse transition rules for operator $\cdot$ in Table \ref{RSETRForBRATC} and $\partial_H$ in Table \ref{RTRForEncapsulation}, we get

      $$\frac{q\xtworightarrow{e_2[m]}e_2\quad(e_2\notin H)}{\partial_H(p\cdot q)\xtworightarrow{e_2[m]}\partial_H(p\cdot e_2)}
      \quad\frac{q\xtworightarrow{e_2[m]}e_2\quad(e_2\notin H)}{\partial_H(p)\cdot \partial_H(q)\xtworightarrow{e_2[m]}\partial_H(p)\cdot\partial_H(e_2)}$$

      $$\frac{q\xtworightarrow{e_2[m]}q'\quad(e_2\notin H)}{\partial_H(p\cdot q)\xtworightarrow{e_2[m]}\partial_H(p\cdot q')}
      \quad\frac{q\xtworightarrow{e_2[m]}q'\quad(e_2\notin H)}{\partial_H(p)\cdot \partial_H(q)\xtworightarrow{e_2[m]}\partial_H(p)\cdot\partial_H(q')}$$

      So, with the assumptions $\partial_H(e_1[m]\cdot q)\sim_s^{fr}\partial_H(e_1[m])\cdot\partial_H(q)$, $\partial_H(p\cdot e_2)\sim_s^{fr}\partial_H(p)\cdot\partial_H(e_2)$, $\partial_H(p'\cdot q)\sim_s^{fr}\partial_H(p')\cdot\partial_H(q)$, $\partial_H(p\cdot q')\sim_s^{fr}\partial_H(p)\cdot\partial_H(q')$, $\partial_H(p\cdot q)\sim_s^{fr}\partial_H(p)\cdot\partial_H(q)$, as desired.
  \item \textbf{Axiom $D6$}. Let $p,q$ be $RAPTC$ processes, and $\partial_H(p\parallel q)=\partial_H(p)\parallel\partial_H(q)$, it is sufficient to prove that $\partial_H(p\parallel q)\sim_s^{fr} \partial_H(p)\parallel\partial_H(q)$. By the forward transition rules for operator $\parallel$ in Table \ref{TRForParallel} and $\partial_H$ in Table \ref{TRForEncapsulation}, we get

      $$\frac{p\xrightarrow{e_1}e_1[m]\quad q\xrightarrow{e_2}e_2[m]\quad(e_1,e_2\notin H)}{\partial_H(p\parallel q)\xrightarrow{\{e_1,e_2\}}\partial_H(e_1[m]\parallel e_2[m])}
      \quad\frac{p\xrightarrow{e_1}e_1[m]\quad q\xrightarrow{e_2}e_2[m]\quad(e_1,e_2\notin H)}{\partial_H(p)\parallel \partial_H(q)\xrightarrow{\{e_1,e_2\}}\partial_H(e_1[m])\parallel\partial_H(e_2[m])}$$

      $$\frac{p\xrightarrow{e_1}p'\quad q\xrightarrow{e_2}e_2[m]\quad(e_1,e_2\notin H)}{\partial_H(p\parallel q)\xrightarrow{\{e_1,e_2\}}\partial_H(p'\parallel e_2[m])}
      \quad\frac{p\xrightarrow{e_1}p'\quad q\xrightarrow{e_2}e_2[m]\quad(e_1,e_2\notin H)}{\partial_H(p)\parallel \partial_H(q)\xrightarrow{\{e_1,e_2\}}\partial_H(p')\parallel\partial_H(e_2[m])}$$

      $$\frac{p\xrightarrow{e_1}e_1[m]\quad q\xrightarrow{e_2}q'\quad(e_1,e_2\notin H)}{\partial_H(p\parallel q)\xrightarrow{\{e_1,e_2\}}\partial_H(e_1[m]\parallel q')}
      \quad\frac{p\xrightarrow{e_1}e_1[m]\quad q\xrightarrow{e_2}q'\quad(e_1,e_2\notin H)}{\partial_H(p)\parallel \partial_H(q)\xrightarrow{\{e_1,e_2\}}\partial_H(e_1[m])\partial_H(q')}$$

      $$\frac{p\xrightarrow{e_1}p'\quad q\xrightarrow{e_2}q'\quad(e_1,e_2\notin H)}{\partial_H(p\parallel q)\xrightarrow{\{e_1,e_2\}}\partial_H(p'\between q')}
      \quad\frac{p\xrightarrow{e_1}p'\quad q\xrightarrow{e_2}q'\quad(e_1,e_2\notin H)}{\partial_H(p)\parallel \partial_H(q)\xrightarrow{\{e_1,e_2\}}\partial_H(p')\between\partial_H(q')}$$

      By the reverse transition rules for operator $\parallel$ in Table \ref{RTRForParallel} and $\partial_H$ in Table \ref{RTRForEncapsulation}, we get

      $$\frac{p\xtworightarrow{e_1[m]}e_1\quad q\xtworightarrow{e_2[m]}e_2\quad(e_1,e_2\notin H)}{\partial_H(p\parallel q)\xtworightarrow{\{e_1[m],e_2[m]\}}\partial_H(e_1\parallel e_2)}
      \quad\frac{p\xtworightarrow{e_1[m]}e_1\quad q\xtworightarrow{e_2[m]}e_2\quad(e_1,e_2\notin H)}{\partial_H(p)\parallel \partial_H(q)\xtworightarrow{\{e_1[m],e_2[m]\}}\partial_H(e_1)\parallel\partial_H(e_2)}$$

      $$\frac{p\xtworightarrow{e_1[m]}p'\quad q\xtworightarrow{e_2[m]}e_2\quad(e_1,e_2\notin H)}{\partial_H(p\parallel q)\xtworightarrow{\{e_1[m],e_2[m]\}}\partial_H(p'\parallel e_2)}
      \quad\frac{p\xtworightarrow{e_1[m]}p'\quad q\xtworightarrow{e_2[m]}e_2\quad(e_1,e_2\notin H)}{\partial_H(p)\parallel \partial_H(q)\xtworightarrow{\{e_1[m],e_2[m]\}}\partial_H(p')\parallel\partial_H(e_2)}$$

      $$\frac{p\xtworightarrow{e_1[m]}e_1\quad q\xtworightarrow{e_2[m]}q'\quad(e_1,e_2\notin H)}{\partial_H(p\parallel q)\xtworightarrow{\{e_1[m],e_2[m]\}}\partial_H(e_1\parallel q')}
      \quad\frac{p\xtworightarrow{e_1[m]}e_1\quad q\xtworightarrow{e_2[m]}q'\quad(e_1,e_2\notin H)}{\partial_H(p)\parallel \partial_H(q)\xtworightarrow{\{e_1[m],e_2[m]\}}\partial_H(e_1)\partial_H(q')}$$

      $$\frac{p\xtworightarrow{e_1[m]}p'\quad q\xtworightarrow{e_2[m]}q'\quad(e_1,e_2\notin H)}{\partial_H(p\parallel q)\xtworightarrow{\{e_1[m],e_2[m]\}}\partial_H(p'\between q')}
      \quad\frac{p\xtworightarrow{e_1[m]}p'\quad q\xtworightarrow{e_2[m]}q'\quad(e_1,e_2\notin H)}{\partial_H(p)\parallel \partial_H(q)\xtworightarrow{\{e_1[m],e_2[m]\}}\partial_H(p')\between\partial_H(q')}$$

      So, with the assumptions $\partial_H(e_1[m]\parallel q')\sim_s^{fr} \partial_H(e_1[m])\parallel\partial_H(q')$, $\partial_H(e_1\parallel q')\sim_s^{fr} \partial_H(e_1)\parallel\partial_H(q')$, $\partial_H(p'\parallel e_2[m])\sim_s^{fr} \partial_H(p')\parallel\partial_H(e_2[m])$, $\partial_H(p'\parallel e_2)\sim_s^{fr} \partial_H(p')\parallel\partial_H(e_2)$, $\partial_H(p'\between q')\sim_s^{fr}\partial_H(p')\between\partial_H(q')$, $\partial_H(p\parallel q)\sim_s^{fr} \partial_H(p)\parallel\partial_H(q)$, as desired.
\end{itemize}
\end{proof}

\begin{theorem}[Soundness of $RAPTC$ modulo FR pomset bisimulation equivalence]\label{SAPTCPBE}
Let $x$ and $y$ be $RAPTC$ terms including encapsulation operator $\partial_H$. If $RAPTC\vdash x=y$, then $x\sim_{p}^{fr} y$.
\end{theorem}

\begin{proof}
Since FR pomset bisimulation $\sim_{p}^{fr}$ is both an equivalent and a congruent relation with respect to the operator $\partial_H$, we only need to check if each axiom in Table \ref{AxiomsForEncapsulation} is sound modulo FR pomset bisimulation equivalence.

From the definition of FR pomset bisimulation (see Definition \ref{PSB}), we know that FR pomset bisimulation is defined by pomset transitions, which are labeled by pomsets. In a pomset transition, the events in the pomset are either within causality relations (defined by $\cdot$) or in concurrency (implicitly defined by $\cdot$ and $+$, and explicitly defined by $\between$), of course, they are pairwise consistent (without conflicts). In Theorem \ref{SAPTCSBE}, we have already proven the case that all events are pairwise concurrent, so, we only need to prove the case of events in causality. Without loss of generality, we take a pomset of $P=\{e_1,e_2:e_1\cdot e_2\}$. Then the pomset transition labeled by the above $P$ is just composed of one single event transition labeled by $e_1$ succeeded by another single event transition labeled by $e_2$, that is, $\xrightarrow{P}=\xrightarrow{e_1}\xrightarrow{e_2}$ or $\xrightarrow{P}=\xtworightarrow{e_2[n]}\xtworightarrow{e_1[m]}$.

Similarly to the proof of soundness of $RAPTC$ modulo FR step bisimulation equivalence (see Theorem \ref{SAPTCSBE}), we can prove that each axiom in Table \ref{AxiomsForEncapsulation} is sound modulo FR pomset bisimulation equivalence, we omit them.
\end{proof}

\begin{theorem}[Soundness of $RAPTC$ modulo FR hp-bisimulation equivalence]\label{SAPTCHPBE}
Let $x$ and $y$ be $RAPTC$ terms including encapsulation operator $\partial_H$. If $RAPTC\vdash x=y$, then $x\sim_{hp}^{fr} y$.
\end{theorem}

\begin{proof}
Since FR hp-bisimulation $\sim_{hp}^{fr}$ is both an equivalent and a congruent relation with respect to the operator $\partial_H$, we only need to check if each axiom in Table \ref{AxiomsForEncapsulation} is sound modulo FR hp-bisimulation equivalence.

From the definition of FR hp-bisimulation (see Definition \ref{HHPB}), we know that FR hp-bisimulation is defined on the posetal product $(C_1,f,C_2),f:C_1\rightarrow C_2\textrm{ isomorphism}$. Two process terms $s$ related to $C_1$ and $t$ related to $C_2$, and $f:C_1\rightarrow C_2\textrm{ isomorphism}$. Initially, $(C_1,f,C_2)=(\emptyset,\emptyset,\emptyset)$, and $(\emptyset,\emptyset,\emptyset)\in\sim_{hp}^{fr}$. When $s\xrightarrow{e}s'$ ($C_1\xrightarrow{e}C_1'$), there will be $t\xrightarrow{e}t'$ ($C_2\xrightarrow{e}C_2'$), and we define $f'=f[e\mapsto e]$. And when $s\xtworightarrow{e[m]}s'$ ($C_1\xtworightarrow{e[m]}C_1'$), there will be $t\xtworightarrow{e[m]}t'$ ($C_2\xtworightarrow{e[m]}C_2'$), and we define $f'=f[e[m]\mapsto e[m]]$. Then, if $(C_1,f,C_2)\in\sim_{hp}^{fr}$, then $(C_1',f',C_2')\in\sim_{hp}^{fr}$.

Similarly to the proof of soundness of $RAPTC$ modulo FR pomset bisimulation equivalence (see Theorem \ref{SAPTCPBE}), we can prove that each axiom in Table \ref{AxiomsForEncapsulation} is sound modulo FR hp-bisimulation equivalence, we just need additionally to check the above conditions on FR hp-bisimulation, we omit them.
\end{proof}

\subsection{Recursion}

\begin{definition}[Weakly guarded recursive expression]
$X$ is weakly guarded in $E$ if each occurrence of $X$ is with some subexpression $\alpha.F$ or $(\alpha_1\parallel\cdots\parallel\alpha_n).F$ or $F.\alpha[m]$ or $F.(\alpha_1[m]\parallel\cdots\parallel\alpha_n[m])$ of $E$.
\end{definition}

\begin{proposition}\label{LUS}
If the variables $\widetilde{X}$ are weakly guarded in $E$, and $E\{\widetilde{P}/\widetilde{X}\}\xrightarrow{\{\alpha_1,\cdots,\alpha_n\}}P'$, or $E\{\widetilde{P}/\widetilde{X}\}\xtworightarrow{\{\alpha_1[m],\cdots,\alpha_n[m]\}}P'$, then $P'$ can not takes the form $E'\{\widetilde{P}/\widetilde{X}\}$ for some expression $E'$.
\end{proposition}
\begin{proof}
It needs to induct on the depth of the inference of $E\{\widetilde{P}/\widetilde{X}\}\xrightarrow{\{\alpha_1,\cdots,\alpha_n\}}P'$ or $E\{\widetilde{P}/\widetilde{X}\}\xtworightarrow{\{\alpha_1[m],\cdots,\alpha_n[m]\}}P'$. We consider $E\{\widetilde{P}/\widetilde{X}\}\xrightarrow{\{\alpha_1,\cdots,\alpha_n\}}P'$.

Case $E\equiv E_1+E_2$. We may have $E_1\xrightarrow{e_1}e_1[m]\cdot E_1'\quad e_1\notin E_2$, $E_1+E_2\xrightarrow{e_1}e_1[m]\cdot E_1'+E_2$, $e_1[m]\cdot E_1'+E_2$ can not takes the form $E'\{\widetilde{P}/\widetilde{X}\}$ for some expression $E'$.

So, there may be not recursive expression for strongly FR truly concurrent bisimulations. For the same reason, there also may be not recursive expression for weakly FR truly concurrent bisimulations.
\end{proof}

\section{Abstraction}\label{abs}

To abstract away from the internal implementations of a program, and verify that the program exhibits the desired external behaviors, the silent step $\tau$ and abstraction operator $\tau_I$ are introduced, where $I\subseteq \mathbb{E}$ denotes the internal events. The transition rule of $\tau$ is shown in Table \ref{TRForTau}. In the following, let the atomic event $e$ range over $\mathbb{E}\cup\{\delta\}\cup\{\tau\}$, and let the communication function $\gamma:\mathbb{E}\cup\{\tau\}\times \mathbb{E}\cup\{\tau\}\rightarrow \mathbb{E}\cup\{\delta\}$, with each communication involved $\tau$ resulting in $\delta$.

\begin{center}
    \begin{table}
        $$\frac{}{\tau\xrightarrow{\tau}\surd}$$
        $$\frac{}{\tau\xtworightarrow{\tau}\surd}$$
        \caption{Transition rule of the silent step}
        \label{TRForTau}
    \end{table}
\end{center}

\begin{theorem}[Conservitivity of $RAPTC$ with silent step]
$RAPTC$ with silent step is a conservative extension of $RAPTC$.
\end{theorem}

\begin{proof}
Since the transition rules of $RAPTC$ are source-dependent, and the transition rules for silent step in Table \ref{TRForTau} contain only a fresh constant $\tau$ in their source, so the transition rules of $RAPTC$ with silent step is a conservative extension of those of $RAPTC$.
\end{proof}

\begin{theorem}[Congruence theorem of $RAPTC$ with silent step]
Rooted branching FR truly concurrent bisimulation equivalences $\approx_{rbp}^{fr}$, $\approx_{rbs}^{fr}$ and $\approx_{rbhp}^{fr}$ are all congruences with respect to $RAPTC$ with silent step.
\end{theorem}

\begin{proof}
It follows the following two facts:
\begin{enumerate}
  \item FR truly concurrent bisimulation equivalences $\sim_{p}^{fr}$, $\sim_s^{fr}$ and $\sim_{hp}^{fr}$ are all congruences with respect to all operators of $RAPTC$, while FR truly concurrent bisimulation equivalences $\sim_{p}^{fr}$, $\sim_s^{fr}$ and $\sim_{hp}^{fr}$ imply the corresponding rooted branching FR truly concurrent bisimulation $\approx{rbp}^{fr}$, $\approx_{rbs}^{fr}$ and $\approx_{rbhp}^{fr}$, so rooted branching FR truly concurrent bisimulation $\approx{rbp}^{fr}$, $\approx_{rbs}^{fr}$ and $\approx_{rbhp}^{fr}$ are all congruences with respect to all operators of $RAPTC$;
  \item While $\mathbb{E}$ is extended to $\mathbb{E}\cup\{\tau\}$, it can be proved that rooted branching FR truly concurrent bisimulation $\approx{rbp}^{fr}$, $\approx_{rbs}^{fr}$ and $\approx_{rbhp}^{fr}$ are all congruences with respect to all operators of $RAPTC$, we omit it.
\end{enumerate}
\end{proof}

\subsection{Algebraic Laws for the Silent Step}

We design the axioms for the silent step $\tau$ in Table \ref{AxiomsForTau}.

\begin{center}
\begin{table}
  \begin{tabular}{@{}ll@{}}
\hline No. &Axiom\\
  $B1$ & $e\cdot\tau=e$\\
  $RB1$ & $\tau\cdot e[m]=e[m]$\\
  $B2$ & $e\cdot(\tau\cdot(x+y)+x)=e\cdot(x+y)$\\
  $RB2$ & $((x+y)\cdot\tau+x)\cdot e[m]=(x+y)\cdot e[m]$\\
  $B3$ & $x\parallel\tau=x$\\
\end{tabular}
\caption{Axioms of silent step}
\label{AxiomsForTau}
\end{table}
\end{center}

\begin{theorem}[Soundness of $RAPTC$ with silent step]\label{SAPTCTAU}
Let $x$ and $y$ be $RAPTC$ with silent step terms. If $RAPTC$ with silent step $\vdash x=y$, then
\begin{enumerate}
  \item $x\approx_{rbs}^{fr} y$;
  \item $x\approx_{rbp}^{fr} y$;
  \item $x\approx_{rbhp}^{fr} y$.
\end{enumerate}
\end{theorem}

\begin{proof}
(1) Soundness of $RAPTC$ with silent step with respect to rooted branching FR step bisimulation $\approx_{rbs}^{fr}$.

Since rooted branching FR step bisimulation $\approx_{rbs}^{fr}$ is both an equivalent and a congruent relation with respect to $RAPTC$ with silent step, we only need to check if each axiom in Table \ref{AxiomsForTau} is sound modulo rooted branching FR step bisimulation equivalence.

Though transition rules in Table \ref{TRForTau} are defined in the flavor of single event, they can be modified into a step (a set of events within which each event is pairwise concurrent), we omit them. If we treat a single event as a step containing just one event, the proof of this soundness theorem does not exist any problem, so we use this way and still use the transition rules in Table \ref{TRForTau}.

\begin{itemize}
  \item \textbf{Axiom $B1$}. Assume that $e\cdot\tau=e$, it is sufficient to prove that $e\cdot\tau\approx_{rbs}^{fr}e$. By the forward transition rules for operator $\cdot$ in Table \ref{STRForBRATC} and $\tau$ in Table \ref{TRForTau}, we get

      $$\frac{e\xrightarrow{e}e[m]}{e\cdot\tau\xrightarrow{e}\xrightarrow{\tau}e[m]}$$

      $$\frac{e\xrightarrow{e}e[m]}{e\xrightarrow{e}e[m]}$$

      By the reverse transition rules for operator $\cdot$ in Table \ref{RSTRForBRATC} and $\tau$ in Table \ref{TRForTau}, there are no transitions.

      So, $e\cdot\tau\approx_{rbs}^{fr}e$, as desired.

  \item \textbf{Axiom $RB1$}. Assume that $\tau \cdot e[m]=e[m]$, it is sufficient to prove that $\tau \cdot e[m]\approx_{rbs}^{fr}e[m]$. By the forward transition rules for operator $\cdot$ in Table \ref{STRForBRATC} and $\tau$ in Table \ref{TRForTau}, there are no transitions.

      By the reverse transition rules for operator $\cdot$ in Table \ref{RSTRForBRATC} and $\tau$ in Table \ref{TRForTau}, we get

      $$\frac{e[m]\xtworightarrow{e[m]}e}{\tau\cdot e[m]\xtworightarrow{e[m]}\xtworightarrow{\tau}e}$$

      $$\frac{e[m]\xtworightarrow{e[m]}e}{e[m]\xtworightarrow{e[m]}e}$$

      So, $\tau \cdot e[m]\approx_{rbs}^{fr}e[m]$, as desired.

  \item \textbf{Axiom $B2$}. Let $p$ and $q$ be $RAPTC$ with silent step processes, and assume that $e\cdot(\tau\cdot(p+q)+p)=e\cdot(p+q)$, it is sufficient to prove that $e\cdot(\tau\cdot(p+q)+p)\approx_{rbs}^{fr}e\cdot(p+q)$. There are several cases, we will not enumerate all. By the forward transition rules for operators $\cdot$ and $+$ in Table \ref{STRForBRATC} and $\tau$ in Table \ref{TRForTau}, we get

      $$\frac{e\xrightarrow{e}e[m]\quad p\xrightarrow{e_1}p'\quad q\xrightarrow{e_1}q'}{e\cdot(\tau\cdot(p+q)+p)\xrightarrow{e}\xrightarrow{\tau}\xrightarrow{e_1}e[m]\cdot((p'+q')+p')}$$

      $$\frac{e\xrightarrow{e}e[m]\quad p\xrightarrow{e_1}p'}{e\cdot(p+q)\xrightarrow{e}\xrightarrow{e_1}e[m]\cdot(p'+q')}$$

      By the reverse transition rules for operators $\cdot$ and $+$ in Table \ref{RSTRForBRATC} and $\tau$ in Table \ref{TRForTau}, there are no transitions.

      So, $e\cdot(\tau\cdot(p+q)+p)\approx_{rbs}^{fr}e\cdot(p+q)$, as desired.

  \item \textbf{Axiom $RB2$}. Let $p$ and $q$ be $RAPTC$ with silent step processes, and assume that $((x+y)\cdot\tau+x)\cdot e[m]=(x+y)\cdot e[m]$, it is sufficient to prove that $((x+y)\cdot\tau+x)\cdot e[m]\approx_{rbs}^{fr}(x+y)\cdot e[m]$. There are several cases, we will not enumerate all. By the forward transition rules for operators $\cdot$ and $+$ in Table \ref{STRForBRATC} and $\tau$ in Table \ref{TRForTau}, there are no transitions.

      By the reverse transition rules for operators $\cdot$ and $+$ in Table \ref{RSTRForBRATC} and $\tau$ in Table \ref{TRForTau}, we get

      $$\frac{e[m]\xtworightarrow{e[m]}e\quad p\xtworightarrow{e_1[n]}p'\quad q\xtworightarrow{e_1[n]}q'}{((p+q)\cdot\tau+p)\cdot e[m]\xtworightarrow{e[m]}\xtworightarrow{\tau}\xtworightarrow{e_1[n]}((p'+q')+p')\cdot e}$$

      $$\frac{e[m]\xtworightarrow{e[m]}e\quad p\xtworightarrow{e_1[n]}p'}{(p+q)\cdot e[m]\xtworightarrow{e[m]}\xtworightarrow{e_1[n]}(p'+q'\cdot e)}$$

      So, $((p+q)\cdot\tau+p)\cdot e[m]\approx_{rbs}^{fr}(p+q)\cdot e[m]$, as desired.

  \item \textbf{Axiom $B3$}. Let $p$ be an $RAPTC$ with silent step, and assume that $p\parallel\tau=p$, it is sufficient to prove that $p\parallel\tau\approx_{rbs}^{fr}p$. By the forward transition rules for operator $\parallel$ in Table \ref{TRForParallel} and $\tau$ in Table \ref{TRForTau}, we get

      $$\frac{p\xrightarrow{e}e[m]}{p\parallel\tau\xRightarrow{e}e[m]}$$

      $$\frac{p\xrightarrow{e}p'}{p\parallel\tau\xRightarrow{e}p'}$$

      By the reverse transition rules for operator $\parallel$ in Table \ref{RTRForParallel} and $\tau$ in Table \ref{TRForTau}, we get

      $$\frac{p\xtworightarrow{e[m]}e}{p\parallel\tau\xTworightarrow{e[m]}e}$$

      $$\frac{p\xtworightarrow{e[m]}p'}{p\parallel\tau\xTworightarrow{e[m]}p'}$$

      So, $p\parallel\tau\approx_{rbs}^{fr}p$, as desired.
\end{itemize}

(2) Soundness of $RAPTC$ with silent step with respect to rooted branching FR pomset bisimulation $\approx_{rbp}^{fr}$.

Since rooted branching FR pomset bisimulation $\approx_{rbp}^{fr}$ is both an equivalent and a congruent relation with respect to $RAPTC$ with silent step, we only need to check if each axiom in Table \ref{AxiomsForTau} is sound modulo rooted branching FR pomset bisimulation $\approx_{rbp}^{fr}$.

From the definition of rooted branching FR pomset bisimulation $\approx_{rbp}^{fr}$ (see Definition \ref{FRRBPSB}), we know that rooted branching FR pomset bisimulation $\approx_{rbp}^{fr}$ is defined by weak pomset transitions, which are labeled by pomsets with $\tau$. In a weak pomset transition, the events in the pomset are either within causality relations (defined by $\cdot$) or in concurrency (implicitly defined by $\cdot$ and $+$, and explicitly defined by $\between$), of course, they are pairwise consistent (without conflicts). In (1), we have already proven the case that all events are pairwise concurrent, so, we only need to prove the case of events in causality. Without loss of generality, we take a pomset of $P=\{e_1,e_2:e_1\cdot e_2\}$. Then the weak pomset transition labeled by the above $P$ is just composed of one single event transition labeled by $e_1$ succeeded by another single event transition labeled by $e_2$, that is, $\xRightarrow{P}=\xRightarrow{e_1}\xRightarrow{e_2}$ or $\xTworightarrow{P}=\xTworightarrow{e_2}\xTworightarrow{e_1}$.

Similarly to the proof of soundness of $RAPTC$ with silent step modulo rooted branching FR step bisimulation $\approx_{rbs}^{fr}$ (1), we can prove that each axiom in Table \ref{AxiomsForTau} is sound modulo rooted branching FR pomset bisimulation $\approx_{rbp}^{fr}$, we omit them.

(3) Soundness of $RAPTC$ with silent step with respect to rooted branching FR hp-bisimulation $\approx_{rbhp}^{fr}$.

Since rooted branching FR hp-bisimulation $\approx_{rbhp}^{fr}$ is both an equivalent and a congruent relation with respect to $RAPTC$ with silent step, we only need to check if each axiom in Table \ref{AxiomsForTau} is sound modulo rooted branching FR hp-bisimulation $\approx_{rbhp}^{fr}$.

From the definition of rooted branching FR hp-bisimulation $\approx_{rbhp}^{fr}$ (see Definition \ref{FRRBHHPB}), we know that rooted branching FR hp-bisimulation $\approx_{rbhp}^{fr}$ is defined on the weakly posetal product $(C_1,f,C_2),f:\hat{C_1}\rightarrow \hat{C_2}\textrm{ isomorphism}$. Two process terms $s$ related to $C_1$ and $t$ related to $C_2$, and $f:\hat{C_1}\rightarrow \hat{C_2}\textrm{ isomorphism}$. Initially, $(C_1,f,C_2)=(\emptyset,\emptyset,\emptyset)$, and $(\emptyset,\emptyset,\emptyset)\in\approx_{rbhp}^{fr}$. When $s\xrightarrow{e}s'$ ($C_1\xrightarrow{e}C_1'$), there will be $t\xRightarrow{e}t'$ ($C_2\xRightarrow{e}C_2'$), and we define $f'=f[e\mapsto e]$. And when $s\xTworightarrow{e[m]}s'$ ($C_1\xTworightarrow{e[m]}C_1'$), there will be $t\xTworightarrow{e[m]}t'$ ($C_2\xTworightarrow{e[m]}C_2'$), and we define $f'=f[e[m]\mapsto e[m]$. Then, if $(C_1,f,C_2)\in\approx_{rbhp}^{fr}$, then $(C_1',f',C_2')\in\approx_{rbhp}^{fr}$.

Similarly to the proof of soundness of $RAPTC$ with silent step modulo rooted branching FR pomset bisimulation equivalence (2), we can prove that each axiom in Table \ref{AxiomsForTau} is sound modulo rooted branching FR hp-bisimulation equivalence, we just need additionally to check the above conditions on rooted branching FR hp-bisimulation, we omit them.
\end{proof}

\subsection{Abstraction}

The unary abstraction operator $\tau_I$ ($I\subseteq \mathbb{E}$) renames all atomic events in $I$ into $\tau$. $RAPTC$ with silent step and abstraction operator is called $RAPTC_{\tau}$. The transition rules of operator $\tau_I$ are shown in Table \ref{TRForAbstraction}.

\begin{center}
    \begin{table}
        $$\frac{x\xrightarrow{e}\surd}{\tau_I(x)\xrightarrow{e}\surd}\quad e\notin I
        \quad\quad\frac{x\xrightarrow{e}x'}{\tau_I(x)\xrightarrow{e}\tau_I(x')}\quad e\notin I$$

        $$\frac{x\xrightarrow{e}\surd}{\tau_I(x)\xrightarrow{\tau}\surd}\quad e\in I
        \quad\quad\frac{x\xrightarrow{e}x'}{\tau_I(x)\xrightarrow{\tau}\tau_I(x')}\quad e\in I$$

        $$\frac{x\xtworightarrow{e[m]}e}{\tau_I(x)\xtworightarrow{e[m]}e}\quad e[m]\notin I
        \quad\quad\frac{x\xtworightarrow{e[m]}x'}{\tau_I(x)\xtworightarrow{e[m]}\tau_I(x')}\quad e[m]\notin I$$

        $$\frac{x\xtworightarrow{e[m]}\surd}{\tau_I(x)\xtworightarrow{\tau}\surd}\quad e[m]\in I
        \quad\quad\frac{x\xtworightarrow{e[m]}x'}{\tau_I(x)\xtworightarrow{\tau}\tau_I(x')}\quad e[m]\in I$$
        \caption{Transition rule of the abstraction operator}
        \label{TRForAbstraction}
    \end{table}
\end{center}

\begin{theorem}[Conservitivity of $RAPTC_{\tau}$]
$RAPTC_{\tau}$ is a conservative extension of $RAPTC$ with silent step.
\end{theorem}

\begin{proof}
Since the transition rules of $RAPTC$ with silent step are source-dependent, and the transition rules for abstraction operator in Table \ref{TRForAbstraction}contain only a fresh operator $\tau_I$ in their source, so the transition rules of $RAPTC_{\tau}$ is a conservative extension of those of $RAPTC$ with silent step.
\end{proof}

\begin{theorem}[Congruence theorem of $RAPTC_{\tau}$]
Rooted branching FR truly concurrent bisimulation equivalences $\approx_{rbp}^{fr}$, $\approx_{rbs}^{fr}$ and $\approx_{rbhp}^{fr}$ are all congruences with respect to $RAPTC_{\tau}$.
\end{theorem}

\begin{proof}

(1) Case rooted branching FR pomset bisimulation equivalence $\approx_{rbp}^{fr}$.

Let $x$ and $y$ be $RAPTC_{\tau}$ processes, and $x\approx_{rbp}^{fr} y$, it is sufficient to prove that $\tau_I(x)\approx_{rbp}^{fr} \tau_I(y)$.

By the transition rules for operator $\tau_I$ in Table \ref{TRForAbstraction}, we can get

$$\tau_I(x)\xrightarrow{X} X[\mathcal{K}] (X\nsubseteq I) \quad \tau_I(y)\xrightarrow{Y} Y[\mathcal{J}] (Y\nsubseteq I)$$

$$\tau_I(x)\xtworightarrow{X[\mathcal{K}]} X (X\nsubseteq I) \quad \tau_I(y)\xtworightarrow{Y[\mathcal{J}]} Y (Y\nsubseteq I)$$

with $X\subseteq x$, $Y\subseteq y$, and $X\sim Y$.

Or, we can get

$$\tau_I(x)\xrightarrow{X} \tau_I(x') (X\nsubseteq I) \quad \tau_I(y)\xrightarrow{Y} \tau_I(y') (Y\nsubseteq I)$$

$$\tau_I(x)\xtworightarrow{X[\mathcal{K}]} \tau_I(x') (X\nsubseteq I) \quad \tau_I(y)\xtworightarrow{Y[\mathcal{J}]} \tau_I(y') (Y\nsubseteq I)$$

with $X\subseteq x$, $Y\subseteq y$, and $X\sim Y$ and the hypothesis $\tau_I(x')\approx_{rbp}^{fr}\tau_I(y')$.

Or, we can get

$$\tau_I(x)\xrightarrow{\tau^*} \surd (X\subseteq I) \quad \tau_I(y)\xrightarrow{\tau^*} \surd (Y\subseteq I)$$

$$\tau_I(x)\xtworightarrow{\tau^*} \surd (X\subseteq I) \quad \tau_I(y)\xtworightarrow{\tau^*} \surd (Y\subseteq I)$$

with $X\subseteq x$, $Y\subseteq y$, and $X\sim Y$.

Or, we can get

$$\tau_I(x)\xrightarrow{\tau^*} \tau_I(x') (X\subseteq I) \quad \tau_I(y)\xrightarrow{\tau^*} \tau_I(y') (Y\subseteq I)$$

$$\tau_I(x)\xtworightarrow{\tau^*} \tau_I(x') (X\subseteq I) \quad \tau_I(y)\xtworightarrow{\tau^*} \tau_I(y') (Y\subseteq I)$$

with $X\subseteq x$, $Y\subseteq y$, and $X\sim Y$ and the hypothesis $\tau_I(x')\approx_{rbp}^{fr}\tau_I(y')$.

So, we get $\tau_I(x)\approx_{rbp}^{fr} \tau_I(y)$, as desired

(2) The cases of rooted branching FR step bisimulation $\approx_{rbs}^{fr}$, rooted branching FR hp-bisimulation $\approx_{rbhp}^{fr}$ can be proven similarly, we omit them.
\end{proof}

We design the axioms for the abstraction operator $\tau_I$ in Table \ref{AxiomsForAbstraction}.

\begin{center}
\begin{table}
  \begin{tabular}{@{}ll@{}}
\hline No. &Axiom\\
  $TI1$ & $e\notin I\quad \tau_I(e)=e$\\
  $RTI1$ & $e[m]\notin I\quad \tau_I(e[m])=e[m]$\\
  $TI2$ & $e\in I\quad \tau_I(e)=\tau$\\
  $RTI2$ & $e[m]\in I\quad \tau_I(e[m])=\tau$\\
  $TI3$ & $\tau_I(\delta)=\delta$\\
  $TI4$ & $\tau_I(x+y)=\tau_I(x)+\tau_I(y)$\\
  $TI5$ & $\tau_I(x\cdot y)=\tau_I(x)\cdot\tau_I(y)$\\
  $TI6$ & $\tau_I(x\parallel y)=\tau_I(x)\parallel\tau_I(y)$\\
\end{tabular}
\caption{Axioms of abstraction operator}
\label{AxiomsForAbstraction}
\end{table}
\end{center}

\begin{theorem}[Soundness of $RAPTC_{\tau}$]\label{SAPTCABS}
Let $x$ and $y$ be $RAPTC_{\tau}$ terms. If $RAPTC_{\tau}\vdash x=y$, then
\begin{enumerate}
  \item $x\approx_{rbs}^{fr} y$;
  \item $x\approx_{rbp}^{fr} y$;
  \item $x\approx_{rbhp}^{fr} y$.
\end{enumerate}
\end{theorem}

\begin{proof}
(1) Soundness of $RAPTC_{\tau}$ with respect to rooted branching FR step bisimulation $\approx_{rbs}^{fr}$.

Since rooted branching FR step bisimulation $\approx_{rbs}^{fr}$ is both an equivalent and a congruent relation with respect to $RAPTC_{\tau}$, we only need to check if each axiom in Table \ref{AxiomsForAbstraction} is sound modulo rooted branching FR step bisimulation equivalence.

Though transition rules in Table \ref{TRForAbstraction} are defined in the flavor of single event, they can be modified into a step (a set of events within which each event is pairwise concurrent), we omit them. If we treat a single event as a step containing just one event, the proof of this soundness theorem does not exist any problem, so we use this way and still use the transition rules in Table \ref{AxiomsForAbstraction}.

We only prove soundness of the non-trivial axioms $TI4-TI6$, and omit the defining axioms $TI1-TI3$.

\begin{itemize}
  \item \textbf{Axiom $TI4$}. Let $p,q$ be $RAPTC_{\tau}$ processes, and $\tau_I(p+ q)=\tau_I(p)+\tau_I(q)$, it is sufficient to prove that $\tau_I(p+ q)\approx_{rbs}^{fr} \tau_I(p)+\tau_I(q)$. By the forward transition rules for operator $+$ in Table \ref{STRForBRATC} and $\tau_I$ in Table \ref{TRForAbstraction}, we get

      $$\frac{p\xrightarrow{e_1}e_1[m]\quad e_1\notin q\quad(e_1\notin I)}{\tau_I(p+ q)\xrightarrow{e_1}\tau_I(e_1[m]+q)}
      \quad\frac{p\xrightarrow{e_1}e_1[m]\quad(e_1\notin I)}{\tau_I(p)+ \tau_I(q)\xrightarrow{e_1}\tau_I(e_1[m])+\tau_I(q)}$$

      $$\frac{q\xrightarrow{e_2}e_2[m]\quad e_2\notin p\quad(e_2\notin I)}{\tau_I(p+ q)\xrightarrow{e_2}\tau_I(p+e_2[m])}
      \quad\frac{q\xrightarrow{e_2}e_2[m]\quad e_2\notin p\quad(e_2\notin I)}{\tau_I(p)+ \tau_I(q)\xrightarrow{e_2}\tau_I(p)+\tau_I(e_1[m])}$$

      $$\frac{p\xrightarrow{e_1}p'\quad e_1\notin q\quad(e_1\notin I)}{\tau_I(p+ q)\xrightarrow{e_1}\tau_I(p'+q)}
      \quad\frac{p\xrightarrow{e_1}p'\quad e_1\notin q\quad(e_1\notin I)}{\tau_I(p)+ \tau_I(q)\xrightarrow{e_1}\tau_I(p')+\tau_I(q)}$$

      $$\frac{q\xrightarrow{e_2}q'\quad e_2\notin p\quad(e_2\notin I)}{\tau_I(p+ q)\xrightarrow{e_2}\tau_I(p+q')}
      \quad\frac{q\xrightarrow{e_2}q'\quad e_2\notin p\quad(e_2\notin I)}{\tau_I(p)+ \tau_I(q)\xrightarrow{e_2}\tau_I(p)+\tau_I(q')}$$

      $$\frac{p\xrightarrow{e_1}p' \quad q\xrightarrow{e_1}q'\quad(e_1\notin I)}{\tau_I(p+ q)\xrightarrow{e_1}\tau_I(p'+q')}
      \quad\frac{p\xrightarrow{e_1}p'\quad q\xrightarrow{e_1}q'\quad(e_1\notin I)}{\tau_I(p)+ \tau_I(q)\xrightarrow{e_1}\tau_I(p')+\tau_I(q')}$$

      $$\frac{p\xrightarrow{e_1}e_1[m]\quad e_1\notin q\quad(e_1\in I)}{\tau_I(p+ q)\xrightarrow{\tau}\tau_I(q)}
      \quad\frac{p\xrightarrow{e_1}e_1[m]\quad e_1\notin q\quad(e_1\in I)}{\tau_I(p)+ \tau_I(q)\xrightarrow{\tau}\tau_I(q)}$$

      $$\frac{q\xrightarrow{e_2}e_2[m]\quad e_2\notin p\quad(e_2\in I)}{\tau_I(p+ q)\xrightarrow{\tau}\tau_I(p)}
      \quad\frac{q\xrightarrow{e_2}e_2[m]\quad e_2\notin p\quad(e_2\in I)}{\tau_I(p)+ \tau_I(q)\xrightarrow{\tau}\tau_I(p)}$$

      $$\frac{p\xrightarrow{e_1}p'\quad e_1\notin q\quad(e_1\in I)}{\tau_I(p+ q)\xrightarrow{\tau}\tau_I(p'+q)}
      \quad\frac{p\xrightarrow{e_1}p'\quad e_1\notin q\quad(e_1\in I)}{\tau_I(p)+ \tau_I(q)\xrightarrow{\tau}\tau_I(p')+\tau_I(q)}$$

      $$\frac{q\xrightarrow{e_2}q'\quad e_2\notin p\quad(e_2\in I)}{\tau_I(p+ q)\xrightarrow{\tau}\tau_I(p+q')}
      \quad\frac{q\xrightarrow{e_2}q'\quad e_2\notin p\quad(e_2\in I)}{\tau_I(p)+ \tau_I(q)\xrightarrow{\tau}\tau_I(p)+\tau_I(q')}$$

      $$\frac{p\xrightarrow{e_1}p'\quad q\xrightarrow{e_1}q'\quad(e_1\in I)}{\tau_I(p+ q)\xrightarrow{\tau}\tau_I(p'+q')}
      \quad\frac{p\xrightarrow{e_1}p'\quad q\xrightarrow{e_1}q'\quad(e_1\in I)}{\tau_I(p)+ \tau_I(q)\xrightarrow{\tau}\tau_I(p')+\tau_I(q')}$$

      By the reverse transition rules for operator $+$ in Table \ref{RSTRForBRATC} and $\tau_I$ in Table \ref{TRForAbstraction}, we get

      $$\frac{p\xtworightarrow{e_1[m]}e_1\quad e_1[m]\notin q\quad(e_1\notin I)}{\tau_I(p+ q)\xtworightarrow{e_1[m]}\tau_I(e_1+q)}
      \quad\frac{p\xtworightarrow{e_1[m]}e_1\quad(e_1[m]\notin I)}{\tau_I(p)+ \tau_I(q)\xtworightarrow{e_1[m]}\tau_I(e_1)+\tau_I(q)}$$

      $$\frac{q\xtworightarrow{e_2[m]}e_2\quad e_2\notin p\quad(e_2[m]\notin I)}{\tau_I(p+ q)\xtworightarrow{e_2[m]}\tau_I(p+e_2)}
      \quad\frac{q\xtworightarrow{e_2[m]}e_2\quad e_2\notin p\quad(e_2[m]\notin I)}{\tau_I(p)+ \tau_I(q)\xtworightarrow{e_2[m]}\tau_I(p)+\tau_I(e_1)}$$

      $$\frac{p\xtworightarrow{e_1[m]}p'\quad e_1\notin q\quad(e_1[m]\notin I)}{\tau_I(p+ q)\xtworightarrow{e_1[m]}\tau_I(p'+q)}
      \quad\frac{p\xtworightarrow{e_1[m]}p'\quad e_1\notin q\quad(e_1[m]\notin I)}{\tau_I(p)+ \tau_I(q)\xtworightarrow{e_1[m]}\tau_I(p')+\tau_I(q)}$$

      $$\frac{q\xtworightarrow{e_2[m]}q'\quad e_2\notin p\quad(e_2[m]\notin I)}{\tau_I(p+ q)\xtworightarrow{e_2[m]}\tau_I(p+q')}
      \quad\frac{q\xtworightarrow{e_2[m]}q'\quad e_2\notin p\quad(e_2[m]\notin I)}{\tau_I(p)+ \tau_I(q)\xtworightarrow{e_2[m]}\tau_I(p)+\tau_I(q')}$$

      $$\frac{p\xtworightarrow{e_1[m]}p' \quad q\xtworightarrow{e_1}q'\quad(e_1[m]\notin I)}{\tau_I(p+ q)\xtworightarrow{e_1[m]}\tau_I(p'+q')}
      \quad\frac{p\xtworightarrow{e_1[m]}p'\quad q\xtworightarrow{e_1}q'\quad(e_1[m]\notin I)}{\tau_I(p)+ \tau_I(q)\xtworightarrow{e_1[m]}\tau_I(p')+\tau_I(q')}$$

      $$\frac{p\xtworightarrow{e_1[m]}e_1\quad e_1\notin q\quad(e_1[m]\in I)}{\tau_I(p+ q)\xtworightarrow{\tau}\tau_I(q)}
      \quad\frac{p\xtworightarrow{e_1[m]}e_1\quad e_1\notin q\quad(e_1[m]\in I)}{\tau_I(p)+ \tau_I(q)\xtworightarrow{\tau}\tau_I(q)}$$

      $$\frac{q\xtworightarrow{e_2[m]}e_2\quad e_2\notin p\quad(e_2[m]\in I)}{\tau_I(p+ q)\xtworightarrow{\tau}\tau_I(p)}
      \quad\frac{q\xtworightarrow{e_2[m]}e_2\quad e_2\notin p\quad(e_2[m]\in I)}{\tau_I(p)+ \tau_I(q)\xtworightarrow{\tau}\tau_I(p)}$$

      $$\frac{p\xtworightarrow{e_1[m]}p'\quad e_1\notin q\quad(e_1[m]\in I)}{\tau_I(p+ q)\xtworightarrow{\tau}\tau_I(p'+q)}
      \quad\frac{p\xtworightarrow{e_1}p'\quad e_1\notin q\quad(e_1[m]\in I)}{\tau_I(p)+ \tau_I(q)\xtworightarrow{\tau}\tau_I(p')+\tau_I(q)}$$

      $$\frac{q\xtworightarrow{e_2[m]}q'\quad e_2\notin p\quad(e_2[m]\in I)}{\tau_I(p+ q)\xtworightarrow{\tau}\tau_I(p+q')}
      \quad\frac{q\xtworightarrow{e_2}q'\quad e_2\notin p\quad(e_2[m]\in I)}{\tau_I(p)+ \tau_I(q)\xtworightarrow{\tau}\tau_I(p)+\tau_I(q')}$$

      $$\frac{p\xtworightarrow{e_1[m]}p'\quad q\xtworightarrow{e_1}q'\quad(e_1[m]\in I)}{\tau_I(p+ q)\xtworightarrow{\tau}\tau_I(p'+q')}
      \quad\frac{p\xtworightarrow{e_1[m]}p'\quad q\xtworightarrow{e_1}q'\quad(e_1[m]\in I)}{\tau_I(p)+ \tau_I(q)\xtworightarrow{\tau}\tau_I(p')+\tau_I(q')}$$

      So, with the assumptions $\tau_I(e_1[m]+ q)\approx_{rbs}^{fr} \tau_I(e_1[m])+\tau_I(q)$, $\tau_I(e_1+ q)\approx_{rbs}^{fr} \tau_I(e_1)+\tau_I(q)$, $\tau_I(p+ e_2[m])\approx_{rbs}^{fr} \tau_I(p)+\tau_I(e_2[m])$, $\tau_I(p+ e_2)\approx_{rbs}^{fr} \tau_I(p)+\tau_I(e_2)$, $\tau_I(p'+ q)\approx_{rbs}^{fr} \tau_I(p')+\tau_I(q)$, $\tau_I(p+ q')\approx_{rbs}^{fr} \tau_I(p)+\tau_I(q')$, $\tau_I(p'+ q')\approx_{rbs}^{fr} \tau_I(p')+\tau_I(q')$ $\tau_I(p+ q)\approx_{rbs}^{fr} \tau_I(p)+\tau_I(q)$, as desired.
  \item \textbf{Axiom $TI5$}. Let $p,q$ be $RAPTC_{\tau}$ processes, and $\tau_I(p\cdot q)=\tau_I(p)\cdot\tau_I(q)$, it is sufficient to prove that $\tau_I(p\cdot q)\approx_{rbs}^{fr} \tau_I(p)\cdot\tau_I(q)$. By forward the transition rules for operator $\cdot$ in Table \ref{STRForBRATC} and $\tau_I$ in Table \ref{TRForAbstraction}, we get

      $$\frac{p\xrightarrow{e_1}e_1[m]\quad(e_1\notin I)}{\tau_I(p\cdot q)\xrightarrow{e_1}\tau_I(e_1[m]\cdot q)}
      \quad\frac{p\xrightarrow{e_1}e_1[m]\quad(e_1\notin I)}{\tau_I(p)\cdot \tau_I(q)\xrightarrow{e_1}\tau_I(e_1[m])\cdot\tau_I(q)}$$

      $$\frac{p\xrightarrow{e_1}p'\quad(e_1\notin I)}{\tau_I(p\cdot q)\xrightarrow{e_1}\tau_I(p'\cdot q)}
      \quad\frac{p\xrightarrow{e_1}p'\quad(e_1\notin I)}{\tau_I(p)\cdot \tau_I(q)\xrightarrow{e_1}\tau_I(p')\cdot\tau_I(q)}$$

      $$\frac{p\xrightarrow{e_1}e_1[m]\quad(e_1\in I)}{\tau_I(p\cdot q)\xrightarrow{\tau}\tau_I(q)}
      \quad\frac{p\xrightarrow{e_1}e_1[m]\quad(e_1\in I)}{\tau_I(p)\cdot \tau_I(q)\xrightarrow{\tau}\tau_I(q)}$$

      $$\frac{p\xrightarrow{e_1}p'\quad(e_1\in I)}{\tau_I(p\cdot q)\xrightarrow{\tau}\tau_I(p'\cdot q)}
      \quad\frac{p\xrightarrow{e_1}p'\quad(e_1\in I)}{\tau_I(p)\cdot \tau_I(q)\xrightarrow{\tau}\tau_I(p')\cdot\tau_I(q)}$$

      By reverse the transition rules for operator $\cdot$ in Table \ref{RSTRForBRATC} and $\tau_I$ in Table \ref{TRForAbstraction}, we get

      $$\frac{q\xtworightarrow{e_2[m]}e_2\quad(e_2\notin I)}{\tau_I(p\cdot q)\xtworightarrow{e_2[m]}\tau_I(p\cdot e_2)}
      \quad\frac{q\xtworightarrow{e_2[m]}e_2\quad(e_2\notin I)}{\tau_I(p)\cdot \tau_I(q)\xtworightarrow{e_2[m]}\tau_I(p)\cdot\tau_I(e_2)}$$

      $$\frac{q\xtworightarrow{e_2[m]}q'\quad(e_2\notin I)}{\tau_I(p\cdot q)\xtworightarrow{e_2[m]}\tau_I(p\cdot q')}
      \quad\frac{q\xtworightarrow{e_2[m]}q'\quad(e_2\notin I)}{\tau_I(p)\cdot \tau_I(q)\xtworightarrow{e_2[m]}\tau_I(p)\cdot\tau_I(q')}$$

      $$\frac{q\xtworightarrow{e_2[m]}e_2\quad(e_2[m]\in I)}{\tau_I(p\cdot q)\xtworightarrow{\tau}\tau_I(p)}
      \quad\frac{q\xtworightarrow{e_1[m]}e_2\quad(e_2[m]\in I)}{\tau_I(p)\cdot \tau_I(q)\xtworightarrow{\tau}\tau_I(p)}$$

      $$\frac{q\xtworightarrow{e_2[m]}q'\quad(e_2[m]\in I)}{\tau_I(p\cdot q)\xtworightarrow{\tau}\tau_I(p\cdot q')}
      \quad\frac{q\xtworightarrow{e_1[m]}q'\quad(e_2[m]\in I)}{\tau_I(p)\cdot \tau_I(q)\xtworightarrow{\tau}\tau_I(p)\cdot\tau_I('q)}$$

      So, with the assumptions $\tau_I(e_1[m]\cdot q)\approx_{rbs}^{fr}\tau_I(e_1[m])\cdot\tau_I(q)$, $\tau_I(p\cdot e_2)\approx_{rbs}^{fr}\tau_I(p)\cdot\tau_I(e_2)$, $\tau_I(p'\cdot q)=\tau_I(p')\cdot\tau_I(q)$, $\tau_I(p\cdot q')=\tau_I(p)\cdot\tau_I(q')$, $\tau_I(p\cdot q)\approx_{rbs}^{fr}\tau_I(p)\cdot\tau_I(q)$, as desired.
  \item \textbf{Axiom $TI6$}. Let $p,q$ be $RAPTC_{\tau}$ processes, and $\tau_I(p\parallel q)=\tau_I(p)\parallel\tau_I(q)$, it is sufficient to prove that $\tau_I(p\parallel q)\approx_{rbs}^{fr} \tau_I(p)\parallel\tau_I(q)$. By the forward transition rules for operator $\parallel$ in Table \ref{TRForParallel} and $\tau_I$ in Table \ref{TRForAbstraction}, we get

      $$\frac{p\xrightarrow{e_1}e_1[m]\quad q\xrightarrow{e_2}e_2[m]\quad(e_1,e_2\notin I)}{\tau_I(p\parallel q)\xrightarrow{\{e_1,e_2\}}\tau_I(e_1[m]\parallel e_2[m])}
      \quad\frac{p\xrightarrow{e_1}e_1[m]\quad q\xrightarrow{e_2}e_2[m]\quad(e_1,e_2\notin I)}{\tau_I(p)\parallel \tau_I(q)\xrightarrow{\{e_1,e_2\}}\tau_I(e_1[m])\parallel \tau_I(e_2[m])}$$

      $$\frac{p\xrightarrow{e_1}p'\quad q\xrightarrow{e_2}e_2[m]\quad(e_1,e_2\notin I)}{\tau_I(p\parallel q)\xrightarrow{\{e_1,e_2\}}\tau_I(p'\parallel e_2[m])}
      \quad\frac{p\xrightarrow{e_1}p'\quad q\xrightarrow{e_2}e_2[m]\quad(e_1,e_2\notin I)}{\tau_I(p)\parallel \tau_I(q)\xrightarrow{\{e_1,e_2\}}\tau_I(p')\parallel \tau_I(e_2[m])}$$

      $$\frac{p\xrightarrow{e_1}e_1[m]\quad q\xrightarrow{e_2}q'\quad(e_1,e_2\notin I)}{\tau_I(p\parallel q)\xrightarrow{\{e_1,e_2\}}\tau_I(e_1[m]\parallel q')}
      \quad\frac{p\xrightarrow{e_1}e_1[m]\quad q\xrightarrow{e_2}q'\quad(e_1,e_2\notin I)}{\tau_I(p)\parallel \tau_I(q)\xrightarrow{\{e_1,e_2\}}\tau_I(e_1[m])\parallel\tau_I(q')}$$

      $$\frac{p\xrightarrow{e_1}p'\quad q\xrightarrow{e_2}q'\quad(e_1,e_2\notin I)}{\tau_I(p\parallel q)\xrightarrow{\{e_1,e_2\}}\tau_I(p'\between q')}
      \quad\frac{p\xrightarrow{e_1}p'\quad q\xrightarrow{e_2}q'\quad(e_1,e_2\notin I)}{\tau_I(p)\parallel \tau_I(q)\xrightarrow{\{e_1,e_2\}}\tau_I(p')\between\tau_I(q')}$$

      $$\frac{p\xrightarrow{e_1}e_1[m]\quad q\xrightarrow{e_2}e_2[m]\quad(e_1\notin I,e_2\in I)}{\tau_I(p\parallel q)\xRightarrow{e_1}\tau_I(e_1[m])}
      \quad\frac{p\xrightarrow{e_1}e_1[m]\quad q\xrightarrow{e_2}e_2[m]\quad(e_1\notin I,e_2\in I)}{\tau_I(p)\parallel \tau_I(q)\xRightarrow{e_1}\tau_I(e_1[m])}$$

      $$\frac{p\xrightarrow{e_1}p'\quad q\xrightarrow{e_2}e_2[m]\quad(e_1\notin I,e_2\in I)}{\tau_I(p\parallel q)\xRightarrow{e_1}\tau_I(p')}
      \quad\frac{p\xrightarrow{e_1}p'\quad q\xrightarrow{e_2}e_2[m]\quad(e_1\notin I,e_2\in I)}{\tau_I(p)\parallel \tau_I(q)\xRightarrow{e_1}\tau_I(p')}$$

      $$\frac{p\xrightarrow{e_1}e_1[m]\quad q\xrightarrow{e_2}q'\quad(e_1\notin I,e_2\in I)}{\tau_I(p\parallel q)\xRightarrow{e_1}\tau_I(e_1[m]\parallel q')}
      \quad\frac{p\xrightarrow{e_1}e_1[m]\quad q\xrightarrow{e_2}q'\quad(e_1\notin I,e_2\in I)}{\tau_I(p)\parallel \tau_I(q)\xRightarrow{e_1}\tau_I(e_1[m])\parallel\tau_I(q')}$$

      $$\frac{p\xrightarrow{e_1}p'\quad q\xrightarrow{e_2}q'\quad(e_1\notin I,e_2\in I)}{\tau_I(p\parallel q)\xRightarrow{e_1}\tau_I(p'\between q')}
      \quad\frac{p\xrightarrow{e_1}p'\quad q\xrightarrow{e_2}q'\quad(e_1\notin I,e_2\in I)}{\tau_I(p)\parallel \tau_I(q)\xRightarrow{e_1}\tau_I(p')\between\tau_I(q')}$$

      $$\frac{p\xrightarrow{e_1}e_1[m]\quad q\xrightarrow{e_2}e_2[m]\quad(e_1\in I,e_2\notin I)}{\tau_I(p\parallel q)\xRightarrow{e_2}\tau_I(e_2[m])}
      \quad\frac{p\xrightarrow{e_1}e_1[m]\quad q\xrightarrow{e_2}e_2[m]\quad(e_1\in I,e_2\notin I)}{\tau_I(p)\parallel \tau_I(q)\xRightarrow{e_2}\tau_I(e_2[m])}$$

      $$\frac{p\xrightarrow{e_1}p'\quad q\xrightarrow{e_2}e_2[m]\quad(e_1\in I,e_2\notin I)}{\tau_I(p\parallel q)\xRightarrow{e_2}\tau_I(p'\parallel e_2[m])}
      \quad\frac{p\xrightarrow{e_1}p'\quad q\xrightarrow{e_2}e_2[m]\quad(e_1\in I,e_2\notin I)}{\tau_I(p)\parallel \tau_I(q)\xRightarrow{e_2}\tau_I(p')\parallel \tau_I(e_2[m])}$$

      $$\frac{p\xrightarrow{e_1}e_1[m]\quad q\xrightarrow{e_2}q'\quad(e_1\in I,e_2\notin I)}{\tau_I(p\parallel q)\xRightarrow{e_2}\tau_I(q')}
      \quad\frac{p\xrightarrow{e_1}e_1[m]\quad q\xrightarrow{e_2}q'\quad(e_1\in I,e_2\notin I)}{\tau_I(p)\parallel \tau_I(q)\xRightarrow{e_2}\tau_I(q')}$$

      $$\frac{p\xrightarrow{e_1}p'\quad q\xrightarrow{e_2}q'\quad(e_1\in I,e_2\notin I)}{\tau_I(p\parallel q)\xRightarrow{e_2}\tau_I(p'\between q')}
      \quad\frac{p\xrightarrow{e_1}p'\quad q\xrightarrow{e_2}q'\quad(e_1\in I,e_2\notin I)}{\tau_I(p)\parallel \tau_I(q)\xRightarrow{e_2}\tau_I(p')\between\tau_I(q')}$$

      $$\frac{p\xrightarrow{e_1}e_1[m]\quad q\xrightarrow{e_2}e_2[m]\quad(e_1,e_2\in I)}{\tau_I(p\parallel q)\xrightarrow{\tau^*}\surd}
      \quad\frac{p\xrightarrow{e_1}e_1[m]\quad q\xrightarrow{e_2}e_2[m]\quad(e_1,e_2\in I)}{\tau_I(p)\parallel \tau_I(q)\xrightarrow{\tau^*}\surd}$$

      $$\frac{p\xrightarrow{e_1}p'\quad q\xrightarrow{e_2}e_2[m]\quad(e_1,e_2\in I)}{\tau_I(p\parallel q)\xrightarrow{\tau^*}\tau_I(p')}
      \quad\frac{p\xrightarrow{e_1}p'\quad q\xrightarrow{e_2}e_2[m]\quad(e_1,e_2\in I)}{\tau_I(p)\parallel \tau_I(q)\xrightarrow{\tau^*}\tau_I(p')}$$

      $$\frac{p\xrightarrow{e_1}e_1[m]\quad q\xrightarrow{e_2}q'\quad(e_1,e_2\in I)}{\tau_I(p\parallel q)\xrightarrow{\tau^*}\tau_I(q')}
      \quad\frac{p\xrightarrow{e_1}e_1[m]\quad q\xrightarrow{e_2}q'\quad(e_1,e_2\in I)}{\tau_I(p)\parallel \tau_I(q)\xrightarrow{\tau^*}\tau_I(q')}$$

      $$\frac{p\xrightarrow{e_1}p'\quad q\xrightarrow{e_2}q'\quad(e_1,e_2\in I)}{\tau_I(p\parallel q)\xrightarrow{\tau^*}\tau_I(p'\between q')}
      \quad\frac{p\xrightarrow{e_1}p'\quad q\xrightarrow{e_2}q'\quad(e_1,e_2\in I)}{\tau_I(p)\parallel \tau_I(q)\xrightarrow{\tau^*}\tau_I(p')\between\tau_I(q')}$$

      By the reverse transition rules for operator $\parallel$ in Table \ref{RTRForParallel} and $\tau_I$ in Table \ref{TRForAbstraction}, we get

      $$\frac{p\xtworightarrow{e_1[m]}e_1\quad q\xtworightarrow{e_2[m]}e_2\quad(e_1[m],e_2[m]\notin I)}{\tau_I(p\parallel q)\xtworightarrow{\{e_1[m],e_2[m]\}}\tau_I(e_1\parallel e_2)}
      \quad\frac{p\xtworightarrow{e_1[m]}e_1\quad q\xtworightarrow{e_2[m]}e_2\quad(e_1[m],e_2[m]\notin I)}{\tau_I(p)\parallel \tau_I(q)\xtworightarrow{\{e_1[m],e_2[m]\}}\tau_I(e_1)\parallel \tau_I(e_2)}$$

      $$\frac{p\xtworightarrow{e_1[m]}p'\quad q\xtworightarrow{e_2[m]}e_2\quad(e_1[m],e_2[m]\notin I)}{\tau_I(p\parallel q)\xtworightarrow{\{e_1[m],e_2[m]\}}\tau_I(p'\parallel e_2)}
      \quad\frac{p\xtworightarrow{e_1[m]}p'\quad q\xtworightarrow{e_2[m]}e_2\quad(e_1[m],e_2[m]\notin I)}{\tau_I(p)\parallel \tau_I(q)\xtworightarrow{\{e_1[m],e_2[m]\}}\tau_I(p')\parallel \tau_I(e_2)}$$

      $$\frac{p\xtworightarrow{e_1[m]}e_1\quad q\xtworightarrow{e_2[m]}q'\quad(e_1[m],e_2[m]\notin I)}{\tau_I(p\parallel q)\xtworightarrow{\{e_1[m],e_2[m]\}}\tau_I(e_1\parallel q')}
      \quad\frac{p\xtworightarrow{e_1[m]}e_1\quad q\xtworightarrow{e_2[m]}q'\quad(e_1[m],e_2[m]\notin I)}{\tau_I(p)\parallel \tau_I(q)\xtworightarrow{\{e_1[m],e_2[m]\}}\tau_I(e_1)\parallel\tau_I(q')}$$

      $$\frac{p\xtworightarrow{e_1[m]}p'\quad q\xtworightarrow{e_2[m]}q'\quad(e_1[m],e_2[m]\notin I)}{\tau_I(p\parallel q)\xtworightarrow{\{e_1[m],e_2[m]\}}\tau_I(p'\between q')}
      \quad\frac{p\xtworightarrow{e_1[m]}p'\quad q\xtworightarrow{e_2[m]}q'\quad(e_1[m],e_2[m]\notin I)}{\tau_I(p)\parallel \tau_I(q)\xtworightarrow{\{e_1[m],e_2[m]\}}\tau_I(p')\between\tau_I(q')}$$

      $$\frac{p\xtworightarrow{e_1[m]}e_1\quad q\xtworightarrow{e_2[m]}e_2\quad(e_1[m]\notin I,e_2[m]\in I)}{\tau_I(p\parallel q)\xTworightarrow{e_1[m]}\tau_I(e_1)}
      \quad\frac{p\xtworightarrow{e_1[m]}e_1\quad q\xtworightarrow{e_2[m]}e_2\quad(e_1[m]\notin I,e_2[m]\in I)}{\tau_I(p)\parallel \tau_I(q)\xTworightarrow{e_1[m]}\tau_I(e_1)}$$

      $$\frac{p\xtworightarrow{e_1[m]}p'\quad q\xtworightarrow{e_2[m]}e_2\quad(e_1[m]\notin I,e_2[m]\in I)}{\tau_I(p\parallel q)\xTworightarrow{e_1[m]}\tau_I(p')}
      \quad\frac{p\xtworightarrow{e_1[m]}p'\quad q\xtworightarrow{e_2[m]}e_2\quad(e_1[m]\notin I,e_2[m]\in I)}{\tau_I(p)\parallel \tau_I(q)\xTworightarrow{e_1[m]}\tau_I(p')}$$

      $$\frac{p\xtworightarrow{e_1[m]}e_1\quad q\xtworightarrow{e_2[m]}q'\quad(e_1[m]\notin I,e_2[m]\in I)}{\tau_I(p\parallel q)\xTworightarrow{e_1[m]}\tau_I(e_1\parallel q')}
      \quad\frac{p\xtworightarrow{e_1[m]}e_1\quad q\xtworightarrow{e_2[m]}q'\quad(e_1[m]\notin I,e_2[m]\in I)}{\tau_I(p)\parallel \tau_I(q)\xTworightarrow{e_1[m]}\tau_I(e_1)\parallel\tau_I(q')}$$

      $$\frac{p\xtworightarrow{e_1[m]}p'\quad q\xtworightarrow{e_2[m]}q'\quad(e_1[m]\notin I,e_2[m]\in I)}{\tau_I(p\parallel q)\xTworightarrow{e_1[m]}\tau_I(p'\between q')}
      \quad\frac{p\xtworightarrow{e_1[m]}p'\quad q\xtworightarrow{e_2[m]}q'\quad(e_1[m]\notin I,e_2[m]\in I)}{\tau_I(p)\parallel \tau_I(q)\xTworightarrow{e_1[m]}\tau_I(p')\between\tau_I(q')}$$

      $$\frac{p\xtworightarrow{e_1[m]}e_1\quad q\xtworightarrow{e_2[m]}e_2\quad(e_1[m]\in I,e_2[m]\notin I)}{\tau_I(p\parallel q)\xTworightarrow{e_2[m]}\tau_I(e_2)}
      \quad\frac{p\xtworightarrow{e_1[m]}e_1\quad q\xtworightarrow{e_2[m]}e_2\quad(e_1[m]\in I,e_2[m]\notin I)}{\tau_I(p)\parallel \tau_I(q)\xTworightarrow{e_2[m]}\tau_I(e_2)}$$

      $$\frac{p\xtworightarrow{e_1[m]}p'\quad q\xtworightarrow{e_2[m]}e_2\quad(e_1[m]\in I,e_2[m]\notin I)}{\tau_I(p\parallel q)\xTworightarrow{e_2[m]}\tau_I(p'\parallel e_2)}
      \quad\frac{p\xtworightarrow{e_1[m]}p'\quad q\xtworightarrow{e_2[m]}e_2\quad(e_1[m]\in I,e_2[m]\notin I)}{\tau_I(p)\parallel \tau_I(q)\xTworightarrow{e_2[m]}\tau_I(p')\parallel \tau_I(e_2)}$$

      $$\frac{p\xtworightarrow{e_1[m]}e_1\quad q\xtworightarrow{e_2[m]}q'\quad(e_1[m]\in I,e_2[m]\notin I)}{\tau_I(p\parallel q)\xTworightarrow{e_2[m]}\tau_I(q')}
      \quad\frac{p\xtworightarrow{e_1[m]}e_1\quad q\xtworightarrow{e_2[m]}q'\quad(e_1[m]\in I,e_2[m]\notin I)}{\tau_I(p)\parallel \tau_I(q)\xTworightarrow{e_2[m]}\tau_I(q')}$$

      $$\frac{p\xtworightarrow{e_1[m]}p'\quad q\xtworightarrow{e_2[m]}q'\quad(e_1[m]\in I,e_2[m]\notin I)}{\tau_I(p\parallel q)\xTworightarrow{e_2[m]}\tau_I(p'\between q')}
      \quad\frac{p\xtworightarrow{e_1[m]}p'\quad q\xtworightarrow{e_2[m]}q'\quad(e_1[m]\in I,e_2[m]\notin I)}{\tau_I(p)\parallel \tau_I(q)\xTworightarrow{e_2[m]}\tau_I(p')\between\tau_I(q')}$$

      $$\frac{p\xtworightarrow{e_1[m]}e_1\quad q\xtworightarrow{e_2[m]}e_2\quad(e_1[m],e_2[m]\in I)}{\tau_I(p\parallel q)\xtworightarrow{\tau^*}\surd}
      \quad\frac{p\xtworightarrow{e_1[m]}e_1\quad q\xtworightarrow{e_2[m]}e_2\quad(e_1[m],e_2[m]\in I)}{\tau_I(p)\parallel \tau_I(q)\xtworightarrow{\tau^*}\surd}$$

      $$\frac{p\xtworightarrow{e_1[m]}p'\quad q\xtworightarrow{e_2[m]}e_2\quad(e_1[m],e_2[m]\in I)}{\tau_I(p\parallel q)\xtworightarrow{\tau^*}\tau_I(p')}
      \quad\frac{p\xtworightarrow{e_1[m]}p'\quad q\xtworightarrow{e_2[m]}e_2\quad(e_1[m],e_2[m]\in I)}{\tau_I(p)\parallel \tau_I(q)\xtworightarrow{\tau^*}\tau_I(p')}$$

      $$\frac{p\xtworightarrow{e_1[m]}e_1\quad q\xtworightarrow{e_2[m]}q'\quad(e_1[m],e_2[m]\in I)}{\tau_I(p\parallel q)\xtworightarrow{\tau^*}\tau_I(q')}
      \quad\frac{p\xtworightarrow{e_1[m]}e_1\quad q\xtworightarrow{e_2[m]}q'\quad(e_1[m],e_2[m]\in I)}{\tau_I(p)\parallel \tau_I(q)\xtworightarrow{\tau^*}\tau_I(q')}$$

      $$\frac{p\xtworightarrow{e_1[m]}p'\quad q\xtworightarrow{e_2[m]}q'\quad(e_1[m],e_2[m]\in I)}{\tau_I(p\parallel q)\xtworightarrow{\tau^*}\tau_I(p'\between q')}
      \quad\frac{p\xtworightarrow{e_1[m]}p'\quad q\xtworightarrow{e_2[m]}q'\quad(e_1[m],e_2[m]\in I)}{\tau_I(p)\parallel \tau_I(q)\xtworightarrow{\tau^*}\tau_I(p')\between\tau_I(q')}$$

      So, with the assumption $\tau_I(p'\between q')=\tau_I(p')\between\tau_I(q')$, $\tau_I(p\parallel q)\approx_{rbs}^{fr} \tau_I(p)\parallel\tau_I(q)$, as desired.
\end{itemize}

(2) Soundness of $RAPTC_{\tau}$ with respect to rooted branching FR pomset bisimulation $\approx_{rbp}^{fr}$.

Since rooted branching FR pomset bisimulation $\approx_{rbp}^{fr}$ is both an equivalent and a congruent relation with respect to $RAPTC_{\tau}$, we only need to check if each axiom in Table \ref{AxiomsForAbstraction} is sound modulo rooted branching FR pomset bisimulation $\approx_{rbp}^{fr}$.

From the definition of rooted branching FR pomset bisimulation $\approx_{rbp}^{fr}$ (see Definition \ref{FRRBPSB}), we know that rooted branching FR pomset bisimulation $\approx_{rbp}^{fr}$ is defined by weak pomset transitions, which are labeled by pomsets with $\tau$. In a weak pomset transition, the events in the pomset are either within causality relations (defined by $\cdot$) or in concurrency (implicitly defined by $\cdot$ and $+$, and explicitly defined by $\between$), of course, they are pairwise consistent (without conflicts). In (1), we have already proven the case that all events are pairwise concurrent, so, we only need to prove the case of events in causality. Without loss of generality, we take a pomset of $P=\{e_1,e_2:e_1\cdot e_2\}$. Then the weak pomset transition labeled by the above $P$ is just composed of one single event transition labeled by $e_1$ succeeded by another single event transition labeled by $e_2$, that is, $\xRightarrow{P}=\xRightarrow{e_1}\xRightarrow{e_2}$  or $\xTworightarrow{P}=\xTworightarrow{e_2}\xTworightarrow{e_1}$.

Similarly to the proof of soundness of $RAPTC_{\tau}$ modulo rooted branching FR step bisimulation $\approx_{rbs}^{fr}$ (1), we can prove that each axiom in Table \ref{AxiomsForAbstraction} is sound modulo rooted branching FR pomset bisimulation $\approx_{rbp}^{fr}$, we omit them.

(3) Soundness of $RAPTC_{\tau}$ with respect to rooted branching FR hp-bisimulation $\approx_{rbhp}^{fr}$.

Since rooted branching FR hp-bisimulation $\approx_{rbhp}^{fr}$ is both an equivalent and a congruent relation with respect to $RAPTC_{\tau}$, we only need to check if each axiom in Table \ref{AxiomsForAbstraction} is sound modulo rooted branching FR hp-bisimulation $\approx_{rbhp}^{fr}$.

From the definition of rooted branching FR hp-bisimulation $\approx_{rbhp}^{fr}$ (see Definition \ref{FRRBHHPB}), we know that rooted branching FR hp-bisimulation $\approx_{rbhp}^{fr}$ is defined on the weakly posetal product $(C_1,f,C_2),f:\hat{C_1}\rightarrow \hat{C_2}\textrm{ isomorphism}$. Two process terms $s$ related to $C_1$ and $t$ related to $C_2$, and $f:\hat{C_1}\rightarrow \hat{C_2}\textrm{ isomorphism}$. Initially, $(C_1,f,C_2)=(\emptyset,\emptyset,\emptyset)$, and $(\emptyset,\emptyset,\emptyset)\in\approx_{rbhp}^{fr}$. When $s\xrightarrow{e}s'$ ($C_1\xrightarrow{e}C_1'$), there will be $t\xRightarrow{e}t'$ ($C_2\xRightarrow{e}C_2'$), and we define $f'=f[e\mapsto e]$. And when $s\xTworightarrow{e[m]}s'$ ($C_1\xTworightarrow{e[m]}C_1'$), there will be $t\xTworightarrow{e[m]}t'$ ($C_2\xTworightarrow{e[m]}C_2'$), and we define $f'=f[e[m]\mapsto e[m]$. Then, if $(C_1,f,C_2)\in\approx_{rbhp}^{fr}$, then $(C_1',f',C_2')\in\approx_{rbhp}^{fr}$.

Similarly to the proof of soundness of $RAPTC_{\tau}$ modulo rooted branching FR pomset bisimulation equivalence (2), we can prove that each axiom in Table \ref{AxiomsForAbstraction} is sound modulo rooted branching FR hp-bisimulation equivalence, we just need additionally to check the above conditions on rooted branching FR hp-bisimulation, we omit them.
\end{proof}

\section{Conclusions}{\label{con}}

We design an axiomatization of reversible truly concurrent process algebra $APTC$ \cite{APTC}. It has algebraic laws of reversible choice, sequence, parallelism, communication, silent step and abstraction, and also the soundness and completeness modulo strongly FR truly concurrent bisimulations and weakly FR truly concurrent bisimulations. It can be used in verification of computer systems with a truly concurrent and reversible flavor.

\newpage

%

\label{lastpage}

\end{document}